\def\spacingNumerator{5}
\def\spacingDenominator{4}

\def\ifundefined#1{\expandafter\ifx\csname#1\endcsname\relax}
\ifundefined{ftmagnification}  \def\ftmagnification{1200} \fi
\ifundefined{spacingNumerator}  \def\spacingNumerator{5} \fi
\ifundefined{spacingDenominator}  \def\spacingDenominator{4} \fi


\magnification\ftmagnification
\tolerance=10000
\hsize=17truecm\vsize=23truecm

\parindent=40pt
\mathsurround=0pt
     \multiply\baselineskip by \spacingNumerator
     \divide \baselineskip by \spacingDenominator 

%
%
\def\today{\ifcase\month\or January\or February\or March\or April\or
     May\or June\or July\or August\or September\or October\or November\or
     December\fi\space\number\day, \number\year}
%
%
\def\dst{\displaystyle}
\def\sst{\scriptstyle}
\def\tst{\textstyle}
%
%
\def\frac#1#2{\dst {#1\over#2}}     
\def\sfrac#1#2{{\tst{#1\over#2}}}   

\def\deqalign#1{\vcenter{\openup1\jot \mathsurround=0pt \ialign{
                \strut\hfil$\displaystyle{##}$&&$\displaystyle{{}##}$\hfil
                \crcr
                #1\crcr}}}         

\def\meqalign#1{\vcenter{\openup1\jot \mathsurround=0pt \ialign{
                &\strut\hfil$\displaystyle{##}$&$\displaystyle{{}##}$\hfil&
                \quad$##$\crcr
                #1\crcr}}}         

%
%
\def\al{\alpha}
\def\be{\beta}
\def\ga{\gamma}
\def\de{\delta}
\def\ep{\epsilon}
\def\ze{\zeta}

\def\th{\theta}
\def\ka{\kappa}
\def\la{\lambda}

\def\si{\sigma}

\def\om{\omega}

\def\De{\Delta}

\def\Si{\Sigma}

%
%
\def\pmb#1{\setbox0=\hbox{#1}       
     \kern-.025em\copy0\kern-\wd0
     \kern.05em\copy0\kern-\wd0
     \kern-.025em\box0}             
\def\0{{\bf 0}}

\def\k{{\bf k}}

\def\t{{\bf t}}

\def\q{{\bf q}}
\def\x{{\bf x}}
\def\y{{\bf y}}

\def\p{{\bf p}}
\def\w{{\bf w}}
\def\z{{\bf z}}

\def\cA{{\cal A}}
\def\cB{{\cal B}}
\def\cE{{\cal E}}
\def\cF{{\cal F}}

\def\cL{{\cal L}}

%
%
\font\tenfrak                 = eufm10
\font\sevenfrak               = eufm7
\font\fivefrak                = eufb5
\newfam\frakfam
     \textfont\frakfam=\tenfrak
     \scriptfont\frakfam=\sevenfrak   
     \scriptscriptfont\frakfam=\fivefrak
\def\frak{\fam\frakfam\tenfrak}
\font \tensans                = cmss10
\font \fivesans               = cmss10 at 5pt
\font \sevensans              = cmss10 at 7pt
\newfam\sansfam
     \textfont\sansfam=\tensans
     \scriptfont\sansfam=\sevensans
     \scriptscriptfont\sansfam=\fivesans
\def\sans{\fam\sansfam\tensans}
%
%
\def\bbbr{{\rm I\!R}}  
\def\bbbn{{\rm I\!N}} 
\def\bbbm{{\rm I\!M}}

\def\bbbc{{\mathchoice {\setbox0=\hbox{$\displaystyle\rm C$}\hbox{\hbox 
to0pt{\kern0.4\wd0\vrule height0.9\ht0\hss}\box0}}
{\setbox0=\hbox{$\textstyle\rm C$}\hbox{\hbox
to0pt{\kern0.4\wd0\vrule height0.9\ht0\hss}\box0}}
{\setbox0=\hbox{$\scriptstyle\rm C$}\hbox{\hbox
to0pt{\kern0.4\wd0\vrule height0.9\ht0\hss}\box0}}
{\setbox0=\hbox{$\scriptscriptstyle\rm C$}\hbox{\hbox
to0pt{\kern0.4\wd0\vrule height0.9\ht0\hss}\box0}}}}
\def\bbbq{{\mathchoice {\setbox0=\hbox{$\displaystyle\rm               
Q$}\hbox{\raise
0.15\ht0\hbox to0pt{\kern0.4\wd0\vrule height0.8\ht0\hss}\box0}}
{\setbox0=\hbox{$\textstyle\rm Q$}\hbox{\raise
0.15\ht0\hbox to0pt{\kern0.4\wd0\vrule height0.8\ht0\hss}\box0}}
{\setbox0=\hbox{$\scriptstyle\rm Q$}\hbox{\raise
0.15\ht0\hbox to0pt{\kern0.4\wd0\vrule height0.7\ht0\hss}\box0}}
{\setbox0=\hbox{$\scriptscriptstyle\rm Q$}\hbox{\raise
0.15\ht0\hbox to0pt{\kern0.4\wd0\vrule height0.7\ht0\hss}\box0}}}}
\def\bbbz{{\mathchoice {\hbox{$\sans\textstyle Z\kern-0.4em Z$}}       
{\hbox{$\sans\textstyle Z\kern-0.4em Z$}}
{\hbox{$\sans\scriptstyle Z\kern-0.3em Z$}}
{\hbox{$\sans\scriptscriptstyle Z\kern-0.2em Z$}}}}
%
%
\def\const{{\rm const}\,}

\def\half{\sfrac{1}{2}}

\def\optbar#1{\vbox{\ialign{##\crcr\hfil${\scriptscriptstyle(}\mkern -1mu
         \vrule height 1.2pt width 3pt depth -.8pt
         {\scriptscriptstyle)}$\hfil\crcr
          \noalign{\kern-1pt\nointerlineskip}$\hfil\displaystyle{#1}\hfil$\crcr}}}
\def\<{\left<}
\def\>{\right>}

\def\smprod{\mathop{\textstyle\prod}}
\def\smsum{\mathop{\textstyle\sum}}
\def\set#1#2{\big\{ \ #1\ \big|\ #2\ \big\}}
\def\eval#1{\big|\lower4pt\hbox{$\displaystyle\sst #1$}}
%
%
\font \tafontt                = cmbx10 scaled\magstep2
\font \tbfontt                = cmbx10 scaled\magstep1
\def\titlea#1{\centerline{\tafontt #1 }\vskip.5truein}
\def\titleb#1{\removelastskip\vskip.3truein%
\noindent{\tbfontt #1 }\vskip.25truein}
\def\titlec#1{\removelastskip\vskip.15truein\noindent{\bf #1 }\vskip.1truein}

%
%
\def\newenvironment#1#2#3#4{\long\def#1##1##2{%
\removelastskip\penalty-100\vskip\baselineskip%
\noindent{#3#2\if!##1!.\else\unskip\ \ignorespaces
##1\unskip\fi\ }{#4\ignorespaces##2\vskip\baselineskip}}}
\newenvironment\lemma{Lemma}{\bf}{\it}
\newenvironment\proposition{Proposition}{\bf}{\it}
\newenvironment\theorem{Theorem}{\bf}{\it}
\newenvironment\corollary{Corollary}{\bf}{\it}
\newenvironment\example{Example}{\bf}{\rm}
\newenvironment\problem{Problem}{\bf}{\rm}
\newenvironment\definition{Definition}{\bf}{\rm}
\newenvironment\remark{Remark}{\bf}{\rm}
\newenvironment\hypothesis{Hypothesis}{\bf}{\it}
\newenvironment\convention{Convention}{\bf}{\it}

\def\Item{\vskip.1in\noindent}

%
%
\long\def\proof#1{\removelastskip\penalty-100\vskip\baselineskip\noindent{\bf
            Proof\if!#1!\else\ \ignorespaces#1\fi:\ }\ \ \ignorespaces}
\long\def\prf{\removelastskip\penalty-100\vskip\baselineskip\noindent{\bf
            Proof:\ }\ \ \ignorespaces}
\def\endproof{\hfill\vrule height .6em width .6em depth 0pt\goodbreak\vskip.25in }

\ifundefined{warnForwardRef}  \def\warnForwardRef{n} \fi
\newcount\chapno
\newcount\sectno
\newcount\equano
\newcount\theono
\newcount\probno

\def\IgNoRe#1{}

\chapno=0
\sectno=0
\equano=0
\theono=0
\probno=0
\def\eqhead{}
\def\frefwarning{\if\warnForwardRef y\immediate\write16{   Forward reference on line \the\inputlineno}\fi}
\def\qqqrefwarning{\immediate\write16{   ??? reference on line \the\inputlineno}}

\def\chap#1{\equano=0\sectno=0\theono=0\probno=0\global\advance\chapno by 1%
\def\eqhead{\ifcase\chapno\or I\or II\or III\or IV\or V\or VI\or VII\or
VIII\or IX\or X\or XI\or XII\or XIII\or XIV\or XV\or XVI\or XVII\or XVIII\or
XIX\or XX\or XXI\or XXII\or XXIII\or XXIV\or XXV\or XXVI\or XXVII\or XXVIII\or XXIX\or XXX\or XXXI\or XXXII\or XXXIII\or XXXIV\or XXXV\or XXXVI\or XXXVII\or XXXVIII\or XXXIX\fi.}%
\titlea{\eqhead \hglue 5pt #1}%
}

\def\sect#1{\global\advance\sectno by 1%
\titleb{\eqhead\number\sectno  \hglue 5pt #1}%
}%

\def\appendix#1#2{\equano=0\sectno=0\theono=0\probno=0\def\eqhead{#1.}
\titlea{Appendix #1: #2}%
}

\def\:#1{\def\temp{\expandafter\IgNoRe\string#1}%
\expandafter\ifx\csname\temp\endcsname\relax%
\expandafter\gdef#1{\qqqrefwarning ???}\fi#1}

\def\Eqn{{\hbox{\global\advance\equano by 1}}%
\eqno ({\rm \eqhead\number\equano})}%

\def\Eqno{{\hbox{\global\advance\equano by 1}}%
 ({\rm \eqhead\number\equano})}%

\def\EQN#1{\Eqn\edef\Zwi{\eqhead\number\equano}%
\global\let #1=\Zwi
}

\def\EQNO#1{\Eqno\edef\Zwi{\eqhead\number\equano}%
\global\let #1=\Zwi
}

\def\STM#1{{\global\advance \theono by 1}%
\eqhead\number\theono
\edef\Zwi{\eqhead\number\theono }
\global\let#1=\Zwi
}

\def\PRB#1{{\global\advance \probno by 1}%
\eqhead\number\probno
\edef\Zwi{\eqhead\number\probno }
\global\let#1=\Zwi
}

\def\PG#1{\def\Zwi{\number\pageno }
\global\let#1=\Zwi
}

\def\Stm{{\global\advance \theono by 1}%
\eqhead\number\theono
}

\def\Prb{{\global\advance \probno by 1}%
\eqhead\number\probno
}

\def\EDEF#1#2{
\def\tEmP{#1}\expandafter\gdef\tEmP{#2}
}



\def\suffix{ps}
\newcount\system
\global\system=3   

\def\ifundefined#1{\expandafter\ifx\csname#1\endcsname\relax}
\ifundefined{figdir}\def\figdir{}\fi
%
%
\newcount\firstline
\newdimen\pswidth  \newdimen\xleft
\newdimen\psheight \newdimen\ytop \newdimen\ybot
\newcount\justx \newcount\justy
\global\justx=0 \global\justy=0
\newdimen\vpos \newtoks\labeL 
\newread\labeLfile \newdimen\xcoord \newdimen\ycoord
\newif\ifdoit 
\newbox\labox
\newdimen\xdvikwid 
\newdimen\xdvikht
\newdimen\pspoints
\newdimen\rwi
\pspoints=1bp
\newcount\temp
\def\readdim#1{\global\read\labeLfile to \temp
\global #1=\temp pt}
%
%
%
%
\def\figcrop#1{\par
\openin\labeLfile=\figdir#1.lbl                                              
\global\read\labeLfile to\firstline\message{#1}               
\global\read\labeLfile to\temp
\readdim{\ybot}
\readdim{\xleft}
\readdim{\ytop}
\global\read\labeLfile to\justx
\global\read\labeLfile to\justy
\global\read\labeLfile to\labeL
\readdim{\pswidth}
\global\advance\pswidth by -\xleft
\readdim{\psheight}
\global\advance\ybot by -\psheight
\global\advance\psheight by -\ytop
\global\read\labeLfile to\justx
\global\read\labeLfile to\justy
\global\read\labeLfile to\labeL
\vbox to\psheight{\vfill
\ifnum\system=1
\fi 
\ifnum\system=2
\fi 
\ifnum\system=3
                                                 \fi         
\ifnum\system=4
\fi         
\ifnum\system=1
\hbox to \pswidth{\kern-\xleft\special{postscriptfile \figdir#1.\suffix }\hfil}\fi
\ifnum\system=2
\hbox to \pswidth{\kern-\xleft\special{ps: plotfile \figdir#1.\suffix }\hfil}\fi
\ifnum\system=3
\hbox to \pswidth{\kern-\xleft\includegraphics{\figdir#1.\suffix}\hfil}\fi
\ifnum\system=4
\hbox to \pswidth{\kern-\xleft\includegraphics{\figdir#1.\suffix}\hfil}\fi
\ifnum\system=5
\hbox to \pswidth{\kern-\xleft\includegraphics{\figdir#1.\suffix}\hfil}\fi 
\ifnum\system=6
   \xdvikwid=\pswidth
   \xdvikht=\psheight
   {\global\divide\xdvikwid by \pspoints}
   {\global\divide\xdvikht by \pspoints}
   \rwi=\xdvikwid
    {\global\multiply\rwi by 10}
\hbox to \pswidth{\kern-\xleft\includegraphics{\figdir#1.\suffix\space}\hfil}\fi                   
\vskip -\baselineskip
\vskip -\ybot 
\vskip-\psheight %
\hbox to\pswidth  {\hss}%
\parindent=0pt\offinterlineskip                                       
\vpos=0 pt%
\loop\readdim{\xcoord}                                 
\ifdim \xcoord < -999pt \doitfalse\else\doittrue\fi                        
\ifdoit \advance \xcoord by -\xleft
\readdim{\ycoord}
\advance \ycoord by -\ytop                              
\global\read\labeLfile to\justx                                       
\global\read\labeLfile to\justy                                       
\global\read\labeLfile to\labeL
\global\setbox\labox=\hbox{\labeL\hskip-0.3em}%
\advance\vpos by-\ycoord                                              
\vskip-\vpos \vpos=\ycoord                                         
\hbox to\pswidth{\hskip\xcoord %
\hbox to 0pt{\ifnum\justx>0\hss\fi%
\vbox to0pt{%
\ifnum\justy<2\vss\fi%
\copy\labox\kern0pt%
\ifnum\justy>0\vss\fi}%
\ifnum\justx<2\hss\fi}%
\hss}%
\repeat%
\advance\vpos by-\psheight%
\vskip-\vpos %
}\closein\labeLfile}
%
%
%
\def\figplace#1#2#3{
\openin\labeLfile=\figdir#1.lbl
\ifeof \labeLfile
       \immediate\write16{***Can't find \figdir#1.lbl; Skipping it.***}
\else  \closein\labeLfile
       \null\hskip#2\raise #3 \hbox{\figcrop{#1}}
\fi
}
%
%
%
%
\def\figput#1{
\openin\labeLfile=\figdir#1.lbl
\ifeof \labeLfile
       \immediate\write16{***Can't find \figdir#1.lbl; Skipping it.***}
\else  \closein\labeLfile
       \hbox{\figcrop{#1}}
\fi
}


    \def\cst#1#2{{\rm const}^{#1}_{#2}\,}
    \def\abcst{{\sst const}}

    \def\dunion{\mathbin{\cup\mkern-8.5mu\cdot}}
    \def\bigdunion{\mathop{\bigcup\mkern-15.5mu\cdot}}

    \def\v{\pmb{$\vert$}}
    \def\V{\pmb{$\big\vert$}}

    \def\tv{\kern8pt\tilde{\kern-8pt\pmb{$\vert$}}}
    \def\tV{\kern8pt\tilde{\kern-8pt\pmb{$\big\vert$}}}
    \def\tVV{\kern8pt\tilde{\kern-8pt\pmb{$\Big\vert$}}}

    \def\tn{|\kern-1pt|\kern-1pt|}
    \def\TN{\big|\kern-1.5pt\big|\kern-1.5pt\big|}
    \def\TTN{\Big|\kern-2pt\Big|\kern-2pt\Big|}

     \def\lc{{\rm loc}}
     \def\tp{{\rm top}}
     \def\bt{{\rm bot}}
     \def\md{{\rm mid}}
     \def\upl{{\rm up}}
     \def\dnl{{\rm dn}}
     \def\li{{\rm l}}
     \def\ci{{\rm c}}
     \def\ri{{\rm r}}
     \def\[{\lbrack\!\lbrack}
     \def\]{\rbrack\!\rbrack}
     \def\squiggle{\raise2pt\hbox{${\scriptstyle\sim}$}}
     \def\dunion{\mathbin{\cup\mkern-8.5mu\cdot}}
     \def\stoday{\number\day\space\ifcase\month\or Jan\or Feb\or 
                      Mar\or Apr\or May\or Jun\or Jul\or Aug\or Sep\or 
                      Oct\or Nov\or Dec\fi, \number\year}

    \def\titled#1{\removelastskip\vskip.15truein\noindent{\it #1 }
                   \vskip.1truein\nobreak}

     \def\fcirc{\circ}

    \def\bka{\pmb{$\ka$}} 
    \def\bpi{{\mathchoice{\pmb{$\pi$}}{\pmb{$\pi$}}
                              {\pmb{$\sst\pi$}}{\pmb{$\scriptscriptstyle\pi$}}}}
    \def\bsi{\pmb{$\si$}} 
    \def\btau{\pmb{$\tau$}} 
     
    \def\cZ{{\cal Z}}

    \def\cM{{\cal M}}
    
    \def\cb{{\frak c}}
    \def\ii{{\rm i}}

    \def\bd{{\bf d}}
    \def\cC{{\cal C}}
    \def\cD{{\cal D}}
    \def\rD{{\rm D}}
    \def\cL{{\cal L}}
    \def\fL{{\frak L}}
    \def\tcL{\tilde{\cal L}}
    \def\cR{{\cal R}}
    \def\cV{{\cal V}}
    \def\fl{{\frak l}}
    \def\veps{{\varepsilon}}
    \def\fv{{\frak v}}
    \def\fK{{\frak K}}
    \def\fN{{\frak N}}
    \def\fX{{\frak X}}
    \def\fY{{\frak Y}}
    \def\fZ{{\frak Z}}
    \def\fV{{\frak V}}
    \def\Ln{{\frak t}}
    \def\Sh{{\frak n}}
    \def\sumprime{\mathop{{\sum}'}}

\def\caproman#1{\ifcase#1\or I\or II\or III\or IV\or V\or VI\or VII\or
VIII\or IX\or X\or XI\or XII\or XIII\or XIV\or XV\or XVI\or XVII\or XVIII\or
XIX\or XX\or XXI\or XXII\or XXIII\or XXIV\or XXV\or XXVI\or XXVII\or XXVIII\or XXIX\or XXX\or XXXI\or XXXII\or XXXIII\or XXXIV\or XXXV\or XXXVI\or XXXVII\or XXXVIII\or XXXIX\fi}%
     \ifundefined{CHwwladders}\def\CHwwladders{II}\fi
     \ifundefined{CHdbubbles}\def\CHwwladders{IV}\fi

   \font\sixrm=cmr6   \font\eightrm=cmr8  
   \font\sixi=cmmi6   \font\eighti=cmmi8  
  \font\sixsy=cmsy6  \font\eightsy=cmsy8 
  \font\sixbf=cmbx6  \font\eightbf=cmbx8 
                     \font\eightit=cmti8 
                     \font\eightsl=cmsl8 
                     \font\eighttt=cmtt8 

\font\eightfrak=eufm7 at 8pt

\def\eightpoint{\def\rm{\fam0\eightrm}
 \textfont0=\eightrm \scriptfont0=\sixrm \scriptscriptfont0=\fiverm
 \textfont1=\eighti \scriptfont1=\sixi \scriptscriptfont1=\fivei
 \textfont2=\eightsy \scriptfont2=\sixsy \scriptscriptfont2=\fivesy
 \textfont3=\tenex \scriptfont3=\tenex \scriptscriptfont3=\tenex
 \textfont\itfam=\eightit \def\it{\fam\itfam\eightit}%
 \textfont\slfam=\eightsl \def\sl{\fam\slfam\eightsl}%
 \textfont\ttfam=\eighttt \def\tt{\fam\ttfam\eighttt}%
 \textfont\frakfam=\eightfrak \def\frak{\fam\frakfam\tenfrak}%
 \textfont\bffam=\eightbf \scriptfont\bffam=\sixbf
 \scriptscriptfont\bffam=\fivebf \def\bf{\fam\bffam\eightbf}%
 \normalbaselineskip=9pt
 \setbox\strutbox=\hbox{\vrule height7pt depth2pt width0pt}%
 \let\sc=\sixrm \let\big=\eightbig \normalbaselines\rm}
\catcode`@=11
\def\footnote#1{\edef\@sf{\spacefactor\the\spacefactor}#1\@sf
     \insert\footins\bgroup\eightpoint
     \interlinepenalty100 \let\par=\endgraf
     \leftskip=0pt \rightskip=0pt
     \splittopskip=10pt plus 1pt minus 1pt \floatingpenalty=20000
     \smallskip\item{#1}\bgroup\strut\aftergroup\@foot\let\next}
\skip\footins=12pt plus 2pt minus 4pt
\dimen\footins=30pc
\catcode`@=12


  \IgNoRe{PG}
  \IgNoRe{EQN}
  \IgNoRe{STM Assertion }
  \IgNoRe{PG}
  \IgNoRe{STM Assertion }
  \IgNoRe{STM Assertion }
 \def\defOSFancynormdomain{\frefwarning II.4} \IgNoRe{STM Assertion }
  \IgNoRe{STM Assertion }
  \IgNoRe{STM Assertion }
  \IgNoRe{STM Assertion }
  \IgNoRe{EQN}
  \IgNoRe{STM Assertion }
  \IgNoRe{STM Assertion }
  \IgNoRe{STM Assertion }
  \IgNoRe{STM Assertion }
  \IgNoRe{STM Assertion }
  \IgNoRe{STM Assertion }
  \IgNoRe{STM Assertion }
  \IgNoRe{STM Assertion }
  \IgNoRe{PG}
  \IgNoRe{STM Assertion }
  \IgNoRe{STM Assertion }
  \IgNoRe{STM Assertion }
  \IgNoRe{STM Assertion }
  \IgNoRe{STM Assertion }
  \IgNoRe{STM Assertion }
  \IgNoRe{STM Assertion }
  \IgNoRe{STM Assertion }
  \IgNoRe{EQN}
  \IgNoRe{PG}
  \IgNoRe{PG}
  \IgNoRe{STM Assertion }
  \IgNoRe{STM Assertion }
  \IgNoRe{STM Assertion }
  \IgNoRe{STM Assertion }
  \IgNoRe{STM Assertion }
  \IgNoRe{STM Assertion }
  \IgNoRe{PG}
  \IgNoRe{EQN}
  \IgNoRe{EQN}
  \IgNoRe{EQN}
  \IgNoRe{EQN}
  \IgNoRe{EQN}
  \IgNoRe{EQN}
  \IgNoRe{STM Assertion }
  \IgNoRe{STM Assertion }
  \IgNoRe{STM Assertion }
  \IgNoRe{STM Assertion }
  \IgNoRe{PG}
  \IgNoRe{STM Assertion }
  \IgNoRe{EQN}
  \IgNoRe{STM Assertion }
  \IgNoRe{STM Assertion }
  \IgNoRe{STM Assertion }
  \IgNoRe{STM Assertion }
  \IgNoRe{PG}
  \IgNoRe{STM Assertion }
 \def\corOSappMonoidIV{\frefwarning A.5} \IgNoRe{STM Assertion }
  \IgNoRe{STM Assertion }
  \IgNoRe{STM Assertion }
  \IgNoRe{PG}
  \IgNoRe{EQN}
  \IgNoRe{EQN}
  \IgNoRe{PG}
  \IgNoRe{STM Assertion }
  \IgNoRe{STM Assertion }
  \IgNoRe{STM Assertion }
  \IgNoRe{EQN}
  \IgNoRe{PG}
  \IgNoRe{STM Assertion }
  \IgNoRe{STM Assertion }
  \IgNoRe{EQN}
  \IgNoRe{STM Assertion }
  \IgNoRe{STM Assertion }
  \IgNoRe{STM Assertion }
  \IgNoRe{STM Assertion }
  \IgNoRe{PG}
  \IgNoRe{STM Assertion }
  \IgNoRe{STM Assertion }
  \IgNoRe{STM Assertion }
  \IgNoRe{STM Assertion }
  \IgNoRe{STM Assertion }
  \IgNoRe{STM Assertion }
  \IgNoRe{STM Assertion }
  \IgNoRe{STM Assertion }
  \IgNoRe{PG}
  \IgNoRe{STM Assertion }
 \def\defOSftcov{\frefwarning IX.3} \IgNoRe{STM Assertion }
  \IgNoRe{STM Assertion }
  \IgNoRe{STM Assertion }
  \IgNoRe{STM Assertion }
  \IgNoRe{EQN}
  \IgNoRe{STM Assertion }
  \IgNoRe{STM Assertion }
  \IgNoRe{PG}
  \IgNoRe{STM Assertion }
  \IgNoRe{STM Assertion }
  \IgNoRe{STM Assertion }
  \IgNoRe{STM Assertion }
  \IgNoRe{STM Assertion }
  \IgNoRe{STM Assertion }
  \IgNoRe{STM Assertion }
  \IgNoRe{STM Assertion }
  \IgNoRe{STM Assertion }
  \IgNoRe{EQN}
  \IgNoRe{STM Assertion }
 \def\defOSsymmetries{\frefwarning B.1} \IgNoRe{STM Assertion }
  \IgNoRe{PG}
  \IgNoRe{STM Assertion }
  \IgNoRe{STM Assertion }
  \IgNoRe{STM Assertion }
  \IgNoRe{STM Assertion }
  \IgNoRe{STM Assertion }
  \IgNoRe{STM Assertion }
  \IgNoRe{STM Assertion }
  \IgNoRe{PG}
  \IgNoRe{STM Assertion }
  \IgNoRe{PG}
  \IgNoRe{PG}
  \IgNoRe{STM Assertion }
  \IgNoRe{STM Assertion }
  \IgNoRe{PG}
 \def\lemOSsectpartunit{\frefwarning XII.3} \IgNoRe{STM Assertion }
  \IgNoRe{STM Assertion }
  \IgNoRe{STM Assertion }
  \IgNoRe{STM Assertion }
  \IgNoRe{STM Assertion }
  \IgNoRe{STM Assertion }
  \IgNoRe{STM Assertion }
  \IgNoRe{STM Assertion }
  \IgNoRe{STM Assertion }
 \def\lemOSNormMom{\frefwarning XII.12} \IgNoRe{STM Assertion }
  \IgNoRe{STM Assertion }
  \IgNoRe{STM Assertion }
  \IgNoRe{STM Assertion }
  \IgNoRe{STM Assertion }
  \IgNoRe{STM Assertion }
  \IgNoRe{STM Assertion }
  \IgNoRe{EQN}
  \IgNoRe{STM Assertion }
 \def\propOSGenDecay{\frefwarning XIII.1} \IgNoRe{STM Assertion }
  \IgNoRe{PG}
  \IgNoRe{STM Assertion }
  \IgNoRe{EQN}
  \IgNoRe{EQN}
 \def\lemOSmorepartunity{\frefwarning XIII.3} \IgNoRe{STM Assertion }
 \def\eqnOSsecpropboundII{\frefwarning XIII.3} \IgNoRe{EQN}
  \IgNoRe{EQN}
  \IgNoRe{STM Assertion }
  \IgNoRe{EQN}
 \def\propOSrealpropbound{\frefwarning XIII.5} \IgNoRe{STM Assertion }
  \IgNoRe{EQN}
  \IgNoRe{STM Assertion }
  \IgNoRe{STM Assertion }
  \IgNoRe{STM Assertion }
  \IgNoRe{STM Assertion }
  \IgNoRe{PG}
  \IgNoRe{STM Assertion }
  \IgNoRe{STM Assertion }
  \IgNoRe{STM Assertion }
  \IgNoRe{STM Assertion }
  \IgNoRe{EQN}
  \IgNoRe{EQN}
  \IgNoRe{STM Assertion }
  \IgNoRe{STM Assertion }
  \IgNoRe{PG}
  \IgNoRe{STM Assertion }
  \IgNoRe{STM Assertion }
  \IgNoRe{STM Assertion }
  \IgNoRe{EQN}
  \IgNoRe{STM Assertion }
  \IgNoRe{EQN}
  \IgNoRe{STM Assertion }
  \IgNoRe{STM Assertion }
  \IgNoRe{EQN}
  \IgNoRe{STM Assertion }
  \IgNoRe{EQN}
  \IgNoRe{EQN}
  \IgNoRe{EQN}
  \IgNoRe{EQN}
  \IgNoRe{STM Assertion }
  \IgNoRe{STM Assertion }
  \IgNoRe{STM Assertion }
  \IgNoRe{STM Assertion }
  \IgNoRe{PG}
  \IgNoRe{STM Assertion }
 \def\defOSsectdiffdecaynorm{\frefwarning XVI.4} \IgNoRe{STM Assertion }
  \IgNoRe{STM Assertion }
  \IgNoRe{STM Assertion }
 \def\defOSsectcheckcF{\frefwarning XVI.7} \IgNoRe{STM Assertion }
  \IgNoRe{STM Assertion }
  \IgNoRe{STM Assertion }
  \IgNoRe{STM Assertion }
  \IgNoRe{STM Assertion }
  \IgNoRe{STM Assertion }
  \IgNoRe{EQN}
  \IgNoRe{EQN}
  \IgNoRe{STM Assertion }
  \IgNoRe{PG}
  \IgNoRe{STM Assertion }
  \IgNoRe{STM Assertion }
  \IgNoRe{EQN}
  \IgNoRe{STM Assertion }
  \IgNoRe{STM Assertion }
  \IgNoRe{EQN}
  \IgNoRe{EQN}
  \IgNoRe{EQN}
  \IgNoRe{EQN}
  \IgNoRe{EQN}
  \IgNoRe{STM Assertion }
  \IgNoRe{STM Assertion }
  \IgNoRe{STM Assertion }
  \IgNoRe{EQN}
  \IgNoRe{EQN}
  \IgNoRe{EQN}
  \IgNoRe{EQN}
  \IgNoRe{EQN}
  \IgNoRe{EQN}
  \IgNoRe{STM Assertion }
  \IgNoRe{STM Assertion }
 \def\lemchannelnorm{\frefwarning D.2} \IgNoRe{STM Assertion }
  \IgNoRe{PG}
  \IgNoRe{STM Assertion }
  \IgNoRe{STM Assertion }
  \IgNoRe{EQN}
  \IgNoRe{EQN}
  \IgNoRe{STM Assertion }
  \IgNoRe{EQN}
  \IgNoRe{EQN}
  \IgNoRe{STM Assertion }
  \IgNoRe{EQN}
  \IgNoRe{EQN}
  \IgNoRe{STM Assertion }
  \IgNoRe{STM Assertion }
  \IgNoRe{PG}
  \IgNoRe{STM Assertion }
  \IgNoRe{STM Assertion }
  \IgNoRe{STM Assertion }
  \IgNoRe{PG}
  \IgNoRe{STM Assertion }
  \IgNoRe{STM Assertion }
  \IgNoRe{STM Assertion }
  \IgNoRe{STM Assertion }
  \IgNoRe{PG}
 \def\propOSresectorI{\frefwarning XIX.4} \IgNoRe{STM Assertion }
  \IgNoRe{STM Assertion }
  \IgNoRe{STM Assertion }
  \IgNoRe{STM Assertion }
  \IgNoRe{STM Assertion }
  \IgNoRe{STM Assertion }
  \IgNoRe{STM Assertion }
  \IgNoRe{STM Assertion }
  \IgNoRe{STM Assertion }
 \def\corOSresectorvanishkzero{\frefwarning XIX.13} \IgNoRe{STM Assertion }
  \IgNoRe{STM Assertion }
  \IgNoRe{STM Assertion }
  \IgNoRe{STM Assertion }
  \IgNoRe{STM Assertion }
  \IgNoRe{PG}
  \IgNoRe{STM Assertion }
  \IgNoRe{PG}
  \IgNoRe{STM Assertion }
  \IgNoRe{STM Assertion }
  \IgNoRe{PG}
  \IgNoRe{STM Assertion }
  \IgNoRe{STM Assertion }
  \IgNoRe{EQN}
  \IgNoRe{PG}
  \IgNoRe{EQN}
  \IgNoRe{STM Assertion }
  \IgNoRe{STM Assertion }
  \IgNoRe{PG}
  \IgNoRe{EQN}
  \IgNoRe{EQN}
  \IgNoRe{EQN}
  \IgNoRe{EQN}
  \IgNoRe{EQN}
  \IgNoRe{STM Assertion }
  \IgNoRe{STM Assertion }
  \IgNoRe{EQN}
  \IgNoRe{STM Assertion }
  \IgNoRe{PG}
  \IgNoRe{PG}
  \IgNoRe{PG}
  \IgNoRe{STM Assertion }
  \IgNoRe{EQN}
  \IgNoRe{STM Assertion }
  \IgNoRe{PG}
  \IgNoRe{STM Assertion }
  \IgNoRe{EQN}
  \IgNoRe{STM Assertion }
  \IgNoRe{STM Assertion }
  \IgNoRe{PG}
  \IgNoRe{EQN}
  \IgNoRe{EQN}
  \IgNoRe{EQN}
  \IgNoRe{STM Assertion }
  \IgNoRe{STM Assertion }
  \IgNoRe{STM Assertion }
  \IgNoRe{EQN}
  \IgNoRe{STM Assertion }
  \IgNoRe{STM Assertion }
  \IgNoRe{STM Assertion }
  \IgNoRe{STM Assertion }
  \IgNoRe{STM Assertion }
  \IgNoRe{STM Assertion }
  \IgNoRe{PG}
  \IgNoRe{STM Assertion }
  \IgNoRe{STM Assertion }
  \IgNoRe{STM Assertion }
  \IgNoRe{STM Assertion }
  \IgNoRe{STM Assertion }
  \IgNoRe{STM Assertion }
  \IgNoRe{STM Assertion }
  \IgNoRe{PG}
  \IgNoRe{PG}


\newcount\CHAPNO
\newcount\APPNO
\CHAPNO=0
\APPNO=1
\def\advCHAPNO{\advance\CHAPNO by 1}
\def\advAPPNO{\advance\APPNO by 1}

\def\caproman#1{\ifcase#1\or I\or II\or III\or IV\or V\or VI\or VII\or
VIII\or IX\or X\or XI\or XII\or XIII\or XIV\or XV\or XVI\or XVII\or XVIII\or
XIX\or XX\or XXI\or XXII\or XXIII\or XXIV\or XXV\or XXVI\or XXVII\or XXVIII\or XXIX\or XXX\or XXXI\or XXXII\or XXXIII\or XXXIV\or XXXV\or XXXVI\or XXXVII\or XXXVIII\or XXXIX\fi}%

\def\capletter#1{\ifcase#1\or A\or B\or C\or D\or E\or F\or G\or
H\or I\or J\or K\or L\or M\or N\or O\or P\or Q\or R\or
S\or T\or U\or V\or W\or X\or Y\or Z\fi}%

\newcount\cHintroI \cHintroI=\CHAPNO \advCHAPNO 
                              
\newcount\cHnorms  \cHnorms=\CHAPNO \advCHAPNO 
                              
\newcount\cHproprengrp \cHproprengrp=\CHAPNO \advCHAPNO 
                              
\newcount\cHcovbounds  \cHcovbounds=\CHAPNO \advCHAPNO 
                              
\newcount\cHinsulator \cHinsulator=\CHAPNO \advCHAPNO

 \advAPPNO

\newcount\cHintroII \cHintroII=\CHAPNO \advCHAPNO 
                              
\newcount\cHamputate \cHamputate=\CHAPNO \advCHAPNO
                              
\newcount\cHscales \cHscales=\CHAPNO \advCHAPNO
                              \edef\CHscales{\caproman\CHAPNO}
\newcount\cHfourier \cHfourier=\CHAPNO \advCHAPNO
                              
\newcount\cHmomentum \cHmomentum=\CHAPNO \advCHAPNO

 \advAPPNO
 \advAPPNO

\newcount\cHintroIII \cHintroIII=\CHAPNO \advCHAPNO
                              
\newcount\cHsectors \cHsectors=\CHAPNO \advCHAPNO
                              \edef\CHsectors{\caproman\CHAPNO}
\newcount\cHsecpropbounds \cHsecpropbounds=\CHAPNO \advCHAPNO
                              
\newcount\cHladdersNotn  \cHladdersNotn=\CHAPNO \advCHAPNO
                              
\newcount\cHestren  \cHestren=\CHAPNO \advCHAPNO
                              
\newcount\cHsecmomnorm \cHsecmomnorm=\CHAPNO \advCHAPNO
                              
\newcount\cHmomestren \cHmomestren=\CHAPNO \advCHAPNO

 \advAPPNO

\newcount\cHintroIV  \cHintroIV=\CHAPNO \advCHAPNO
                              
\newcount\cHcomparison   \cHcomparison=\CHAPNO \advCHAPNO
                              
\newcount\cHsumsmom  \cHsumsmom=\CHAPNO \advCHAPNO
                              
\newcount\cHsectorsmom   \cHsectorsmom=\CHAPNO \advCHAPNO
                              
\newcount\cHppladsect    \cHppladsect=\CHAPNO \advCHAPNO

 \advAPPNO


  \IgNoRe{STM Assertion }
  \IgNoRe{PG}
  \IgNoRe{EQN}
  \IgNoRe{EQN}
  \IgNoRe{EQN}
  \IgNoRe{EQN}
  \IgNoRe{EQN}
 \def\defNPscales{\frefwarning I.2} \IgNoRe{STM Assertion }
  \IgNoRe{STM Assertion }
  \IgNoRe{STM Assertion }
  \IgNoRe{STM Assertion }
  \IgNoRe{STM Assertion }
  \IgNoRe{STM Assertion }
  \IgNoRe{STM Assertion }
  \IgNoRe{STM Assertion }
 \def\defNPstrongasymm{\frefwarning I.10} \IgNoRe{STM Assertion }
  \IgNoRe{STM Assertion }
  \IgNoRe{STM Assertion }
  \IgNoRe{PG}
  \IgNoRe{PG}
  \IgNoRe{PG}
  \IgNoRe{PG}
  \IgNoRe{EQN}
  \IgNoRe{EQN}
  \IgNoRe{EQN}
  \IgNoRe{EQN}
  \IgNoRe{PG}
  \IgNoRe{PG}
  \IgNoRe{PG}
  \IgNoRe{EQN}
  \IgNoRe{PG}
  \IgNoRe{PG}
  \IgNoRe{EQN}
  \IgNoRe{EQN}
  \IgNoRe{EQN}
  \IgNoRe{EQN}
  \IgNoRe{EQN}
  \IgNoRe{EQN}
  \IgNoRe{EQN}
  \IgNoRe{EQN}
  \IgNoRe{EQN}
  \IgNoRe{EQN}
  \IgNoRe{PG}
  \IgNoRe{PG}
  \IgNoRe{EQN}
  \IgNoRe{EQN}
  \IgNoRe{EQN}
  \IgNoRe{STM Assertion }
  \IgNoRe{PG}
  \IgNoRe{EQN}
  \IgNoRe{EQN}
  \IgNoRe{EQN}
  \IgNoRe{EQN}
  \IgNoRe{STM Assertion }
  \IgNoRe{STM Assertion }
  \IgNoRe{STM Assertion }
  \IgNoRe{STM Assertion }
  \IgNoRe{STM Assertion }
  \IgNoRe{STM Assertion }
  \IgNoRe{STM Assertion }
  \IgNoRe{STM Assertion }
  \IgNoRe{STM Assertion }
  \IgNoRe{EQN}
  \IgNoRe{EQN}
  \IgNoRe{EQN}
  \IgNoRe{EQN}
  \IgNoRe{STM Assertion }
  \IgNoRe{EQN}
  \IgNoRe{STM Assertion }
  \IgNoRe{EQN}
 \def\exNPsectorizebound{\frefwarning A.1} \IgNoRe{STM Assertion }
  \IgNoRe{PG}
  \IgNoRe{EQN}
  \IgNoRe{STM Assertion }
  \IgNoRe{STM Assertion }
  \IgNoRe{EQN}
  \IgNoRe{PG}
  \IgNoRe{PG}
  \IgNoRe{STM Assertion }
 \def\defNPFancynormdomain{\frefwarning V.2} \IgNoRe{STM Assertion }
  \IgNoRe{PG}
  \IgNoRe{STM Assertion }
  \IgNoRe{STM Assertion }
  \IgNoRe{STM Assertion }
  \IgNoRe{STM Assertion }
  \IgNoRe{STM Assertion }
  \IgNoRe{STM Assertion }
  \IgNoRe{STM Assertion }
  \IgNoRe{STM Assertion }
  \IgNoRe{PG}
  \IgNoRe{STM Assertion }
  \IgNoRe{STM Assertion }
  \IgNoRe{STM Assertion }
  \IgNoRe{STM Assertion }
  \IgNoRe{EQN}
 \def\defNPsectnorm{\frefwarning VI.6} \IgNoRe{STM Assertion }
  \IgNoRe{STM Assertion }
  \IgNoRe{STM Assertion }
  \IgNoRe{EQN}
  \IgNoRe{STM Assertion }
  \IgNoRe{STM Assertion }
  \IgNoRe{STM Assertion }
  \IgNoRe{STM Assertion }
  \IgNoRe{PG}
 \def\defbubbleprop{\frefwarning VII.1} \IgNoRe{STM Assertion }
  \IgNoRe{STM Assertion }
  \IgNoRe{STM Assertion }
 \def\defParicleHoleDecomp{\frefwarning VII.4} \IgNoRe{STM Assertion }
  \IgNoRe{STM Assertion }
  \IgNoRe{STM Assertion }
  \IgNoRe{STM Assertion }
  \IgNoRe{STM Assertion }
  \IgNoRe{STM Assertion }
  \IgNoRe{EQN}
  \IgNoRe{EQN}
  \IgNoRe{EQN}
  \IgNoRe{STM Assertion }
  \IgNoRe{PG}
  \IgNoRe{STM Assertion }
  \IgNoRe{STM Assertion }
  \IgNoRe{STM Assertion }
  \IgNoRe{EQN}
  \IgNoRe{EQN}
  \IgNoRe{EQN}
  \IgNoRe{STM Assertion }
  \IgNoRe{STM Assertion }
  \IgNoRe{EQN}
  \IgNoRe{EQN}
  \IgNoRe{EQN}
  \IgNoRe{STM Assertion }
  \IgNoRe{PG}
  \IgNoRe{PG}
  \IgNoRe{STM Assertion }
  \IgNoRe{STM Assertion }
  \IgNoRe{PG}
  \IgNoRe{STM Assertion }
  \IgNoRe{STM Assertion }
  \IgNoRe{EQN}
  \IgNoRe{EQN}
  \IgNoRe{EQN}
  \IgNoRe{EQN}
  \IgNoRe{EQN}
  \IgNoRe{EQN}
  \IgNoRe{EQN}
  \IgNoRe{EQN}
  \IgNoRe{EQN}
  \IgNoRe{EQN}
  \IgNoRe{EQN}
  \IgNoRe{EQN}
  \IgNoRe{EQN}
  \IgNoRe{EQN}
  \IgNoRe{STM Assertion }
  \IgNoRe{EQN}
  \IgNoRe{PG}
  \IgNoRe{EQN}
  \IgNoRe{STM Assertion }
  \IgNoRe{EQN}
  \IgNoRe{STM Assertion }
  \IgNoRe{STM Assertion }
  \IgNoRe{EQN}
  \IgNoRe{EQN}
  \IgNoRe{EQN}
  \IgNoRe{STM Assertion }
  \IgNoRe{EQN}
  \IgNoRe{EQN}
  \IgNoRe{EQN}
  \IgNoRe{EQN}
  \IgNoRe{EQN}
  \IgNoRe{EQN}
  \IgNoRe{EQN}
  \IgNoRe{EQN}
  \IgNoRe{EQN}
  \IgNoRe{EQN}
  \IgNoRe{EQN}
  \IgNoRe{EQN}
  \IgNoRe{EQN}
  \IgNoRe{EQN}
  \IgNoRe{EQN}
  \IgNoRe{EQN}
  \IgNoRe{EQN}
  \IgNoRe{EQN}
  \IgNoRe{EQN}
  \IgNoRe{EQN}
  \IgNoRe{EQN}
  \IgNoRe{STM Assertion }
  \IgNoRe{PG}
  \IgNoRe{PG}
  \IgNoRe{STM Assertion }
  \IgNoRe{PG}
  \IgNoRe{EQN}
  \IgNoRe{EQN}
  \IgNoRe{EQN}
  \IgNoRe{EQN}
  \IgNoRe{STM Assertion }
  \IgNoRe{EQN}
  \IgNoRe{PG}
  \IgNoRe{EQN}
  \IgNoRe{EQN}
  \IgNoRe{EQN}
  \IgNoRe{EQN}
  \IgNoRe{EQN}
  \IgNoRe{EQN}
  \IgNoRe{EQN}
  \IgNoRe{EQN}
  \IgNoRe{EQN}
  \IgNoRe{EQN}
  \IgNoRe{EQN}
  \IgNoRe{EQN}
  \IgNoRe{STM Assertion }
  \IgNoRe{STM Assertion }
  \IgNoRe{EQN}
  \IgNoRe{EQN}
  \IgNoRe{PG}
  \IgNoRe{PG}
  \IgNoRe{STM Assertion }
  \IgNoRe{EQN}
  \IgNoRe{STM Assertion }
  \IgNoRe{PG}
  \IgNoRe{EQN}
  \IgNoRe{EQN}
  \IgNoRe{EQN}
  \IgNoRe{STM Assertion }
  \IgNoRe{STM Assertion }
  \IgNoRe{EQN}
  \IgNoRe{EQN}
  \IgNoRe{EQN}
  \IgNoRe{EQN}
  \IgNoRe{EQN}
  \IgNoRe{STM Assertion }
  \IgNoRe{EQN}
  \IgNoRe{EQN}
  \IgNoRe{EQN}
  \IgNoRe{EQN}
  \IgNoRe{STM Assertion }
  \IgNoRe{STM Assertion }
  \IgNoRe{EQN}
  \IgNoRe{STM Assertion }
  \IgNoRe{STM Assertion }
  \IgNoRe{STM Assertion }
  \IgNoRe{STM Assertion }
  \IgNoRe{PG}
  \IgNoRe{STM Assertion }
  \IgNoRe{STM Assertion }
  \IgNoRe{STM Assertion }
  \IgNoRe{STM Assertion }
  \IgNoRe{STM Assertion }
  \IgNoRe{STM Assertion }
  \IgNoRe{STM Assertion }
 \def\defNPsectdiffdecaynorm{\frefwarning XIII.12} \IgNoRe{STM Assertion }
  \IgNoRe{STM Assertion }
  \IgNoRe{STM Assertion }
  \IgNoRe{STM Assertion }
  \IgNoRe{STM Assertion }
  \IgNoRe{STM Assertion }
  \IgNoRe{STM Assertion }
  \IgNoRe{STM Assertion }
  \IgNoRe{PG}
  \IgNoRe{STM Assertion }
  \IgNoRe{STM Assertion }
  \IgNoRe{STM Assertion }
 \def\defNPdisjointfnspaces{\frefwarning XIV.6} \IgNoRe{STM Assertion }
  \IgNoRe{STM Assertion }
  \IgNoRe{STM Assertion }
  \IgNoRe{STM Assertion }
  \IgNoRe{STM Assertion }
  \IgNoRe{STM Assertion }
  \IgNoRe{STM Assertion }
  \IgNoRe{STM Assertion }
  \IgNoRe{STM Assertion }
  \IgNoRe{EQN}
  \IgNoRe{STM Assertion }
  \IgNoRe{STM Assertion }
  \IgNoRe{STM Assertion }
  \IgNoRe{STM Assertion }
  \IgNoRe{STM Assertion }
  \IgNoRe{STM Assertion }
  \IgNoRe{EQN}
  \IgNoRe{STM Assertion }
  \IgNoRe{PG}
  \IgNoRe{PG}
  \IgNoRe{STM Assertion }
  \IgNoRe{STM Assertion }
  \IgNoRe{STM Assertion }
  \IgNoRe{EQN}
  \IgNoRe{STM Assertion }
  \IgNoRe{PG}
  \IgNoRe{EQN}
  \IgNoRe{STM Assertion }
  \IgNoRe{STM Assertion }
  \IgNoRe{EQN}
  \IgNoRe{EQN}
  \IgNoRe{EQN}
  \IgNoRe{EQN}
  \IgNoRe{EQN}
  \IgNoRe{EQN}
  \IgNoRe{EQN}
  \IgNoRe{EQN}
  \IgNoRe{EQN}
  \IgNoRe{EQN}
  \IgNoRe{EQN}
  \IgNoRe{EQN}
  \IgNoRe{EQN}
  \IgNoRe{EQN}
  \IgNoRe{EQN}
  \IgNoRe{EQN}
  \IgNoRe{EQN}
  \IgNoRe{EQN}
  \IgNoRe{EQN}
  \IgNoRe{EQN}
  \IgNoRe{STM Assertion }
  \IgNoRe{STM Assertion }
  \IgNoRe{PG}
  \IgNoRe{EQN}
  \IgNoRe{EQN}
  \IgNoRe{EQN}
  \IgNoRe{EQN}
  \IgNoRe{EQN}
  \IgNoRe{EQN}
  \IgNoRe{EQN}
  \IgNoRe{EQN}
  \IgNoRe{EQN}
  \IgNoRe{EQN}
  \IgNoRe{EQN}
  \IgNoRe{EQN}
  \IgNoRe{EQN}
  \IgNoRe{EQN}
  \IgNoRe{EQN}
  \IgNoRe{EQN}
  \IgNoRe{EQN}
  \IgNoRe{EQN}
  \IgNoRe{EQN}
  \IgNoRe{EQN}
  \IgNoRe{EQN}
 \def\lemNPhoelder{\frefwarning C.1} \IgNoRe{STM Assertion }
  \IgNoRe{PG}
 \def\defmodcompLadder{\frefwarning D.1} \IgNoRe{STM Assertion }
 \def\theoremmodcompLadder{\frefwarning D.2} \IgNoRe{STM Assertion }
  \IgNoRe{EQN}
  \IgNoRe{EQN}
  \IgNoRe{PG}
  \IgNoRe{EQN}
  \IgNoRe{EQN}
  \IgNoRe{EQN}
  \IgNoRe{EQN}
  \IgNoRe{EQN}
  \IgNoRe{EQN}
  \IgNoRe{EQN}
  \IgNoRe{EQN}
  \IgNoRe{STM Assertion }
  \IgNoRe{STM Assertion }
  \IgNoRe{STM Assertion }
  \IgNoRe{STM Assertion }
 \def\corcompLadder{\frefwarning D.7} \IgNoRe{STM Assertion }
  \IgNoRe{PG}
  \IgNoRe{PG}


\newcount\CHAPNO
\newcount\APPNO
\CHAPNO=0
\APPNO=1
\def\advCHAPNO{\advance\CHAPNO by 1}
\def\advAPPNO{\advance\APPNO by 1}

\def\caproman#1{\ifcase#1\or I\or II\or III\or IV\or V\or VI\or VII\or
VIII\or IX\or X\or XI\or XII\or XIII\or XIV\or XV\or XVI\or XVII\or XVIII\or
XIX\or XX\or XXI\or XXII\or XXIII\or XXIV\or XXV\or XXVI\or XXVII\or XXVIII\or XXIX\or XXX\or XXXI\or XXXII\or XXXIII\or XXXIV\or XXXV\or XXXVI\or XXXVII\or XXXVIII\or XXXIX\fi}%

\def\capletter#1{\ifcase#1\or A\or B\or C\or D\or E\or F\or G\or
H\or I\or J\or K\or L\or M\or N\or O\or P\or Q\or R\or
S\or T\or U\or V\or W\or X\or Y\or Z\fi}%

\newcount\cHintroI \cHintroI=\CHAPNO \advCHAPNO 
\newcount\cHintroOverview  \cHintroOverview=\CHAPNO \advCHAPNO 
                              \edef\CHintroOverview{\caproman\CHAPNO}  
\newcount\cHrenmap \cHrenmap=\CHAPNO \advCHAPNO 

 \advAPPNO

\newcount\cHintroII \cHintroII=\CHAPNO \advCHAPNO 
                              
\newcount\cHfirstscale \cHfirstscale=\CHAPNO \advCHAPNO
                              
\newcount\cHnewsectors \cHnewsectors=\CHAPNO \advCHAPNO
                              \edef\CHnewsectors{\caproman\CHAPNO}
\newcount\cHphladders \cHphladders=\CHAPNO \advCHAPNO
                              \edef\CHphladders{\caproman\CHAPNO}
\newcount\cHfinitescale \cHfinitescale=\CHAPNO \advCHAPNO
                              
\newcount\cHstep \cHstep=\CHAPNO \advCHAPNO
                              
\newcount\cHrecurs \cHrecurs=\CHAPNO \advCHAPNO
                              
 \advAPPNO

\newcount\cHintroIII \cHintroIII=\CHAPNO \advCHAPNO
                              
\newcount\cHtildefinitescale \cHtildefinitescale=\CHAPNO \advCHAPNO
                              
\newcount\cHtildenewsectors \cHtildenewsectors=\CHAPNO \advCHAPNO
                              
\newcount\cHtildephladders \cHtildephladders=\CHAPNO \advCHAPNO
                              
\newcount\cHtildestep  \cHtildestep=\CHAPNO \advCHAPNO

 \advAPPNO
 \advAPPNO


\def\APpropbnd{A}
\def\APmodelbnd{B}
\def\APbubsec{C}
 \def\pgLADI{\frefwarning 1} \IgNoRe{PG}
 \def\pgLADIa{\frefwarning 1} \IgNoRe{PG}
 \def\lemLADprimitivemanfred{\frefwarning I.1} \IgNoRe{STM Assertion }
 \def\eqnLADpartofunity{\frefwarning I.1} \IgNoRe{EQN}
 \def\defLADscales{\frefwarning I.2} \IgNoRe{STM Assertion }
 \def\pgLADIb{\frefwarning 4} \IgNoRe{PG}
 \def\defLADsectors{\frefwarning I.3} \IgNoRe{STM Assertion }
 \def\eqnLADfourdunion{\frefwarning I.2} \IgNoRe{EQN}
 \def\defLADtransinv{\frefwarning I.4} \IgNoRe{STM Assertion }
 \def\eqnLADbemu{\frefwarning I.3} \IgNoRe{EQN}
 \def\defLADfourtrans{\frefwarning I.5} \IgNoRe{STM Assertion }
 \def\defLADsectorized{\frefwarning I.6} \IgNoRe{STM Assertion }
 \def\remLADsectorizedreduction{\frefwarning I.7} \IgNoRe{STM Assertion }
 \def\defLADphladder{\frefwarning I.8} \IgNoRe{STM Assertion }
 \def\pgLADIc{\frefwarning 8} \IgNoRe{PG}
 \def\remLADhardsoft{\frefwarning I.9} \IgNoRe{STM Assertion }
 \def\remLADphreduction{\frefwarning I.10} \IgNoRe{STM Assertion }
 \def\defLADlonelinfty{\frefwarning I.11} \IgNoRe{STM Assertion }
 \def\defLADlonelinftyII{\frefwarning I.12} \IgNoRe{STM Assertion }
 \def\pgLADId{\frefwarning 9} \IgNoRe{PG}
 \def\defLADdiffdecay{\frefwarning I.13} \IgNoRe{STM Assertion }
 \def\defLADlonelinftyIV{\frefwarning I.14} \IgNoRe{STM Assertion }
 \def\eqnLADcj{\frefwarning I.4} \IgNoRe{EQN}
 \def\defLADnormdomainnorm{\frefwarning I.15} \IgNoRe{STM Assertion }
 \def\lemLADNormcomparison{\frefwarning I.16} \IgNoRe{STM Assertion }
 \def\defLADresectoriz{\frefwarning I.17} \IgNoRe{STM Assertion }
 \def\pgLADIe{\frefwarning 12} \IgNoRe{PG}
 \def\pgLADIf{\frefwarning 13} \IgNoRe{PG}
 \def\remLADreresectoriz{\frefwarning I.18} \IgNoRe{STM Assertion }
 \def\eqnLADflipped{\frefwarning I.5} \IgNoRe{EQN}
 \def\defLADcompoundphladder{\frefwarning I.19} \IgNoRe{STM Assertion }
 \def\thmLADmodcompLadder{\frefwarning I.20} \IgNoRe{STM Assertion }
 \def\pgLADIg{\frefwarning 14} \IgNoRe{PG}
 \def\remLADmodcompLadder{\frefwarning I.21} \IgNoRe{STM Assertion }
 \def\thmLADmodcompLaddercont{\frefwarning I.22} \IgNoRe{STM Assertion }
\def\CHwwladders{II}
 \def\convLADresectorbullet{\frefwarning II.1} \IgNoRe{STM Assertion }
 \def\defLADcompoundphladderscalej{\frefwarning II.2} \IgNoRe{STM Assertion }
 \def\pgLADII{\frefwarning 17} \IgNoRe{PG}
 \def\pgLADIIa{\frefwarning 17} \IgNoRe{PG}
 \def\propLADaltcmpladders{\frefwarning II.3} \IgNoRe{STM Assertion }
 \def\lemLADaltcmpladdersII{\frefwarning II.4} \IgNoRe{STM Assertion }
 \def\lemLADaltcmpladdersIII{\frefwarning II.5} \IgNoRe{STM Assertion }
 \def\defLADspinIndependent{\frefwarning II.6} \IgNoRe{STM Assertion }
 \def\remLADspinIndependent{\frefwarning II.7} \IgNoRe{STM Assertion }
 \def\lemLADchargespin{\frefwarning II.8} \IgNoRe{STM Assertion }
 \def\pgLADIIb{\frefwarning 21} \IgNoRe{PG}
 \def\remLADchargespin{\frefwarning II.9} \IgNoRe{STM Assertion }
 \def\lemLADchargespinflip{\frefwarning II.10} \IgNoRe{STM Assertion }
 \def\lemLADchargespinII{\frefwarning II.11} \IgNoRe{STM Assertion }
 \def\corLADaltcmpladders{\frefwarning II.12} \IgNoRe{STM Assertion }
 \def\convLADlr{\frefwarning II.13} \IgNoRe{STM Assertion }
 \def\secrepnorm{\frefwarning II.14} \IgNoRe{STM Assertion }
 \def\eqnLADDelta{\frefwarning II.1} \IgNoRe{EQN}
 \def\pgLADIIc{\frefwarning 24} \IgNoRe{PG}
 \def\remLADscaledtonormdomain{\frefwarning II.15} \IgNoRe{STM Assertion }
 \def\resecrepnorm{\frefwarning II.16} \IgNoRe{STM Assertion }
 \def\lemLADresectornorm{\frefwarning II.17} \IgNoRe{STM Assertion }
 \def\eqnLADchisbnd{\frefwarning II.2} \IgNoRe{EQN}
 \def\topbottom{\frefwarning II.18} \IgNoRe{STM Assertion }
 \def\bbound{\frefwarning II.19} \IgNoRe{STM Assertion }
 \def\dbbound{\frefwarning II.20} \IgNoRe{STM Assertion }
 \def\pgLADIId{\frefwarning 27} \IgNoRe{PG}
 \def\leibniz{\frefwarning II.21} \IgNoRe{STM Assertion }
 \def\bboundcor{\frefwarning II.22} \IgNoRe{STM Assertion }
 \def\dbboundcor{\frefwarning II.23} \IgNoRe{STM Assertion }
 \def\bdbbound{\frefwarning II.24} \IgNoRe{STM Assertion }
 \def\eqnLADFbnd{\frefwarning II.3} \IgNoRe{EQN}
 \def\eqnLADbnd{\frefwarning II.4} \IgNoRe{EQN}
 \def\eqnLADladindbnd{\frefwarning II.5} \IgNoRe{EQN}
 \def\eqnLADGbnd{\frefwarning II.6} \IgNoRe{EQN}
 \def\remLADLbnd{\frefwarning II.25} \IgNoRe{STM Assertion }
 \def\ladderinduction{\frefwarning II.26} \IgNoRe{STM Assertion }
 \def\eqnLADqbnd{\frefwarning II.7} \IgNoRe{EQN}
 \def\lemCombinedCors{\frefwarning II.27} \IgNoRe{STM Assertion }
 \def\eqnLADKbnd{\frefwarning II.8} \IgNoRe{EQN}
 \def\eqnliA{\frefwarning II.9} \IgNoRe{EQN}
 \def\eqnliB{\frefwarning II.10} \IgNoRe{EQN}
 \def\eqnliC{\frefwarning II.11} \IgNoRe{EQN}
 \def\eqnliD{\frefwarning II.12} \IgNoRe{EQN}
 \def\pgLADIIe{\frefwarning 42} \IgNoRe{PG}
 \def\lemLADlebupbnd{\frefwarning II.28} \IgNoRe{STM Assertion }
 \def\lemLADtermwisecont{\frefwarning II.29} \IgNoRe{STM Assertion }
 \def\eqnMBPmab{\frefwarning II.13} \IgNoRe{EQN}
\def\CHbubbles{III}
 \def\eqnLADfullbubdecomp{\frefwarning III.1} \IgNoRe{EQN}
 \def\eqnBubIa{\frefwarning III.2} \IgNoRe{EQN}
 \def\pgLADIII{\frefwarning 46} \IgNoRe{PG}
 \def\eqnBubI{\frefwarning III.3} \IgNoRe{EQN}
 \def\eqnBubII{\frefwarning III.4} \IgNoRe{EQN}
 \def\defLADbubblenorm{\frefwarning III.1} \IgNoRe{STM Assertion }
 \def\lemLADbubblenorm{\frefwarning III.2} \IgNoRe{STM Assertion }
 \def\eqnLADbubblenormA{\frefwarning III.5} \IgNoRe{EQN}
 \def\eqnLADbubblenormB{\frefwarning III.6} \IgNoRe{EQN}
 \def\eqnBubbleTensor{\frefwarning III.7} \IgNoRe{EQN}
 \def\eqnVI{\frefwarning III.8} \IgNoRe{EQN}
 \def\defLADtrasfermomcutoff{\frefwarning III.3} \IgNoRe{STM Assertion }
 \def\lemXIII{\frefwarning III.4} \IgNoRe{STM Assertion }
 \def\remLADOR{\frefwarning III.5} \IgNoRe{STM Assertion }
 \def\defLADshortnorm{\frefwarning III.6} \IgNoRe{STM Assertion }
 \def\defLADKspace{\frefwarning III.7} \IgNoRe{STM Assertion }
 \def\remLADKnorm{\frefwarning III.8} \IgNoRe{STM Assertion }
 \def\rebbound{\frefwarning III.9} \IgNoRe{STM Assertion }
 \def\defLADtdiff{\frefwarning III.10} \IgNoRe{STM Assertion }
 \def\lemXIV{\frefwarning III.11} \IgNoRe{STM Assertion }
 \def\remLADtdiff{\frefwarning III.12} \IgNoRe{STM Assertion }
 \def\remXIVector{\frefwarning III.13} \IgNoRe{STM Assertion }
 \def\lemXV{\frefwarning III.14} \IgNoRe{STM Assertion }
 \def\eqnLADcsn{\frefwarning III.9} \IgNoRe{EQN}
 \def\eqnLADlonepropbnd{\frefwarning III.10} \IgNoRe{EQN}
 \def\eqnLADlinftypropbnd{\frefwarning III.11} \IgNoRe{EQN}
 \def\eqnLADderivlinftypropbnd{\frefwarning III.12} \IgNoRe{EQN}
 \def\rebboundII{\frefwarning III.15} \IgNoRe{STM Assertion }
 \def\eqnLADbboundZ{\frefwarning III.13} \IgNoRe{EQN}
 \def\propLADlarget{\frefwarning III.16} \IgNoRe{STM Assertion }
 \def\eqnLADsmalli{\frefwarning III.14} \IgNoRe{EQN}
 \def\eqnLADctocprime{\frefwarning III.15} \IgNoRe{EQN}
 \def\eqnLADassume{\frefwarning III.16} \IgNoRe{EQN}
 \def\eqLemXV{\frefwarning III.17} \IgNoRe{EQN}
 \def\eqnVII{\frefwarning III.18} \IgNoRe{EQN}
 \def\remXXII{\frefwarning III.17} \IgNoRe{STM Assertion }
 \def\lemLADzprops{\frefwarning III.18} \IgNoRe{STM Assertion }
 \def\propLADzdiff{\frefwarning III.19} \IgNoRe{STM Assertion }
 \def\lemXXIV{\frefwarning III.20} \IgNoRe{STM Assertion }
 \def\lemXXV{\frefwarning III.21} \IgNoRe{STM Assertion }
 \def\eqnLADspatcnt{\frefwarning III.19} \IgNoRe{EQN}
 \def\eqnVIII{\frefwarning III.20} \IgNoRe{EQN}
 \def\propXXVIII{\frefwarning III.22} \IgNoRe{STM Assertion }
 \def\lemXXIX{\frefwarning III.23} \IgNoRe{STM Assertion }
 \def\eqnLADwsum{\frefwarning III.21} \IgNoRe{EQN}
 \def\eqnIX{\frefwarning III.22} \IgNoRe{EQN}
 \def\propXXX{\frefwarning III.24} \IgNoRe{STM Assertion }
 \def\eqnX{\frefwarning III.23} \IgNoRe{EQN}
 \def\lemXXXI{\frefwarning III.25} \IgNoRe{STM Assertion }
 \def\lemLADrectangle{\frefwarning III.26} \IgNoRe{STM Assertion }
 \def\eqnLADmeqmrho{\frefwarning III.24} \IgNoRe{EQN}
 \def\propXXXII{\frefwarning III.27} \IgNoRe{STM Assertion }
 \def\eqnLADmrhotoms{\frefwarning III.25} \IgNoRe{EQN}
 \def\eqnLADmsup{\frefwarning III.26} \IgNoRe{EQN}
 \def\eqnLADmsupmax{\frefwarning III.27} \IgNoRe{EQN}
 \def\eqnLADgMh{\frefwarning III.28} \IgNoRe{EQN}
 \def\lemBubTnonzero{\frefwarning III.28} \IgNoRe{STM Assertion }
 \def\pgLADIIIa{\frefwarning 77} \IgNoRe{PG}
 \def\lemLADtnonzero{\frefwarning III.29} \IgNoRe{STM Assertion }
 \def\propMPsamelim{\frefwarning III.30} \IgNoRe{STM Assertion }
 \def\pgLADIIIb{\frefwarning 79} \IgNoRe{PG}
 \def\corMPsamelim{\frefwarning III.31} \IgNoRe{STM Assertion }
\def\CHdbubbles{IV}
 \def\pgLADIV{\frefwarning 83} \IgNoRe{PG}
 \def\lemXXXIV{\frefwarning IV.1} \IgNoRe{STM Assertion }
 \def\eqnXIa{\frefwarning IV.1} \IgNoRe{EQN}
 \def\eqnXIb{\frefwarning IV.2} \IgNoRe{EQN}
 \def\eqnLADdupdown{\frefwarning IV.3} \IgNoRe{EQN}
 \def\lemXVbis{\frefwarning IV.2} \IgNoRe{STM Assertion }
 \def\corLADgDgWh{\frefwarning IV.3} \IgNoRe{STM Assertion }
 \def\redbbound{\frefwarning IV.4} \IgNoRe{STM Assertion }
 \def\redbboundII{\frefwarning IV.5} \IgNoRe{STM Assertion }
 \def\eqnLADdbboundZ{\frefwarning IV.4} \IgNoRe{EQN}
 \def\propLADdlarget{\frefwarning IV.6} \IgNoRe{STM Assertion }
 \def\eqnLADdsmalli{\frefwarning IV.5} \IgNoRe{EQN}
 \def\eqnLADdctocprime{\frefwarning IV.6} \IgNoRe{EQN}
 \def\eqnLADdassume{\frefwarning IV.7} \IgNoRe{EQN}
 \def\lemXXXVI{\frefwarning IV.7} \IgNoRe{STM Assertion }
 \def\lemXXXVII{\frefwarning IV.8} \IgNoRe{STM Assertion }
 \def\propXXXVIII{\frefwarning IV.9} \IgNoRe{STM Assertion }
 \def\eqnXII{\frefwarning IV.8} \IgNoRe{EQN}
\def\APpropbnd{A}
 \def\lemLADeprime{\frefwarning A.1} \IgNoRe{STM Assertion }
 \def\pgLADA{\frefwarning 97} \IgNoRe{PG}
 \def\lemLADsimplepropbnd{\frefwarning A.2} \IgNoRe{STM Assertion }
 \def\eqnLADcspI{\frefwarning A.1} \IgNoRe{EQN}
 \def\eqnLADcspII{\frefwarning A.2} \IgNoRe{EQN}
 \def\lemLADdeltabnd{\frefwarning A.3} \IgNoRe{STM Assertion }
 \def\eqnLADdepartbnds{\frefwarning A.3} \IgNoRe{EQN}
\def\APmodelbnd{B}
 \def\lemLadBmom{\frefwarning B.1} \IgNoRe{STM Assertion }
 \def\eqnMana{\frefwarning B.1} \IgNoRe{EQN}
 \def\eqnManb{\frefwarning B.2} \IgNoRe{EQN}
 \def\pgLADB{\frefwarning 102} \IgNoRe{PG}
 \def\eqnMancm{\frefwarning B.3} \IgNoRe{EQN}
 \def\eqnMand{\frefwarning B.4} \IgNoRe{EQN}
 \def\eqnMane{\frefwarning B.5} \IgNoRe{EQN}
 \def\eqnManc{\frefwarning B.6} \IgNoRe{EQN}
 \def\eqnMante{\frefwarning B.7} \IgNoRe{EQN}
 \def\thmLADBposn{\frefwarning B.2} \IgNoRe{STM Assertion }
 \def\eqnManaa{\frefwarning B.8} \IgNoRe{EQN}
 \def\eqnManaabis{\frefwarning B.9} \IgNoRe{EQN}
 \def\eqnManbb{\frefwarning B.10} \IgNoRe{EQN}
 \def\eqnManrho{\frefwarning B.11} \IgNoRe{EQN}
 \def\eqnMancc{\frefwarning B.12} \IgNoRe{EQN}
 \def\lemLadABdiff{\frefwarning B.3} \IgNoRe{STM Assertion }
 \def\eqnLADBminusBp{\frefwarning B.13} \IgNoRe{EQN}
 \def\eqnLADIbnd{\frefwarning B.14} \IgNoRe{EQN}
 \def\eqnMEa{\frefwarning B.15} \IgNoRe{EQN}
 \def\eqnMEb{\frefwarning B.16} \IgNoRe{EQN}
 \def\eqnMEc{\frefwarning B.17} \IgNoRe{EQN}
 \def\eqnMEd{\frefwarning B.18} \IgNoRe{EQN}
 \def\eqnMEe{\frefwarning B.19} \IgNoRe{EQN}
 \def\eqnMEf{\frefwarning B.20} \IgNoRe{EQN}
 \def\lemLadBmodbub{\frefwarning B.4} \IgNoRe{STM Assertion }
\def\APbubsec{C}
 \def\lemAO{\frefwarning C.1} \IgNoRe{STM Assertion }
 \def\pgLADC{\frefwarning 117} \IgNoRe{PG}
 \def\eqnAOI{\frefwarning C.1} \IgNoRe{EQN}
 \def\eqnAOII{\frefwarning C.2} \IgNoRe{EQN}
 \def\lemAI{\frefwarning C.2} \IgNoRe{STM Assertion }
 \def\eqnAI{\frefwarning C.3} \IgNoRe{EQN}
 \def\eqnAIp{\frefwarning C.4} \IgNoRe{EQN}
 \def\lemAIII{\frefwarning C.3} \IgNoRe{STM Assertion }
 \def\eqnAII{\frefwarning C.5} \IgNoRe{EQN}
 \def\lemAIV{\frefwarning C.4} \IgNoRe{STM Assertion }
 \def\eqnAIII{\frefwarning C.6} \IgNoRe{EQN}
 \def\eqnLADappCn{\frefwarning C.7} \IgNoRe{EQN}
 \def\pgLADref{\frefwarning 125} \IgNoRe{PG}
 \def\pgLADnot{\frefwarning 127} \IgNoRe{PG}

\EDEF\APpropbnd{A}
\EDEF\APmodelbnd{B}
\EDEF\APbubsec{C}


{\nopagenumbers
\multiply\baselineskip by \spacingDenominator\divide \baselineskip by\spacingNumerator

\null\vskip3truecm

%
%
\centerline{\tafontt Particle--Hole Ladders}

\vskip0.75in
\centerline{Joel Feldman{\parindent=.15in\footnote{$^{*}$}{Research supported 
in part by the
 Natural Sciences and Engineering Research Council of Canada and the Forschungsinstitut f\"ur Mathematik, ETH Z\"urich}}}
\centerline{Department of Mathematics}
\centerline{University of British Columbia}
\centerline{Vancouver, B.C. }
\centerline{CANADA\ \   V6T 1Z2}
\centerline{feldman@math.ubc.ca}
\centerline{http:/\hskip-3pt/www.math.ubc.ca/\squiggle
feldman/}
\vskip0.3in
\centerline{Horst Kn\"orrer, Eugene Trubowitz}
\centerline{Mathematik}
\centerline{ETH-Zentrum}
\centerline{CH-8092 Z\"urich}
\centerline{SWITZERLAND}
\centerline{knoerrer@math.ethz.ch, trub@math.ethz.ch}
\centerline{http:/\hskip-3pt/www.math.ethz.ch/\squiggle
knoerrer/}

\vskip0.75in
\noindent
%
{\bf Abstract.\ \ \ } 
A self contained analysis demonstrates that the sum of all particle-hole 
ladder contributions for a two dimensional, weakly coupled fermion gas 
with a strictly convex Fermi curve at temperature zero is bounded. This
is used in our construction of two dimensional Fermi liquids.

\vfill
\eject


\titleb{Table of Contents}
\halign{\hfill#\ &\hfill#\ &#\hfill&\ p\ \hfil#&\ p\ \hfil#\cr
\noalign{\vskip.1in}
\S I&\omit Introduction                             \span&\:\pgLADI&\omit\cr
&&Ladders in Momentum Space                              &\omit&\:\pgLADIa\cr
&&Scales and Sectors                                     &\omit&\:\pgLADIb\cr
&&Particle--Hole Ladders                                 &\omit&\:\pgLADIc\cr
&&Norms                                                  &\omit&\:\pgLADId\cr
&&The Propagators                                        &\omit&\:\pgLADIe\cr
&&Resectorization                                        &\omit&\:\pgLADIf\cr
&&Compound Particle--Hole Ladders                        &\omit&\:\pgLADIg\cr
\noalign{\vskip.1in}
\S II&\omit Reduction to Bubble Estimates           \span&\:\pgLADII\cr
&&Combinatorial Structure of Compound Ladders          &\omit&\:\pgLADIIa\cr
&&Spin Independence                                   &\omit&\:\pgLADIIb\cr
&&Scaled Norms                                        &\omit&\:\pgLADIIc\cr
&&Bubble and Double Bubble Bounds                     &\omit&\:\pgLADIId\cr
&&The Infrared Limit                                  &\omit&\:\pgLADIIe\cr
\noalign{\vskip.1in}
\S III&\omit Bubbles                                 \span&\:\pgLADIII\cr
&&The Infrared Limit -- Nonzero Transfer Momentum     &\omit&\:\pgLADIIIa\cr
&&The Infrared Limit -- Reduction to Factorized Cutoffs  &\omit&\:\pgLADIIIb\cr
\noalign{\vskip.1in}
\S IV&\omit Double Bubbles                                 \span&\:\pgLADIV\cr
\noalign{\vskip.1in}
{\bf Appendices}\span\cr
\noalign{\vskip.1in}
\S A&\omit Bounds on Propagators                       \span&\:\pgLADA\cr
\noalign{\vskip.1in}
\S B&\omit Bound on the Generalized Model Bubble       \span&\:\pgLADB\cr
\noalign{\vskip.1in}
\S C&\omit Sector Counting with Specified Transfer Momentum\span&\:\pgLADC\cr
\noalign{\vskip.1in}
 &\omit References                                    \span&\:\pgLADref \cr
\noalign{\vskip.1in}
 &\omit Notation                                      \span&\:\pgLADnot \cr
}
\vfill\eject
\multiply\baselineskip by \spacingNumerator\divide \baselineskip by\spacingDenominator}
\pageno=1


\chap{ Introduction}\PG\pgLADI

In this paper we study the contributions of generalized particle-hole
ladders to the four--point Green's function of a many fermion system.
Formally, the amputated four--point Green's function, 
$G_4({\sst(p_1,\si_1),(p_2,\si_2),(p_3,\si_3),(p_4,\si_4)})$ with incoming particles of momenta $p_1,p_4\in\bbbr\times\bbbr^d$ and spins $\si_1,\si_4\in
\{\uparrow,\downarrow\}$ and outgoing particles of momenta $p_2,p_3$ and spins 
$\si_2,\si_3$, can be written as a sum of values of Feynman diagrams with 
four external legs. The propagator of these diagrams is
$C(k)=\sfrac{1}{\imath k_0-e(\k)}$, where $k=(k_0,\k)\in\bbbr\times\bbbr^d$ 
and the dispersion relation $e(\k)$ 
(into which the chemical potential has been absorbed) characterizes the independent fermion approximation. The interaction of the model determines the diagram vertices, $V({\sst(k_1,\si_1),(k_2,\si_2),(k_3,\si_3),(k_4,\si_4)})$,
$k_1+k_4=k_2+k_3$. Here, the incoming momenta are $k_1,k_4$ and the outgoing momenta are $k_2,k_3$.

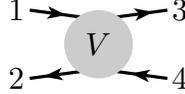
\centerline{\figput{vertex}}
\vskip.2in
\goodbreak\titlec{Ladders in Momentum Space}\PG\pgLADIa

The most important contributions to this four--point function are 
ladders. The contribution of the particle--hole ladder with $\ell+1$ rungs

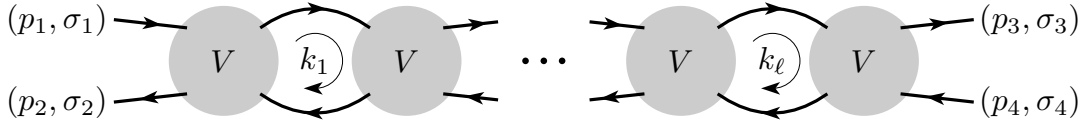
\centerline{\figput{PHladder}}

\noindent is
$$\eqalign{
\sum_{\tau_{i,1},\tau_{i,2}\in\{\uparrow,\downarrow\}\atop i=1,\cdots,\ell}
\int\sfrac{d^{d+1}k_1}{(2\pi)^{d+1}}\cdots\sfrac{d^{d+1}k_\ell}{(2\pi)^{d+1}}
\ &V({\sst(p_1,\si_1),(p_2,\si_2),(p_1+k_1,\tau_{1,1}),(p_2+k_1,\tau_{1,2})})
C({\sst p_1+k_1})C({\sst p_2+k_1})\cr
\noalign{\vskip-.2in}
&V({\sst(p_1+k_1,\tau_{1,1}),(p_2+k_1,\tau_{1,2}),\cdots})\cdots
V({\sst\cdots,(p_1+k_\ell,\tau_{\ell,1}),(p_2+k_\ell,\tau_{\ell,2})})\cr
\noalign{\vskip-.05in}
&C({\sst p_1+k_\ell})C({\sst p_2+k_\ell})
V({\sst(p_1+k_\ell,\tau_{\ell,1}),(p_2+k_\ell,\tau_{\ell,2}),
         (p_3,\si_3),(p_4,\si_4)})
}$$
 The contribution of the particle--particle ladder with $\ell+1$ rungs

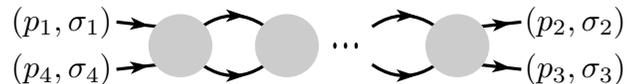
\centerline{\figput{PPladder}}

\noindent is
$$\eqalign{
\sum_{\tau_{i,1},\tau_{i,2}\in\{\uparrow,\downarrow\}\atop i=1,\cdots,\ell}
\int\sfrac{d^{d+1}k_i}{(2\pi)^{d+1}}\cdots\sfrac{d^{d+1}k_\ell}{(2\pi)^{d+1}}
\ &V({\sst(p_1,\si_1),(p_1+k_1,\tau_{1,1}),(p_4-k_1,\tau_{1,2}), (p_4,\si_4)})
C({\sst p_1+k_1})C({\sst p_4-k_1})\cr
\noalign{\vskip-.2in}
&V({\sst(p_1+k_1,\tau_{1,1}),\cdots,(p_4-k_1,\tau_{1,2})})\cdots
V({\sst\cdots,(p_1+k_\ell,\tau_{\ell,1}),(p_4-k_\ell,\tau_{\ell,2}),\cdots})\cr
\noalign{\vskip-.05in}
&C({\sst p_1+k_\ell})C({\sst p_4-k_\ell})
V({\sst(p_1+k_\ell,\tau_{\ell,1}),
         (p_2,\si_2),(p_3,\si_3),(p_4-k_\ell,\tau_{\ell,2})})
}$$
Ladders with two rungs are called bubbles. The values of the bubbles 
with dispersion relation $e(\k)=\sfrac{|\k|^2}{2m}-\mu$ and interaction 
$V({\sst(p_1,\si_1),(p_2,\si_2),(p_3,\si_3),(p_4,\si_4)})
=\la\big(\de_{\si_1,\si_2}\de_{\si_3,\si_4}
-\de_{\si_1,\si_3}\de_{\si_2,\si_4}\big)$ are well--known for $d=2,3$ [FHN].
The particle--particle bubble has a logarithmic singularity [FKST, 
Proposition II.1b] at transfer momentum $p_1+p_4=0$ which is 
responsible for the formation of Cooper pairs and the onset of superconductivity. This singularity persists in models having dispersion relations that are symmetric about the origin, i.e. $e(\k)=e(-\k)$. On the other hand, if $e(\k)$ is strongly asymmetric in the sense of Definition
\defNPstrongasymm\ of [FKTf1] 
then the particle--particle bubble remains continuous and, in particular, 
bounded [FKLT1, page 297].

For the particle--hole bubble with $d=2$ and $e(\k)=\sfrac{|\k|^2}{2m}-\mu$
$$\eqalign{
&\int_{\bbbr^3} \sfrac{d^3 k}{(2\pi)^3}\ C(k+p_1)\,C(k+p_2)
=\int_{\bbbr^3} \sfrac{d^3 k}{(2\pi)^3}   
\sfrac{1}{ i(k_0+t_0/2)-e(\k+\t/2)}  \sfrac{1}{i(k_0-t_0/2)-e(\k-\t/2)}\cr
&\hskip.5in
= \cases{
-\sfrac{m}{2\pi}+\sfrac{m}{2\pi|\t|^2}\, {\rm Re}
\sqrt{{\sst|\t|^2(|\t|^2-4k_F^2)-4m^2t_0^2-4\imath mt_0|\t|^2}} 
& if $t_0,|\t|\ne 0$ or $|\t|\ge 2k_F$\cr
\noalign{\vskip.05in}
-\sfrac{m}{2\pi} & if $t_0= 0$ and $0<|\t|\le 2k_F$\cr
\noalign{\vskip.05in}
0& if $t_0\ne 0$ and $\t=0$\cr}
}$$
where $t=p_1-p_2$ is the transfer momentum, $k_F=\sqrt{2m\mu}$ is the radius of 
the Fermi surface and $\sqrt{\ }$ is the square root with nonnegative real
part and cut along the negative real axis. See, for example, 
[FHN (2.22) or FKST, Proposition II.1a]. This is $C^\infty$ on
$\set{t\in\bbbr\times~\bbbr^2}{t_0\ne~0\hbox{ or }|\t|>2k_F}$, is H\"older
continuous of degree 1 in a neighbourhood of any $t$ with $t_0=0,\ 0<|\t|<2k_F$
and is H\"older continuous of degree $\half$ in a neighbourhood of any $t$ with $t_0=0,\ |\t|=2k_F$, but cannot be continuously extended to $t=0$. However its restriction to $t_0=0$ does have a $C^\infty$ extension at the point $\t=0$. The discontinuity at $t=0$ persists for general, even
strongly asymmetric, $e(\k)$. For this reason, bounds on particle--hole ladders in position space are not straight forward.

That the restriction of the particle--hole bubble to $t_0=0$ does have a $C^\infty$ extension for a large class of smooth dispersion relations may be seen by the following argument, which was shown to us by Manfred Salmhofer
[S].
A generalization of this argument is used in Proposition \:\propXXXII.
\lemma{\STM\lemLADprimitivemanfred}{ 
Choose a  ``scale parameter'' $M>1$ and a function 
$\nu\in C^\infty_0\big(\big[\sfrac{1}{M},\,2M\big]\big)$ that 
takes values in $[0,1]$, is identically 1 on $\big[\sfrac{2}{M},M\big]$
is monotone on $\big[\sfrac{1}{M},\sfrac{2}{M}\big]$ and $[M,2M]$ and obeys
$$
\sum_{j=0}^\infty \nu\big(M^{2j}x\big) = 1
\EQN\eqnLADpartofunity$$
for $0<x<1$. Set
$\nu_0^{[0,j]}(k_0)=\sum_{\ell=0}^j\nu(M^{2\ell}k_0^2)$
and let $u(k,\t)$ be a bounded $C^\infty$ function with
compact support in $\k$ and bounded derivatives. Let $e(\k)$ be a
$C^\infty$ function that obeys $\lim\limits_{|\k|\rightarrow\infty}e(\k)=+\infty$.
Assume that the gradient of $e(\k)$ does not vanish on the Fermi surface $F=\set{\k\in\bbbr^d}{e(\k)=0}$.
Then
$$
B(\t)=\lim_{j\rightarrow\infty}
\int dk\ \frac{\nu^{[0,j]}_0(k_0)u(k,\t)}
{[ik_0-e(\k)][ik_0-e(\k+\t)]}
$$
is $C^\infty$ for $\t$ in a neighbourhood of $0$.
}
\prf Write
$$\deqalign{
B_j(\t)
&=\int\! dk\ \sfrac{\nu^{[0,j]}_0(k_0)u(k,\t)}
{[ik_0-e(\k)][ik_0-e(\k+\t)]}
&=\int\! dk\
\sfrac{\nu^{[0,j]}_0(k_0)u(k,\t)}{e(\k)-e(\k+\t)}
\Big[\sfrac{1}{ik_0-e(\k)}-\sfrac{1}{ik_0-e(\k+\t)}\Big]\cr
&=\int dk\ \sfrac{\nu^{[0,j]}_0(k_0)u(k,\t)}{e(\k)-e(\k+\t)}
\int_0^1\!\!ds\ \sfrac{d\hfill}{ds}\sfrac{1}{ik_0-E(\k,\t,s)}
&=\int dk\int_0^1ds\ 
\sfrac{\nu^{[0,j]}_0(k_0)u(k,\t)}
{[ik_0-E(\k,\t,s)]^2}\cr
}$$
where
$$
E(\k,\t,s)=s e(\k)+(1-s)e(\k+\t)
$$
Make, for each fixed $s$ and $k_0$, the change of variables from $\k$ to $E$ 
and $d-1$ variables $\th$ on $F$. Denote by $J(E,\t,\th,s)$ the Jacobian of 
this change of variables and set 
$$
f(k_0,E,\th,\t,s)= u\big((k_0,\k(E,\th,\t,s)),\t\big)J(E,\th,\t,s)
$$
Because $u$ has compact support in $\k$, $f$ vanishes unless $|E|\le\cE$,
for some finite $\cE$. Thus
$$
B_j(\t)
=\int_0^1 ds\int\, d\th\int\, dk_0\int_{-\cE}^\cE dE\ 
\sfrac{\nu^{[0,j]}_0(k_0)f(k_0,E,\th,\t,s)}{[ik_0-E]^2}
$$
Set
$$
B'_j(\t)
=\int_0^1 ds\int d\th\int dk_0\int_{-\cE}^\cE dE\ 
\sfrac{\nu^{[0,j]}_0(k_0)f(k_0,0,\th,\t,s)}{[ik_0-E]^2}
$$
Since
$$
\Big|\partial_\t^\al \big[\sfrac{\nu^{[0,j]}_0(k_0)f(k_0,E,\th,\t,s)}{[ik_0-E]^2}
-\sfrac{\nu^{[0,j]}_0(k_0)f(k_0,0,\th,\t,s)}{[ik_0-E]^2}\big]\!\Big|
\le\cst{}{\al}\!\sfrac{|E|}{k_0^2+E^2}
$$
is integrable on $\bbbr\times[-\cE,\cE]$, 
$\lim\limits_{j\rightarrow\infty}B_j(\t)-B'_j(\t)$ exists and
is $C^\infty$ by the Lebesgue dominated convergence theorem.
So it suffices to consider
$$
B'_j(\t)
=-2\cE\int_0^1 ds\int d\th\int dk_0\ 
\sfrac{\nu^{[0,j]}_0(k_0)f(k_0,0,\th,\t,s)}{k_0^2+\cE^2}
$$
Since
$$
\Big|\partial_\t^\al \sfrac{\nu^{[0,j]}_0(k_0)f(k_0,0,\th,\t,s)}{k_0^2+\cE^2}\Big|
\le\cst{}{\al}\sfrac{1}{k_0^2+\cE^2}
$$
is integrable on $\bbbr$, 
$\lim\limits_{j\rightarrow\infty}B'_j(\t)$ exists and
is $C^\infty$ by the Lebesgue dominated convergence theorem.
\endproof

\goodbreak\titlec{Scales and Sectors}\PG\pgLADIb

In this paper, we derive position space bounds for generalized particle--hole ladders in two space dimensions as they arise in a multiscale analysis.
The main result is Theorem \:\thmLADmodcompLadder, which is used in 
[FKTf2], under the name Theorem \theoremmodcompLadder, to help construct 
a Fermi liquid. We assume that the dispersion relation $e(\k)$ is 
$C^{r_e+3}$ for some $r_e\ge 6$,
that its gradient does not vanish on the Fermi curve $F=\set{\k\in\bbbr^2}{e(\k)=0}$ and that the Fermi curve is nonempty, connected, compact and
strictly convex (meaning that its curvature does not vanish anywhere). We also fix the number $r_0\ge 6$ of derivatives in $k_0$
that we wish to control.

We introduce scales as in 
[FKTf1, Definition \defNPscales] and [FKTo2, \S\CHscales]:

\definition{\STM\defLADscales}{
\Item i) 
For $j\ge 1$, the $j^{\rm th}$ scale function on $\bbbr\times\bbbr^2$ is defined as
$$ 
\nu^{(j)}(k)=\nu\left(M^{2j}(k_0^2+e(\k)^2)\right) 
$$
where $\nu$ is the function of (\eqnLADpartofunity). It may be constructed
by choosing a function $\varphi\in C_0^\infty\big((-2,2)\big)$ that is
identically one on $[-1,1]$ and setting $\nu(x)=\varphi(x/M)-\varphi(Mx)$
for $x>0$ and zero otherwise.
By construction, $\nu^{(j)}$ is identically one on
$$
\set{k=(k_0,\k) \in \bbbr\times\bbbr^2}
{\sqrt{\sfrac{2}{M}}\, \sfrac{1}{M^j}\le |ik_0-e(\k)|\le \sqrt{M} \sfrac{1}{M^j} }
$$
The support of $\nu^{(j)}$ is called the $j^{\rm th}$ shell. By construction, it is contained in
$$
\set{k\in \bbbr\times\bbbr^2}
{\sfrac{1}{\sqrt{M}}\, \sfrac{1}{M^j}\le |ik_0-e(\k)|\le \sqrt{2M} \sfrac{1}{M^j} }
$$
The momentum $k$ is said to be of scale $j$ if $k$ lies in the $j^{\rm th}$ shell.
\Item ii) 
For $j\ge 1$, set
$$
\nu^{(\ge j)}(k)=\smsum_{i\ge j}\nu^{(i)}(k)
$$
for $|ik_0-e(\k)|>0$ and $\nu^{(\ge j)}(k)=1$ for $|ik_0-e(\k)|=0$.
Equivalently, $\nu^{(\ge j)}(k)=\varphi\big(M^{2j-1}(k_0^2+e(\k)^2)\big)$.
By construction, $\nu^{(\ge j)}$ is identically 1 on 
$$
\set{k \in \bbbr\times\bbbr^2} {|ik_0-e(\k)|\le \sqrt{M} \sfrac{1}{M^j} }
$$ 
The support of $\nu^{(\ge j)}$ is called the $j^{\rm th}$ neighbourhood of the Fermi surface. By construction, it  is contained in
$$
\set{k\in \bbbr\times\bbbr^2}{|ik_0-e(\k)|\le \sqrt{2M} \sfrac{1}{M^j} }
$$
The support of $\varphi\big(M^{2j-2}(k_0^2+e(\k)^2)\big)$ is called the 
$j^{\rm th}$ extended neighbourhood. It  is contained in
$$
\set{k\in \bbbr\times\bbbr^d}{|ik_0-e(\k)|\le \sqrt{2}M\sfrac{1}{M^{j}} }
$$
}
\goodbreak

To estimate functions in position space and still make use of conservation of 
momentum, we use sectorization. See [FKTf1, Example \exNPsectorizebound].
The following definition  is also made in [FKTf2, \S\CHnewsectors] and 
[FKTo3, \S\CHsectors].

\definition{\STM\defLADsectors (Sectors and sectorizations)}{ 
\Item i)
Let $I$ be an interval on the Fermi surface $F$ and $j\ge 1$. Then
$$
s=\set{k\ {\rm in\ the\ } j^{\rm th}\ {\rm neighbourhood}}{\pi_F(k)\in I}
$$
is called a sector of length $|I|$ at scale $j$. 
Here $k\mapsto\pi_F(k)$ is a projection on the Fermi surface. 
Two different sectors $s$ and $s'$ are called neighbours if  
$s'\cap s\ne \emptyset$.
\Item ii)
A sectorization of length $\fl$ at scale $j$ is a set $\Si$
of sectors of length $\fl$ at scale $j$ that obeys 
\item{-} the set $\Si$ of sectors covers the Fermi surface
\item{-} each sector in $\Si$ has precisely two neighbours in $\Si$, 
one to its left and one to its right
\item{-} if $s,\ s'\in\Si$ are neighbours then 
$\sfrac{1}{16}\fl\le |s\cap s'\cap F|\le\sfrac{1}{8}\fl$

\noindent Observe that there are at most 
$2\,{\rm length}(F)/\fl$ sectors in $\Si$.

}

In the renormalization group map of [FKTf1] and [FKTo3], 
we integrate over fields whose arguments $(x,\si,s)$ lie in $\cB^\updownarrow\times\Si$, where $\cB^\updownarrow=(\bbbr\times\bbbr^2)\times
\{\uparrow,\downarrow\}$ is the set of all ``(positions, spins)''. On the
other hand, we are interested in the dependence of the two and four--point
functions on external momenta. To distinguish between the set of all positions and the set of all momenta, we denote by $\bbbm=\bbbr\times\bbbr^2$, the set of 
all possible momenta. The set of all possible positions shall still be denoted $\bbbr\times\bbbr^2$. Thus the external variables 
 $(k,\si)$ lie in  $\check\cB^\updownarrow=\bbbm\times
\{\uparrow,\downarrow\}$. In total, legs of four--legged kernels may lie in 
the disjoint union $\fY^\updownarrow_\Si=\check\cB^\updownarrow\dunion(\cB^\updownarrow\times\Si)$ 
for some sectorization $\Si$.
The four--legged kernels over $\fY^\updownarrow_\Si$ that we consider here  arise in [FKTf2, \S\CHphladders] as particle--hole reductions 
(as in Definition \defParicleHoleDecomp\ of [FKTf2])
of four--legged kernels on $\fX_\Si=\check\cB\dunion\,(\cB\times\Si)$ where
$\check\cB=\check\cB^\updownarrow\times\{0,1\}$ and $\cB=\cB^\updownarrow\times\{0,1\}$ and
$\{0,1\}$ is the set of creation/annihilation indices.
Particle--hole reduction sets the creation/annihilation index to zero for
legs number one and four and to one for legs number two and three.
To simplify the notation in this paper, we shall eliminate the spin 
variables so that the legs lie in 
$$
\fY_\Si=\bbbm\dunion\big((\bbbr\times\bbbr^2)\times\Si\big)
$$
Sometimes a four--legged kernel will have different sectorizations $\Si,\Si'$ 
on its two left hand legs and on its two right hand legs. Therefore, we introduce the space
$$
\fY^{(4)}_{\Si,\Si'}=\fY^2_\Si\times \fY^2_{\Si'}
$$

Since $\fY_\Si$ is the disjoint union of $\bbbm$ and 
$(\bbbr\times\bbbr^2)\times\Si$, the space $\fY^{(4)}_{\Si,\Si'}$ is the disjoint union 
$$
\fY^{(4)}_{\Si,\Si'}=
\bigdunion_{i_1,i_2,i_3,i_4\in\{0,1\}}
\fY_{i_1,\Si}\times \fY_{i_2,\Si}\times \fY_{i_3,\Si'}\times \fY_{i_4,\Si'}
\EQN\eqnLADfourdunion$$
where $\fY_{0,\Si}=\bbbm$ and 
$\fY_{1,\Si}=(\bbbr\times\bbbr^2)\times\Si$. If $f$ is a function on
$\fY^{(4)}_{\Si,\Si'}$, we denote by $f\big|_{(i_1,\cdots,i_4)}$ its restriction
to $\fY_{i_1,\Si}\times\fY_{i_2,\Si}\times \fY_{i_3,\Si'}\times \fY_{i_4,\Si'}$ 
under the identification (\eqnLADfourdunion).

\definition{\STM\defLADtransinv (Translation invariance)}{
Let $\Si$ and $\Si'$ be sectorizations.
\Item{i)} 
Let $y\in\fY_\Si$ and $t\in\bbbr\times\bbbr^2$. We set
$$
T_ty=\cases{ k & if $y=k\in\bbbm$\cr
       (x+t,s) & if $y=(x,s)\in\big(\bbbr\times\bbbr^2\big)\times\Si$\cr}
$$
\Item{ii)} Let $i_1,\cdots,i_4\in\{0,1\}$.
A function $f$ on 
$\fY_{i_1,\Si}\times\fY_{i_2,\Si}\times \fY_{i_3,\Si'}\times \fY_{i_4,\Si'}$ is 
called translation invariant, if for all $t \in \bbbr \times \bbbr^2$
$$
f(T_ty_1,\cdots,T_ty_4) = \Big(\prod_{1\le\mu\le 4\atop i_\mu=0}
e^{\imath(-1)^{b_\mu}<y_\mu,t>_-}\Big)f(y_1,\cdots,y_4)
$$
where
$$
b_\mu=\cases{0& if $\mu=1,4$\cr
             1& if $\mu=2,3$\cr}
\EQN\eqnLADbemu$$
and $<\!k,x\!>_-=-k_0x_0+\k_1\x_1+\k_2\x_2$. This choice of $b_\mu$ 
reflects our image of $f$ as a particle--hole kernel, with first and fourth,
resp. second and third, arguments being creation, resp. annihilation, 
arguments. 
\Item{iii)} 
A function $f$ on $\fY^{(4)}_{\Si,\Si'}$ is translation invariant if
$f\big|_{(i_1,\cdots,i_4)}$ is translation invariant for all
$i_1,\cdots,i_4\in\{0,1\}$.

\noindent
A function $f$ on $\big(\fY^\updownarrow_\Si\big)^4$ is translation invariant if
$f({\sst
(\,\cdot\,,\si_1),(\,\cdot\,,\si_2),(\,\cdot\,,\si_3),(\,\cdot\,,\si_4)})$ is translation invariant for all
$\si_1,\cdots,\si_4\in\{\uparrow,\downarrow\}$.

}
\definition{\STM\defLADfourtrans (Fourier transform)}{
Let $\Si$, $\Si'$ be sectorizations. Set $\fY_{2,\Si}=\bbbm\times\Si$.
\Item{i)}  Let $i_1,\cdots,i_4\in\{0,1,2\}$ and $1\le \mu\le 4$ such that
$i_\mu=1$. The Fourier transform of a function $f$ on 
$\fY_{i_1,\Si}\times\fY_{i_2,\Si}\times \fY_{i_3,\Si'}\times \fY_{i_4,\Si'}$
with respect to the $\mu^{\rm th}$ variable is the function on 
$\fY_{i'_1,\Si}\times\fY_{i'_2,\Si}\times \fY_{i'_3,\Si'}\times \fY_{i'_4,\Si'}$
with
$$
i'_\nu=\cases{i_\nu & if $\nu\ne \mu$\cr
              2     & if $\nu=\mu$\cr}
$$
defined by
$$
(\Phi_\mu f)({\sst y_1,\cdots,y_{\mu-1},(k,s),y_{\mu+1},\cdots,y_4}) 
=\int 
e^{\imath (-1)^{b_\mu}<k,x>_-}
f({\sst y_1,\cdots,y_{\mu-1},(x,s),y_{\mu+1},\cdots,y_4})\ d^3x 
$$ 
\Item{ii)}  Let $i_1,\cdots,i_4\in\{0,1\}$ with $i_\mu=1$ for at least one 
$1\le\mu\le 4$. The total Fourier transform $\check f$ of a translation
invariant function $f$ on 
$\fY_{i_1,\Si}\times\fY_{i_2,\Si}\times \fY_{i_3,\Si'}\times \fY_{i_4,\Si'}$
is defined by
$$
\check f(y_1,y_2,y_3,y_4)\ (2\pi)^{3}\de(k_1-k_2-k_3+k_4) 
=\Big(\smprod_{1\le\mu\le 4\atop i_\mu=1}\hskip-7pt\Phi_\mu\ f\Big)(y_1,y_2,y_3,y_4) 
$$ 
where
$y_\mu=k_\mu$ when $i_\mu=0$ and $y_\mu=(k_\mu,s_\mu)$ when $i_\mu=1$.
$\check f$ is defined on the set of all $ (y_1,y_2,y_3,y_4)\in
\fY_{2i_1,\Si}\times\fY_{2i_2,\Si}\times \fY_{2i_3,\Si'}\times \fY_{2i_4,\Si'}$
for which $k_1-k_2=k_3-k_4$.
}
\definition{\STM\defLADsectorized (Sectorized Functions)}{
Let $\Si$ and $\Si'$ be sectorizations.
\Item{i)} Let $i_1,\cdots,i_4\in\{0,1\}$. A translation invariant function
$f$  on 
$\fY_{i_1,\Si}\times\fY_{i_2,\Si}\times \fY_{i_3,\Si'}\times \fY_{i_4,\Si'}$
is sectorized if, for each $1\le\mu\le 4$ with $i_\mu=1$, the total
Fourier transform 
$\check f({\sst y_1,\cdots,y_{\mu-1},(k,s),y_{\mu+1},\cdots,y_4})$
vanishes unless $k$ is in the $j^{\rm th}$ extended neighbourhood
and $\pi_F(k)\in s$.
\Item{ii)} 
A translation invariant function $f$ on $\fY^{(4)}_{\Si,\Si'}$ is  sectorized
if $f\big|_{(i_1,\cdots,i_4)}$ is sectorized for all $i_1,\cdots,i_4\in\{0,1\}$.

\noindent
A translation invariant function $f$ on $\big(\fY^\updownarrow_\Si\big)^4$ is sectorized if
$f({\sst
(\,\cdot\,,\si_1),(\,\cdot\,,\si_2),(\,\cdot\,,\si_3),(\,\cdot\,,\si_4)})$ is sectorized for all
$\si_1,\cdots,\si_4\in\{\uparrow,\downarrow\}$.

}
\remark{\STM\remLADsectorizedreduction}{
If $f$ is a function in the space $\check\cF_{4,\Si}$ of Definition \defNPdisjointfnspaces\ 
of [FKTf2] (or Definition \defOSsectcheckcF.iii of [FKTo3]),
then its particle--hole reduction is a sectorized function on 
$\big(\fY^\updownarrow_\Si\big)^4$.

}
\goodbreak\titlec{Particle--Hole Ladders}\PG\pgLADIc
\definition{\STM\defLADphladder}{ 
\Item{i)} A (spin independent) propagator is a translation invariant function 
on $\big(\bbbr\times\bbbr^2\big)^2$. If $A(x,x')$ is a propagator, then its transpose is $A^t(x,x')=A(x',x)$.
\Item{ii)} A (spin independent) bubble propagator is a translation invariant function on $\big(\bbbr\times\bbbr^2\big)^4$. If $A$ and $B$ are propagators,
we define the bubble propagator 
$$
A\otimes B(x_1,x_2,x_3,x_4)=A(x_1,x_3)B(x_2,x_4)
$$
We set
$$\eqalign{
\cC(A,B) &=\kern 10pt (A+B)\otimes(A+B)^t-B\otimes B^t\cr
&= \kern 10pt A\otimes A^t\kern 6pt +\kern 6pt A\otimes B^t 
\kern 7pt+\kern 7pt B\otimes A^t\cr
&=\figplace{bubPropLad}{0 in}{-0.02 in}
}$$
\vskip.1in
\Item{ iii)} Let $\Si,\Si',\Si''$ be sectorizations, $P$ be a bubble propagator
and $F$ be a function on  
$\fY_{i_1,\Si''}\times\fY_{i_2,\Si''}\times{(\bbbr\times\bbbr^2)}^2$. 
If $K$ is a function on 
$\fY_{\Si}\times\fY_{\Si}\times \fY_{1,\Si'}\times \fY_{1,\Si'}$, we set
$$\eqalign{
(K\bullet P)({\sst y_1,y_2;x_3,x_4}) 
&=  \smsum_{s'_1,s'_2\in \Si'} \int {\sst dx'_1 dx'_2}\ 
K({\sst y_1,y_2,(x'_1,s'_1),(x'_2,s'_2)})\ 
P({\sst x'_1,x'_2;x_3,x_4}) \cr
}$$
If $K$ is a function on 
$\fY_{1,\Si}\times\fY_{1,\Si}\times \fY_{i_3,\Si'}\times \fY_{i_4,\Si'}$, 
we set, when $i_1,i_2,i_3,i_4$ are not all 0,
$$\eqalign{
(F\bullet K)({\sst y_1,y_2,y_3,y_4}) 
&=  \smsum_{s_1,s_2\in \Si}\int {\sst dx_1 dx_2 }\ 
F({\sst y_1,y_2;x_1,x_2})
K({\sst (x_1,s_1),(x_2,s_2),y_3,y_4}) \cr
}$$
and when $i_1,i_2,i_3,i_4=0$,
$$\eqalign{
(F\bullet K)({\sst k_1,k_2,k_3,k_4}) {\sst (2\pi)^3\de(k_1-k_2-k_3+k_4)}
&=  \smsum_{s_1,s_2\in \Si}\int {\sst dx_1 dx_2 }\ 
F({\sst k_1,k_2;x_1,x_2})
K({\sst (x_1,s_1),(x_2,s_2),k_3,k_4})\cr
}$$
Observe that $K\bullet P$ is a function on 
$\fY^2_\Si\times(\bbbr\times\bbbr^2)^2$
and $F\bullet K$ is a function on $\fY^{(4)}_{\Si,\Si'}$. 
If $K'$ is a function on $(\fY^\updownarrow_\Si)^{4}$ and $F'$ is a function on 
$(\fY^\updownarrow_\Si)^{2}\times(\cB^\updownarrow)^2$ we set
$$\eqalign{
(K'\bullet P)({\sst (\,\cdot\,,\si_1),(\,\cdot\,,\si_2),(\,\cdot\,,\si_3),(\,\cdot\,,\si_4)})
&=K'({\sst (\,\cdot\,,\si_1),(\,\cdot\,,\si_2),(\,\cdot\,,\si_3),(\,\cdot\,,\si_4)})
\bullet P   \cr
}$$ 
and
$$\eqalign{
&(F'\bullet K')({\sst
(\,\cdot\,,\si_1),(\,\cdot\,,\si_2),(\,\cdot\,,\si_3),(\,\cdot\,,\si_4)})\cr
&\hskip1in=  \smsum_{\tau_1,\tau_2\in \{\uparrow,\downarrow\}}
F'({\sst (\,\cdot\,,\si_1),(\,\cdot\,,\si_2),(\,\cdot\,,\tau_1),(\,\cdot\,,\tau_2)})
\bullet
K'({\sst (\,\cdot\,,\tau_1),(\,\cdot\,,\tau_2),(\,\cdot\,,\si_3),(\,\cdot\,,\si_4)}) \cr
}$$

\Item iv) Let $\ell \ge 1\,$. Let, for $1\le i\le \ell+1$, 
$\Si^{(i)},\ \Si^{' (i)}$ be sectorizations  and  
$K_i$ a function on $\fY^{(4)}_{\Si^{(i)},\ \Si^{' (i)}}$. Furthermore, let 
$P_1,\cdots,P_\ell$ be  bubble propagators.
The ladder with rungs $K_1,\cdots,K_{\ell+1}$ and 
bubble propagators $P_1,\cdots,P_\ell$ is defined to be
$$
K_1\bullet P_1 \bullet K_2 \bullet P_2\bullet \cdots \bullet
 K_{\ell}\bullet P_\ell\bullet K_{\ell+1}
$$
If $\Si$ is  a sectorization and $K'_1,\ \cdots,\ K'_{\ell+1}$
are functions on $\big(\fY^\updownarrow_\Si\big)^4$, the ladder with rungs $K'_1,\cdots,K'_{\ell+1}$ and 
bubble propagators $P_1,\cdots,P_\ell$ is defined to be
$$
K'_1\bullet P_1 \bullet K'_2 \bullet P_2\bullet \cdots \bullet
 K'_{\ell}\bullet P_\ell\bullet K'_{\ell+1}
$$

}

\remark{\STM\remLADhardsoft}{
We typically use $\cC(A,B)$ with $A$ being the part, $\nu^{(j)}(k)C(k)$,
of the propagator, $C(k)$, having momentum in the $j^{\rm th}$ shell and 
$B$ being the part, $\nu^{(\ge j+1)}(k)C(k)$,
of the propagator having momentum in the $(j+1)^{\rm st}$ neighbourhood.
The bubble propagator $\cC(A,B)$ always contains at least one ``hard line'' $A$
and may or may not contain one ``soft line'' $B$. The latter are created
by Wick ordering. See [FKTf1, \S\CHintroOverview, subsection 9].

}

\remark{\STM\remLADphreduction}{
If $F_1,\ F_2$ are functions on $\big(\fX_\Si\big)^4$ and $A,\ B$ 
are propagators over $\cB$ in the sense of Definition \defbubbleprop.i
of [FKTf2], then the particle--hole reduction of 
$F_1\bullet\cC(A,B)\bullet F_2$ (with the $\cC(A,B)$ of 
Definition \defbubbleprop.i of [FKTf2]) is equal to 
$$
-F_1^{\rm ph}\bullet \cC\big(  A({\sst(\,\cdot\, 1), (\,\cdot\, 0)}),
                               B({\sst(\,\cdot\, 1), (\,\cdot\, 0)})\big)
\bullet F_2^{\rm ph}
$$
(with the $\cC$ of Definition  \defLADphladder) since $B({\sst(x,\si,0), (x',\si', 1)})
=-B({\sst(\,\cdot\, 1), (\,\cdot\, 0)})^t({\sst(x,\si), (x',\si')})$.

}

\goodbreak\titlec{Norms}\PG\pgLADId

In the momentum space variables, we take suprema of the function and 
its derivatives. In the position space variables, we will apply the 
$L^1$--$L^\infty$ norm of Definition \:\defLADlonelinfty, below, to the 
function and to  the function multiplied by various coordinate differences.
\definition{\STM\defLADlonelinfty}{ Let $f$ be a function on $\big(\bbbr\times
\bbbr^2\big)^n$. Its $L^1$--$L^\infty$ norm is
$$
\tn f\tn_{1,\infty} 
=\max\limits_{1\le j_0 \le n}\ 
\sup\limits_{x_{j_0} \in \bbbr\times
\bbbr^2}\  
\int \prod\limits_{j=1,\cdots, n \atop j\ne j_0}\!\!\! dx_j\ 
| f( x_1,\cdots,x_n) | 
$$

}
\noindent Multiple derivatives are labeled by a multiindex 
$\de=(\de_0,\de_1,\de_2)\in\bbbn_0\times\bbbn_0^2$. For such a multiindex,
we set $|\de|=\de_0+\de_1+\de_2$, $\de!=\de_0!\,\de_1!\,\de_2!$
 and $x^\de=x_0^{\de_0}x_1^{\de_1}x_2^{\de_2}$
for $x\in\bbbr\times\bbbr^2$.

\definition{\STM\defLADlonelinftyII}{ Let $\Si$ be a sectorization and 
$A$ a function on $\big((\bbbr\times\bbbr^2)\times\Si\big)^2$. 
For a multiindex $\de\in\bbbn_0\times\bbbn_0^2$,
we define
$$
\v A\v^\de_{1,\Si} 
=\max_{i=1,2}\ \max_{s_i\in\Si}\sum_{s_{3-i}\in\Si}
\TN (x-y)^{\de}\,A\big((x,s_1),(y,s_2)\big)\TN_{1,\infty}
$$

}

Variables for four--point functions may be momenta or
position/sector pairs. Therefore we introduce differential--decay operators 
that differentiate momentum space variables and multiply position space 
variables by coordinate differences. We again use the identification
$$
\fY^{(4)}_{\Si,\Si'}=\bigdunion_{i_1,i_2,i_3,i_4\in\{0,1\}}
\fY_{i_1,\Si}\times \fY_{i_2,\Si}\times \fY_{i_3,\Si'}\times \fY_{i_4,\Si'}
$$
of (\eqnLADfourdunion).

\definition{\STM\defLADdiffdecay (Differential--decay operators)}{
Let $\Si$ and $\Si'$ be sectorizations, 
$\de=(\de_0,\de_1,\de_2)\in\bbbn_0\times\bbbn_0^2$ a multiindex and 
$\mu,\mu'\in\{1,2,3,4\}$ with $\mu\ne\mu'$.

\Item i)
Let $i_1,\cdots,i_4\in\{0,1\}$ and $f$ be a function on 
$\fY_{i_1,\Si}\times\fY_{i_2,\Si}\times \fY_{i_3,\Si'}\times\fY_{i_4,\Si'}$.
If $i_\mu=0$, multiplication
by the $\de^{\rm th}$ power of the position variable dual to $k_\mu$
(see Definition \defLADfourtrans) is implemented by
$$\eqalign{
&\rD^\de_\mu
f(\cdots,k_\mu,\cdots) 
= (-1)^{\de_1+\de_2}(-1)^{b_\mu|\de|}\imath^{|\de|}\ 
\sfrac{\partial^{\de_0}\hfill}{\partial k_{\mu,0}^{\de_0}}\,
\sfrac{\partial^{\de_1}\hfill}{\partial \k_{\mu,1}^{\de_1}} 
\sfrac{\partial^{\de_2}\hfill}{\partial \k_{\mu,2}^{\de_2}} 
f(\cdots,k_\mu,\cdots) 
}$$
In general, set
$$
\rD_{\mu;\mu'}^\de f=
\cases{
\big(\rD_\mu^\de-\rD_{\mu'}^\de\big) f &if $i_\mu=i_{\mu'}=0$\cr
\noalign{\vskip.05in}
\big(\rD_\mu^\de-x_{\mu'}^\de\big) f &if $i_\mu=0, i_{\mu'}=1$\cr
\noalign{\vskip.05in}
\big(x_\mu^\de-\rD_{\mu'}^\de\big) f &if $i_\mu=1,\ i_{\mu'}=0$\cr
\noalign{\vskip.05in}
\big(x_\mu^\de-x_{\mu'}^\de\big) f &if $i_\mu=i_{\mu'}=1$\cr
}
$$
Here, when $i_\mu=1$, the $\mu^{\rm th}$ argument of $f$ is $(x_\mu,s_\mu)$.

\Item{ii)}
If $f$ is a function on $\fY^{(4)}_{\Si,\Si'}$, then
$\big(\rD_{\mu;\mu'}^\de f\big)\big|_{(i_1,\cdots,i_4)}
=\rD_{\mu;\mu'}^\de \big(f\big|_{(i_1,\cdots,i_4)}\big)$ 
for all $i_1,\cdots,i_4\in\{0,1\}$.
}

\definition{\STM\defLADlonelinftyIV}{ Let $\Si,\Si'$ be sectorizations.
\Item i) Let $i_1,\cdots,i_4\in\{0,1\}$ and $f$ be a  function on 
$\fY_{i_1,\Si}\times\fY_{i_2,\Si}\times \fY_{i_3,\Si'}\times \fY_{i_4,\Si'}$. 
For multiindices $\de_\li,\ \de_\ci,\ \de_\ri\in\bbbn_0\times\bbbn_0^2$,
we define
$$
\v f\v^{(\de_\li,\de_\ci,\de_\ri)}_{\Si,\Si'} 
=\max_{s_\nu\in\Si\atop {\nu=1,2\atop{\rm\ with\ }i_\nu=1}}
\max_{s_\nu\in\Si'\atop {\nu=3,4\atop{\rm\ with\ }i_\nu=1}}
\sup_{k_\nu\in\bbbm\atop {\nu=1,2,3,4\atop{\rm\ with\ }i_\nu=0}}
\max_{\mu=1,2\atop\mu'=3,4}\ 
\TN \rD_{1;2}^{\de_\li}\rD_{\mu;\mu'}^{\de_\ci}\rD_{3;4}^{\de_\ri}\,f\TN_{1,\infty}
$$
Here, the $\nu^{\rm th}$ argument of $f$ is $k_\nu$ when $i_\nu=0$ and
$(x_\nu,s_\nu)$ when $i_\nu=1$. The $\tn\ \cdot\ \tn_{1,\infty}$ of 
Definition \defLADlonelinfty\ is applied to all spatial arguments of
$\rD_{1;2}^{\de_\li}\rD_{\mu;\mu'}^{\de_\ci}\rD_{3;4}^{\de_\ri}\,f$. 

\Item ii) If $f$ is a  function on $\fY^{(4)}_{\Si,\Si'}$,
we define
$$
\v f\v^{(\de_\li,\de_\ci,\de_\ri)}_{\Si,\Si'} 
=\sum_{i_1,i_2,i_3,i_4\in\{0,1\}}
\V f\big|_{(i_1,\cdots,i_4)}\V^{(\de_\li,\de_\ci,\de_\ri)}_{\Si,\Si'}
$$ 

}
\noindent In this Definition, the system $(\de_\li,\de_\ci,\de_\ri)$ 
of multiindices indicates, roughly speaking, that one takes
$\de_\li$ derivatives with respect to the momentum flowing between the two 
left legs, $\de_\ri$ derivatives with respect to the momentum flowing between
the two right legs and $\de_\ci$ derivatives with respect to momenta
flowing from the left hand side to the right hand side.

In [FKTf1--f3] and [FKTo1--o4], we combine the norms of all derivatives
of a function in a formal power series. We denote by $\fN_3$ the set of all formal power series $X=\sum\limits_{\de\in\bbbn_0\times\bbbn_0^2} X_\de t^\de$ 
in the
variables $t=(t_0,t_1,t_2)$ with coefficients $X_\de\in\bbbr_+\cup\{\infty\}$.
See Definition \defNPFancynormdomain\ of [FKTf2] or 
Definition \defOSFancynormdomain\ of [FKTo1].

\noindent
A quantity in $\fN_3$ characteristic of the power counting for derivatives
in scale $j$ is
$$
\cb_j=\sum_{\de_1+\de_2\le r_e\atop |\de_0|\le r_0}  M^{j|\de|}\,t^\de
+\sum_{\de_1+\de_2> r_e\atop {\rm or\ }|\de_0|> r_0}\infty\, t^\de
\EQN\eqnLADcj$$

\definition{\STM\defLADnormdomainnorm}{ Let $\Si$ be a sectorization.
\Item i) For a function $A$ on $\big((\bbbr\times\bbbr^2)\times\Si\big)^2$,
we define
$$
\v A\v_{1,\Si} 
=\sum_{\de\in\bbbn_0\times\bbbn_0^2}\sfrac{1}{\de!}\v A\v^\de_{1,\Si}\ t^\de 
$$

\Item ii) For a function $f$ on $\fY^4_\Si=\fY^{(4)}_{\Si,\Si}$,
we define
$$
\v f\v_{\Si} 
=\sum_{\de\in\bbbn_0\times\bbbn_0^2}\sfrac{1}{\de!}
\Big(\max_{\de_\li+\de_\ci+\de_\ri=\de}\v f\v^{(\de_\li,\de_\ci,\de_\ri)}_{\Si,\Si}
\Big)t^\de
$$ 

\Item iii) For a function $f$ on $\big(\fY^\updownarrow_\Si\big)^4$,
we define
$$
\v f\v_{\Si} 
=\sum_{\si_1,\cdots,\si_4\in\{\uparrow,\downarrow\}}
\v f({\sst(\,\cdot\,,\si_1),(\,\cdot\,,\si_2),(\,\cdot\,,\si_3),(\,\cdot\,,\si_4)})\v_{\Si} 
$$ 
}
\noindent
The following Lemma, whose proof follows immediately from the 
various definitions and Lemma \lemchannelnorm.ii of [FKTo3], compares 
these norms with the norms of Definition \defNPsectnorm\ of [FKTf2].
\goodbreak
\lemma{\STM\lemLADNormcomparison}{
 Let $\Si$ be a sectorization.
\Item i) Let $f$ be a sectorized, translation invariant function 
on $\big(\fY^\updownarrow_\Si\big)^4$ and $V_{\rm ph}(f)$ its particle--hole 
value as in Definition \defParicleHoleDecomp\ of [FKTf2]. 
Let  $\v \ \cdot\ \tv_{3,\Si}$
         be the norm of Definition \defNPsectdiffdecaynorm\ of [FKTf3]
(or Definition \defOSsectdiffdecaynorm\ of [FKTo3]).
Then there is a constant $\abcst$, that depends only on $r_0$ and $r$, such that
$$
\V V_{\rm ph}(f)\tV_{3,\Si}\le \abcst\, \v f\v_{\Si}
+\sum_{\de_1+\de_2> r\atop{\rm or\ }\de_0> r_0}\infty\,t^\de
$$

\Item ii) 
Let $g$ be a function in the space $\check\cF_{4,\Si}$ of Definition \defNPdisjointfnspaces\ 
of [FKTf2] (or Definition \defOSsectcheckcF.iii of [FKTo3])
and $g^{\rm ph}$ its particle--hole reduction
as in Definition \defParicleHoleDecomp\ of [FKTf2].
Then there is a universal $\abcst$  such that
$$
\v g^{\rm ph}\v_{\Si} \le \abcst\, \V g\tV_{3,\Si}
$$

}

\goodbreak\titlec{The Propagators}\PG\pgLADIe

The propagators we use in the multiscale analysis of [FKTf1--f3] are of the form 
$$
C_v^{(j)}(k)=\sfrac{\nu^{(j)}(k)}{ik_0-e(\k)-v(k)}\qquad
C_v^{(\ge j)}(k)=\sfrac{\nu^{(\ge j)}(k)}{ik_0-e(\k)-v(k)}
$$
with functions $v(k)$ satisfying $|v(k)|\le\half |\imath k_0-e(\k)|$.
Their Fourier transforms are
$$\meqalign{
C_v^{(j)}(x,y)
   &= \int\sfrac{d^3k}{(2\pi)^3}\, e^{\imath<k,x-y>_-}C_v^{(j)}(k) &&
C_v^{(\ge j)}(x,y)
   &= \int\sfrac{d^3k}{(2\pi)^3}\, e^{\imath<k,x-y>_-}C_v^{(\ge j)}(k)\cr
C_v^{(j)}(y)
   &= \int\sfrac{d^3k}{(2\pi)^3}\, e^{-\imath<k,y>_-}C_v^{(j)}(k) &&
C_v^{(\ge j)}(y)
   &= \int\sfrac{d^3k}{(2\pi)^3}\, e^{-\imath<k,y>_-}C_v^{(\ge j)}(k)\cr
}$$
The function $v(k)$ will be the sum of Fourier transforms of sectorized,
translation invariant functions $p\big((x,s),(x,s')\big)$ on $\Big(\big(\bbbr\times\bbbr^2\big)\times\Si\Big)^2$ for various 
sectorizations $\Si$. The Fourier transform of such a function is defined as
$$\eqalign{
\check p(k) = \sum_{s,s'\in\Si}\int d^3x\  e^{\imath<k,x>_-}\,
p({\sst (0,s),(x,s') }) 
\cr 
}$$ 

\goodbreak\titlec{Resectorization}\PG\pgLADIf

We now fix $\half<\aleph<\sfrac{2}{3}$ and set $\fl_j=\sfrac{1}{M^{\aleph j}}$.
Furthermore, we select, for each $j\ge 1$, a sectorization $\Si_j$ of length
$\fl_j$ at scale $j$ and a partition of unity $\set{\chi_s}{s\in\Si_j}$
of the $j^{\rm th}$ neighbourhood  which fulfills 
Lemma \lemOSsectpartunit\ of [FKTo3] with $\Si=\Si_j$. The Fourier
transform of $\chi_s$ is
$$
\hat\chi_s(x) 
=\int e^{-\imath<k,x>_-}\,\chi_s(k)\,
\sfrac{d^3k}{(2\pi)^3} 
$$
\definition{\STM\defLADresectoriz (Resectorization)}{ 
Let $j,j',j_\li,j_\li',j_\ri,j_\ri'\ge 1$. 
\Item i)  Let $p$ be a sectorized, translation invariant function on $\Big(\big(\bbbr\times\bbbr^2\big)\times\Si_j\Big)^2$. Then, for $j'\ne j$, the 
$j'$--resectorization of $p$ is 
$$
p_{\Si_{j'}}({\sst\,(x_1,s_1),(x_2,s_2)} )
=\smsum_{s'_1,s'_{2} \in \Si_j} \int {\sst dx'_1\, dx'_2}\ 
\hat\chi_{s_1}({\sst x_1-x'_1})\ 
p({\sst(x'_1,s'_1),(x'_2,s'_2)} )\ 
\hat\chi_{s_2}({\sst x'_2-x_2}) 
$$
It is a sectorized, translation invariant 
function on $\Big(\big(\bbbr\times\bbbr^2\big)\times\Si_{j'}\Big)^2$.
If $j=j'$, we set $p_{\Si_{j'}}=p$.
\Item ii)
Let $i_1,\cdots,i_4\in\{0,1\}$ and $f$ be a function on 
$\fY_{i_1,\Si_{j_\li}}\times\fY_{i_2,\Si{j_\li}}\times \fY_{i_3,\Si{j_\ri}}\times\fY_{i_4,\Si{j_\ri}}$ that is sectorized and
 translation invariant. Then the 
$(j'_\li,j'_\ri)$--resectorization of $f$ is the sectorized, 
translation invariant function on 
$\fY_{i_1,\Si_{j'_\li}}\times\fY_{i_2,\Si{j'_\li}}\times \fY_{i_3,\Si{j'_\ri}}\times\fY_{i_4,\Si{j'_\ri}}$
defined by
$$
f_{\Si_{j_\li'},\Si_{j_\ri'}}({\sst\,y_1,y_2,y_3,y_4} )
=\smsum_{s'_\mu\in \Si_{j_\li}\atop\mu\in\{1,2\}\cap S}
\smsum_{s'_\mu\in \Si_{j_\ri}\atop\mu\in\{3,4\}\cap S}
\int \smprod_{\mu\in S}\Big( {\sst dx'_\mu}\ 
\hat\chi_{s_\mu}({\sst (-1)^{b_\mu}(x_\mu-x'_\mu)})\Big)\ 
f({\sst\,y'_1,y'_2,y'_3,y'_4} )\ 
$$
where
$$
S=\set{\mu}{i_\mu=1}\cap\cases{
          \{1,2,3,4\}& if $j'_\li\ne j_\li,\ j'_\ri\ne j_\ri$\cr
          \noalign{\vskip.05in}
          \{1,2\}& if $j'_\li\ne j_\li,\ j'_\ri= j_\ri$\cr
          \noalign{\vskip.05in}
          \{3,4\}& if $j'_\li= j_\li,\ j'_\ri\ne j_\ri$\cr
          \noalign{\vskip.05in}
          \emptyset& if $j'_\li= j_\li,\ j'_\ri= j_\ri$\cr
}
$$
and $y'_\mu=y_\mu$ for $\mu\notin S$ and, for $\mu\in S$,
$$
y_\mu=(x_\mu,s_\mu)\qquad y'_\mu=(x'_\mu,s'_\mu) 
$$
\Item iii)
If $f$ is a  sectorized, 
translation invariant function on $\fY^{(4)}_{\Si_{j_\li},\Si_{j_\ri}}$, then
$\big(f_{\Si_{j_\li'},\Si_{j_\ri'}}\big)\big|_{(i_1,\cdots,i_4)}
=\big(f\big|_{(i_1,\cdots,i_4)}\big)_{\Si_{j_\li'},\Si_{j_\ri'}}$ 
for all $i_1,\cdots,i_4\in\{0,1\}$. If $j'_\li=j'_\ri=j'$, we set
$f_{\Si_{j'}}=f_{\Si_{j'},\Si_{j'}}$.
\Item iv)
If $f$ is a  sectorized, translation invariant function on 
$\big(\fY^\updownarrow_{\Si_j}\big)^4$, then 
$$
f_{\Si_{j'}}({\sst
(\,\cdot\,,\si_1),(\,\cdot\,,\si_2),(\,\cdot\,,\si_3),(\,\cdot\,,\si_4)})
=\Big(f({\sst
(\,\cdot\,,\si_1),(\,\cdot\,,\si_2),(\,\cdot\,,\si_3),(\,\cdot\,,\si_4)})
\Big)_{\Si_{j'}}
$$
 for all
$\si_1,\cdots,\si_4\in\{\uparrow,\downarrow\}$.

}
\remark{\STM\remLADreresectoriz}{ 
Let $K$ and $H$ be sectorized translation invariant functions on
$\fY^{(4)}_{\Si_{i_\li},\Si_{j_\li}}$ and
 $\fY^{(4)}_{\Si_{i_\ri},\Si_{j_\ri}}$ respectively. Let $P$ be a bubble 
propagator. If the Fourier transform 
$$
\int \smprod_{\mu=1}^4dx_\mu\ \smprod_{\mu=1}^4e^{-\imath(-1)^{b_\mu}<k_\mu,x_\mu>_-}\ 
P(x_1,x_2,x_3,x_4)
$$
of $P$ is supported on the 
$\max\{j'_\li,i'_\ri\}^{\rm th}$ neighbourhood, then
$$
\Big[K\bullet P\bullet H\Big]_{\Si_{i'_\li},\Si_{j'_\ri}}
=K_{\Si_{i'_\li},\Si_{j'_\li}}\bullet P\bullet H_{\Si_{i'_\ri},\Si_{j'_\ri}}
$$

}
\goodbreak\titlec{Compound Particle--Hole Ladders}\PG\pgLADIg

Define, for any set $\cZ$ and any function $K$ on $\cZ^4$,
the flipped function 
$$
K^f(z_1,z_2,z_3,z_4)=-K(z_1,z_3,z_2,z_4)
\EQN\eqnLADflipped$$
\goodbreak
\definition{\STM\defLADcompoundphladder}{
Let $\vec F=\big(F^{(2)},F^{(3)},\cdots \big)$ be a sequence of 
sectorized, translation invariant functions $F^{(i)}$ on $\big(\fY^\updownarrow_{\Si_i}\big)^4$
and $v(k)$ a function on $\bbbm$ such that
 $|v(k)| \le \half|\imath k_0 -e(\k)|$.
We define, recursively on 
$0\le j <\infty$, the compound particle--hole (or wrong way) ladders up
to scale $j$, denoted by $\,\cL^{(j)}=\cL^{(j)}_v(\vec F)\,$, as
$$\eqalign{
\cL^{(0)}&=0 \cr
\cL^{(j+1)}&= \cL^{(j)}_{\Si_j}
+ \smsum_{\ell=1}^\infty 
\big(F +\cL^{(j)}_{\Si_j}+\cL^{(j)\,f}_{\Si_j} \big)
\bullet \cC^{(j)}\bullet \cdots \cC^{(j)}
\bullet
\big(F +\cL^{(j)}_{\Si_j}+\cL^{(j)\,f}_{\Si_j} \big)\cr
}$$
where  $F = \sum_{i=2}^{j}{F}_{\Si_j}^{(i)}$ and the $\ell^{\rm th}$ term
has $\ell$ bubble propagators $\cC^{(j)} = \cC\big(C^{(j)}_v, C^{(\ge j+1)}_v \big)$. Observe that $\cL^{(1)}=\cL^{(2)}=0$.

}
\vskip.25in
\theorem{\STM\thmLADmodcompLadder}{
For every $\veps>0$ there are constants $\rho_0,\const$\footnote{$^{(1)}$}{Throughout this paper we use ``$\const$''
to denote unimportant constants that depend only on the dispersion relation $e(\k)$ and the scale parameter $M$. In particular, they do not depend on the scale $j$.} such that 
the following holds.  
Let $\vec F=\big(F^{(2)},F^{(3)},\cdots \big)$ be a sequence of 
sectorized, translation invariant spin 
independent\footnote{$^{(2)}$}{``Spin independence'' is formally defined in
Definition \:\defLADspinIndependent.} functions $F^{(i)}$ on $\big(\fY^\updownarrow_{\Si_i}\big)^4$
and $\vec p=\big(p^{(2)},p^{(3)},\cdots \big)$ be a sequence of 
sectorized, translation invariant functions
$p^{(i)}$ on $\Big(\big(\bbbr\times\bbbr^2\big)\times\Si_i\Big)^2$.
Assume that there is $\rho \le \rho_0$ such that for $i\ge 2$
$$
\v F^{(i)}\v_{\Sigma_i} \le \sfrac{\rho}{M^{\veps i}}\cb_i\qquad\quad 
\v p^{(i)}\v_{1,\Sigma_i} \le \sfrac{\rho\,\fl_i}{M^i}\cb_i\qquad
\check p^{(i)}(0,\k)=0
$$
Set 
$\,v(k) =\smsum_{i=2}^\infty \check p^{(i)}(k)\,$. Then for all $j\ge 1$
$$
\V \cL^{(j+1)}_v(\vec F)\V_{\Si_j} 
\le \const \rho^2 \,\cb_j
$$
}
\remark{\STM\remLADmodcompLadder}{
Theorem \thmLADmodcompLadder\ and Theorem \theoremmodcompLadder\ of 
[FKTf3] are equivalent. If one replaces the functions $F^{(i)}$
of Theorem \theoremmodcompLadder\ of [FKTf3] by 24 times their 
particle--hole reductions, then, by Corollary \corcompLadder\ of  [FKTf3] and 
Remark \remLADphreduction, the concepts of compound ladders of 
Definition \defLADcompoundphladder\ and Definition \defmodcompLadder\  of 
[FKTf3] coincide. Hence Theorem \thmLADmodcompLadder\ and Theorem 
\theoremmodcompLadder\ of [FKTf3] are equivalent by Lemma \lemLADNormcomparison.
}

Theorem \thmLADmodcompLadder\ will be proven following Corollary \:\bdbbound.
The core of the proof consists of bounds on two types of ladder fragments,
that look like 

\vskip.1in
\centerline{\figput{wwbubble5}\qquad
\figplace{dbubble4}{0 in}{-0.2 in}}

\noindent
and are  called
particle--hole bubbles and double bubbles, and a combinatorial result,
Corollary \:\corLADaltcmpladders, that enables one to express general ladders
in terms of these fragments. The most subtle part of the bound, Theorem
\:\bbound, on particle--hole bubbles is a generalization of 
Lemma \lemLADprimitivemanfred. The bound, Theorem
\:\dbbound, on double bubbles also exploits ``volume improvement
due to overlapping loops''. A simple introduction to this phenomenon is 
provided at the beginning of \S\:\CHdbubbles. 

Ladders with external momenta have an infrared limit that behaves 
much like the model bubble of Lemma \lemLADprimitivemanfred.

\theorem{\STM\thmLADmodcompLaddercont}{
Under the hypotheses of Theorem \thmLADmodcompLadder, the limit
$$
\fL(q,q',t,\si_1,\cdots\si_4)
=\lim_{j\rightarrow\infty}
\cL^{(j)}_v(\vec F)\big|
_{i_1,i_2,i_3,i_4=0}
({\sst(q+{t\over 2},\si_1),(q-{t\over 2},\si_2),
(q'+{t\over 2},\si_3),(q'-{t\over 2},\si_4)})
$$
exists for transfer momentum $t\ne 0$ and is continuous in $(q,q',t)$ for 
$t\ne 0$. The restrictions to $\t=0$ and to $t_0=0$, namely,
$\fL(q,q',(t_0,\0),\si_1,\cdots\si_4)$ and
$\fL(q,q',(0,\t),\si_1,\cdots\si_4)$,
have continuous extensions to $t=0$.
}

\noindent This Theorem is proven following Lemma \:\lemLADtermwisecont.
Notation tables are provided at the end of the paper.

\vfill\eject
    
\chap{Reduction to Bubble Estimates}\PG\pgLADII

\EDEF\CHwwladders{\caproman\chapno}

For the rest of the paper, we fix a sequence 
$\vec F=\big(F^{(2)},F^{(3)},\cdots \big)$  of sectorized, translation invariant, spin independent functions $F^{(i)}$ on $\big(\fY^\updownarrow_{\Si_i}\big)^4$
and a sequence $\vec p=\big(p^{(2)},p^{(3)},\cdots \big)$  of 
sectorized, translation invariant functions
$p^{(i)}$ on $\Big(\big(\bbbr\times\bbbr^2\big)\times\Si_i\Big)^2$ as in Theorem
\thmLADmodcompLadder\ and we set $v(k)=\sum_{i=2}^\infty\check p^{(i)}(k)$.
Denote $\cL^{(j+1)}=\cL^{(j+1)}_v(\vec F)$ and define
the particle--hole bubble propagator of scale $j$ by
$$
\cC^{(j)}=\cC\big(C^{(j)}_v,C^{(\ge j+1)}_v\big)
=\sum_{i_1,i_2\ge 1 \atop { \min(i_1,i_2)=j}}
C_v^{(i_1)}\otimes C_v^{(i_2)\,t}
$$ 
and set
$$
\cC^{[j_1,j_2]}
= \sum_{j=j_1}^{j_2} {\cal C}^{(j)}
= C_v^{(\ge j_1)}\otimes C_v^{(\ge j_1)\,t} - 
      C_v^{(\ge j_2+1)}\otimes C_v^{(\ge j_2+1)\,t}
$$
\titlec{Combinatorial Structure of Compound Ladders}\PG\pgLADIIa
In this section, we use the following
\convention{\STM\convLADresectorbullet}{
Let $K$ and $K'$ be functions on $\big(\fY_{\Si_{j}}\big)^4$
and $\big(\fY_{\Si_{j'}}\big)^4$, respectively. Then the notation
$K+K'$ denotes the function $K_{\Si_{\max\{j,j'\}}}+K'_{\Si_{\max\{j,j'\}}}$
on $\big(\fY_{\Si_{\max\{j,j'\}}}\big)^4$. The same convention is used 
when $K$ and $K'$ are functions on $\big(\fY^\updownarrow_{\Si_{j}}\big)^4$
and $\big(\fY^\updownarrow_{\Si_{j'}}\big)^4$.

}
\definition{\STM\defLADcompoundphladderscalej}{
We define, recursively on 
$0\le j <\infty$, sectorized, translation invariant, spin independent functions
$\,L^{(j)}\,$, on $\big(\fY^\updownarrow_{\Si_{j-1}}\big)^4$ by
$$\eqalign{
L^{(0)}&=L^{(1)}=L^{(2)}=0 \cr
L^{(j+1)}&= \sum_{\ell=1}^\infty \,
\sumprime_{i_1,\cdots,i_{\ell+1} \ge 2 \atop{ j_1,\cdots,j_\ell\ge 0}}\hskip-5pt
\Big[\big(F^{(i_1)}\!+\!L^{(i_1)\,f} \big) \bullet {\cal C}^{(j_1)} 
\bullet \cdots \bullet {\cal C}^{(j_\ell)} 
\bullet \big(F^{(i_{\ell+1})}\!+\!L^{(i_{\ell+1})\,f} \big)\Big]_{\Si_j} \cr
}$$
where the sum $\sumprime$ imposes the constraints
$$\eqalign{
&\max \{j_1,\cdots,j_\ell \} = j \cr
&i_m \le \min \{j_{m-1},j_m\} \ \ \ \ {\rm for\ all} \ \ 1\le m\le \ell+1 \cr
}$$
When $m=1$, $\min \{j_{m-1},j_m\}=j_1$ and when $m=\ell+1$, $\min \{j_{m-1},j_m\}=j_\ell$.

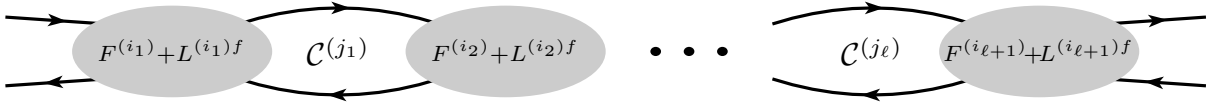
\centerline{\figplace{CmpLadder}{-.1 in}{0 in} }
\goodbreak\noindent
Observe that $\,L^{(j)}\,$ depends only on the components 
$\,F^{(2)},\cdots,F^{(j-1)}\,$ of $\,\vec F\,$. 

}
\proposition{\STM\propLADaltcmpladders}{
$$\leqalignno{
{\cal L}^{(j+1)} &= \sum\limits_{i=0}^{j+1} L^{(i)}_{\Si_{j}}
& i)\cr
{\cal L}^{(j+1)} 
&= \sum_{\ell=1}^\infty \  \sum_{i_1,\cdots,i_{\ell+1}=2}^{j} \  
\Big[\big(F^{(i_1)}+L^{(i_1)\,f} \big)  \bullet 
{\cal C}^{[\max \{i_1,i_2\},j]} \bullet
\big(F^{(i_2)}+L^{(i_2)\,f} \big) \bullet \cdots & ii)\cr
&\hskip1.9in \cdots \bullet 
{\cal C}^{[\max \{i_\ell,i_{\ell+1}\},j]} \bullet
\big(F^{(i_{\ell+1})}+L^{(i_{\ell+1})\,f} \big)\Big]_{\Si_j}\cr
L^{(j+1)} &= 
\Big( \sum_{i=2}^{j} F^{(i)}_{\Si_j} +{\cal L}^{(j)\,f }_{\Si_j} +{\cal L}^{(j)}_{\Si_j} \Big)
\bullet {\cal C}^{(j)} \bullet
\Big( \sum_{i=2}^{j} F^{(i)}_{\Si_j} +{\cal L}^{(j)\,f }_{\Si_j} +{\cal L}^{(j+1)} \Big)
& iii)\cr
}$$
}
\vskip.1in
To prove Proposition \propLADaltcmpladders, we define
$$
\tcL^{(j+1)} = \sum\limits_{i=0}^{j+1} L^{(i)}_{\Si_{j}}
$$
and verify, in Lemmas \:\lemLADaltcmpladdersII\ and \:\lemLADaltcmpladdersIII,
 parts (ii) and (iii) of the Proposition, but with $\cL^{(k)}$
replaced by $\tcL^{(k)}$. Then we prove that $\tcL^{(k)}=\cL^{(k)}$.
\lemma{\STM\lemLADaltcmpladdersII}{
$$\eqalign{
\tcL^{(j+1)} 
&= \sum_{\ell=1}^\infty \  \sum_{i_1,\cdots,i_{\ell+1}=2}^{j} \  
\Big[\big(F^{(i_1)}+L^{(i_1)\,f} \big)  \bullet 
{\cal C}^{[\max \{i_1,i_2\},j]} \bullet
\big(F^{(i_2)}+L^{(i_2)\,f} \big) \bullet \cdots \cr
&\hskip1.9in \cdots \bullet 
{\cal C}^{[\max \{i_\ell,i_{\ell+1}\},j]} \bullet
\big(F^{(i_{\ell+1})}+L^{(i_{\ell+1})\,f} \big)\Big]_{\Si_j}\cr
}$$
}
\prf 
$$\deqalign{
\tcL^{(j+1)} 
&= \sum_{\ell=1}^\infty \  \sum_{j_1,\cdots,j_\ell=0}^{j} \ 
&\sum_{i_m=2 \atop{1\le m \le \ell+1}}^{\min\{j_{m-1},j_m\}}\ \ \ 
\Big[\big(F^{(i_1)}+ & L^{(i_1)\,f} \big) 
\bullet 
{\cal C}^{(j_1)} \bullet
\big(F^{(i_2)}+L^{(i_2)\,f} \big) \bullet \cdots \cr
& &     &\ \ \ \ \ \cdots \bullet 
{\cal C}^{(j_\ell)} \bullet
\big(F^{(i_{\ell+1})}+L^{(i_{\ell+1})\,f} \big)\Big]_{\Si_j} \cr
&= \sum_{\ell=1}^\infty \  \sum_{i_1,\cdots,i_{\ell+1}=2}^{j} \ 
&\sum_{j_m=\max\{i_m,i_{m+1}\} \atop{1\le m \le \ell}}^{j}
\Big[\big(F^{(i_1)}+ & L^{(i_1)\,f} \big) 
\bullet 
{\cal C}^{(j_1)} \bullet
\big(F^{(i_2)}+L^{(i_2)\,f} \big) \bullet \cdots \cr
& &    &\ \ \ \ \ \cdots \bullet 
{\cal C}^{(j_\ell)} \bullet
\big(F^{(i_{\ell+1})}+L^{(i_{\ell+1})\,f} \big)\Big]_{\Si_j} \cr
}$$
\endproof
\lemma{\STM\lemLADaltcmpladdersIII}{
$$\leqalignno{
L^{(j+1)} &= 
\Big(\sum_{i=2}^{j} F^{(i)}_{\Si_j} +\tcL^{(j)\,f }_{\Si_j} +\tcL^{(j)}_{\Si_j} \Big)
\bullet {\cal C}^{(j)} \bullet
\Big( \sum_{i=2}^{j} F^{(i)}_{\Si_j} +\tcL^{(j)\,f }_{\Si_j} +\tcL^{(j+1)} \Big)
& i)\cr
L^{(j+1)} &=\sum_{\ell=1}^\infty 
\bigg[
\Big(\sum_{i=2}^{j} F^{(i)}_{\Si_j} +\tcL^{(j)\,f }_{\Si_j} +\tcL^{(j)}_{\Si_j} \Big)
\bullet {\cal C}^{(j)}\bigg]^\ell \bullet
\Big( \sum_{i=2}^{j} F^{(i)}_{\Si_j} +\tcL^{(j)\,f }_{\Si_j} +\tcL^{(j)}_{\Si_j} \Big)
& ii)\cr
}$$
}
\prf i)
$$\eqalign{
L^{(j+1)} 
&= \sum_{\ell=1}^\infty \ \sum_{\ell'=1}^\ell 
\sum_{j_1,\cdots,j_\ell=0 
\atop{j_1,\cdots,j_{\ell'-1}\le j-1 \atop{j_{\ell'}=j }}}^{j}\hskip-5.5pt
\sum_{i_m=2 \atop{1\le m \le \ell+1}}^{\min\{j_{m-1},j_m\}}
\Big[\big(F^{(i_1)}\!+\!L^{(i_1)\,f} \big)  \bullet 
{\cal C}^{(j_1)} \bullet
\big(F^{(i_2)}+L^{(i_2)\,f} \big) \bullet \cdots \cr
&\hskip3in\cdots \bullet 
{\cal C}^{(j_\ell)} \bullet
\big(F^{(i_{\ell+1})}+L^{(i_{\ell+1})\,f} \big)\Big]_{\Si_j} \cr
&=\figplace{wwladder3}{-0.1 in}{-.3in}\cr
}$$
\vskip.05in
\noindent
Splitting up the sum according to whether $\ell'=1$, $1<\ell'<\ell$ or 
$\ell'=\ell$, we have
\goodbreak
$$\eqalign{
L^{(j+1)}
&=\bigg[ \sum_{i=2}^{j}
\big(F^{(i)}_{\Si_j} +L^{(i)\,f}_{\Si_j} \big) \bigg]
\bullet {\cal C}^{(j)} \bullet
\bigg[ \sum_{i=2}^{j} \big(F^{(i)}_{\Si_j} +L^{(i)\,f}_{\Si_j} \big)  \bigg] \cr
&\ \ + \sum_{\ell=2}^\infty  
\sum_{j_1,\cdots,j_\ell=0 
 \atop{j_1=j }}^{j} \ 
\sum_{i=2}^{j}\ \sum_{i_m=2 \atop{2\le m \le \ell+1}}^{\min\{j_{m-1},j_m\}}
\big(F^{(i)}_{\Si_j}+L^{(i)\,f}_{\Si_j} \big) \bullet {\cal C}^{(j)}\bullet \cr
& \hskip1.5in \bigg[\big(F^{(i_2)}+L^{(i_2)\,f} \big) 
\bullet {\cal C}^{(j_2)} \bullet \cdots \bullet 
{\cal C}^{(j_\ell)} \bullet
\big(F^{(i_{\ell+1})}+L^{(i_{\ell+1})\,f} \big)\bigg]_{\Si_j}\cr
\noalign{\goodbreak}
&\ \ +\sum_{\ell=3}^\infty 
\ \sum_{\ell'=2}^{\ell-1}
\bigg[ 
\sum_{j_1,\cdots,j_{\ell'-1}=0}^{j-1} \hskip-6.5pt
\sum_{i_1,\cdots,i_{\ell'}\ge 2 
      \atop{ i_m\le \min\{j_{m-1},j_m\}
       \atop{ {\rm for\ } m=1,\cdots,\ell'-1
        \atop{i_{\ell'} \le j_{\ell'-1} }}}}
\hskip-23pt\big(F^{(i_1)}\!+\!L^{(i_1)\,f} \big) 
\bullet {\cal C}^{(j_1)} \cdots 
\big(F^{(i_{\ell'})}\!+\!L^{(i_{\ell'})\,f} \big) 
\bigg]_{\Si_j}\!\bullet {\cal C}^{(j)} \cr
&  \hskip.2in\bullet
\bigg[ 
\sum_{j_{\ell'+1},\cdots,j_\ell=0}^{j} 
\sum_{i_{\ell'+1},\cdots,i_{\ell+1}\ge 2 
      \atop{ i_m\le \min\{j_{m-1},j_m\}
       \atop{ {\rm for\ } m=\ell'+2,\cdots,\ell+1
        \atop{i_{\ell'+1} \le j_{\ell'+1} }}}}
\hskip-22pt\big(F^{(i_{\ell'+1})}\!+\!L^{(i_{\ell'+1})\,f} \big) 
\bullet {\cal C}^{(j_{\ell'+1})} \cdots 
\big(F^{(i_{\ell+1})}\!+\!L^{(i_{\ell+1})\,f} \big) 
\bigg]_{\Si_j} \cr
&\ \ + \sum_{\ell=2}^\infty  \ 
\sum_{j_1,\cdots,j_{\ell-1}=0 }^{j-1}\hskip-12.5pt
\sum_{i_m=2 \atop{1\le m \le \ell}}^{\min\{j_{m-1},j_m\}}\sum_{i=2}^{j}
\bigg[\big(F^{(i_1)}\!+\!L^{(i_1)\,f} \big) 
\!\bullet {\cal C}^{(j_1)} \cdots  
{\cal C}^{(j_{\ell-1})}\! \bullet
\big(F^{(i_\ell)}\!+\!L^{(i_\ell)\,f} \big)\bigg]_{\Si_j} \cr
& \hskip3.8in \bullet {\cal C}^{(j)}\bullet  
\big(F^{(i)}_{\Si_j}+L^{(i)\,f}_{\Si_j} \big) \cr
\noalign{\goodbreak}
&=  \bigg[ \sum_{i=2}^{j}\! 
\big(F^{(i)}_{\Si_j} +L^{(i)\,f}_{\Si_j} \big)\! \bigg]\!\!
\bullet {\cal C}^{(j)} \bullet
\!\!\bigg[ \sum_{i=2}^{j}\! \big(F^{(i)}_{\Si_j} +L^{(i)\,f}_{\Si_j} \big)  \bigg]\! 
+\!\bigg[ \sum_{i=2}^{j}\! 
\big(F^{(i)}_{\Si_j} +L^{(i)\,f}_{\Si_j}\big)\! \bigg]\!\!
\bullet {\cal C}^{(j)} \bullet
\tcL^{(j+1)}
\cr
& \hskip1.3in+ \tcL^{(j)}_{\Si_j}  \bullet  {\cal C}^{(j)} \bullet\tcL^{(j+1)} \ 
+ \ \tcL^{(j)}_{\Si_j}  \bullet  {\cal C}^{(j)} \bullet 
\bigg[ \sum_{i=2}^{j} \big(F^{(i)}_{\Si_j} +L^{(i)\,f}_{\Si_j} \big) \bigg] \cr
\noalign{\goodbreak}
&= \bigg[\sum_{i=2}^{j}  F^{(i)}_{\Si_j} +\tcL^{(j)\,f }_{\Si_j} +\tcL^{(j)}_{\Si_j}\bigg]
\bullet {\cal C}^{(j)} \bullet
\bigg[\sum_{i=2}^{j} F^{(i)}_{\Si_j} +\tcL^{(j)\,f }_{\Si_j} +\tcL^{(j+1)} \bigg]
}$$
\Item ii) Substituting
$
\tcL^{(j+1)}=\tcL^{(j)}_{\Si_j}+L^{(j+1)}
$
into part (i) gives
$$\eqalign{
L^{(j+1)} &= 
\Big(\sum_{i=2}^{j}F^{(i)}_{\Si_j}+\tcL^{(j)\,f }_{\Si_j}
+\tcL^{(j)}_{\Si_j}\Big)\bullet {\cal C}^{(j)} \bullet
\Big( \sum_{i=2}^{j} F^{(i)}_{\Si_j} +\tcL^{(j)\,f }_{\Si_j}+\tcL^{(j)}_{\Si_j}\Big)\cr
&\hskip2in+\Big(\sum_{i=2}^{j}F^{(i)}_{\Si_j}+\tcL^{(j)\,f }_{\Si_j}
+\tcL^{(j)}_{\Si_j}\Big)\bullet {\cal C}^{(j)} \bullet
L^{(j+1)}
}$$
Now just iterate.
\endproof
\proof{ of Proposition \propLADaltcmpladders}
By Lemma \lemLADaltcmpladdersIII.ii,
$$\eqalign{
\tcL^{(j+1)} &= \tcL^{(j)}_{\Si_j} + L^{(j+1)} \cr
&=\tcL^{(j)}_{\Si_j}
+ \smsum_{\ell=1}^\infty 
\big(F +\tcL^{(j)}_{\Si_j}+\tcL^{(j)\,f}_{\Si_j} \big)
\bullet \cC^{(j)}\bullet \cdots \cC^{(j)}
\bullet
\big(F +\tcL^{(j)}_{\Si_j}+\tcL^{(j)\,f}_{\Si_j} \big)\cr
}$$
where  $F = \sum_{i=2}^{j}{F}_{\Si_j}^{(i)}$ and the $\ell^{\rm th}$ term
has $\ell$ bubble propagators $\cC^{(j)}$.
Thus $\tcL^{(j)}$ obeys the same initial condition and recursion relation
as that defining $\cL^{(j)}$ in Definition \defLADcompoundphladder. Therefore,
they are equal. Hence the Proposition follows from Lemma 
\lemLADaltcmpladdersII\ and Lemma \lemLADaltcmpladdersIII.i. 
\endproof

\titlec{Spin Independence}\PG\pgLADIIb

The following discussion shows how spin independent functions on $\big(\fY^\updownarrow_\Si\big)^4$ are related to functions on $\fY_\Si^4$.

\definition{\STM\defLADspinIndependent (Spin Independence)}{ 
Let $\fZ_\li$ and $\fZ_\ri$ be sets and let $f$ 
be a function on $\big(\fZ_\li\times\{\uparrow,\downarrow\}\big)^2
\times \big(\fZ_\ri\times\{\uparrow,\downarrow\}\big)^2$. Set, for each $A\in SU(2)$,
$$
f^A\big((\cdot,\si_1),\cdots,(\cdot,\si_4)\big)
=\sum_{\tau_1,\cdots\tau_4}
f\big((\cdot,\tau_1),\cdots,(\cdot,\tau_4)\big)\ 
A_{\tau_1,\si_1}\bar A_{\tau_2,\si_2}\bar A_{\tau_3,\si_3}A_{\tau_4,\si_4}
$$
$f$ is called (particle--hole) spin independent if $f=f^A$ for all $A\in SU(2)$.
}
\remark{\STM\remLADspinIndependent}{
Let $F$ be a four--legged kernel on $\fX_\Si$. If $F$ is spin independent
in the sense of Definition \defOSsymmetries.S\ of [FKTo2], then its particle--hole reduction is spin independent in the sense of Definition \defLADspinIndependent.

}
\lemma{\STM\lemLADchargespin (Charge Spin Representation)}{
Let $\fZ_\li$ and $\fZ_\ri$ be sets and let $f$ 
be a spin independent function 
on $\big(\fZ_\li\times\{\uparrow,\downarrow\}\big)^2
\times \big(\fZ_\ri\times\{\uparrow,\downarrow\}\big)^2$.
Then, there are functions $f_C$ and $f_S$ on $\fZ_\li^2\times\fZ_\ri^2$ 
such that
$$\eqalign{
f\big((z_1,\si_1),(z_2,\si_2),(z_3,\si_3),(z_4,\si_4)\big)
&=\sfrac{1}{2}f_C(z_1,z_2,z_3,z_4)\de_{\si_1,\si_2}\de_{\si_3,\si_4}\cr
&\hskip.5in+f_S(z_1,z_2,z_3,z_4)\big[\de_{\si_1,\si_3}\de_{\si_2,\si_4}
-\half \de_{\si_1,\si_2}\de_{\si_3,\si_4}
\big]
}$$

}
\prf
The statement is essentially [N, (1--7)]. The proof is outlined in [N] between
(3--40) and (3--41). For the readers convenience, we include a detailed proof.

The $z$'s play no role,
so we suppress them. Then the function $f(\si_1,\si_2,\si_3,\si_4)$
can be viewed as an element 
of $\bbbc^{16}=\bbbc^2\otimes\bbbc^2\otimes\bbbc^2\otimes\bbbc^2$ and 
$M_A:f\mapsto f^A$ is a linear map on $\bbbc^{16}$.
The map $A\mapsto M_A$ is a representation of $SU(2)$ on $\bbbc^{16}$. 
Denote by $S_n$ the
standard $(2n+1)$ dimensional  ``spin $n$'' irreducible representation of $SU(2)$.
In particular, the identity representation $A\mapsto A$ is $S_{1/2}$.
Since the representation $A\mapsto \bar A$ is unitarily equivalent to
$S_{1/2}$, the representation $A\mapsto M_A$ is unitarily equivalent to
$
S_{1/2}\otimes S_{1/2}\otimes S_{1/2}\otimes S_{1/2}
\cong
(S_0\oplus S_1)\otimes(S_0\oplus S_1)
\cong 2S_0\oplus 3S_1\oplus S_2
$. Thus the dimension of the subspace $\set{f\in\bbbc^{16}}{f=f^A\ \forall
A\in SU(2)}$ is exactly two. Since $f(\si_1,\si_2,\si_3,\si_4)
=\de_{\si_1,\si_2}\de_{\si_3,\si_4}$ and $f(\si_1,\si_2,\si_3,\si_4)
=\de_{\si_1,\si_3}\de_{\si_2,\si_4}-\half \de_{\si_1,\si_2}\de_{\si_3,\si_4}$ 
are two independent elements of that subspace, every $f\in\bbbc^{16}$
obeying $f=f^A$ for all $A\in SU(2)$ is a linear combination of 
$\de_{\si_1,\si_2}\de_{\si_3,\si_4}$ and
$\de_{\si_1,\si_3}\de_{\si_2,\si_4}-\half \de_{\si_1,\si_2}\de_{\si_3,\si_4}$.
\endproof
\remark{\STM\remLADchargespin}{
$$\eqalign{
f_C&=f\big((\,\cdot\,,\uparrow),(\,\cdot\,,\uparrow),(\,\cdot\,,\uparrow),(\,\cdot\,,\uparrow)\big)
+f\big((\,\cdot\,,\uparrow),(\,\cdot\,,\uparrow),(\,\cdot\,,\downarrow),(\,\cdot\,,\downarrow)\big)\cr
&=f\big((\,\cdot\,,\downarrow),(\,\cdot\,,\downarrow),(\,\cdot\,,\downarrow),(\,\cdot\,,\downarrow)\big)
+f\big((\,\cdot\,,\downarrow),(\,\cdot\,,\downarrow),(\,\cdot\,,\uparrow),(\,\cdot\,,\uparrow)\big)\cr
f_S&=f\big((\,\cdot\,,\uparrow),(\,\cdot\,,\downarrow),(\,\cdot\,,\uparrow),(\,\cdot\,,\downarrow)\big)
=f\big((\,\cdot\,,\downarrow),(\,\cdot\,,\uparrow),(\,\cdot\,,\downarrow),(\,\cdot\,,\uparrow)\big)\cr
}$$

}
\lemma{\STM\lemLADchargespinflip}{
If $K$ is a spin independent function on $(\fZ\times\{\uparrow,\downarrow\})^{4}$, then 
$$
\big(K^f\big)_C=\half\big(K_C+3K_S\big)^f\qquad
\big(K^f\big)_S=\half \big(K_C-K_S\big)^f
$$
where $K^f$ is the flipped function of (\eqnLADflipped).
}
\prf 
$$\eqalign{
&K^f\big((z_1,\si_1),(z_2,\si_2),(z_3,\si_3),(z_4,\si_4)\big)
=-K\big((z_1,\si_1),(z_3,\si_3),(z_2,\si_2),(z_4,\si_4)\big)\cr
&\hskip.5in
=-\sfrac{1}{2}K_C(z_1,z_3,z_2,z_4)\de_{\si_1,\si_3}\de_{\si_2,\si_4}
-K_S(z_1,z_3,z_2,z_4)\big[\de_{\si_1,\si_2}\de_{\si_3,\si_4}
-\half \de_{\si_1,\si_3}\de_{\si_2,\si_4}\big]\cr
&\hskip.5in
=K^f_S(z_1,z_2,z_3,z_4)\de_{\si_1,\si_2}\de_{\si_3,\si_4}
+\sfrac{1}{2}\big(K^f_C-K^f_S\big)(z_1,z_2,z_3,z_4)\de_{\si_1,\si_3}\de_{\si_2,\si_4}
\cr
&\hskip.5in
=\sfrac{1}{4}\big(K^f_C+3K^f_S\big)(z_1,z_2,z_3,z_4)\de_{\si_1,\si_2}\de_{\si_3,\si_4}\cr
&\hskip1in
+\sfrac{1}{2}\big(K^f_C-K^f_S\big)(z_1,z_2,z_3,z_4)
\big[\de_{\si_1,\si_3}\de_{\si_2,\si_4}-\half \de_{\si_1,\si_2}\de_{\si_3,\si_4}
\big]
\cr
}$$
\endproof
\lemma{\STM\lemLADchargespinII}{
If $H'$ and $K'$ are spin independent functions on $(\fY^\updownarrow_\Si)^{4}$
and $P$ is a bubble propagator, then 
$$\eqalign{
(H'\bullet P\bullet K')_C&=H'_C\bullet P\bullet K'_C\cr
(H'\bullet P\bullet K')_S&=H'_S\bullet P\bullet K'_S\cr
}$$

}
\prf This Lemma follows directly from Remark \remLADchargespin.
\endproof
Parts (ii) and (iii) of Proposition \propLADaltcmpladders, 
Lemma \lemLADchargespinflip\ and Lemma \lemLADchargespinII\ give a coupled
system of recursion relations for $\cL^{(j)}_C$,  $\cL^{(j)}_S$,
$L^{(j)}_C$ and $L^{(j)}_S$.
\corollary{\STM\corLADaltcmpladders}{
$$\leqalignno{
{\cal L}^{(j+1)}_C 
&= \sum_{\ell=1}^\infty \  \sum_{i_1,\cdots,i_{\ell+1}=2}^{j} \  
\Big[\big(F^{(i_1)}_C+\half L^{(i_1)\,f}_C+\sfrac{3}{2} L^{(i_1)\,f}_S \big)  
\bullet {\cal C}^{[\max \{i_1,i_2\},j]} \bullet
& i)\cr
&\hskip1.2in \cdots \bullet 
{\cal C}^{[\max \{i_\ell,i_{\ell+1}\},j]} \bullet
\big(F^{(i_{\ell+1})}_C+\half L^{(i_{\ell+1})\,f}_C+\sfrac{3}{2} L^{(i_{\ell+1})\,f}_S \big)\Big]_{\Si_j}\cr
{\cal L}^{(j+1)}_S
&= \sum_{\ell=1}^\infty \  \sum_{i_1,\cdots,i_{\ell+1}=2}^{j} \  
\Big[\big(F^{(i_1)}_S+\half L^{(i_1)\,f}_C-\half L^{(i_1)\,f}_S \big)  \bullet 
{\cal C}^{[\max \{i_1,i_2\},j]} \bullet
\cr
&\hskip1.2in \cdots \bullet 
{\cal C}^{[\max \{i_\ell,i_{\ell+1}\},j]} \bullet
\big(F^{(i_{\ell+1})}_S+\half L^{(i_{\ell+1})\,f}_C-\half L^{(i_{\ell+1})\,f}_S \big)\Big]_{\Si_j}\cr
L^{(j+1)}_C &= 
\Big( \sum_{i=2}^{j} F^{(i)}_{C\,\Si_j} 
+\half {\cal L}^{(j)\,f }_{C\,\Si_j}
+\sfrac{3}{2} {\cal L}^{(j)\,f }_{S\,\Si_j} 
+{\cal L}^{(j)}_{C\,\Si_j} \Big)
\bullet {\cal C}^{(j)} & ii) \cr
&\hskip1.5in\bullet \Big( \sum_{i=2}^{j} F^{(i)}_{C\,\Si_j} 
+\half {\cal L}^{(j)\,f }_{C\,\Si_j}
+\sfrac{3}{2} {\cal L}^{(j)\,f }_{S\,\Si_j} 
 +{\cal L}^{(j+1)}_C \Big)
\cr
L^{(j+1)}_S &= 
\Big( \sum_{i=2}^{j} F^{(i)}_{S\,\Si_j} 
+\half {\cal L}^{(j)\,f }_{C\,\Si_j} 
-\half {\cal L}^{(j)\,f }_{S\,\Si_j} 
+{\cal L}^{(j)}_{S\,\Si_j} \Big)
\bullet {\cal C}^{(j)} \cr
&\hskip1.5in \bullet
\Big( \sum_{i=2}^{j} F^{(i)}_{S\,\Si_j} 
+\half {\cal L}^{(j)\,f }_{C\,\Si_j} 
-\half {\cal L}^{(j)\,f }_{S\,\Si_j} 
+{\cal L}^{(j+1)}_S \Big)
\cr
}$$

}
\vskip.25in

Theorem \thmLADmodcompLadder\ will be proven by bounding each term on the right
hand side of Corollary \corLADaltcmpladders.i. Each such term is a 
particle--hole ladder of the form 
$$
\big(G^{(i_1)}+K^{(i_1)\,f} \big)  \bullet 
{\cal C}^{[\max \{i_1,i_2\},j]} \bullet \cdots \bullet 
{\cal C}^{[\max \{i_\ell,i_{\ell+1}\},j]} \bullet
\big(G^{(i_{\ell+1})}+K^{(i_{\ell+1})\,f} \big)
$$ 
where $G^{(i)}$ is either $F_C^{(i)}$ or $F_S^{(i)}$ and
$K^{(i)}$ is a linear combination of $L_C^{(i)}$ and $L_S^{(i)}$.
This ladder has rungs $\big(G^{(i_\nu)}+K^{(i_\nu)\,f} \big)$
which are connected by particle--hole propagators ${\cal C}^{[i,j]}$.
The induction step will consist in adding an additional rung to the left
of the ladder. More precisely, we will prove a bound on 
$$
\big(G^{(i_1)}+K^{(i_1)\,f} \big)  \bullet {\cal C}^{[i,j]} \bullet H
$$
with 
$\ 
H=\big(G^{(i_2)}+K^{(i_2)\,f} \big)  \bullet 
{\cal C}^{[\max \{i_2,i_3\},j]} \bullet \cdots  \bullet
\big(G^{(i_{\ell+1})}+K^{(i_{\ell+1})\,f} \big)
\ $, assuming bounds on $H$. The expression
$$
G^{(i_1)}  \bullet {\cal C}^{[i,j]} \bullet H
$$
is a particle--hole bubble

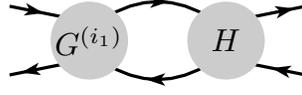
\centerline{\figplace{wwbubble5}{0.2 in}{0 in}}

\noindent
We will derive the necessary bounds on general particle--hole bubbles
in Theorem \:\bbound. By Corollary \corLADaltcmpladders.ii,
$$
K^{(i_1)\,f}  \bullet {\cal C}^{[i,j]} \bullet H
=\Big(G_1^{(i_1)}\bullet\cC^{(i_1-1)}\bullet G_2^{(i_1)}\Big)^f
 \bullet {\cal C}^{[i,j]} \bullet H
$$
with $G_1^{(i)}$ and $G_2^{(i)}$ linear combinations of 
$\sum\limits_{k=2}^{i-1} F^{(k)}_C$, $\sum\limits_{k=2}^{i-1} F^{(k)}_S$,
$\cL^{(i-1)}_C $, $\cL^{(i-1)}_S $, $\cL^{(i-1)\,f }_C$, $\cL^{(i-1)\,f }_S$,
$\cL^{(i)}_C $, $\cL^{(i)}_S $. It
is a double bubble

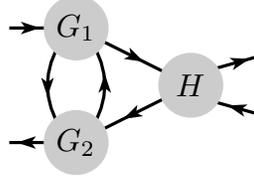
\centerline{\figplace{dbubble4}{0 in}{0 in}}

\noindent
Bounds on double bubbles will be obtained in Theorem \:\dbbound.

\titlec{Scaled Norms}\PG\pgLADIIc

In the induction procedure outlined above the various ladders naturally
have different sectorization scales at their left and right hand ends.
This was the motivation for Definition \defLADlonelinftyIV.

\convention{\STM\convLADlr}{Introduce, for scales $\ell,r$, the short hand
notation
$$
\fY_{\ell,r}=\fY^{(4)}_{\Si_\ell,\Si_r}
$$

}
\definition{\STM\secrepnorm}{ For a function $f$ on $\fY_{\ell,r}$ and
multiindices $\de_\li,\de_\ci,\de_\ri\in\bbbn_0\times\bbbn_0^2$, set
$$\eqalign{
\|f\|^{(\de_\li,\de_\ci,\de_\ri)}_{\ell,r}
&=\sfrac{1}{M^{\ell|\de_\li|+|\de_\ci|\max(\ell,r)+r|\de_\ri|}}
\v f\v^{(\de_\li,\de_\ci,\de_\ri)}_{\Si_\ell,\Si_r}\cr
|f|^{[\de_\li,\de_\ci,\de_\ri]}_{\ell,r}
&=\max_{\de'_\li\le\de_\li
        \atop{\de'_\ci\le\de_\ci
        \atop \de'_\ri\le\de_\ri
        }}
            \|f\|^{(\de'_\li,\de'_\ci,\de'_\ri)}_{\ell,r}
\cr
}$$
The norm $\v\ \cdot\ \v^{(\de_\li,\de_\ci,\de_\ri)}_{\Si_\ell,\Si_r}$
was defined in Definition \defLADlonelinftyIV.
If $\ell=r=j$, set
$$
|f|^{\[\de\]}_j
=\max_{\de_\li,\de_\ci,\de_\ri\in\bbbn_0\times\bbbn_0^2\atop
        \de_\li+\de_\ci+\de_\ri\le\de}
            \|f\|^{(\de_\li,\de_\ci,\de_\ri)}_{j,j}
$$
}

Set
$$\eqalign{
\De&=\set{\de\in\bbbn_0\times\bbbn_0^2}{\de_{0}\le r_0,\ \de_1+\de_2\le r_e}\cr 
\vec\De&=\set
    {\vec\de=(\de_\li,\de_\ci,\de_\ri)\in\big(\bbbn_0\times\bbbn_0^2\big)^3}
    {\de_\li+\de_\ci+\de_\ri\in\De}\cr
}\EQN\eqnLADDelta$$
where $r_e+3$ is the degree of differentiability of the dispersion relation $e(\k)$ and $r_0$ is the number of $k_0$ derivatives that we wish to control.
The numbers $r_e$ and $r_0$ also determine the number of finite coefficients in the formal power series $\cb_j$ of (\eqnLADcj). The following remark relates the formal power series norms of Definition \defLADnormdomainnorm.ii
to the norms of Definition \secrepnorm.

\remark{\STM\remLADscaledtonormdomain}{ 
There is a constant $\abcst$, depending only on $r_e$ and $r_0$ such that the 
following holds.
Let $f$ be a sectorized, translation invariant function on $\fY^4_{\Si_j}$. 
\Item i)
$$
\v f\v_{\Si_j}\le \Big[\max_{\de\in\De}|f|^{\[\de\]}_{j}\Big]\ \cb_j
$$
\Item ii) If there is a number $\ga$ such that $\v f\v_{\Si_j}\le\ga\cb_j$,
then
$$
|f|^{\[\de\]}_{j}
\le \abcst\,\ga\qquad
\hbox{ for all }\de\in\De
$$
}

\noindent Thus to prove Theorem \thmLADmodcompLadder, it suffices to prove that
$$
\max_{\de\in\De}|\cL_C^{(j+1)}|^{\[\de\]}_{j} \le \const \rho^2\qquad
\max_{\de\in\De}|\cL_S^{(j+1)}|^{\[\de\]}_{j} \le \const \rho^2
$$

\definition{\STM\resecrepnorm (Norms and Resectorization)}{ 
Let $\ell,\ell',r,r'\ge 0$. For a sectorized, translation invariant,
function $f$ on $\fY_{\ell',r'}$ and
multiindices $\vec\de\in{\big(\bbbn_0\times\bbbn_0^2\big)}^3$, set
$$\eqalign{
|f|^{[\vec\de]}_{\ell,r}
&=\big|f_{\Si_\ell,\Si_r}\big|^{[\vec\de]}_{\ell,r}
\cr
}$$
If $\ell=r=j$ and $\de\in\bbbn_0\times\bbbn_0^2$, set
$$
|f|^{\[\de\]}_j
=\max_{\de_\li,\de_\ci,\de_\ri\in\bbbn_0\times\bbbn_0^2\atop
        \de_\li+\de_\ci+\de_\ri\le\de}
            |f|^{[\de_\li,\de_\ci,\de_\ri]}_{j,j}
$$
}
As in Proposition \propOSresectorI\ of [FKTo4], one proves 
\lemma{\STM\lemLADresectornorm}{
Let $\ell\ge\ell'\ge1$ and $r\ge r'\ge 1$. 
Let $f$ be a sectorized, translation invariant, function on $\fY_{\ell',r'}$ 
and let $\vec\de=(\de_\li,\de_\ci,\de_\ri)\in\vec\De$. Then
$$\eqalign{
|f|^{[\vec\de]}_{\ell,r}
&\le \abcst\Big\{ \sfrac{1}{M^{\ell-\ell'}}\sfrac{1}{M^{r-r'}}|f|^{[\vec\de]}_{\ell',r'}
+\sfrac{1}{M^{\ell-\ell'}}|f|^{[\de_\li,\de_\ci,0]}_{\ell',r'}
+\sfrac{1}{M^{r-r'}}|f|^{[0,\de_\ci,\de_\ri]}_{\ell',r'}
+|f|^{[0,\de_\ci,0]}_{\ell',r'}\Big\}
\cr
&\le \abcst |f|^{[\vec\de]}_{\ell',r'}
}$$
The constant $\abcst$ depends only on $\De$.
}
\prf
Let $f$ be a function on 
$\fY_{i_1,\Si_{\ell'}}\times\fY_{i_2,\Si_{\ell'}}\times \fY_{i_3,\Si{r'}}\times\fY_{i_4,\Si_{r'}}$.
We consider the case $i_1=i_2=i_3=i_4=1$ and $\ell'<\ell$, $r'<r$. The
other cases are similar, but easier.
Recall from Definition \defLADresectoriz\ that, 
$$\eqalign{
&f_{\Si_{\ell},\Si_{r}}({\sst(x_1,s_1),(x_2,s_2),(x_3,s_3),(x_4,s_4)} )\cr
&\hskip.9in=\smsum_{s'_\nu\in \Si_{\ell'}\atop\nu\in\{1,2\}}
\smsum_{s'_\nu\in \Si_{r'}\atop\nu\in\{3,4\}}
\int \smprod_{\nu=1}^4\Big( {\sst dx'_\nu}\ 
\hat\chi_{s_\nu}({\sst (-1)^{b_\nu}(x_\nu-x'_\nu)})\Big)\ 
f({\sst(x'_1,s'_1),(x'_2,s'_2),(x'_3,s'_3),(x'_4,s'_4)} )\cr
}$$
First observe that, for any fixed $s_1,\cdots,s_4$, there are at most 
$3^4$ choices of $(s'_1,\cdots,s'_4)$ for which the integral 
$\int \smprod_{\nu=1}^4\Big( {\sst dx'_\nu}\ 
\hat\chi_{s_\nu}(\cdots)\Big)\ f(\cdots )$
fails to vanish identically, because $f$ is sectorized and 
$\ell'<\ell$, $r'<r$. So it suffices to consider any fixed 
$s'_1,\cdots,s'_4$. Hence by Leibniz's Rule (Lemma \:\leibniz), 
$\|f_{\Si_\ell,\Si_r}\|^{(\de_\li,\de_\ci,\de_\ri)}_{\ell,r}$ is bounded 
by a constant, which depends only on
$\De$, times the maximum of
$$\eqalign{
&\sfrac{1}{M^{\ell|\de_\li|+|\de_\ci|\max(\ell,r)+r|\de_\ri|}}
\int\smprod_{\nu=2}^4 dx_\nu\int \smprod_{\nu=1}^4\Big( {\sst dx'_\nu}\ 
\big|(x_\nu-x'_\nu)^{\be_\nu}\hat\chi_{s_\nu}({\sst (-1)^{b_\nu}(x_\nu-x'_\nu)})\big|
\Big)\cr
&\hskip2.5in\big|\rD_{1;2}^{\al_\li}\rD_{\mu;\mu'}^{\al_\ci}\rD_{3;4}^{\al_\ri}
f({\sst(x'_1,s'_1),(x'_2,s'_2),(x'_3,s'_3),(x'_4,s'_4)} )\big|\cr
&\le \sfrac{1}{M^{\ell|\de_\li|+|\de_\ci|\max(\ell,r)+r|\de_\ri|}}
\Big(\smprod_{\nu=1}^4\big\|x_\nu^{\be_\nu}\hat\chi_{s_\nu}(x_\nu)\big\|_{L^1}
\Big)\V f\V_{\Si_{\ell'},\Si_{r'}}^{(\al_\li,\al_\ci,\al_\ri)}\cr
&= \sfrac{M^{\ell'|\al_\li|+|\al_\ci|\max(\ell',r')+r'|\al_\ri|}}
{M^{\ell|\de_\li|+|\de_\ci|\max(\ell,r)+r|\de_\ri|}}
\Big(\smprod_{\nu=1}^4\big\|x_\nu^{\be_\nu}\hat\chi_{s_\nu}(x_\nu)\big\|_{L^1}
\Big)\big\| f\big\|_{\Si_{\ell'},\Si_{r'}}^{(\al_\li,\al_\ci,\al_\ri)}\cr
}$$
over $x_1,s_1,\cdots,s_4,s'_1,\cdots,s'_4$ and $\mu\in\{1,2\},\ \mu'\in\{3,4\}$
and $\al_\li,\al_\ci,\al_\ri$ and 
$$
\be_\nu=\be_{\nu,\li}+\be_{\nu,\ci}+\be_{\nu,\ri}\qquad\nu=1,\cdots,4
$$
obeying
$$\eqalign{
&\be_{1,\li}+\al_\li+\be_{2,\li}=\de_\li\quad
\be_{\mu,\ci}+\al_\ci+\be_{\mu',\ci}=\de_\ci\quad
\be_{3,\ri}+\al_\ri+\be_{4,\ri}=\de_\ri\cr
&\be_{1,\ri}=\be_{2,\ri}=\be_{3,\li}=\be_{4,\li}=\be_{\nu,\ci}=0
\qquad\hbox{for }\nu\ne\mu,\mu'
}$$
In particular
$$
\ell|\de_\li|+|\de_\ci|\max(\ell,r)+r|\de_\ri|
\ge \ell|\al_\li+\be_1+\be_2|+|\al_\ci|\max(\ell,r)+r|\al_\ri+\be_3+\be_4|
$$
By  Lemma \lemOSsectpartunit\ of [FKTo3]
$$
\big\|x_\nu^{\be_\nu}\hat\chi_{s_\nu}(x_\nu)\big\|_{L^1}
\le\abcst\cases{M^{|\be_\nu|\ell}& if $\nu\in\{1,2\}$\cr
                M^{|\be_\nu|r}& if $\nu\in\{3,4\}$\cr}
\EQN\eqnLADchisbnd$$
so that
$$\eqalign{
&\sfrac{M^{\ell'|\al_\li|+|\al_\ci|\max(\ell',r')+r'|\al_\ri|}}
{M^{\ell|\de_\li|+|\de_\ci|\max(\ell,r)+r|\de_\ri|}}
\smprod_{\nu=1}^4\big\|x_\nu^{\be_\nu}\hat\chi_{s_\nu}(x_\nu)\big\|_{L^1}\cr
&\hskip1in\le \abcst\sfrac{M^{\ell'|\al_\li|+|\al_\ci|\max(\ell',r')+r'|\al_\ri|}}
{M^{\ell|\de_\li|+|\de_\ci|\max(\ell,r)+r|\de_\ri|}}
M^{\ell|\be_1+\be_2|+r|\be_3+\be_4|}\cr
&\hskip1in\le \abcst\sfrac{1}{M^{(\ell-\ell')|\al_\li|+(r-r')|\al_\ri|}}
}$$
and
$$\eqalign{
\|f_{\Si_\ell,\Si_r}\|^{(\de_\li,\de_\ci,\de_\ri)}_{\ell,r}
&\le  \abcst\max_{\al_\li\le\de_\li\atop{\al_\ci\le\de_\ci\atop\al_\ri\le\de_\ri}}
\sfrac{1}{M^{(\ell-\ell')|\al_\li|+(r-r')|\al_\ri|}}
\big\| f\big\|_{\Si_{\ell'},\Si_{r'}}^{(\al_\li,\al_\ci,\al_\ri)}
}$$
and the Lemma follows.

\endproof

\goodbreak
\titlec{Bubble and Double Bubble Bounds}\PG\pgLADIId

\definition{\STM\topbottom}{ Let $i\le j$. Then
$$
\cC^{[i,j]}=\cC_{\tp}^{[i,j]}+\cC_{\md}^{[i,j]}+\cC_{\bt}^{[i,j]}
$$
where
$$\eqalign{
\cC_{\tp}^{[i,j]}&=\sum_{i\le i_t\le j\atop i_b>j}C^{(i_t)}_v\otimes C^{(i_b)\,t}_v\cr
\cC_{\md}^{[i,j]}&=\sum_{i\le i_t\le j\atop i\le i_b\le j}C^{(i_t)}_v
\otimes C^{(i_b)\,t}_v\cr
\cC_{\bt}^{[i,j]}&=\sum_{i_t> j\atop i\le i_b\le j}C^{(i_t)}_v
\otimes C^{(i_b)\,t}_v\cr
}$$

}

\theorem{\STM\bbound\ (Bubble Bound)}{ Let $1\le i,\ell \le j$ and 
$\de_\li,\de_\ri\in\De$. Let $g$ and $h$ be sectorized,
 translation invariant functions on $\fY_{\ell,i}$ and $\fY_{i,j}$ 
respectively. Then\hfill\break
a)
$$
\big|g\bullet\cC^{[i,j]}\bullet h\big|_{\ell,j}^{[\de_\li,0,\de_\ri]}
\le\const\ i\  \max_{\al_\ri,\al_\li\in\bbbn_0\times\bbbn_0^2
            \atop|\al_\ri|+|\al_\li|\le 3}
\big|g\big|_{\ell,i}^{[\de_\li,0,\al_\ri]}\big|h\big|_{i,j}^{[\al_\li,0,\de_\ri]}
$$ 
\noindent b)
For any $\be\in\De$
$$\eqalign{
\sfrac{1}{M^{|\be|j}}\big\|g\bullet \rD_{1;3}^{\be}\cC^{[i,j]}_{\tp}
\bullet h\big\|_{\ell,j}^{(\de_\li,0,\de_\ri)}
&\le\const \big\|g\big\|_{\ell,i}^{(\de_\li,0,0)}
\big\|h\big\|_{i,j}^{(0,0,\de_\ri)}\cr
\sfrac{1}{M^{|\be|j}}\big\|g\bullet \rD_{2;4}^{\be}\cC^{[i,j]}_{\bt}
\bullet h\big\|_{\ell,j}^{(\de_\li,0,\de_\ri)}
&\le\const \big\|g\big\|_{\ell,i}^{(\de_\li,0,0)}
\big\|h\big\|_{i,j}^{(0,0,\de_\ri)}\cr
}$$
\noindent 
c) 
$$\eqalign{
\big\|g\bullet\cC^{[i,j]}_\md\bullet h\big\|_{\ell,j}^{(\de_\li,0,\de_\ri)}
&\le\const |j-i+1|
\big\|g\big\|_{\ell,i}^{(\de_\li,0,0)}\big\|h\big\|_{i,j}^{(0,0,\de_\ri)}\cr
}$$
and for any $\be\in\De$ with $|\be|\ge 1$ and
$(\mu,\mu')=(1,3),(2,4)$
$$\eqalign{
\sfrac{1}{M^{|\be|j}}\big\|g\bullet \rD_{\mu;\mu'}^{\be}\cC^{[i,j]}_{\md}
\bullet h\big\|_{\ell,j}^{(\de_\li,0,\de_\ri)}
&\le\const \big\|g\big\|_{\ell,i}^{(\de_\li,0,0)}
\big\|h\big\|_{i,j}^{(0,0,\de_\ri)}\cr
}$$
}
\noindent This Theorem is proven in \S III.

\theorem{\STM\dbbound\ (Double Bubble Bound)}{ 
Let $1\le \ell\le i\le j$, $\nu\in\bbbn_0\times\bbbn_0^2$ 
 and $\de_\li,\de_\ri\in\De$.
Let $g_1,\ g_2$ and $h$ be sectorized, translation invariant functions on $\fY_{\ell,\ell}$,  $\fY_{\ell,\ell}$ and $\fY_{i,j}$  respectively. 
Let $\cD$ be either
$$
\cD^{(\ell)}_{\nu,\upl}(x_1,x_2,x_3,x_4)
=\sfrac{1}{M^{|\nu| \ell}}
\sum_{m=\ell}^\infty \rD^{\nu}_{1;3}C_v^{(\ell)}(x_1,x_3)C_v^{(m)}(x_4,x_2)
$$
or
$$
\cD^{(\ell)}_{\nu,\dnl}(x_1,x_2,x_3,x_4)
=\sfrac{1}{M^{|\nu| \ell}}
\sum_{m=\ell+1}^\infty C_v^{(m)}(x_1,x_3)\rD^{\nu}_{2;4}C_v^{(\ell)}(x_4,x_2)
$$
a) If $\nu+\de_\li+\al\in\De$ for all $|\al|\le 3$, then
$$
\big|(g_1\bullet \cD\bullet g_2)^f\bullet\cC^{[i,j]}\bullet h\big|_{\ell,j}
          ^{[\de_\li,0,\de_\ri]}
\le\const\ i\ \sqrt{\fl_\ell}\max_{\al_\upl,\al_\dnl,\al_\li\in\bbbn_0\times\bbbn_0^2
            \atop|\al_\upl|+|\al_\dnl|+|\al_\li|\le 3} \big|g_1\big|_{\ell}^{\[\de_\li+\al_\upl\]} \big|g_2\big|_{\ell}^{\[\de_\li+\al_\dnl\]}
\big|h\big|_{i,j}^{[\al_\li,0,\de_\ri]}
$$ 
\noindent
b) If $\nu+\de_\li\in\De$, then for any $\be\in\De$
$$\eqalign{
\sfrac{1}{M^{|\be|j}}\big\|(g_1\bullet \cD\bullet g_2)^f
\bullet \rD_{1;3}^{\be}\cC^{[i,j]}_{\tp}
\bullet h\big\|_{\ell,j}^{(\de_\li,0,\de_\ri)}
&\le\const \sqrt{\fl_\ell} \big|g_1\big|_{\ell,\ell}^{[0,\de_\li,0]}
\big|g_2\big|_{\ell,\ell}^{[0,\de_\li,0]} \big\|h\big\|_{i,j}^{(0,0,\de_\ri)}\cr
\sfrac{1}{M^{|\be|j}}\big\|(g_1\bullet \cD\bullet g_2)^f
\bullet \rD_{2;4}^{\be}\cC^{[i,j]}_{\bt}
\bullet h\big\|_{\ell,j}^{(\de_\li,0,\de_\ri)}
&\le\const \sqrt{\fl_\ell} \big|g_1\big|_{\ell,\ell}^{[0,\de_\li,0]}
\big|g_2\big|_{\ell,\ell}^{[0,\de_\li,0]} \big\|h\big\|_{i,j}^{(0,0,\de_\ri)}\cr
}$$

\noindent 
c) If $\nu+\de_\li\in\De$, then
$$\eqalign{
\big\|(g_1\bullet \cD\bullet g_2)^f \bullet\cC^{[i,j]}_\md\bullet h\big\|_{\ell,j}^{(\de_\li,0,\de_\ri)}
&\le\const|j-i+1| \sqrt{\fl_\ell} \big|g_1\big|_{\ell,\ell}^{[0,\de_\li,0]} \big|g_2\big|_{\ell,\ell}^{[0,\de_\li,0]} \big\|h\big\|_{i,j}^{(0,0,\de_\ri)}\cr
}$$ 
and for any $\be\in\De$ with $|\be|\ge 1$ and
$(\mu,\mu')=(1,3),(2,4)$
$$\eqalign{
\sfrac{1}{M^{|\be|j}}\big\|(g_1\bullet \cD\bullet g_2)^f
\bullet \rD_{\mu;\mu'}^{\be}\cC^{[i,j]}_{\md}
\bullet h\big\|_{\ell,j}^{(\de_\li,0,\de_\ri)}
&\le\const \sqrt{\fl_\ell} \big|g_1\big|_{\ell,\ell}^{[0,\de_\li,0]}
\big|g_2\big|_{\ell,\ell}^{[0,\de_\li,0]} \big\|h\big\|_{i,j}^{(0,0,\de_\ri)}\cr
}$$
}
\noindent This Theorem is proven in \S IV.

\remark{}{
Observe that
$$
\cC^{(\ell)}=\cD^{(\ell)}_{0,\upl}+\cD^{(\ell)}_{0,\dnl}
$$
}

We use Leibniz's rule to convert Theorems \bbound\ and \dbbound\ into bounds
on derivatives of $g\bullet\cC^{[i,j]}\bullet h$
and $(g_1\bullet\cC^{(\ell)}\bullet g_2)^f\bullet\cC^{[i,j]}\bullet h$
with respect to transfer momenta. These bounds are stated in Corollaries
\:\bboundcor, \:\dbboundcor\ and \:\bdbbound, below.

\lemma{\STM\leibniz  (Leibniz's Rule)}{
Let $\ell_1,r_1,\ell_2,r_2\ge 1$, $P$ a bubble propagator and
$K_1,\ K_2$ sectorized, translation invariant functions on $\fY_{\ell_1,r_1}$ 
and $\fY_{\ell_2,r_2}$, respectively. 
Let $\mu,\nu\in\{1,2\}$, $\mu',\nu'\in\{3,4\}$
 and $\de\in\bbbn_0\times\bbbn_0^2$. 
Then,
$$
\rD^{\de}_{\nu;\nu'}(K_1\bullet P\bullet K_2)
=\sum_{\be_1,\be_2,\be_3\in\bbbn_0\times\bbbn_0^2
\atop\be_1+\be_2+\be_3=\de}
{\tst{\de\choose\be_1,\be_2,\be_3}}
(\rD^{\be_1}_{\nu;\mu+2}K_1)\bullet(\rD^{\be_2}_{\mu;\mu'}P)
\bullet (\rD^{\be_3}_{\mu'-2;\nu'}K_2)
$$
Here ${\tst{\de\choose\be_1,\be_2,\be_3}}=\sfrac{\de!}{\be_1!\be_2!\be_3!\ }$.

}
\prf The proof is trivial.
\endproof

\corollary{\STM\bboundcor}{ Let $1\le \ell\le i \le j$ and
$\de_\li, \de_\ci, \de_\ri\in\De$. 
Let $g$ and $h$ be sectorized, translation invariant functions on $\fY_{\ell,\ell}$ and $\fY_{i,j}$  respectively. 
\Item a)
$$\eqalign{
\big\|g\bullet \cC^{[i,j]}_{\tp}\bullet h\big\|_{\ell,j}^{(\de_\li,\de_\ci,\de_\ri)}
&\le\const \big|g\big|_{\ell,\ell}^{[\de_\li,\de_\ci,0]}
\big|h\big|_{i,j}^{[0,\de_\ci,\de_\ri]}\cr
\big\|g\bullet \cC^{[i,j]}_{\bt}\bullet h\big\|_{\ell,j}^{(\de_\li,\de_\ci,\de_\ri)}
&\le\const \big|g\big|_{\ell,\ell}^{[\de_\li,\de_\ci,0]}
\big|h\big|_{i,j}^{[0,\de_\ci,\de_\ri]}\cr
}$$
\Item b) For
$\mu\in\{1,2\}$ and $\mu'\in\{3,4\}$
$$\eqalign{
\rD_{\mu;\mu'}^{\de_\ci}\big(g\bullet \cC^{[i,j]}_{\md}\bullet h\big)
=\sum_{\be_1,\be_2,\be_3\in\bbbn_0\times\bbbn_0^2\atop \be_1+\be_2+\be_3=\de_c}
{\tst {\de_c\choose\be_1,\be_2,\be_3}}
\rD_{\mu;3}^{\be_1}g\bullet \rD_{1;3}^{\be_2}\cC^{[i,j]}_{\md}
\bullet \rD_{1;\mu'}^{\be_3}h
}$$
and, for all $\be_1+\be_2+\be_3=\de_\ci$,
$$\eqalign{
&\sfrac{1}{M^{|\de_\ci| j}}\big\|\rD_{\mu;3}^{\be_1}g\bullet 
\rD_{1;3}^{\be_2}\cC^{[i,j]}_{\md}
\bullet \rD_{1;\mu'}^{\be_3}h
\big\|_{\ell,j}^{(\de_\li,0,\de_\ri)}\cr
&\hskip1in\le\sfrac{\const}{M^{|\be_1|(j-\ell)}}\cases{
(j-i+1)
\big\|g\big\|_{\ell,\ell}^{(\de_\li,\be_1,0)}
\big\|h\big\|_{i,j}^{(0,\be_3,\de_\ri)}& if $\be_2=0$ \cr
\noalign{\vskip.1in}
i\ \max\limits_{|\al_\ri+\al_\li|\le 3}
\big|g\big|_{\ell,\ell}^{[\de_\li,\be_1,\al_\ri]}
\big|h\big|_{i,j}^{[\al_\li,\be_3,\de_\ri]}& if $\be_2=0$ \cr
\noalign{\vskip.1in}
\phantom{(j-i+1)}
\big\|g\big\|_{\ell,\ell}^{(\de_\li,\be_1,0)}
\big\|h\big\|_{i,j}^{(0,\be_3,\de_\ri)}& if $\be_2\ne 0$ \cr
}}$$
}
\prf a) We consider the case of $\tp$.  By Leibniz, 
$$\eqalign{
\rD_{\mu;\mu'}^{\de_c}\big(g\bullet \cC^{[i,j]}_{\tp}\bullet h\big)
=\sum_{\be_1,\be_2,\be_3\in\bbbn_0\times\bbbn_0^2\atop \be_1+\be_2+\be_3=\de_c}
{\tst {\de_c\choose\be_1,\be_2,\be_3}}
\rD_{\mu;3}^{\be_1}g\bullet \rD_{1;3}^{\be_2}\cC^{[i,j]}_{\tp}
\bullet \rD_{1;\mu'}^{\be_3}h
}$$
The desired inequality follows by the triangle inequality, Theorem \bbound b
and Lemma \lemLADresectornorm,
with $\rD_{\mu;3}^{\be_1}g$ in place of $g$  
and $\rD_{1;\mu'}^{\be_3}h$ in place of $h$.

\Item b) The first statement is again Leibniz's rule.
By the first statement of Theorem \bbound.c, with $\rD_{\mu;3}^{\be_1}g$ in
place of $g$ and $\rD_{1;\mu'}^{\be_3}h$ in place of $h$,
$$\eqalign{
\sfrac{1}{M^{|\de_\ci| j}}\big\|\rD_{\mu;3}^{\be_1}g\bullet \cC^{[i,j]}_{\md}
\bullet \rD_{1;\mu'}^{\be_3}h
\big\|_{\ell,j}^{(\de_\li,0,\de_\ri)}
&\le\const\sfrac{j-i+1}{M^{|\de_\ci| j}}
\big\|\rD_{\mu;3}^{\be_1}g\big\|_{\ell,i}^{(\de_\li,0,0)}
\big\|\rD_{1;\mu'}^{\be_3}h\big\|_{i,j}^{(0,0,\de_\ri)}\cr
&\hskip-.5in\le\const\sfrac{j-i+1}{M^{|\de_\ci| j}}
\big\|\rD_{\mu;3}^{\be_1}g\big\|_{\ell,\ell}^{(\de_\li,0,0)}
\big\|\rD_{1;\mu'}^{\be_3}h\big\|_{i,j}^{(0,0,\de_\ri)}\cr
&\hskip-.5in\le\const\sfrac{j-i+1}{M^{|\de_\ci| j}}
M^{|\be_1|\ell}\big\|g\big\|_{\ell,\ell}^{(\de_\li,\be_1,0)}
M^{|\be_3|j}\big\|h\big\|_{i,j}^{(0,\be_3,\de_\ri)}\cr
&\hskip-.5in\le\const\sfrac{j-i+1}{M^{|\be_1|(j-\ell)}}
\big\|g\big\|_{\ell,\ell}^{(\de_\li,\be_1,0)}
\big\|h\big\|_{i,j}^{(0,\be_3,\de_\ri)}\cr
}$$
For the second inequality, we used the variant
$$
\big\|\rD_{\mu;3}^{\be_1}g\big\|_{\ell,i}^{(\de_\li,0,0)}
=\sfrac{1}{M^{|\de_\li|\ell}}\big|\rD_{1;2}^{\de_\li}\rD_{\mu;3}^{\be_1}g
                \big|_{\ell,i}^{[0,0,0]}
\le\abcst\, \sfrac{1}{M^{|\de_\li|\ell}}\big|\rD_{1;2}^{\de_\li}\rD_{\mu;3}^{\be_1}g
                \big|_{\ell,\ell}^{[0,0,0]}
=\abcst\,\big\|\rD_{\mu;3}^{\be_1}g\big\|_{\ell,\ell}^{(\de_\li,0,0)}
$$
of Lemma \lemLADresectornorm.
The proof of the second case is similar, but with
$$\eqalign{
\big\|g\bullet\cC_\md^{[i,j]}\bullet h\big\|_{\ell,j}^{(\de_\li,0,\de_\ri)}
&\le \big|g\bullet\cC^{[i,j]}\bullet h\big|_{\ell,j}^{[\de_\li,0,\de_\ri]}
\!+\big|g\bullet\cC_\tp^{[i,j]}\bullet h\big|_{\ell,j}^{[\de_\li,0,\de_\ri]}
\!+\big|g\bullet\cC_\bt^{[i,j]}\bullet h\big|_{\ell,j}^{[\de_\li,0,\de_\ri]}\cr
&\le\const \ i\ 
\max_{\al_\ri,\al_\li\in\bbbn_0\times\bbbn_0^2
            \atop|\al_\ri|+|\al_\li|\le 3}
\big|g\big|_{\ell,i}^{[\de_\li,0,\al_\ri]}\big|h\big|_{i,j}^{[\al_\li,0,\de_\ri]}
}$$ 
(by Theorem \bbound.a,b)
used in place of the first statement of Theorem \bbound.c. The proof of the
third case is again similar, but with the second statement of Theorem \bbound.c
used in place of the first statement of Theorem \bbound.c.
\endproof

\corollary{\STM\dbboundcor}{ 
 Let $1\le \ell\le i \le j$ and
$\de_\li, \de_\ci, \de_\ri\in\De$. 
Let $g_1,\ g_2$ and $h$ be sectorized, translation invariant functions on $\fY_{\ell,\ell}$, $\fY_{\ell,\ell}$ and $\fY_{i,j}$  respectively. Let 
$\mu\in\{1,2\}$, $\mu'\in\{3,4\}$ and
$$
g=(g_1\bullet\cC^{(\ell)}\bullet g_2)^f
$$
\Item a) If $\de_\li+\de_\ci\in\De$, then
$$\eqalign{
\big\|g\bullet \cC^{[i,j]}_{\tp}\bullet h
               \big\|_{\ell,j}^{(\de_\li,\de_\ci,\de_\ri)}
&\le\const\sqrt{\fl_\ell}\ \big|g_1\big|_{\ell}^{\[\de_\li+\de_\ci\]}
\big|g_2\big|_{\ell}^{\[\de_\li+\de_\ci\]}
\big|h\big|_{i,j}^{[0,\de_\ci,\de_\ri]}\cr
\big\|g\bullet \cC^{[i,j]}_{\bt}\bullet h
            \big\|_{\ell,j}^{(\de_\li,\de_\ci,\de_\ri)}
&\le\const\sqrt{\fl_\ell}\ \big|g_1\big|_{\ell,\ell}^{\[\de_\li+\de_\ci\]}
\big|g_2\big|_{\ell}^{\[\de_\li+\de_\ci\]}
\big|h\big|_{i,j}^{[0,\de_\ci,\de_\ri]}\cr
}$$
\Item b) 
Let $\be_1+\be_2+\be_3=\de_\ci$. If 
$\de_\li+\be_1\in\De$, then
$$\eqalign{
&\sfrac{1}{M^{|\de_\ci| j}}\big\|\rD_{\mu;3}^{\be_1}g\bullet 
\rD_{1;3}^{\be_2}\cC^{[i,j]}_{\md}
\bullet \rD_{1;\mu'}^{\be_3}h
\big\|_{\ell,j}^{(\de_\li,0,\de_\ri)}\cr
&\hskip.5in\le\sfrac{\const}{M^{|\be_1|(j-\ell)}}\sqrt{\fl_\ell}\cases{
(j-i+1)
\big|g_1\big|_{\ell}^{\[\de_\li+\be_1\]}
\big|g_2\big|_{\ell}^{\[\de_\li+\be_1\]}
\big\|h\big\|_{i,j}^{(0,\be_3,\de_\ri)}& if $\be_2=0$ \cr
\noalign{\vskip.1in}
\phantom{(j-i+1)}
\big|g_1\big|_{\ell}^{\[\de_\li+\be_1\]}
\big|g_2\big|_{\ell}^{\[\de_\li+\be_1\]}
\big\|h\big\|_{i,j}^{(0,\be_3,\de_\ri)}& if $\be_2\ne 0$ \cr
}}$$
If $\be_2=0$ and
$\de_\li+\be_1+\al\in\De$ for all $|\al|\le 3$, then
$$\eqalign{
&\sfrac{1}{M^{|\de_\ci| j}}\big\|\rD_{\mu;3}^{\be_1}g\bullet 
\rD_{1;3}^{\be_2}\cC^{[i,j]}_{\md}
\bullet \rD_{1;\mu'}^{\be_3}h
\big\|_{\ell,j}^{(\de_\li,0,\de_\ri)}\cr
&\hskip.5in\le\sfrac{\const}{M^{|\be_1|(j-\ell)}}\ i\ \sqrt{\fl_\ell}
\max\limits_{\al_\upl,\al_\dnl,\al_\li\in\bbbn_0\times\bbbn_0^2
            \atop|\al_\upl|+|\al_\dnl|+|\al_\li|\le 3} \big|g_1\big|_{\ell}^{\[\de_\li+\be_1+\al_\upl\]} \big|g_2\big|_{\ell}^{\[\de_\li+\be_1+\al_\dnl\]}
\big|h\big|_{i,j}^{[\al_\li,\be_3,\de_\ri]}
}$$
}\goodbreak
\prf a) We consider the case of $\tp$.  By Leibniz, 
$$\eqalign{
\rD_{1;\mu'}^{\de_\ci}\big(g\bullet \cC^{[i,j]}_{\tp}\bullet h\big)
&=\sum_{\be_1,\be_2,\be_3\in\bbbn_0\times\bbbn_0^2\atop \be_1+\be_2+\be_3=\de_\ci}
{\tst {\de_\ci\choose\be_1,\be_2,\be_3}}
\rD_{1;3}^{\be_1}(g_1\bullet\cC^{(\ell)}\bullet g_2)^f\bullet \rD_{1;3}^{\be_2}\cC^{[i,j]}_{\tp}
\bullet \rD_{1;\mu'}^{\be_3}h\cr
\rD_{2;\mu'}^{\de_\ci}\big(g\bullet \cC^{[i,j]}_{\tp}\bullet h\big)
&=\sum_{\be_1,\be_2,\be_3\in\bbbn_0\times\bbbn_0^2\atop \be_1+\be_2+\be_3=\de_\ci}
{\tst {\de_\ci\choose\be_1,\be_2,\be_3}}
\rD_{2;3}^{\be_1}(g_1\bullet\cC^{(\ell)}\bullet g_2)^f\bullet \rD_{1;3}^{\be_2}\cC^{[i,j]}_{\tp}
\bullet \rD_{1;\mu'}^{\be_3}h\cr
}$$
Substitute 
$\cC^{(\ell)}=\cD^{(\ell)}_{0,\upl}+\cD^{(\ell)}_{0,\dnl}$.  
We consider the case of $\upl$. Then
$$\eqalign{
&\rD_{1;3}^{\be_1}(g_1\bullet\cD^{(\ell)}_{0,\upl}\bullet g_2)^f\bullet \rD_{1;3}^{\be_2}\cC^{[i,j]}_{\tp}
\bullet \rD_{1;\mu'}^{\be_3}h
=(\rD_{1;2}^{\be_1}g_1\bullet\cD^{(\ell)}_{0,\upl}\bullet g_2)^f\bullet \rD_{1;3}^{\be_2}\cC^{[i,j]}_{\tp}
\bullet \rD_{1;\mu'}^{\be_3}h
\cr
&\rD_{2;3}^{\be_1}(g_1\bullet\cD^{(\ell)}_{0,\upl}\bullet g_2)^f\bullet \rD_{1;3}^{\be_2}\cC^{[i,j]}_{\tp}
\bullet \rD_{1;\mu'}^{\be_3}h
=\big(\rD_{3;2}^{\be_1}(g_1\bullet\cD^{(\ell)}_{0,\upl}
\bullet g_2)\big)^f\bullet \rD_{1;3}^{\be_2}\cC^{[i,j]}_{\tp}
\bullet \rD_{1;\mu'}^{\be_3}h\cr
&\hskip.5in=(-1)^{|\be_1|}\hskip-20pt
\sum_{\ga_1,\ga_2,\ga_3\in\bbbn_0\times\bbbn_0^2\atop \ga_1+\ga_2+\ga_3=\be_1}M^{|\ga_2|\ell}
{\tst {\be_1\choose\ga_1,\ga_2,\ga_3}}
\big(\rD_{2;3}^{\ga_1}g_1\bullet\cD^{(\ell)}_{\ga_2,\upl}
\bullet \rD_{1;3}^{\ga_3}g_2\big)^f\bullet \rD_{1;3}^{\be_2}\cC^{[i,j]}_{\tp}
\bullet \rD_{1;\mu'}^{\be_3}h\cr
}$$
The $\sfrac{1}{M^{|\de_\ci| j}}\big\|\ \cdot\ \big\|_{\ell,j}^{(\de_\li,0,\de_\ri)}$
norm of each term is bounded by Theorem \dbbound.b.

\Item b) 
As above, we must estimate the $\sfrac{1}{M^{|\de_\ci| j}}\big\|\ \cdot\
\big\|_{\ell,j}^{(\de_\li,0,\de_\ri)}$ norm of terms like
$$
\big(\rD_{1;2}^{\be_1}g_1\bullet\cD^{(\ell)}_{0,\upl}
\bullet g_2\big)^f\bullet \rD_{1;3}^{\be_2}\cC^{[i,j]}_{\md}
\bullet \rD_{1;\mu'}^{\be_3}h
$$
and
$$
M^{|\ga_2|\ell}
\big(\rD_{2;3}^{\ga_1}g_1\bullet\cD^{(\ell)}_{\ga_2,\upl}
\bullet \rD_{1;3}^{\ga_3}g_2\big)^f\bullet \rD_{1;3}^{\be_2}\cC^{[i,j]}_{\md}
\bullet \rD_{1;\mu'}^{\be_3}h
$$
with $\ga_1+\ga_2+\ga_3=\be_1$.
This is done using Theorem \dbbound. (In the last case, we write 
$\cC^{[i,j]}_\md=\cC^{[i,j]}-\cC^{[i,j]}_\tp-\cC^{[i,j]}_\bt$.)
\endproof
\goodbreak

\corollary{\STM\bdbbound}{
 Let $1\le \ell\le i \le j$, $1\le r\le j$ and
$\de_\li, \de_\ci, \de_\ri\in\De$. 
 Let $\mu\in\{1,2\}$ and $\mu'\in\{3,4\}$.
Let $h$ be a sectorized, translation invariant function on  $\fY_{i,r}$
and let $h'=h_{\Si_i,\Si_j}$ be its resectorization 
as in Definition \defLADresectoriz.i.
\Item a) Let $g$ be a sectorized, translation invariant function on $\fY_{\ell,i}$. Then
$$\eqalign{
\sfrac{1}{M^{j|\de_\ci|}}\big\|g\bullet\cC^{[i,j]}_\md\bullet \rD^{\de_\ci}_{\mu;\mu'}
 h'\big\|_{\ell,j}^{(\de_\li,0,\de_\ri)}
&\le\const\hskip-7pt \max_{\al_\ri,\al_\li\in\bbbn_0\times\bbbn_0^2
            \atop|\al_\ri|+|\al_\li|\le 3}
\big|g\big|_{\ell,i}^{[\de_\li,0,\al_\ri]}
\Big(\sfrac{j-i+1}{M^{j-i}}\big|h\big|_{i,r}^{[0,\de_\ci,\de_\ri]}
+i\ \big|h\big|_{i,r}^{[\al_\li,0,0]}\Big)\cr
}$$
\Item b) Let $g_1$ and $ g_2$  be sectorized, translation invariant functions on $\fY_{\ell,\ell}$. If $\de_\li+\al\in\De$ for all $|\al|\le 3$, then
$$\eqalign{
&\sfrac{1}{M^{j|\de_\ci|}}\big\|(g_1\bullet\cC^{(\ell)}\bullet g_2)^f
\bullet\cC^{[i,j]}_\md\bullet \rD^{\de_\ci}_{\mu;\mu'}
 h'\big\|_{\ell,j}^{(\de_\li,0,\de_\ri)}\cr
&\hskip0.5in\le\const \sqrt{\fl_\ell}
\max_{\al_\upl,\al_\dnl,\al_\li\in\bbbn_0\times\bbbn_0^2
            \atop|\al_\upl|+|\al_\dnl|+|\al_\li|\le 3}
 \big|g_1\big|_{\ell}^{\[\de_\li+\al_\upl\]} 
 \big|g_2\big|_{\ell}^{\[\de_\li+\al_\dnl\]}
\Big(\sfrac{j-i+1}{M^{j-i}}\big|h\big|_{i,r}^{[0,\de_\ci,\de_\ri]}
+i\ \big|h\big|_{i,r}^{[\al_\li,0,0]}\Big)\cr
}$$

}
\prf We prove part a. 
First suppose that $h$ is a function on 
$\fY_{\Si_i}^2\times\big((\bbbr\times\bbbr^2)\times\Si_r\big)^2$. Then,
 for $s_3,s_4\in\Si_j$,
$$
h'\big(\,\cdot\,,\,\cdot\,,(\,\cdot\,,s_3),(\,\cdot\,,s_4)\big)
=h\bullet\Big(\hat\chi_{s_3}\otimes\hat\chi_{s_4}^t\Big)
$$
 We have
$$
\sfrac{1}{M^{j|\de_\ci|}}\big\|g\bullet\cC^{[i,j]}_\md\bullet \rD^{\de_\ci}_{\mu;\mu'}h'\big\|_{\ell,j}^{(\de_\li,0,\de_\ri)}
=\sfrac{1}{M^{j(|\de_\ci|+|\de_\ri|)}}\big\|g\bullet\cC^{[i,j]}_\md\bullet \rD^{\de_\ci}_{\mu;\mu'}\rD^{\de_\ri}_{3;4} h'\big\|_{\ell,j}^{(\de_\li,0,0)}
$$
Apply Leibniz to 
$\ 
\rD^{\de_\ci}_{\mu;\mu'}\rD^{\de_\ri}_{3;4}\big( h\bullet (\hat\chi_{s_3}\otimes
\hat\chi_{s_4}^t)\big),$ yielding a sum of terms of the form 
$$
\rD^{\be_1}_{\mu;\mu'}\rD^{\ga_1}_{3;4} h\bullet 
\rD^{\be_2+\ga_2}_{1;3}\rD^{\be_3+\ga_3}_{2;4}
(\hat\chi_{s_3}\otimes\hat\chi_{s_4}^t)
$$ 
with $\be_1+\be_2+\be_3=\de_\ci$ and $\ga_1+\ga_2+\ga_3=\de_\ri$.
If $\be_1+\ga_1=0$ we apply Theorem \bbound
a and otherwise we apply Theorem \bbound
c. The Lemma follows from
$$\eqalign{
\big| h\bullet 
\rD^{\be_2+\ga_2}_{1;3}\rD^{\be_3+\ga_3}_{2;4}
(\hat\chi_{s_3}\otimes\hat\chi_{s_4}^t)\big)\big|_{i,j}^{[\al_\li,0,0]}
&\le M^{j(|\de_\ci+\de_\ri|)}\big|h\big|_{i,r}^{[\al_\li,0,0]}\cr
\big\|\rD^{\be_1}_{\mu;\mu'}\rD^{\ga_1}_{3;4} h\bullet 
\rD^{\be_2+\ga_2}_{1;3}\rD^{\be_3+\ga_3}_{2;4}
(\hat\chi_{s_3}\otimes\hat\chi_{s_4}^t)\big)\big\|^{(0,0,0)}_{i,j}
&\le M^{j(|\be_2+\ga_2+\be_3+\ga_3|)}
M^{|\be_1|i+|\ga_1|r}
\big|h\big|_{i,r}^{[0,\de_\ci,\de_\ri]}\cr
&\hskip-0.5in\le M^{j(|\de_\ci+\de_\ri|-1)}M^{i}\big|h\big|_{i,r}^{[0,\de_\ci,\de_\ri]}
\quad\hbox{if $|\be_1+\ga_1|\ge 1$}\cr
}$$
which is proven in the same way as Lemma \lemLADresectornorm\ and, 
in particular, uses (\eqnLADchisbnd) with $r=j$.
If one of the third or fourth arguments of $h$ lie in momentum space,
$\bbbm$, the argument is similar, except that the corresponding
$\hat\chi_{s_3}$ or $\hat\chi_{s_4}$ is omitted.
The proof of part b is similar with Theorem \dbbound\ 
used in place of Theorem \bbound.
\endproof
\vskip.25in
\noindent{\bf Proof  of Theorem \thmLADmodcompLadder} (assuming Theorems \bbound\ and \dbbound): \hfill\break
\noindent
Let $\de\in\De$.
By the hypothesis of the Theorem and Remark \remLADscaledtonormdomain.ii,
there is a constant $c_F$ such that
$$
\max_{\de\in\De}|F_C^{(i)}|^{\[\de\]}_{i} \le \sfrac{c_F}{M^{\veps i}}\rho\qquad
\max_{\de\in\De}|F_S^{(i)}|^{\[\de\]}_{i} \le \sfrac{c_F}{M^{\veps i}} \rho
\EQN\eqnLADFbnd$$
We prove by induction on $j$ that
$$
\max_{\de\in\De}|\cL_C^{(i)}|^{\[\de\]}_{i-1} \le c_\cL\ \rho^2\qquad
\max_{\de\in\De}|\cL_S^{(i)}|^{\[\de\]}_{i-1} \le c_\cL\ \rho^2\qquad
\hbox{for all $i\le j$}
\EQN\eqnLADbnd$$
with a constant $c_\cL$, independent of $j$.
See Remark \remLADscaledtonormdomain.
By construction
 $\cL^{(0)}=\cL^{(1)}=\cL^{(2)}=0$. Now assume that (\eqnLADbnd) holds for some $j\ge2$. We prove that
$$
\max_{\de\in\De}|\cL_S^{(j+1)}|^{\[\de\]}_{j} \le c_\cL\ \rho^2
\EQN\eqnLADladindbnd$$
The bound on $\cL_C^{(j+1)}$ is similar.

For $\,i\le j\,$ we have, by Corollary \corLADaltcmpladders.ii,
$$\eqalign{
 L_C^{(i)}
        &=G_{C,1}^{(i-1)}\bullet{\cal C}^{(i-1)}\bullet G_{C,2}^{(i-1)}\cr
 L_S^{(i)} 
       &= G_{S,1}^{(i-1)} \bullet {\cal C}^{(i-1)} \bullet G_{S,2}^{(i-1)}\cr
}$$
with
$$\eqalign{
G_{C,1}^{(i-1)} &=\sum_{i'=2}^{i-1} F_{C\,\Si_{i-1}}^{(i')} 
  +\half\cL_{C\,\Si_{i-1}}^{(i-1)\,f }
  +\sfrac{3}{2}\cL_{S\,\Si_{i-1}}^{(i-1)\,f }
  +\cL_{C\,\Si_{i-1}}^{(i-1)}\cr 
G_{C,2}^{(i-1)} &=\sum_{i'=2}^{i-1} F_{C\,\Si_{i-1}}^{(i')} 
  +\half\cL_{C\,\Si_{i-1}}^{(i-1)\,f }
  +\sfrac{3}{2}\cL_{S\,\Si_{i-1}}^{(i-1)\,f }
  +\cL_{C}^{(i)}\cr 
G_{S,1}^{(i-1)} &=\sum_{i'=2}^{i-1} F_{S\,\Si_{i-1}}^{(i')} 
  +\half\cL_{C\,\Si_{i-1}}^{(i-1)\,f }
  -\half\cL_{S\,\Si_{i-1}}^{(i-1)\,f }
  +\cL_{S\,\Si_{i-1}}^{(i-1)}\cr 
G_{S,2}^{(i-1)} &=\sum_{i'=2}^{i-1} F_{S\,\Si_{i-1}}^{(i')} 
  +\half\cL_{C\,\Si_{i-1}}^{(i-1)\,f }
  -\half\cL_{S\,\Si_{i-1}}^{(i-1)\,f }
  +\cL_{S}^{(i)}\cr 
}$$
The hypotheses (\eqnLADFbnd) on $\vec F$ and the induction hypotheses 
(\eqnLADbnd) imply, via Lemma \lemLADresectornorm,
 that, when $\rho$ is small enough and $M^\veps$ is large enough,
$$
\max_{\de\in\De}\big|G_{C,\nu}^{(i-1)}\big|^{\[\de\]}_{i-1}
 \le c_F\rho \qquad
\max_{\de\in\De}\big|G_{S,\nu}^{(i-1)}\big|^{\[\de\]}_{i-1}
 \le c_F\rho  
\EQN\eqnLADGbnd$$
for $i\le j,\ \nu=1,2$.

\remark{\STM\remLADLbnd}{
For $i\le j$
$$
\max_{\de\in\De}\big|L_C^{(i)}\big|^{\[\de\]}_{i-1}\le \const c_F^2\rho^2\qquad
\max_{\de\in\De}\big|L_S^{(i)}\big|^{\[\de\]}_{i-1}\le \const c_F^2\rho^2
$$
where $\const$ is $(2+3^{r_0+2r_e})$ times the constant of 
Corollary \bboundcor.
}
\prf
We prove the Remark for $L_C^{(i)}$. Fix $(\de_\li,\de_\ci,\de_\ri)\in\vec\De$.
Decomposing
$$
{\cal C}^{[i-1,i-1]}=\big({\cal C}_\tp^{[i-1,i-1]}+{\cal C}_\bt^{[i-1,i-1]}\big)
+{\cal C}_\md^{[i-1,i-1]}
$$ 
and applying Corollary \bboundcor, parts a and b respectively, with $\ell,i,j$
 all replaced by $i-1$, we have
$$\eqalign{
\big|G_{C,1}^{(i-1)} \bullet \cC^{[i-1,i-1]} \bullet G_{C,2}^{(i-1)}\big|_{i-1,i-1}^{[\de_\li,\de_\ci,\de_\ri]}
&\le\const \big((1+1)+3^{|\de_c|}\big)
\big|G_{C,1}^{(i-1)}\big|_{i-1,i-1}^{[\de_\li,\de_\ci,0]}\ 
\big|G_{C,2}^{(i-1)}\big|_{i-1,i-1}^{[0,\de_\ci,\de_\ri]}\cr
&\le\const (2+3^{r_0+2r_e}) c_F^2\rho^2
}$$
\endproof
Set, for each $i\ge 1$, 
$$
\fv_i=i^2\max\big\{\sqrt{\fl_i},\sfrac{1}{M^{\veps i}}\big\}
$$
and
$$
K^{(i)}=F^{(i)}_S+\half L^{(i)\,f}_{C\,\Si_i}-\half L^{(i)\,f}_{S\,\Si_i}
$$ 
Then, by Corollary \corLADaltcmpladders.i,
$$
{\cal L}^{(j+1)}_S
= \sum_{\ell=1}^\infty \  \sum_{i_1,\cdots,i_{\ell+1}=2}^{j} \  
\Big[K^{(i_1)}  \bullet {\cal C}^{[\max \{i_1,i_2\},j]} \bullet
K^{(i_2)} \bullet  \cdots \bullet 
{\cal C}^{[\max \{i_\ell,i_{\ell+1}\},j]} \bullet
K^{(i_{\ell+1})}\Big]_{\Si_j}
$$
We put the main estimates required to complete the proof of Theorem
 \thmLADmodcompLadder\ in

\lemma{\STM\ladderinduction}{ 
Let $\ell\ge 1$ and $\,i_1,\cdots,i_{\ell+1}\le j\,$. 
\Item a) For $|\al|\le 3$ and $\de\in\De$,
$$
\Big| 
K^{(i_1)} \bullet 
{\cal C}^{[\max \{i_1,i_2\},j]} \bullet 
K^{(i_2)} \bullet  
 \cdots \bullet 
K^{(i_{\ell+1})} 
\Big|_{i_1,j}^{[\al,0,\de]} 
\le \cst{\ell}{} \big(c_F\rho \big)^{\ell+1}\ \sfrac{i_{\ell+1}}{i_1}\ 
\fv_{i_1}\cdots \fv_{i_{\ell}} 
$$
\Item b) 
For $(\de_\li,0,\de_\ri)\in\vec\De$
$$\eqalign{
\Big|
K^{(i_1)} \bullet 
{\cal C}^{[\max \{i_1,i_2\},j]} \bullet &
K^{(i_2)} \bullet  
 \cdots \bullet 
K^{(i_{\ell+1})} 
\Big|_{j,j}^{[\de_\li,0,\de_\ri]} 
\le \cst{\ell}{} \big(c_F\rho \big)^{\ell+1}\fv_{i_2}\cdots \fv_{i_{\ell}} 
\min \big\{ \fv_{i_1},\fv_{i_{\ell+1}}\big\}\cr  
}$$
\Item c) 
For $0\ne\de\in\De$, $\mu\in\{1,2\}$ and $\mu'\in\{3,4\}$,
there are sectorized, translation invariant functions $k',\,k''$  on $\fY_{i_1,j}$ such that
$$
\sfrac{1}{M^{j|\de|}} \rD^{\de}_{\mu;\mu'} \Big[ 
\big(K^{(i_1)} \bullet 
{\cal C}^{[\max \{i_1,i_2\},j]} \bullet  \cdots \bullet 
K^{(i_{\ell+1})}\big)_{\Si_{i_1},\Si_j}\Big]
= k'+k''
$$
and, for all $|\al|\le 3$ and all $\ga$ with $\ga+\de\in\De$,
$$\eqalign{
\big| k' \big|_{i_1,j}^{[\al,0,\ga]} &
\le \cst{\ell}{} \big(c_F\rho \big)^{\ell+1}\ \sfrac{i_{\ell+1}}{i_1}\  
 \fv_{i_1}\cdots \fv_{i_{\ell}}  \cr
\big| k'' \big|_{i_1,j}^{[0,0,\ga]} &
\le \sfrac{j-i_1+1}{M^{j-i_1}}\, \cst{\ell}{} \big(c_F\rho \big)^{\ell+1}\ 
\sfrac{i_{\ell+1}}{i_1}\  
 \fv_{i_1}\cdots \fv_{i_{\ell}} \cr
}$$
\Item d) 
For $(\de_\li,\de_\ci,\de_\ri)\in\vec\De$ with $|\de_\ci|\ge 1$,
$$\eqalign{
\Big|  K^{(i_1)} \bullet 
{\cal C}^{[\max \{i_1,i_2\},j]} \bullet 
 \cdots \bullet 
K^{(i_{\ell+1})}
 \Big|_{j,j}^{[\de_\li,\de_\ci,\de_\ri]}
\le \cst{\ell}{} \big(c_F\rho \big)^{\ell+1}\ 
\fv_{i_2}\cdots \fv_{i_{\ell}}\,
\min \big\{ \fv_{i_1},\fv_{i_{\ell+1}}\big\}\cr
}$$
}

\noindent
 Set $i=\max \{i_1,i_2\}\,$ and write,
for $(\al_\li,\al_\ci,\al_\ri)\in\vec\De$ and $\al_\upl,\al_\dnl\in\De$,
$$\eqalign{
q(\al_\li,\al_\ci,\al_\ri;\al_\upl,\al_\dnl)
=\ &
i\,\big|F^{(i_1)}_S\big|_{i_1,i_1}^{[\al_\li,\al_\ci,\al_\ri]}\cr
&\!\!+i\sqrt{\fl_{i_1-1}}\Big[
\big|G^{(i_1-1)}_{C,1}\big|_{i_1-1}^{\[\al_\upl\]}
\big|G^{(i_1-1)}_{C,2}\big|_{i_1-1}^{\[\al_\dnl\]}
\!+\!\big|G^{(i_1-1)}_{S,1}\big|_{i_1-1}^{\[\al_\upl\]}
\big|G^{(i_1-1)}_{S,2}\big|_{i_1-1}^{\[\al_\dnl\]}\Big]
}$$
By (\eqnLADFbnd) and (\eqnLADGbnd)
$$
q(\al_\li,\al_\ci,\al_\ri;\al_\upl,\al_\dnl)
\le i\sfrac{c_F\rho}{M^{\veps i_1}}+ i\, 2c_F^2\rho^2\sqrt{\fl_{i_1-1}}
\le  2 c_F \rho  \,\fv_{i_1}\,\sfrac{i_2}{i_1}
\EQN\eqnLADqbnd$$
for $\rho$ sufficiently small. The proof of Lemma \ladderinduction\ follows
\lemma{\STM\lemCombinedCors}{
 Let $1\le i_1,i_2\le i \le j$, $1\le r\le j$ and
$\de_\li, \de_\ci, \de_\ri\in\De$. 
 Let $\mu\in\{1,2\}$ and $\mu'\in\{3,4\}$ and $\lc\in\{\tp,\bt,\md\}$.
If $\lc\in\{\tp,\md\}$, set $(\nu,\nu')=(1,3)$.
If $\lc=\bt$, set $(\nu,\nu')=(2,4)$.
Let $H$  be a sectorized, translation invariant function on  $\fY_{i_2,r}$.
\Item a) Let $\be_2+\be_3=\de$ with $\be_2\ne 0$
 and either $|\de_\li|\le 3$, $\de+\de_\ri\in\De$ 
or $\de_\li+\de_\ri+\de\in\De$. Then
$$\eqalign{
\sfrac{1}{M^{j|\de|}}\big|K^{(i_1)}\bullet
\rD^{\be_2}_{\nu;\nu'}\cC^{[i,j]}_\lc\bullet 
\rD^{\be_3}_{1;\mu'}H_{\Si_{i_2},\Si_j}\big|_{i_1,j}^{[\de_\li,0,\de_\ri]}
&\le\const\ q({\sst\de_\li,0,0;\de_\li,\de_\li})\  \big|H\big|_{i_2,j}^{[0,\be_3,\de_\ri]}
\cr
}$$
\Item b) Let  $|\de_\li|\le 3$ and $\de+\de_\ri\in\De$. Then
$$\eqalign{
&\sfrac{1}{M^{j|\de|}}\big|K^{(i_1)}\bullet
\cC^{[i,j]}_\lc\bullet 
\rD^{\de}_{1;\mu'}H_{\Si_{i_2},\Si_j}\big|_{i_1,j}^{[\de_\li,0,\de_\ri]}\cr
&\hskip0.5in\le\const
\max_{\al_\upl,\al_\dnl,\al_\li\in\bbbn_0\times\bbbn_0^2
            \atop|\al_\upl|+|\al_\dnl|+|\al_\li|\le 3}
 q({\sst\de_\li,0,\al_\upl+\al_\dnl;\de_\li+\al_\upl,\de_\li+\al_\dnl})\ 
\Big(\big|H\big|_{i_2,j}^{[0,\de,\de_\ri]}
+\big|H\big|_{i_2,j}^{[\al_\li,0,0]}\Big)\cr
}$$
\Item c)  Let $\be_1+\be_2+\be_3=\de$ and  
$\be_1+\de_\li,\de,\de_\ri\in\De$. Then
$$\eqalign{
&\sfrac{1}{M^{j|\de|}}\big|\rD^{\be_1}_{\mu;3}K^{(i_1)}\bullet
\rD^{\be_2}_{\nu;\nu'}\cC^{[i,j]}_\lc\bullet 
\rD^{\be_3}_{1;\mu'}H_{\Si_{i_2},\Si_j}\big|_{i_1,j}^{[\de_\li,0,\de_\ri]}\cr
&\hskip1.5in\le\const\ 
\sfrac{j-i+1}{M^{|\be_1|(j-i_1)}}\ 
q({\sst\de_\li,\be_1,0;\de_\li+\be_1,\de_\li+\be_1})\  \big|H\big|_{i_2,j}^{[0,\be_3,\de_\ri]}\cr
}$$

}
\prf Subbing in the definition of $K^{(i_1)}$ and applying Lemma
\lemLADresectornorm,
$$\eqalign{
&\big|\rD^{\be_1}_{\mu;3} K^{(i_1)}\bullet
\rD^{\be_2}_{\nu;\nu'}\cC^{[i,j]}_\lc\bullet 
\rD^{\be_3}_{1;\mu'}H_{\Si_{i_2},\Si_j}\big|_{i_1,j}^{[\de_\li,0,\de_\ri]}\cr
&\hskip.5in\le \big|\rD^{\be_1}_{\mu;3}F^{(i_1)}_S\bullet
\rD^{\be_2}_{\nu;\nu'}\cC^{[i,j]}_\lc\bullet 
\rD^{\be_3}_{1;\mu'}H_{\Si_{i},\Si_j}\big|_{i_1,j}^{[\de_\li,0,\de_\ri]}\cr
&\hskip.9in+\const 
\big|\rD^{\be_1}_{\mu;3}\big(G_{C,1}^{(i_1-1)}\bullet{\cal C}^{(i_1-1)}\bullet G_{C,2}^{(i_1-1)}\big)^f\bullet
\rD^{\be_2}_{\nu;\nu'}\cC^{[i,j]}_\lc\bullet 
\rD^{\be_3}_{1;\mu'}H_{\Si_{i},\Si_j}\big|_{i_1-1,j}^{[\de_\li,0,\de_\ri]}\cr
&\hskip.9in+\const 
\big|\rD^{\be_1}_{\mu;3}\big(G_{S,1}^{(i_1-1)}\bullet{\cal C}^{(i_1-1)}\bullet G_{S,2}^{(i_1-1)}\big)^f\bullet
\rD^{\be_2}_{\nu;\nu'}\cC^{[i,j]}_\lc\bullet 
\rD^{\be_3}_{1;\mu'}H_{\Si_{i},\Si_j}\big|_{i_1-1,j}^{[\de_\li,0,\de_\ri]}\cr
}$$
We were able to replace $H_{\Si_{i_2},\Si_j}$ by $H_{\Si_{i},\Si_j}$
without changing the $\bullet$ products because $i\ge i_2$ and 
$\cC^{[i,j]}_\lc$ is supported in the $i^{\rm th}$ neighbourhood.
\Item a)
By  Corollary \bboundcor.b with $\be_1=0$ and $\de_\ci=\de$
(and, when $\lc=\tp,\bt$, Theorem \bbound.b with $h=\sfrac{1}{M^{j|\be_3|}}\rD^{\be_3}_{1;\mu'}H_{\Si_{i},\Si_j}$)
$$\eqalign{
\sfrac{1}{M^{j|\de|}}\big|F^{(i_1)}_S\bullet
\rD^{\be_2}_{\nu;\nu'}\cC^{[i,j]}_\lc\bullet 
\rD^{\be_3}_{1;\mu'}H_{\Si_{i},\Si_j}\big|_{i_1,j}^{[\de_\li,0,\de_\ri]}
\le \const \big|F^{(i_1)}_S\big|_{i_1}^{[\de_\li,0,0]}
\big|H\big|_{i_2,j}^{[0,\be_3,\de_\ri]}
}$$
provided $\de_\li,\de_\ri,\de\in\De$. For $T=C,S$,
$$\eqalign{
&\sfrac{1}{M^{j|\de|}}\big|\big(G_{T,1}^{(i_1-1)}\bullet{\cal C}^{(i_1-1)}\bullet G_{T,2}^{(i_1-1)}\big)^f\bullet
\rD^{\be_2}_{\nu;\nu'}\cC^{[i,j]}_\lc\bullet 
\rD^{\be_3}_{1;\mu'}H_{\Si_{i},\Si_j}\big|_{i_1-1,j}^{[\de_\li,0,\de_\ri]}\cr
&\hskip1in\le\const\sqrt{\fl_{i_1-1}}\ 
\big|G_{T,1}^{(i_1-1)}\big|_{i_1-1}^{\[\de_\li\]}
\big|G_{T,2}^{(i_1-1)}\big|_{i_1-1}^{\[\de_\li\]}
\big|H\big|_{i_2,j}^{[0,\be_3,\de_\ri]}
}$$
by  Corollary \dbboundcor.b with $\ell=i_1-1$, $\be_1=0$ and $\de_\ci=\de$
(and Theorem \dbbound.b when $\lc=\tp,\bt$),
provided $\de_\li,\de_\ri,\de\in\De$.
\goodbreak
\Item b)
By  Corollary \bdbbound.a with $\de_\ci=\de$, $\ell=i_1$ and $r=j$
 (and Corollary \bboundcor.a when $\lc=\tp,\bt$)
$$\eqalign{
&\sfrac{1}{M^{j|\de|}}\big|F^{(i_1)}_S\bullet\cC^{[i,j]}_\lc\bullet 
\rD^{\de}_{1;\mu'}H_{\Si_{i},\Si_j}\big|_{i_1,j}^{[\de_\li,0,\de_\ri]}\cr
&\hskip1in\le\const  
\max_{\al_\li,\al_\ri\in\bbbn_0\times\bbbn_0^2
            \atop|\al_\li|+|\al_\ri|\le 3}
         \big|F^{(i_1)}_S\big|_{i_1}^{[\de_\li,0,\al_\ri]}
\Big(\big|H\big|_{i_2,j}^{[0,\de,\de_\ri]}
+i\big|H\big|_{i_2,j}^{[\al_\li,0,0]}\Big)
}$$
provided $\de_\li,\de_\ri,\de\in\De$. For $T=C,S$,
$$\eqalign{
&\sfrac{1}{M^{j|\de|}}\big|\big(G_{T,1}^{(i_1-1)}\bullet{\cal C}^{(i_1-1)}\bullet G_{T,2}^{(i_1-1)}\big)^f\bullet\cC^{[i,j]}_\lc\bullet 
\rD^{\de}_{1;\mu'}H_{\Si_{i},\Si_j}\big|_{i_1-1,j}^{[\de_\li,0,\de_\ri]}\cr
&\hskip.2in\le\const \sqrt{\fl_{i_1-1}} \hskip-6pt
\max_{\al_\upl,\al_\dnl,\al_\li\in\bbbn_0\times\bbbn_0^2
            \atop|\al_\upl|+|\al_\dnl|+|\al_\li|\le 3}
\big|G_{T,1}^{(i_1-1)}\big|_{i_1-1}^{\[\de_\li+\al_\upl\]}
\big|G_{T,2}^{(i_1-1)}\big|_{i_1-1}^{\[\de_\li+\al_\dnl\]}
\Big(\big|H\big|_{i_2,j}^{[0,\de,\de_\ri]}
+i\big|H\big|_{i_2,j}^{[\al_\li,0,0]}\Big)
}$$
by  Corollary \bdbbound.b with $\de_\ci=\de$ and $\ell=i_1-1$
 (and Corollary \dbboundcor.a when $\lc=\tp,\bt$) provided $\de_\li+\al\in\De$
for all $|\al|\le 3$ (which is certainly the case when $|\de_\li|\le 3$)
and $\de_\ri,\de\in\De$.
\Item c)
By  Corollary \bboundcor.b, with $\de_\ci=\de$ (and Theorem \bbound.b 
with $\be=\be_2$, $g=\sfrac{1}{M^{i_1|\be_1|}}\rD^{\be_1}_{\mu;3}F^{(i_1)}_S$
and $h=\sfrac{1}{M^{j|\be_3|}}\rD^{\be_3}_{1;\mu'}H_{\Si_{i},\Si_j}$,
when $\lc=\tp,\bt$)
$$\eqalign{
&\sfrac{1}{M^{j|\de|}}\big|\rD^{\be_1}_{\mu;3}F^{(i_1)}_S\bullet
\rD^{\be_2}_{\nu;\nu'}\cC^{[i,j]}_\lc\bullet 
\rD^{\be_3}_{1;\mu'}H_{\Si_{i},\Si_j}\big|_{i_1,j}^{[\de_\li,0,\de_\ri]}\cr
&\hskip1in\le\const  
\sfrac{j-i+1}{M^{|\be_1|(j-i_1)}}         
\big|F^{(i_1)}_S\big|_{i_1}^{[\de_\li,\be_1,0]}
\big|H\big|_{i_2,j}^{[0,\be_3,\de_\ri]}
}$$
provided $\de_\li,\de_\ri,\de\in\De$.
For $T=C,S$,
$$\eqalign{
&\sfrac{1}{M^{j|\de|}}
\big|\rD^{\be_1}_{\mu;3}\big(G_{T,1}^{(i_1-1)}\bullet{\cal C}^{(i_1-1)}\bullet G_{T,2}^{(i_1-1)}\big)^f\bullet
\rD^{\be_2}_{\nu;\nu'}\cC^{[i,j]}_\lc\bullet 
\rD^{\be_3}_{1;\mu'}H_{\Si_{i},\Si_j}\big|_{i_1-1,j}^{[\de_\li,0,\de_\ri]}\cr
&\hskip1in\le\const  
\sfrac{j-i+1}{M^{|\be_1|(j-i_1)}} \sqrt{\fl_{i_1-1}}
\big|G_{T,1}^{(i_1-1)}\big|_{i_1-1}^{\[\de_\li+\be_1\]}
\big|G_{T,2}^{(i_1-1)}\big|_{i_1-1}^{\[\de_\li+\be_1\]}
\big|H\big|_{i_2,j}^{[0,\be_3,\de_\ri]}
}$$
by  Corollary \dbboundcor.b with $\de_\ci=\de$ and $\ell=i_1-1$ 
(and Theorem \dbbound.b when $\lc=\tp,\bt$ -- the $\rD^{\be_1}_{\mu;3}$ 
is treated as in the proof of Corollary \dbboundcor.a)
provided $\de_\li+\be_1,\de_\ri,\de\in\De$.
\endproof

\proof{of Lemma \ladderinduction}
The proof is by induction on $\ell$.
We begin the induction at $\ell=1$. Observe that, by (\eqnLADFbnd),
Remark \remLADLbnd\ and Lemma \lemLADresectornorm,
$$
\max_{\de\in\De}|K^{(i_2)}|^{\[\de\]}_{i_2} \le c_F\rho
\EQN\eqnLADKbnd$$
for all $\de\in\De$, if $\rho$ is sufficiently small.

\noindent
a) 
By Lemma \lemCombinedCors.b, with $\de=0$, $\de_\li=\al$ and $\de_\ri=\de$,
(\eqnLADqbnd), (\eqnLADKbnd) and Lemma \lemLADresectornorm,
$$\eqalign{
&\Big|K^{(i_1)}
 \bullet \cC_\lc^{[i,j]} \bullet K^{(i_2)} \Big|_{i_1,j}^{[\al,0,\de]}\cr
&\hskip.3in\le\const 
\max_{\al_\upl,\al_\dnl,\al_\li\in\bbbn_0\times\bbbn_0^2
            \atop|\al_\upl|+|\al_\dnl|+|\al_\li|\le 3}
q({\sst \al,0,\al_\upl+\al_\dnl;\al+\al_\upl,\al+\al_\dnl})\Big[
\big|K^{(i_2)} \big|_{i_2}^{[0,0,\de]}
+\big|K^{(i_2)} \big|_{i_2}^{[\al_\li,0,0]}
\Big] 
\cr
 &\hskip.3in\le  \const c^2_F \rho^2  \,\fv_{i_1}\,\sfrac{i_2}{i_1}
}$$
for all of $\lc=\md,\tp,\bt$.
Observe that $\al+\al_\upl+\al_\dnl+\al_\li\in\De$, since, by hypothesis, 
$r_e,r_0\ge 6$.

\Item b) By symmetry, we may assume, without loss of generality that
$i_1\ge i_2$. Then $\fv_{i_2}\cdots \fv_{i_{\ell}}\,
\min \big\{\fv_{i_1}\ ,\ \fv_{i_{\ell+1}}\big\}$ reduces to $\fv_{i_1}$.
By Lemma \lemLADresectornorm,
Lemma \lemCombinedCors.c, with $\be_1=\be_2=\be_3=0$, 
(\eqnLADqbnd), (\eqnLADKbnd) and part a of this Lemma with $\ell=1$,
$$\eqalign{
&\Big|K^{(i_1)}
 \bullet {\cal C}^{[i,j]} \bullet K^{(i_2)} \Big|_{j,j}^{[\de_\li,0,\de_\ri]}
\cr
&\hskip.3in\le\sfrac{\const}{M^{j-i}}
\Big|K^{(i_1)}
 \bullet {\cal C}^{[i,j]} \bullet K^{(i_2)} \Big|_{i,j}^{[\de_\li,0,\de_\ri]}
+\const
\Big|K^{(i_1)}
 \bullet {\cal C}^{[i,j]} \bullet K^{(i_2)} \Big|_{i,j}^{[0,0,\de_\ri]}\cr
&\hskip.3in\le\sfrac{\const(j-i+1)}{M^{j-i}}
 q({\sst \de_\li,0,0;\de_\li,\de_\li})
\big|K^{(i_2)} \big|_{i_2}^{[0,0,\de_\ri]}
+\const\Big|K^{(i_1)}
 \bullet {\cal C}^{[i,j]} \bullet K^{(i_2)} \Big|_{i,j}^{[0,0,\de_\ri]}\cr
&\hskip.3in \le \const c^2_F \rho^2  \,\fv_{i_1}\,\sfrac{i_2}{i_1}
\le \const c^2_F \rho^2  \,\fv_{i_1} \cr
}$$

\Item c) 
Substitute $\cC^{[i,j]}=\cC_\tp^{[i,j]}+\cC_\md^{[i,j]}+\cC_\bt^{[i,j]}$
into
$$
\rD^{\de}_{\mu;\mu'} \Big[\big( 
K^{(i_1)} \bullet {\cal C}^{[i,j]} \bullet K^{(i_2)}\big)_{\Si_{i_1},\Si_j}\Big]
=\rD^{\de}_{\mu;\mu'} \Big[
K^{(i_1)} \bullet {\cal C}^{[i,j]} \bullet K^{(i_2)}_{\Si_{i_2},\Si_j}\Big]
$$
and apply Leibniz's rule (Lemma \leibniz) using the routing which gives
$\rD^{\be_2}_{1;3}\cC_\tp^{[i,j]}$,
$\rD^{\be_2}_{1;3}\cC_\md^{[i,j]}$
and $\rD^{\be_2}_{2;4}\cC_\bt^{[i,j]}$. We define $k'$ to be $\sfrac{1}{M^{j|\de|}}$ times the sum of all resulting terms having no
derivatives acting on $K^{(i_1)}$ and  $k''$ to be 
$\sfrac{1}{M^{j|\de|}}$ times the sum of all terms having at least 
one derivative acting on $K^{(i_1)}$. 
Fix any $\al$, $\al'$ and $\ga,\ga'$ obeying, $|\al|\le3$,
$\ga+\de\in\De$ and $\al'+\ga'+\de\in\De$.
We show
$$\eqalign{
\big| k' \big|_{i_1,j}^{[\al,0,\ga]} &
\le \const c^2_F\rho^2\,\fv_{i_1}\,\sfrac{i_2}{i_1} \cr
\big| k'' \big|_{i_1,j}^{[0,0,\ga]} &
\le\const c^2_F \sfrac{j-i_1+1}{M^{j-i_1}}\, \rho^2
 \,\fv_{i_1}\,\sfrac{i_2}{i_1}\cr
}\EQN\eqnliA$$
and
$$
\big|k'\big|_{j,j}^{[\al',0,\ga']}
+\big|k''\big|_{j,j}^{[\al',0,\ga']}\le \const c^2_F \rho^2 \,\fv_{i_1}
\,\sfrac{i_2}{i_1}
\EQN\eqnliB$$
Let $\lc\in\{\md,\tp,\bt\}$. If $\lc\in\{\tp,\md\}$, set $(\nu,\nu')=(1,3)$.
If $\lc=\bt$, set $(\nu,\nu')=(2,4)$.
The contributions to $k'$ and $k''$ coming from $\cC_\lc^{[i,j]}$ are
$$\eqalign{
k'_\lc&=\sfrac{1}{M^{j|\de|}}
\sum_{\be_2,\be_3\in\bbbn_0\times\bbbn_0^2\atop \be_2+\be_3=\de}
{\tst{\de\choose\be_2,\be_3}}
K^{(i_1)}
\bullet \rD_{\nu;\nu'}^{\be_2}\cC^{[i,j]}_{\lc}
\bullet \rD_{1;\mu'}^{\be_3}K^{(i_2)}_{\Si_{i_2},\Si_j}\cr
k''_\lc&=\sfrac{1}{M^{j|\de|}}
\sum_{{\be_1,\be_2,\be_3\in\bbbn_0\times\bbbn_0^2\atop \be_1+\be_2+\be_3=\de}\atop|\be_1|>0}
{\tst{\de\choose\be_1,\be_2,\be_3}}
\rD_{\mu;3}^{\be_1}K^{(i_1)}
\bullet \rD_{\nu;\nu'}^{\be_2}\cC^{[i,j]}_{\lc}
\bullet \rD_{1;\mu'}^{\be_3}K^{(i_2)}_{\Si_{i_2},\Si_j}\cr
}$$

We first bound $k'_\lc$. Fix $\be_2+\be_3=\de$.
First consider $\be_2\ne 0$.
Let $(\de_\li,\de_\ri)=(\al,\ga)$ or $(\al',\ga')$.
By Lemma \lemCombinedCors.a,
$$\eqalign{
\sfrac{1}{M^{j|\de|}}
\Big|K^{(i_1)}\bullet \rD_{\nu;\nu'}^{\be_2}\cC^{[i,j]}_{\lc}
\bullet \rD_{1;\mu'}^{\be_3} K^{(i_2)}_{\Si_{i_2},\Si_j}\Big|_{i_1,j}^{[\de_\li,0,\de_\ri]}
&\le\const
 q({\sst \de_\li,0,0;\de_\li,\de_\li})
 \big|K^{(i_2)}\big|_{i_2}^{[0,\be_3,\de_\ri]}\cr
&\le\const c^2_F\rho^2\fv_{i_1}\,\sfrac{i_2}{i_1}\cr
}$$
Next consider $\be_2= 0$. 
By Lemma \lemCombinedCors.b, with $\de_\li=\al$ and $\de_\ri=\ga$,
$$\eqalign{
&\sfrac{1}{M^{j|\de|}}
\Big|K^{(i_1)}\bullet \rD_{\nu;\nu'}^{\be_2}\cC^{[i,j]}_{\lc}
\bullet \rD_{1;\mu'}^{\be_3} K^{(i_2)}_{\Si_{i_2},\Si_j}\Big|_{i_1,j}^{[\al,0,\ga]}\cr
&\hskip.2in=\sfrac{1}{M^{j|\de|}}
\Big|K^{(i_1)}\bullet \cC^{[i,j]}_{\lc}
\bullet \rD_{1;\mu'}^{\de} K^{(i_2)}_{\Si_{i_2},\Si_j}\Big|_{i_1,j}^{[\al,0,\ga]}\cr
&\hskip.2in
\le\const
\max_{\al_\upl,\al_\dnl,\al_\li\in\bbbn_0\times\bbbn_0^2
            \atop|\al_\upl|+|\al_\dnl|+|\al_\li|\le 3}
 q({\sst \al,0,\al_\upl+\al_\dnl;\al+\al_\upl,\al+\al_\dnl})
\Big[\big|K^{(i_2)}\big|_{i_2}^{[0,\de,\ga]}+
\big|K^{(i_2)}\big|_{i_2}^{[\al_\li,0,0]}\Big]\cr
&\hskip.2in\le\const c^2_F\rho^2\fv_{i_1}\,\sfrac{i_2}{i_1}\cr
}$$
and 
$$\eqalign{
&\sfrac{1}{M^{j|\de|}}
\Big|K^{(i_1)}\bullet
\rD_{\nu;\nu'}^{\be_2}\cC^{[i,j]}_{\lc}
\bullet \rD_{1;\mu'}^{\be_3} K^{(i_2)}_{\Si_{i_2},\Si_j}\Big|_{j,j}^{[\al',0,\ga']}\cr
&\hskip.8in\le \const\sfrac{1}{M^{j|\de|}}
\Big|K^{(i_1)}\bullet \cC^{[i,j]}_{\lc}
\bullet \rD_{1;\mu'}^{\de} K^{(i_2)}_{\Si_{i_2},\Si_j}\Big|_{i_1,j}^{[0,0,\ga']}\cr
&\hskip1.5in+\const\sfrac{1}{M^{j-i_1}} \sfrac{1}{M^{j|\de|}}
\Big|K^{(i_1)}\bullet\cC^{[i,j]}_{\lc}
\bullet \rD_{1;\mu'}^{\de} K^{(i_2)}_{\Si_{i_2},\Si_j}\Big|_{i_1,j}^{[\al',0,\ga']}
\cr
&\hskip.8in\le \const\sfrac{1}{M^{j|\de|}}
\Big|K^{(i_1)}\bullet \cC^{[i,j]}_{\lc}
\bullet \rD_{1;\mu'}^{\de} K^{(i_2)}_{\Si_{i_2},\Si_j}\Big|_{i_1,j}^{[0,0,\ga']}\cr
&\hskip1.5in
+\const\sfrac{j-i+1}{M^{j-i_1}}
 q({\sst \al',0,0;\al',\al'})
\big|K^{(i_2)}\big|_{i_2,j}^{[0,\de,\ga']}\cr
&\hskip.8in\le\const c^2_F\rho^2\fv_{i_1}\,\sfrac{i_2}{i_1}\cr
}$$
In the first step we applied Lemma \lemLADresectornorm. In the second, we 
applied Lemma \lemCombinedCors.c with $\be_1=\be_2=0$, $\be_3=\de$, 
$\de_\li=\al'$ and $\de_\ri=\ga'$.
 In the third step we applied
the conclusion of the last estimate and Lemma \lemLADresectornorm.

To bound $\big| k''_\lc \big|_{i_1,j}^{[\de_\li,0,\de_\ri]}$ with 
$\de_\li+\de_\ri+\de\in\De$ observe that,
by Lemma \lemCombinedCors.c,
 for all $\be_1+\be_2+\be_3=\de$ with $\be_1\ne 0$,
$$\eqalign{
&\sfrac{1}{M^{j|\de|}}
\Big|\rD_{\mu;3}^{\be_1}K^{(i_1)}\bullet \rD_{\nu;\nu'}^{\be_2}\cC^{[i,j]}_{\lc}
\bullet \rD_{1;\mu'}^{\be_3} K^{(i_2)}_{\Si_{i_2},\Si_j}\Big|_{i_1,j}^{[\de_\li,0,\de_\ri]}\cr
&\hskip.15in\le\const\sfrac{j-i+1}{M^{|\be_1|(j-i_1)}}
 q({\sst \de_\li,\be_1,0;\de_\li+\be_1,\de_\li+\be_1})
 \big|K^{(i_2)}\big|_{i_2,j}^{[0,\be_3,\de_\ri]}\cr
&\hskip.15in\le\const\sfrac{j-i_1+1}{M^{j-i_1}}
c^2_F\rho^2\fv_{i_1}\,\sfrac{i_2}{i_1}\cr
}$$
Setting $(\de_\li,\de_\ri)=(0,\ga)$, we get the $k''$ estimate of (\eqnliA).
Setting $(\de_\li,\de_\ri)=(\al',\ga')$ and using Lemma \lemLADresectornorm,
we get the $k''$ estimate of (\eqnliB).

\noindent 
When $\ell=1$, part c follows from (\eqnliA).

\Item d) Again, we may assume, without loss of generality that
$i_1\ge i_2$. By part c and (\eqnliB)
$$\eqalign{
\sfrac{1}{M^{j|\de_\ci|}} 
\Big| \rD^{\de_\ci}_{\mu;\mu'} \Big[ K^{(i_1)} \bullet \cC^{[i,j]} \bullet 
K^{(i_2)}_{\Si_{i_2},\Si_j} \Big]  \Big|_{j,j}^{[\de_\li,0,\de_\ri]}
&\le \big|k'\big|_{j,j}^{[\de_\li,0,\de_\ri]}
+\big|k''\big|_{j,j}^{[\de_\li,0,\de_\ri]}\cr
&\le \const c^2_F \rho^2 \,\fv_{i_1}
}$$

This finishes the case $\ell=1$.

\noindent
{ \bf Induction step}: We assume that the Lemma holds for $\ell-1$. Write
$$\eqalign{
K^{(i_1)} \bullet 
{\cal C}^{[i,j]} \bullet 
K^{(i_2)} \bullet  
 \cdots \bullet 
K^{(i_{\ell+1})}
= K^{(i_1)}\bullet 
{\cal C}^{[i,j]} \bullet H
}$$
with
$$
H=K^{(i_2)} \bullet
{\cal C}^{[\max \{i_2,i_3\},j]} \bullet  
 \cdots \bullet 
K^{(i_{\ell+1})}
$$
Set
$$
\fV= \cst{\ell-1}{}(c_F\rho)^\ell\ \fv_{i_2}\cdots \fv_{i_{\ell}} 
$$
The induction hypothesis applies to $H$. So, for all $|\al|\le 3$ 
and $\de\in\De$
$$
\big| H \big|_{i_2,j}^{[\al,0,\de]}\ \le\ \fV\,\sfrac{i_{\ell+1}}{i_2}
$$
Furthermore, for each $0\ne\de\in\De$,
$\mu\in\{1,2\}$ and $\mu'\in\{3,4\}$, there is a decomposition
$$
\sfrac{1}{M^{j|\de|}} \rD^{\de}_{\mu;\mu'} H_{\Si_{i_2},\Si_j}
=h_{\de,\mu,\mu'}'+h_{\de,\mu,\mu'}''
$$ 
with, for all $|\al|\le 3$ and all $\ga$ with $\ga+\de\in\De$,
$$\eqalign{
\big| h_{\de,\mu,\mu'}' \big|_{i_2,j}^{[\al,0,\ga]} &
\le \fV\,\sfrac{i_{\ell+1}}{i_2}\cr
\big| h_{\de,\mu,\mu'}'' \big|_{i_2,j}^{[0,0,\ga]} &
\le \sfrac{j-i_2+1}{M^{j-i_2}}\, \fV\,\sfrac{i_{\ell+1}}{i_2}\cr
}$$
In particular,
$$
\big| H \big|_{i_2,j}^{[0,\de_\ci,\de_\ri]}\ \le\ 2\fV\,\sfrac{i_{\ell+1}}{i_2}
$$
for $\de_\ci+\de_\ri\in\De$. 
\Item a)
By Lemma \lemCombinedCors.b, with $\de=0$, $\de_\li=\al$ and $\de_\ri=\de$,
(\eqnLADqbnd) and Lemma \lemLADresectornorm,
$$\eqalign{
\Big|K^{(i_1)}
 \bullet {\cal C}^{[i,j]} \bullet H \Big|_{i_1,j}^{[\al,0,\de]}
&\le \const\max_{\al_\upl,\al_\dnl,\al_\li\in\bbbn_0\times\bbbn_0^2
            \atop|\al_\upl|+|\al_\dnl|+|\al_\li|\le 3}
q({\sst \al,0,\al_\upl+\al_\dnl;\al+\al_\upl,\al+\al_\dnl})
|H|_{i_2,j}^{[\al_\li,0,\de]}\cr
 &\le  \const c_F\rho  \ \fv_{i_1}\sfrac{i_2}{i_1}
\,\fV\,\sfrac{i_{\ell+1}}{i_2}
 \le  \const c_F\rho  \ \fv_{i_1}
\,\fV\,\sfrac{i_{\ell+1}}{i_1}
}$$

\Item b) 
 Again, we may assume, without loss of generality that
$i_1\ge i_{\ell+1}$. Then the factor $ \cst{\ell}{}(c_F\rho)^{\ell+1}\fv_{i_2}\cdots \fv_{i_{\ell}}\,
\min \big\{\fv_{i_1},\fv_{i_{\ell+1}}\big\}$ in
the right hand side of  the statement 
reduces to $\const c_F\rho\, \fv_{i_1}\fV$.
The remainder of the proof is virtually identical to that for $\ell=1$.

\Item c)
Substitute $\cC^{[i,j]}=\cC_\tp^{[i,j]}+\cC_\md^{[i,j]}+\cC_\bt^{[i,j]}$
into
$$
\rD^{\de}_{\mu;\mu'} \Big[\big( 
K^{(i_1)} \bullet {\cal C}^{[i,j]} \bullet H\big)_{\Si_{i_1},\Si_j}\Big]
=\rD^{\de}_{\mu;\mu'} \Big[
K^{(i_1)} \bullet {\cal C}^{[i,j]} \bullet H_{\Si_{i_2},\Si_j}\Big]
$$
and apply Leibniz's rule (Lemma \leibniz) using the routing which gives
$\rD^{\be_2}_{1;3}\cC_\tp^{[i,j]}$,
$\rD^{\be_2}_{1;3}\cC_\md^{[i,j]}$
and $\rD^{\be_2}_{2;4}\cC_\bt^{[i,j]}$. We define $k'$ to be $\sfrac{1}{M^{j|\de|}}$ times the sum of all resulting terms having no
derivatives acting on $K^{(i_1)}$ and  $k''$ to be 
$\sfrac{1}{M^{j|\de|}}$ times the sum of all terms having at least 
one derivative acting on $K^{(i_1)}$. 
Fix any $\al$, $\al'$ and $\ga,\ga'$ obeying, $|\al|\le3$,
$\ga+\de\in\De$ and $\al'+\ga'+\de\in\De$.
We show
$$\eqalign{
\big| k' \big|_{i_1,j}^{[\al,0,\ga]} &
\le \const c_F\rho\,\fv_{i_1}\ \fV\,\sfrac{i_{\ell+1}}{i_1} \cr
\big| k'' \big|_{i_1,j}^{[0,0,\ga]} &
\le\const c_F \sfrac{j-i_1+1}{M^{j-i_1}}\, \rho \,\fv_{i_1}\ \fV\,\sfrac{i_{\ell+1}}{i_1}\cr
}\EQN\eqnliC$$
and
$$
\big|k'\big|_{j,j}^{[\al',0,\ga']}
+\big|k''\big|_{j,j}^{[\al',0,\ga']}\le \const c_F \rho \,\fv_{i_1}\ \fV
\,\sfrac{i_{\ell+1}}{i_1}
\EQN\eqnliD$$
Let $\lc\in\{\md,\tp,\bt\}$. If $\lc\in\{\tp,\md\}$, set $(\nu,\nu')=(1,3)$.
If $\lc=\bt$, set $(\nu,\nu')=(2,4)$. The contributions to $k'$ and $k''$
coming from $\cC_\lc^{[i,j]}$ are
$$\eqalign{
k'_\lc&=\sfrac{1}{M^{j|\de|}}
\sum_{\be_2,\be_3\in\bbbn_0\times\bbbn_0^2\atop \be_2+\be_3=\de}
{\tst{\de\choose\be_2,\be_3}}
K^{(i_1)}
\bullet \rD_{\nu;\nu'}^{\be_2}\cC^{[i,j]}_{\lc}
\bullet \rD_{1;\mu'}^{\be_3}H_{\Si_{i_2},\Si_j}\cr
k''_\lc&=\sfrac{1}{M^{j|\de|}}
\sum_{{\be_1,\be_2,\be_3\in\bbbn_0\times\bbbn_0^2\atop \be_1+\be_2+\be_3=\de}\atop|\be_1|>0}
{\tst{\de\choose\be_1,\be_2,\be_3}}
\rD_{\mu;3}^{\be_1}K^{(i_1)}
\bullet \rD_{\nu;\nu'}^{\be_2}\cC^{[i,j]}_{\lc}
\bullet \rD_{1;\mu'}^{\be_3}H_{\Si_{i_2},\Si_j}\cr
}$$

We first bound $k'_\lc$. Fix $\be_2+\be_3=\de$.
First consider $\be_2\ne 0$.
Let $(\de_\li,\de_\ri)=(\al,\ga)$ or $(\al',\ga')$.
By Lemma \lemCombinedCors.a,
$$\eqalign{
\sfrac{1}{M^{j|\de|}}
\Big|K^{(i_1)}\bullet \rD_{\nu;\nu'}^{\be_2}\cC^{[i,j]}_{\lc}
\bullet \rD_{1;\mu'}^{\be_3} H_{\Si_{i_2},\Si_j}\Big|_{i_1,j}^{[\de_\li,0,\de_\ri]}
&\le\const
 q({\sst \de_\li,0,0;\de_\li,\de_\li})
 |H|_{i_2,j}^{[0,\be_3,\de_\ri]}\cr
&\le \const c_F\rho  \ \fv_{i_1}\sfrac{i_2}{i_1}
\,\fV\,\sfrac{i_{\ell+1}}{i_2}\cr
& \le  \const c_F\rho  \ \fv_{i_1}
\,\fV\,\sfrac{i_{\ell+1}}{i_1}\cr
}$$
Next consider $\be_2= 0$. By (\eqnLADqbnd),
$$\eqalign{
&\sfrac{1}{M^{j|\de|}}
\Big|K^{(i_1)}\bullet \rD_{\nu;\nu'}^{\be_2}\cC^{[i,j]}_{\lc}
\bullet \rD_{1;\mu'}^{\be_3} H_{\Si_{i_2},\Si_j}\Big|_{i_1,j}^{[\al,0,\ga]}\cr
&\hskip.8in=\sfrac{1}{M^{j|\de|}}
\Big|K^{(i_1)}\bullet \cC^{[i,j]}_{\lc}
\bullet \rD_{1;\mu'}^{\de} H_{\Si_{i_2},\Si_j}\Big|_{i_1,j}^{[\al,0,\ga]}\cr
&\hskip.8in\le
\Big|K^{(i_1)}\bullet \cC^{[i,j]}_{\lc}
\bullet h'_{\de,1,\mu'}\Big|_{i_1,j}^{[\al,0,\ga]}
+
\Big|K^{(i_1)}\bullet \cC^{[i,j]}_{\lc}
\bullet h''_{\de,1,\mu'}\Big|_{i_1,j}^{[\al,0,\ga]}
\cr
&\hskip.8in
\le\const
\max_{\al_\upl,\al_\dnl,\al_\li\in\bbbn_0\times\bbbn_0^2
            \atop|\al_\upl|+|\al_\dnl|+|\al_\li|\le 3}
 q({\sst \al,0,\al_\upl+\al_\dnl;\al+\al_\upl,\al+\al_\dnl})
\big|h'_{\de,1,\mu'}\big|_{i_2,j}^{[\al_\li,0,\ga]}\cr
&\hskip2in
+\const q({\sst \al,0,0;\al,\al})
(j-i+1) \big|h''_{\de,1,\mu'}\big|_{i_2,j}^{[0,0,\ga]}\cr
&\hskip.8in
 \le  \const c_F\rho  \ \fv_{i_1}
\,\fV\,\sfrac{i_{\ell+1}}{i_1}\cr
}$$
since $\sfrac{(j-i+1)(j-i_2-1)}{M^{j-i_2}}\le\const$.
The term with $h'_{\de,1,\mu'}$ was bounded using Lemma \lemCombinedCors.b
with $\de=0$. 
The term with $h''_{\de,1,\mu'}$ was bounded using Lemma \lemCombinedCors.c
with $\be_1=\be_2=\be_3=0$. 
Again, with $\be_2=0$,
$$
\sfrac{1}{M^{j|\de|}}
\Big|K^{(i_1)}\bullet \rD_{\nu;\nu'}^{\be_2}\cC^{[i,j]}_{\lc}
\bullet \rD_{1;\mu'}^{\be_3} H_{\Si_{i_2},\Si_j}\Big|_{j,j}^{[\al',0,\ga']}
\le \const c_F\rho  \ \fv_{i_1}
\,\fV\,\sfrac{i_{\ell+1}}{i_1}
$$
as in the proof of part (c) when $\ell=1$.

We bound $\big| k''_\lc \big|_{i_1,j}^{[\de_\li,0,\de_\ri]}$ with 
$\de_\li+\de_\ri+\de\in\De$ as for $\ell=1$. Observe that,
by Lemma \lemCombinedCors.c,
 for all $\be_1+\be_2+\be_3=\de$ with $\be_1\ne 0$,
$$\eqalign{
&\sfrac{1}{M^{j|\de|}}
\Big|\rD_{\mu;3}^{\be_1}K^{(i_1)}\bullet \rD_{1,3}^{\be_2}\cC^{[i,j]}_{\lc}
\bullet \rD_{1;\mu'}^{\be_3} H_{\Si_{i_2},\Si_j}\Big|_{i_1,j}^{[\de_\li,0,\de_\ri]}\cr
&\hskip.15in\le\const\sfrac{j-i+1}{M^{|\be_1|(j-i_1)}}
 q({\sst \de_\li,\be_1,0;\de_\li+\be_1,\de_\li+\be_1})
 |H|_{i_2,j}^{[0,\be_3,\de_\ri]}\cr
&\hskip.15in\le\const\sfrac{j-i_1+1}{M^{j-i_1}}
c_F\rho\,\fv_{i_1}\ \fV\,\sfrac{i_{\ell+1}}{i_1}\cr
}$$
Setting $(\de_\li,\de_\ri)=(0,\ga)$, we get the $k''$ estimate of (\eqnliC).
Setting $(\de_\li,\de_\ri)=(\al',\ga')$ and using Lemma \lemLADresectornorm,
we get the $k''$ estimate of (\eqnliD).

\Item d) Part (d) follows from part (c) and (\eqnliD) as in the case $\ell=1$.

\endproof

\noindent{\bf Completion of the proof of Theorem \thmLADmodcompLadder:}

We prove (\eqnLADladindbnd).
Let $ \vec\de=(\de_\li,\de_\ci,\de_\ri) \in\vec\De$. 
By parts b) and d) of the Lemma above, for $\ell\ge 1$,
$$\eqalign{
\Big|
K^{(i_1)} \bullet 
{\cal C}^{[\max \{i_1,i_2\},j]} \bullet &
K^{(i_2)} \bullet  
 \cdots \bullet 
K^{(i_{\ell+1})} 
\Big|_{j,j}^{[\vec\de]} 
\le \cst{\ell}{} (c_F\rho)^{\ell+1}\ 
\fv_{i_2}\cdots\fv_{i_\ell}\,
\min \big\{\fv_{i_1},\fv_{i_{\ell+1}}\big\}\cr
}$$
Therefore, by Corollary \corLADaltcmpladders.i,
$$\eqalign{
\big| \cL_S^{(j+1)} \Big|_{j,j}^{[\vec\de]}
& \le \smsum_{\ell=1}^\infty \  \smsum_{i_1,\cdots,i_{\ell+1}=2}^{j}
 \cst{\ell}{} (c_F\rho)^{\ell+1}\ 
\fv_{i_2}\cdots\fv_{i_\ell}\,
\min \big\{\fv_{i_1},\fv_{i_{\ell+1}}\big\}\cr
& \le \const c^2_F \rho^2\,\smsum_{\ell=1}^\infty \ 
\Big( (\const c_F \rho)^{\ell-1} \hskip -12pt \smsum_{i_2,\cdots,i_{\ell}=2}^{\infty}
\fv_{i_2}\cdots\fv_{i_\ell}\Big)\,
\Big( \smsum_{i_1,i_{\ell+1}=2}^{\infty}
\min \big\{\fv_{i_1},\fv_{i_{\ell+1}}\big\}
\Big) \cr
& \le \const c^2_F \rho^2\,\smsum_{\ell=1}^\infty \ 
 (\const c_F \rho)^{\ell-1} 
\Big( \smsum_{i_1 \ge i_{\ell+1}} \fv_{i_1}
 + \smsum_{i_1 < i_{\ell+1}}  \fv_{i_{\ell+1}} \Big) \cr
& \le \const c^2_F\rho^2\, \smsum_{i=2}^\infty (i-1) \fv_i
\ \le\ \const c^2_F \rho^2\  
\ =\ c_\cL\rho^2\cr
}$$
when $\rho$ is small enough.
This concludes the induction step in the proof of Theorem \thmLADmodcompLadder.
\endproof

\titlec{The Infrared Limit}\PG\pgLADIIe
Define, for each $j\ge 2$, $\ell\ge  1$ and $\ii_1,\cdots,\ii_{\ell+1}\ge 2$,
the function 
$$
\fL^{(j)}_{\ell,\ii_1,\cdots,\ii_{\ell+1}}
:\bbbm^3\times\{\uparrow,\downarrow\}^4\rightarrow\bbbc
$$ 
by
$$\eqalign{
&\fL^{(j)}_{\ell,\ii_1,\cdots,\ii_{\ell+1}}(q,q',t,\si_1,\cdots\si_4)\cr
&\hskip.1in=
\Big[\big(F^{(\ii_1)}+L^{(\ii_1)\,f} \big)  \bullet 
{\cal C}^{[\max \{\ii_1,\ii_2\},j]}\!\bullet\!
\big(F^{(\ii_2)}+L^{(\ii_2)\,f} \big) \!\bullet \cdots \cr
&\hskip.3in\cdots \bullet\! 
{\cal C}^{[\max \{\ii_\ell,\ii_{\ell+1}\},j]}\! \bullet\!
\big(F^{(\ii_{\ell+1})}+L^{(\ii_{\ell+1})\,f} \big)\Big]_{i_1,i_2,i_3,i_4=0}
({\sst(q+{t\over 2},\si_1),(q-{t\over 2},\si_2),
(q'+{t\over 2},\si_3),(q'-{t\over 2},\si_4)})\cr
}$$
By Proposition \propLADaltcmpladders.ii
$$\eqalign{
&\cL^{(j+1)}\Big|_{i_1,i_2,i_3,i_4=0}
({\sst(q+{t\over 2},\si_1),(q-{t\over 2},\si_2),
(q'+{t\over 2},\si_3),(q'-{t\over 2},\si_4)})\cr
&\hskip1in=\sum_{\ell=1}^\infty \  \sum_{\ii_1,\cdots,\ii_{\ell+1}=2}^{j} \  
\fL^{(j)}_{\ell,\ii_1,\cdots,\ii_{\ell+1}}(q,q',t,\si_1,\cdots\si_4)\cr
}$$
\lemma{\STM\lemLADlebupbnd}{
$$
\sum_{\ell=1}^\infty \  \sum_{\ii_1,\cdots,\ii_{\ell+1}=2}^\infty \  
\sup_{j\ge\max\{\ii_1,\cdots,\ii_{\ell+1}\}}\ 
\sup_{q,q',t\in\bbbm\atop\si_i\in\{\uparrow,\downarrow\}}
\Big|\fL^{(j)}_{\ell,\ii_1,\cdots,\ii_{\ell+1}}(q,q',t,\si_1,\cdots\si_4)\Big|
<\infty
$$
}
\prf
By Lemma \ladderinduction.ii, with $\de_\li=\de_\ri=0$, and the analogous
bound for $\cL^{(j+1)}_C$,
$$
\sup_{q,q',t\in\bbbm\atop\si_i\in\{\uparrow,\downarrow\}}
\Big|\fL^{(j)}_{\ell,\ii_1,\cdots,\ii_{\ell+1}}(q,q',t,\si_1,\cdots\si_4)\Big|
\le \cst{\ell}{} \big(c_F\rho \big)^{\ell+1}\fv_{\ii_2}\cdots \fv_{\ii_{\ell}} 
\min \big\{ \fv_{\ii_1},\fv_{\ii_{\ell+1}}\big\}
$$
uniformly in $j$. Hence, as in the final part of the 
proof of Theorem \thmLADmodcompLadder,
$$\eqalign{
&\sum_{\ell=1}^\infty \  \sum_{\ii_1,\cdots,\ii_{\ell+1}=2}^\infty \  
\sup_{j\ge\max\{\ii_1,\cdots,\ii_{\ell+1}\}}\ 
\sup_{q,q',t\in\bbbm\atop\si_i\in\{\uparrow,\downarrow\}}
\Big|\fL^{(j)}_{\ell,\ii_1,\cdots,\ii_{\ell+1}}(q,q',t,\si_1,\cdots\si_4)\Big|\cr
&\hskip.5in \le \const c^2_F \rho^2\,\smsum_{\ell=1}^\infty \ 
\Big( (\const c_F \rho)^{\ell-1} \hskip -12pt \smsum_{\ii_2,\cdots,\ii_{\ell}=2}^{\infty}
\fv_{\ii_2}\cdots\fv_{\ii_\ell}\Big)\,
\Big( \smsum_{\ii_1,\ii_{\ell+1}=2}^{\infty}
\min \big\{\fv_{\ii_1},\fv_{\ii_{\ell+1}}\big\}
\Big) \cr
&\hskip.5in \le \const c^2_F \rho^2\,\smsum_{\ell=1}^\infty \ 
 (\const c_F \rho)^{\ell-1} 
\Big( \smsum_{\ii_1 \ge \ii_{\ell+1}} \fv_{\ii_1}
 + \smsum_{\ii_1 < \ii_{\ell+1}}  \fv_{\ii_{\ell+1}} \Big) \cr
&\hskip.5in \le \const c^2_F\rho^2\, \smsum_{\ii=2}^\infty (\ii-1) \fv_\ii
\ <\ \infty\cr
}$$
when $\rho$ is small enough.
\endproof
\lemma{\STM\lemLADtermwisecont}{
For $t\ne 0$, the limit
$$
\fL_{\ell,\ii_1,\cdots,\ii_{\ell+1}}(q,q',t,\si_1,\cdots\si_4)
=\lim_{j\rightarrow\infty}
\fL^{(j)}_{\ell,\ii_1,\cdots,\ii_{\ell+1}}(q,q',t,\si_1,\cdots\si_4)
$$
exists. The limit is continuous in $(q,q',t)$ for $t\ne 0$.
The restrictions to $\t=0$ and to $t_0=0$, namely,
$\fL_{\ell,\ii_1,\cdots,\ii_{\ell+1}}(q,q',(t_0,\0),\si_1,\cdots\si_4)$ and
$\fL_{\ell,\ii_1,\cdots,\ii_{\ell+1}}(q,q',(0,\t),\si_1,\cdots\si_4)$,
have continuous extensions to $t=0$.

}
\prf
It suffices to consider separately the spin and charge parts,
in the sense of Lemma \lemLADchargespin,  of $\fL^{(j)}_{\ell,\ii_1,\cdots,\ii_{\ell+1}}(q,q',t,\si_1,\cdots\si_4)$.
We denote them $\fL^{(j)}_{X,\ell,\ii_1,\cdots,\ii_{\ell+1}}(q,q',t)$ 
with $X=S,C$. 
The existence and continuity of the limits when $t\ne0$ shall be proven
in Lemma \:\lemLADtnonzero. 

Recall that the bubble propagator $\cC^{[i,j]}$
has momentum space kernel
$$\eqalign{
\cC^{[i,j]}(p,k)
&=\sfrac{\nu^{(\ge i)}(p)\nu^{(\ge i)}(k)
-\nu^{(\ge j+1)}(p)\nu^{(\ge j+1)}(k)}
{[ip_0-e'(p)][ik_0-e'(k)]}\cr
}$$
where $e'(k)=e(\k)-v(k)$. 
Define the model particle--hole bubble propagators
$$\eqalign{
\cA_{i,j}(p,k)&=\sfrac{\nu^{(\ge i)}(p)\nu^{(\ge i)}(k)
[1-\nu_j(e(\p))\nu_j(e(\k))]}
{[ip_0-e'(p)][ik_0-e'(k)]}\cr
\cB_{i,j}(p,k)&=\sfrac{\nu^{(\ge i)}(p)\nu^{(\ge i)}(k)
[1-\nu_j(p_0)\nu_j(k_0)]}
{[ip_0-e'(p)][ik_0-e'(k)]}\cr
}\EQN\eqnMBPmab$$
where
$$\eqalign{
\nu_j(\om)&=\sum_{m=j}^\infty\nu\big(M^{2m}\om^2\big)\cr
}$$
with $\nu$ being the single scale cutoff introduced in Definition
\defLADscales. 
Let
$$\eqalign{
&A^{(j)}_{X,\ell,\ii_1,\cdots,\ii_{\ell+1}}(q,q',t_0)\cr
&=
\big[K_X^{(\ii_{\ell+1})}\!\bullet\! 
\cA_{\max \{\ii_{\ell+1},\ii_\ell\},j}\!\bullet\! K_X^{(\ii_\ell)} 
\!\bullet \cdots 
\bullet\! \cA_{\max \{\ii_2,\ii_1\},j}\! \bullet\!
K_X^{(\ii_1)}\big]_{i_1,i_2,i_3,i_4=0}
({\sst q+{t\over 2}, q-{t\over 2},
q'+{t\over 2},q'-{t\over 2}})\big|_{\t=0}\cr
}$$
and
$$\eqalign{
&B^{(j)}_{X\ell,\ii_1,\cdots,\ii_{\ell+1}}(q,q',\t)\cr
&=\!
\big[K_X^{(\ii_{\ell+1})}\!\bullet\! 
\cB_{\max \{\ii_{\ell+1},\ii_\ell\},j}\!\bullet\! K_X^{(\ii_\ell)} 
\!\bullet \cdots 
\bullet\! \cB_{\max \{\ii_2,\ii_1\},j}\! \bullet\!
K_X^{(\ii_1)}\big]_{i_1,i_2,i_3,i_4=0}
({\sst q+{t\over 2}, q-{t\over 2},
q'+{t\over 2},q'-{t\over 2}})\big|_{t_0=0}\cr
}$$
where  $K_X^{(i)}$ is  
$F^{(i)}_S+\half L^{(i)\,f}_{C\,\Si_i}-\half L^{(i)\,f}_{S\,\Si_i}$ when
$X=S$ and 
$F^{(i)}_C+\half L^{(i)\,f}_{C\,\Si_i}+\sfrac{3}{2} L^{(i)\,f}_{S\,\Si_i}$
when $X=C$.
By Corollary \:\corMPsamelim\ the differences 
$$
\cL^{(j)}_{X,\ell,\ii_1,\cdots,\ii_{\ell+1}}(q,q',(t_0,\0))
-A^{(j)}_{X\ell,\ii_{\ell+1},\cdots,\ii_1}(q,q',t_0)
$$
and 
$$
\cL^{(j)}_{X,\ell,\ii_1,\cdots,\ii_{\ell+1}}(q,q',(0,\t))
-B^{(j)}_{X\ell,\ii_{\ell+1},\cdots,\ii_1}(q,q',\t)
$$
both converge to zero for all $t\ne 0$. The bounds on the $|K_X^{(\ii_m)}|_{\ii_m,\ii_m} $'s
required by Corollary \:\corMPsamelim\ are provided by (\eqnLADKbnd) 
with $\de=0$.

That 
$\lim\limits_{j\rightarrow\infty}
      A^{(j)}_{X\ell,\ii_1,\cdots,\ii_{\ell+1}}(q,q',t_0)$
and 
$\lim\limits_{j\rightarrow\infty}
      B^{(j)}_{X\ell,\ii_1,\cdots,\ii_{\ell+1}}(q,q',\t)$
exist and are continuous at $t=0$ is proven using
 Lemma \:\lemLadABdiff\  inductively on $\ell$, with $I=F$, the full Fermi
surface. For the induction step from $\ell-1$ to $\ell$, set $z=(q,q')$
and use
$$\eqalign{
u_j(k,t_0,z)&=K_X^{(i_{\ell+1})}(q+\sfrac{t}{2}, q-\sfrac{t}{2},
k+\sfrac{t}{2},k-\sfrac{t}{2})\big|_{\t=0}
\nu^{(\ge i)}(k+t)\nu^{(\ge i)}(k)
A^{(j)}_{\ell-1,\ii_1,\cdots,\ii_\ell}(k,q',t_0)
\cr
v_j(k,\t,z)&=K_X^{(i_{\ell+1})}(q+\sfrac{t}{2}, q-\sfrac{t}{2},
k+\sfrac{t}{2},k-\sfrac{t}{2})\big|_{t_0=0}
\nu^{(\ge i)}(k+t)\nu^{(\ge i)}(k)
B^{(j)}_{\ell-1,\ii_1,\cdots,\ii_\ell}(k,q',\t)
\cr
n_j(\om)&=\nu_{i-1}(\om)\big[1-\nu_j(\om)^2\big]\cr
i&=\max\{\ii_{\ell+1},\ii_\ell\}\cr
}$$
Also fix some $0<\aleph''<\aleph$, and use 
$\tilde\aleph = \aleph''+\sfrac{1}{2^{\ell}}(\aleph-\aleph'')$ and 
$\aleph' = \aleph''+\sfrac{1}{2^{\ell+1}}(\aleph-\aleph'')$.
\endproof

\proof{of Theorem \thmLADmodcompLaddercont}
By the Lebesgue dominated convergence theorem and the uniform bounds
of Lemma \lemLADlebupbnd, the existence of the limit
$\lim\limits_{j\rightarrow\infty}$ and its continuity for $t\ne 0$,
 as well as the existence and continuity of the limits
$\lim\limits_{t_0\rightarrow 0}\lim\limits_{j\rightarrow\infty}$ 
and $\lim\limits_{\t\rightarrow 0}\lim\limits_{j\rightarrow\infty}$
applied to $\cL^{(j+1)}\Big|_{i_1,i_2,i_3,i_4=0}
({\sst(q+{t\over 2},\si_1),(q-{t\over 2},\si_2),
(q'+{t\over 2},\si_3),(q'-{t\over 2},\si_4)})$ follow
from the corresponding properties of
$\fL^{(j)}_{\ell,\ii_1,\cdots,\ii_{\ell+1}}(q,q',t,\si_1,\cdots\si_4)$, for
$\ell\ge  1$ and $\ii_1,\cdots,\ii_{\ell+1}\ge 2$. These were proven in Lemma
\lemLADtermwisecont.

\endproof

\vfill\eject

\chap{Bubbles}\PG\pgLADIII

\EDEF\CHbubbles{\caproman\chapno}

The bulk of this section is devoted to the proof of Theorem \bbound. 
Parts b and c, reformulated as Theorem \:\rebbound,  are relatively easy 
to prove. To do so, we fully decompose
$$
\cC^{[i,j]}=\sum_{m=i}^j\sum_{m_1,m_2\in\bbbn_0\atop \min\{m_1,m_2\}=m}
C_v^{(m_1)}\otimes C_v^{(m_2)\,t}
\EQN\eqnLADfullbubdecomp$$ 
and bound each term naively to achieve ordinary power counting.
The factor $j-i+1=\sum\limits_{m=i}^j1$ in the first statement of part c is a reflection of
the marginality of four--legged diagrams in naive power counting. When power
counting bubbles with propagator $\rD_{\mu,\mu'}^{\be}\cC^{[i,j]}$, $|\be|\ge 1$,
the sum
$\sum\limits_{m=i}^j1$ is replaced by $\sum\limits_{m=i}^jM^{|\be|m}
\le\const M^{|\be|j}$, which is cancelled by the factors 
$\sfrac{1}{M^{|\be|j}}$ on the left hand sides of parts b and c.
In the $\be=0$ statement of part b, naive power counting gives
$\sum\limits_{i_t=i}^j\sum\limits_{i_b>j}M^{-(i_b-i_t)}\le\const$. 

The proof of Theorem \bbound a, which follows Theorem \:\rebboundII,
 relies on two distinct phenomena, volume improvement
for large transfer momentum and a sign cancellation in momentum space for
small transfer momentum. The mechanism underlying the sign cancellation
has been illustrated in the model Lemma \lemLADprimitivemanfred\ and is 
fully implemented in Theorem \:\rebboundII. 

We now sketch the idea behind volume improvement. To unravel the sector sums
of the $\bullet$ product of Definition \defLADphladder, we define,
for any translation invariant functions $K$ on 
$\fY_\Si^2\times\big(\bbbr\times\bbbr^2\big)$, 
$K'$ on $\big(\bbbr\times\bbbr^2\big)\times\fY_{\Si'}^2$ 
and bubble propagator $P$,
$$\eqalign{
K\fcirc P( y_1,y_2,x_3,x_4)&= \int dx_1 dx_2\ 
K(y_1,y_2,x_1,x_2) P( x_1,x_2,x_3,x_4)\cr
P\fcirc K'(x_1,x_2,y_3,y_4)&= \int dx_3 dx_4\ 
P( x_1,x_2,x_3,x_4) K'( x_3,x_4,y_3,y_4)\cr
}$$
If at least one of $y_1,y_2,y_3,y_4$ is in $\big(\bbbr\times\bbbr^2)\times\Si$
or $\big(\bbbr\times\bbbr^2)\times\Si'$
$$
K\fcirc K'( y_1,y_2,y_3,y_4)= \int dx_1 dx_2\ 
K( y_1,y_2,x_1,x_2) K'( x_1,x_2,y_3,y_4)
$$
On the other hand, if all of $k_1,k_2,k_3,k_4$ are in $\bbbm$,
$K\fcirc K'(k_1,k_2,k_3,k_4)$ is determined by
$$
K  \fcirc K'(k_1,k_2,k_3,k_4)\  (2\pi)^3\de(k_1-k_2-k_3+k_4)
=  \int  dx_1 dx_2\ K(k_1,k_2,x_1,x_2)\,K'(x_1,x_2,k_3,k_4)
$$
or equivalently, by
$$
K\fcirc K'(k_1,k_2,k_3,k_4)
=\int  dx_n\ \
K(k_1,k_2,x_1,x_2)\,K'(x_1,x_2,k_3,k_4)\big|_{x_{3-n}=0}
$$
for $n\in\{1,2\}$.
Then, for the functions $g$ and $h$ of the Theorem,
$$\eqalign{
\big(g\bullet\cC^{[i,j]}\bullet h\big)
({\sst y_1,y_2,y_3,y_4})
&=\sum_{u_1,u_2\in\Si_i\atop v_1,v_2\in\Si_i}
g({\sst y_1,y_2,(\,\cdot\,,u_1),(\,\cdot\,,u_2)})
\fcirc\cC^{[i,j]}\fcirc
 h({\sst(\,\cdot\,,v_1),(\,\cdot\,,v_2),y_3,y_4})\cr
}\EQN\eqnBubIa$$
Consider the case in which all external arguments $y_1,\cdots,y_4$ are momenta
$k_1,\cdots,k_4$. Then
$$\eqalign{
&\big(g\bullet\cC^{[i,j]}\bullet h\big)
({\sst k_1,k_2,k_3,k_4})\cr
&\hskip.3in=\sfrac{1}{(2\pi)^3}\int {\sst d^3p d^3k}
 \sum_{u_1,u_2\in\Si_i\atop v_1,v_2\in\Si_i} \de({\sst k_1-k_2-p+k})\,
\check g({\sst k_1,k_2,(p,u_1),(k,u_2)})
\,\cC^{[i,j]}(p,k)\,
\check h({\sst(p,v_1),(k,v_2),k_3,k_4})
}\EQN\eqnBubI$$
where
$$
\cC^{[i,j]}(p,k)=\sum_{m=i}^j\sum_{m_1,m_2\in\bbbn_0\atop \min\{m_1,m_2\}=m}
C_v^{(m_1)}(p) C_v^{(m_2)}(k)
$$
In order for $\cC^{[i,j]}(p,k)$ to be nonzero, one must have 
$p$ and $k$ in the $i^{\rm th}$ neighbourhood. In particular, $\p$ and $\k$ must lie within
a distance $\sfrac{\const}{M^i}$ of the Fermi curve $F$. Furthermore, by conservation
of momentum at the vertex $g$, the ``transfer momentum''
$$
t=k_1-k_2
$$
is equal to $p-k$. Thus, the set of pairs $(p,k)$ for which the 
integrand of (\eqnBubI) does not vanish is contained in
$$
\set{(k,p)\in \big({\rm supp}\, C^{(\ge i)}\big)^2}{p-k=t}
$$
For each fixed large $\t$, the volume of 
$$
\set{\p\in {\rm supp}\, C^{(\ge i)}}{\p-\t\in {\rm supp}\, C^{(\ge i)}}
={\rm supp}\, C^{(\ge i)}\cap\big(\t+{\rm supp}\, C^{(\ge i)}\big)
\EQN\eqnBubII$$
is very small compared to the volume of ${\rm supp}\, C^{(\ge i)}$, 
as the following figure illustrates.
\vskip.2in
\centerline{\figplace{voleffecttrans}{-.5 in}{0 in}}
\vskip-.2in

\noindent In naive power counting, the volume of the set (\eqnBubII)
is bounded by the volume of ${\rm supp}\, C^{(\ge i)}$, yielding a relatively
loose bound. There is a similar volume improvement, when, for example,
the external arguments $y_1=(x_1,s_1)$ and $y_2=(x_2,s_2)$ lie
$\big(\bbbr\times\bbbr^2\big)\times\Si_\ell$ and the sectors $s_1$ and $s_2$ are widely separated. For a more detailed discussion of this volume 
improvement in perturbation theory see [FKLT2].

We now give a somewhat more detailed technical outline of the contents of this section. By sector counting and relatively simple propagator estimates, the volume improvement effect can be implemented for all summands
$$
\cC^{(m)}=\sum_{m_1,m_2\in\bbbn_0\atop \min\{m_1,m_2\}=m}
C_v^{(m_1)}\otimes C_v^{(m_2)\,t}
$$ 
of (\eqnLADfullbubdecomp) for which $\sfrac{1}{M^m}$ is small compared
to the transfer momentum. Sector counting is made precise in Remark 
\:\remLADtdiff.ii and Lemma \:\lemAI. The basic propagator estimates are stated
in Appendix \APpropbnd\ and are adapted to the present situation in 
Lemma \:\lemXV.
Lemma \:\lemXIV\ shows how one can combine sector counting and propagator 
estimates on quantities like $g\bullet\cC^{(m)}\bullet h$. The resulting 
estimates turn out to be summable over those $m$'s for which $\sfrac{1}{M^m}$ 
is smaller than the transfer momentum. This is used to prove parts (b) and 
(c) of Theorem \bbound\ (which are reformulated as Theorem \rebbound)
and to reduce the statement of part (a) of Theorem \bbound\ 
to the situation of transfer momentum smaller than $\fl_j$.

The situation of small transfer momentum is treated in Theorem \:\rebboundII.
To estimate $g\bullet\cC^{[i,j]}\bullet h$ when the transfer momentum
is small compared to $\fl_j$, we replace $\cC^{[i,j]}$ with a model bubble
propagator $\cM$ with a factorized cutoff similar to that of Lemma \lemLADprimitivemanfred.
In Proposition \:\propXXXII, we use a position space bound on $\cM$ (which is 
proven in Appendix \APmodelbnd) to estimate $g\bullet\cM\bullet h$. 
Propositions \:\propLADzdiff, \:\propXXVIII\ and \:\propXXX\ use 
sector counting and simple propagator estimates as above to bound 
$g\bullet(\cC^{[i,j]}-\cM)\bullet h$.

The results Lemma \:\lemBubTnonzero\ through Corollary
\corMPsamelim, of the final two subsections, are used in the proof, in Lemma
\lemLADtermwisecont, of the existence and continuity properties
of the infrared limit of ladders. 

\vskip.25in

Before we implement the program outlined above, we introduce some 
notation, prove some utility Lemmata and reformulate Theorem  \bbound\ in terms 
of the new notation.

Let
$$
\fY=\bbbm\dunion\big(\bbbr\times\bbbr^2\big)
$$
be the disjoint union of the set, $\bbbm$, of all possible momenta
and the set, $\bbbr\times\bbbr^2$, of all possible positions.
We consider $\fY$ as the special case of the space $\fY_\Si$
of the introduction, with the set of sectors $\Si=\Si_0$ where $\Si_0$ contains
only a single element, namely all momentum space, $\bbbm$.
In particular, as in (\eqnLADfourdunion), $\fY^4$ is the disjoint union
$$
\fY^4=
\bigdunion_{i_1,i_2,i_3,i_4\in\{0,1\}}
\fY_{i_1}\times \fY_{i_2}\times \fY_{i_3}\times \fY_{i_4}
$$
where $\fY_{0}=\bbbm$ and $\fY_{1}=\bbbr\times\bbbr^2$. For a translation
invariant function $f$ on $\fY^4$, we define 
$$
\tn f\tn=\V f\V_{\Si_0,\Si_0}^{(0,0,0)}
$$ 
using the norm $\V \ \cdot\ \V_{\Si,\Si'}^{(0,0,0)}$ of 
Definition \defLADlonelinftyIV. Concretely,
$$
\tn f\tn=\sum_{i_1,i_2,i_3,i_4\in\{0,1\}}
\sup_{k_\nu\in\bbbm\atop {\nu=1,2,3,4\atop{\rm\ with\ }i_\nu=0}}\ 
\TN f\big|_{(i_1,\cdots,i_4)}\TN_{1,\infty}
$$
Here, the $\nu^{\rm th}$ argument of $f$ is $k_\nu$ when $i_\nu=0$ and
$x_\nu$ when $i_\nu=1$. The $\tn\ \cdot\ \tn_{1,\infty}$ norm of 
Definition \defLADlonelinfty\ is applied to all spatial arguments of
$f\big|_{(i_1,\cdots,i_4)}$. 

\definition{\STM\defLADbubblenorm}{
We define the bubble operator norm of any translation invariant bubble propagator $P(x_1,x_2,x_3,x_4)$
by
$$
\|P\|_{\rm bubble}=\sup_{G,H}
\frac{\tn G\fcirc P\fcirc H\tn}{\tn G\tn\,\tn H\tn}
$$
where the $\sup$ is over nonzero, translation invariant functions on
$\fY^4$.

}
\lemma{\STM\lemLADbubblenorm}{
Let $P$ be a translation invariant bubble propagator. Then
$$\eqalign{
\|P\|_{\rm bubble}\le\min\Big\{
&\min_{n=1,2}\sup_{x_1,x_2}\int dy_n\sup_{y_{\bar n}}
|P(x_1,x_2,y_1,y_2)|,\cr
&\min_{n=1,2}\sup_{y_1,y_2}\int dx_n\sup_{x_{\bar n}}
|P(x_1,x_2,y_1,y_2)|
\Big\}
}$$
where $\bar n=2$ if $n=1$ and $\bar n=1$ if $n=2$. 
}
\prf
Let $c_P$ be the right hand side of the claim. 
We must prove that
$$
\tn G\fcirc P\fcirc H\tn\le c_P\ \tn G\tn\,\tn H\tn
$$
for all translation invariant functions, $G,\ H$ on $\fY^4$.
It suffices to consider $G$ and $H$ obeying
$$
G=G\big|_{(i_1,i_2,1,1)}\qquad
H=H\big|_{(1,1,i_3,i_4)}
$$
for some ${i_1,i_2,i_3,i_4\in\{0,1\}}$.

\goodbreak
First consider the case $i_1=i_2=i_3=i_4=0$.  By definition
$$\eqalignno{
&| G \fcirc P \fcirc H(k_1,k_2,k_3,k_4)|\cr
&\hskip.5in\le  \int  d^3x_2\, d^3y_1\,  d^3y_2\ 
|G(k_1,k_2,0,x_2)|\,|P(0,x_2,y_1,y_2)|\,|H(y_1,y_2,k_3,k_4)|\cr
&\hskip.5in\le \tn G\tn\, \sup_{x_2}\int dy_1\,dy_2\ 
|P(0,x_2,y_1,y_2)|\,|H(y_1,y_2,k_3,k_4)|&\EQNO\eqnLADbubblenormA\cr
&\hskip.5in\le \tn G\tn\, \sup_{x_2}\int dy_n\ 
\big[\sup_{y_{\bar n}}|P(0,x_2,y_1,y_2)|\big]
\int dy_{\bar n}\ |H(y_1,y_2,k_3,k_4)|\cr
&\hskip.5in\le \tn G\tn\,\tn H\tn\, \sup_{x_2}\int dy_n\ 
\sup_{y_{\bar n}}|P(0,x_2,y_1,y_2)|\cr
}$$
\goodbreak
The other bound is achieved in a similar fashion, starting from
$$\eqalign{
&| G \fcirc P \fcirc H(k_1,k_2,k_3,k_4)|\cr
&\hskip.5in\le  \int  d^3x_1\, d^3x_2\,  d^3y_2\ 
|G(k_1,k_2,x_1,x_2)|\,|P(x_1,x_2,0,y_2)|\,|H(0,y_2,k_3,k_4)|\cr
}$$

Now consider the case in which at least one of $i_1,\ i_2,\ i_3,\ i_4$
is one. Pick any $\ell\in\{1,2,3,4\}$ with $i_\ell=1$. Then, by
translation invariance,
$$\eqalignno{
&\sup_{y_\ell}
\sup_{y_\nu\in\bbbm\atop {\nu=1,2,3,4\atop{\rm\ with\ }i_\nu=0}}\ 
\int \smprod_{\nu=1,2,3,4\atop{{\rm\ with\ }i_\nu=1\atop{\rm and\ }\nu\ne\ell}}dy_\nu\ 
\big|G \fcirc P\fcirc H(y_1,y_2,y_3,y_4)\big|\cr
&\hskip.2in\le
\sup_{y_\ell}
\sup_{y_\nu\in\bbbm\atop {\nu=1,2,3,4\atop{\rm\ with\ }i_\nu=0}}\ 
\hskip-7pt\int\hskip-7pt \smprod_{\nu=1,2,3,4\atop{{\rm\ with\ }i_\nu=1\atop{\rm and\ }\nu\ne\ell}}
\hskip-10pt dy_\nu\ 
\smprod_{\nu=1,2,3,4}\hskip-10pt dx_\nu\ 
\big|G(y_1,y_2,x_1,x_2)P(x_1,x_2,x_3,x_4)H(x_3,x_4,y_3,y_4)\big|\cr
&\hskip.2in=
\sup_{y_\nu\in\bbbm\atop {\nu=1,2,3,4\atop{\rm\ with\ }i_\nu=0}}\ 
\hskip-7pt\int\hskip-7pt \smprod_{\nu=1,2,3,4\atop{{\rm\ with\ }i_\nu=1\atop{\rm and\ }\nu\ne\ell}}
\hskip-10pt dy_\nu\ 
\smprod_{\nu=1,2,3,4}\hskip-10pt dx_\nu\ 
\big|G(y_1,y_2,x_1,x_2)P(x_1,x_2,x_3,x_4)H(x_3,x_4,y_3,y_4)\big|_{y_\ell=0}\cr
&\hskip.2in=
\sup_{y_\nu\in\bbbm\atop {\nu=1,2,3,4\atop{\rm\ with\ }i_\nu=0}}\ 
\hskip-7pt\int\hskip-7pt \smprod_{\nu=1,2,3,4\atop{{\rm\ with\ }i_\nu=1\atop{\rm and\ }\nu\ne\ell}}
\hskip-10pt dy_\nu\ 
\smprod_{\nu=1,2,3,4}\hskip-10pt dx_\nu\ 
\big|G(y_1,y_2,0,x_2)P(0,x_2,x_3,x_4)H(x_3,x_4,y_3,y_4)\big|_{y_\ell=-x_1}\cr
&\hskip.2in=
\sup_{y_\nu\in\bbbm\atop {\nu=1,2,3,4\atop{\rm\ with\ }i_\nu=0}}\ 
\hskip-7pt\int\hskip-7pt \smprod_{\nu=1,2,3,4\atop{\rm\ with\ }i_\nu=1}
\hskip-10pt dy_\nu\ 
\smprod_{\nu=2,3,4}\hskip-7pt dx_\nu\ 
\big|G(y_1,y_2,0,x_2)P(0,x_2,x_3,x_4)H(x_3,x_4,y_3,y_4)\big|\cr
&\hskip.2in\le\tn G\tn
\sup_{y_\nu\in\bbbm\atop {\nu=3,4\atop{\rm\ with\ }i_\nu=0}}\hskip-4pt
\sup_{x_2}
\int\hskip-7pt \smprod_{\nu=3,4\atop{\rm\ with\ }i_\nu=1}
\hskip-10pt dy_\nu\ 
\smprod_{\nu=3,4}\hskip-3pt dx_\nu\ 
\big|P(0,x_2,x_3,x_4)H(x_3,x_4,y_3,y_4)\big|
&\EQNO\eqnLADbubblenormB\cr
\noalign{\goodbreak}
&\hskip.2in\le\tn G\tn
\sup_{y_\nu\in\bbbm\atop {\nu=3,4\atop{\rm\ with\ }i_\nu=0}}\hskip-4pt
\sup_{x_2}
\int dx_3\, dx_4\, \big|P(0,x_2,x_3,x_4)\big|
\int\hskip-7pt \smprod_{\nu=3,4\atop{\rm\ with\ }i_\nu=1}
\hskip-10pt dy_\nu\ \big|H(x_3,x_4,y_3,y_4)\big|
\cr
}$$
For the second equality, we made the change of variables $y_\nu\rightarrow y_\nu+x_1$,
for each $\nu\ne\ell$ with $i_\nu=1$ and the change of variables 
$x_\nu\rightarrow x_\nu+x_1$, for each $\nu=2,3,4$ and then used translation
invariance of the three kernels. This replaces ``$y_\ell=0$'' 
by ``$y_\ell=-x_1$''. For the third equality, we made the change of 
variables $x_1\rightarrow -y_\ell$. Now we may continue as in the case $i_1=i_2=i_3=i_4=0$.

\endproof
\goodbreak

Our bubble propagators are typically of the form
$P=A\otimes B^t$ with translation invariant propagators $A$ and $B$.
If $A$ is a translation invariant propagator, we write $A(y-x)$ in place of 
$A(x,y)$. With this convention the $L^1$--$L^\infty$ norm of Definition
\defLADlonelinfty\ reduces to the $L^1$ norm $\|A\|_{L^1}=\int |A(y)|\ d^3y$.
If $P=A\otimes B^t$, then
$$P(x_1,x_2,y_1,y_2)=A(y_1-x_1)B(x_2-y_2)
= \figplace{wwBubProp2}{-.1 in}{-.22 in}
$$ 
and, by Lemma \lemLADbubblenorm,
$$
\|P\|_{\rm bubble}\le\min\big\{\|A\|_{L^\infty}\|B\|_{L^1},
\|A\|_{L^1}\|B\|_{L^\infty}\big\}
\EQN\eqnBubbleTensor$$

Given any function $W(p,k)$ on $\bbbm^2$, we associate to it the particle--hole bubble propagator
$$\eqalign{
W(x_1,x_2,y_1,y_2)
&=\int\sfrac{d^3p}{(2\pi)^3}\sfrac{d^3k}{(2\pi)^3}\ W(p,k)\,
e^{\imath <p,x_1-y_1>_-}e^{\imath <k,y_2-x_2>_-}\cr
&=\int\sfrac{d^3t}{(2\pi)^3}\sfrac{d^3k}{(2\pi)^3}\ W(k+t,k)
e^{\imath <k,x_1-y_1+y_2-x_2>_-}e^{\imath <t,x_1-y_1>_-}\cr
}\EQN\eqnVI$$
\centerline{\figplace{wwBubProp3}{0 in}{0 in}}\hfill\break
Here $k$ is the loop momentum and $t=p-k$ is the transfer momentum.

Motivated by the introduction to this section, we often treat small and 
large transfer momenta differently. To isolate a specific set of 
transfer momenta,
 we use a function $R(t)$ on $\bbbm$ that is supported there.
\definition{\STM\defLADtrasfermomcutoff}{
For any function $W(x_1,x_2,y_1,y_2)$ and any function $R(t)$, with Fourier
transform $\hat R(z)$, we set
$$
W_R(x_1,x_2,y_1,y_2)=\int dz\ W(x_1,x_2,y_1-z,y_2-z)\,\hat R(z)
$$
If $W(x_1,x_2,y_1,y_2)$ is associated with $W(p,k)$ as in (\eqnVI), then
$W_R(x_1,x_2,y_1,y_2)$ is associated with
$$
W_R(p,k)=W(p,k)R(p-k)
$$

}

\lemma{\STM\lemXIII}{
Let $A$ and $B$ be translation invariant propagators and $R(t)$ a function on 
$\bbbm$. Then
$$
\|(A\otimes B^t)_R\|_{\rm bubble}\le\|\hat R(x)\|_{L^1}
\min\big\{\|A(x)\|_{L^\infty}\|B(x)\|_{L^1},
\|A(x)\|_{L^1}\|B(x)\|_{L^\infty}\big\}
$$
}
\prf By Definition \defLADtrasfermomcutoff,
$$\eqalign{
(A\otimes B^t)_R(x_1,x_2,y_1,y_2)
&=\int dz\ A(y_1-x_1-z)B(x_2-y_2+z)\,\hat R(z)
}$$
By Lemma \lemLADbubblenorm,
$$
\|(A\otimes B^t)_R\|_{\rm bubble}\le \min_{n=1,2}\ 
\sup_{x_1,x_2}\int dy_n\sup_{y_{\bar n}}|(A\otimes B^t)_R(x_1,x_2,y_1,y_2)|
$$
We treat $n=1$. The other case is similar.
$$\eqalign{
 \sup_{x_1,x_2}\int dy_1\sup_{y_2}|(A\otimes B^t)_R(x_1,x_2,y_1,y_2)|
&\le \int dy_1\sup_{y_2}\int dz\ |A(y_1-z)B(-y_2-z)\hat R(z)|\cr
&\le \|B(x)\|_{L^\infty}\int dy_1\, dz\ |A(y_1-z)\hat R(z)|\cr
&=\|B(x)\|_{L^\infty}\|A(x)\|_{L^1}\|\hat R(z)\|_{L^1}
}$$
\endproof
\remark{\STM\remLADOR}{
Define, for any function $\hat R(x)$, the bubble operator
$$
O_R(x_1,x_2,y_1,y_2)=\hat R(y_1-x_1)\de(x_2-y_2+y_1-x_1)
$$
Then, for any bubble propagator $W$,
$$
W\circ O_R=W_R
$$
}

\vskip.25in

Replacing $g$ by $\sfrac{1}{M^{\ell|\de_\li|}}\rD^{\de_\li}_{1,2}g$ and 
$h$ by $\sfrac{1}{M^{j|\de_\ri|}}\rD^{\de_\ri}_{3,4}h$ in Theorem
\bbound\ reduces consideration of the norm 
$\big|g\bullet\cC^{[i,j]}\bullet h\big|_{\ell,j}^{[\de_\li,0,\de_\ri]}$ to a
$\big|\ \cdot\ \big|_{\ell,j}^{[0,0,0]}$ norm. Therefore, we introduce the 
short hand notation
\definition{\STM\defLADshortnorm}{ For $f$ a function on $\fY_{i_\li,i_\ri}$,
set
$$
\big|f\big|_{i_\li,i_\ri}=\big|f\big|_{i_\li,i_\ri}^{[0,0,0]}
$$
}

With the reduction to $\de_\li=\de_\ri=0$, indicated above, Theorem \bbound\ 
becomes bounds on the $\big|\ \cdot\ \big|_{\ell,j}$ norm
of quantities like $g\bullet\cC^{[i,j]}\bullet h$.  For the rest of this 
section, we fix $\ell\ge 1$ and consider, more generally,  
$\big|\ \cdot\ \big|_{\ell,r}$ norms with $r\ge j$. The 
$\big|\ \cdot\ \big|_{\ell,r}$ norm of a function $f$ is obtained by fixing all arguments that lie in $\bbbm$ and the sectors of all arguments that lie 
in $\big(\bbbr\times\bbbr^2\big)\times\Si_\ell$ or
$\big(\bbbr\times\bbbr^2\big)\times\Si_r$ and taking the $\tn\ \cdot\ \tn_{1,\infty}$ of the result. The transfer momentum $t$ is determined by the
momenta and sectors of the last two arguments of $f$. This motivates the 
following
\definition{\STM\defLADKspace}{
\Item i) 
Let $\fK_r=\bbbm\dunion\Si_r$ be the disjoint union of the set $\bbbm$ 
of external momenta and the set $\Si_r$ of sectors of scale $r$. 
\Item ii) Let $\ka_1,\ka_2\in\fK_r$. The subset $\ka_1-\ka_2$ of $\bbbm$ is defined by
$$
\ka_1-\ka_2=\cases{
\{\ka_1-\ka_2\}& if $\ka_1,\ka_2\in\bbbm$\cr
\noalign{\vskip.05in}
\set{\ka_1-k_2}{k_2\in\ka_2}& if $\ka_1\in\bbbm$, $\ka_2\in\Si_r$\cr
\noalign{\vskip.05in}
\set{k_1-\ka_2}{k_1\in\ka_1}& if $\ka_1\in\Si_r$, $\ka_2\in\bbbm$\cr
\noalign{\vskip.05in}
\set{k_1-k_2}{k_1\in\ka_1,k_2\in\ka_2}& if $\ka_1,\ka_2\in\Si_r$\cr
}
$$
\Item ii) Let $f$ be a function on $\fY_{\ell,r}$ and $\ka_1,\ka_2\in\fK_r$.
 Then
$$
\big\|f\big\|_{\ka_1,\ka_2}\hskip-5pt
=\cases{
\sum\limits_{i_1,i_2\in\{0,1\}}
\max\limits_{s_\nu\in\Si_\ell\atop {\rm\ if\ }i_\nu=1}\hskip-5pt
\sup\limits_{k_\nu\in\bbbm\atop {\rm\ if\ }i_\nu=0}
\TN f\big|_{(i_1,i_2,0,0)}({\sst y_1,y_2,\ka_1,\ka_2})\TN_{1,\infty}
&\hskip-2pt if $\ka_1,\ka_2\in\bbbm$\cr
\noalign{\vskip.05in}
\sum\limits_{i_1,i_2\in\{0,1\}}
\max\limits_{s_\nu\in\Si_\ell\atop {\rm\ if\ }i_\nu=1}\hskip-5pt
\sup\limits_{k_\nu\in\bbbm\atop {\rm\ if\ }i_\nu=0}
\TN f\big|_{(i_1,i_2,0,1)}({\sst y_1,y_2,\ka_1,(x_4,\ka_2)})\TN_{1,\infty}
&\hskip-2pt if $\ka_1\in\bbbm$, $\ka_2\in\Si_r$\cr
\noalign{\vskip.05in}
\sum\limits_{i_1,i_2\in\{0,1\}}
\max\limits_{s_\nu\in\Si_\ell\atop {\rm\ if\ }i_\nu=1}\hskip-5pt
\sup\limits_{k_\nu\in\bbbm\atop {\rm\ if\ }i_\nu=0}
\TN f\big|_{(i_1,i_2,1,0)}({\sst y_1,y_2,(x_3,\ka_1),\ka_2})\TN_{1,\infty}
&\hskip-2pt if $\ka_1\in\Si_r$, $\ka_2\in\bbbm$\cr
\noalign{\vskip.05in}
\sum\limits_{i_1,i_2\in\{0,1\}}
\max\limits_{s_\nu\in\Si_\ell\atop {\rm\ if\ }i_\nu=1}\hskip-5pt
\sup\limits_{k_\nu\in\bbbm\atop {\rm\ if\ }i_\nu=0}
\TN f\big|_{(i_1,i_2,1,1)}({\sst y_1,y_2,(x_3,\ka_1),(x_4,\ka_2)})\TN_{1,\infty}
&\hskip-2pt if $\ka_1,\ka_2\in\Si_r$\cr
}
$$
Here, we use the decomposition of (\eqnLADfourdunion) and,
for $\nu=1,2$, $y_\nu=k_\nu$ if $i_\nu=0$ and $y_\nu=(x_\nu,s_\nu)$
if $i_\nu=1$.
}
\remark{\STM\remLADKnorm}{
For a function $f$ on $\fY_{\ell,r}$
$$
\big|f\big|_{\ell,r}
\le 4\Big\{
\sup_{k_1,k_2\in\bbbm}\big\|f\big\|_{k_1,k_2}
+ \sup_{k_1\in\bbbm\atop \si_2\in\Si_r}\big\|f\big\|_{k_1,\si_2}
+ \sup_{\si_1\in\Si_r\atop k_2\in\bbbm}\big\|f\big\|_{\si_1,k_2}
+ \sup_{\si_1,\si_2\in\bbbm}\big\|f\big\|_{\si_1,\si_2}
\Big\}
$$
}

We now state the reformulation of parts (b) and (c) of Theorem \bbound.
 Recall the decomposition
$$
\cC^{[i,j]}=\cC_{\tp}^{[i,j]}+\cC_{\md}^{[i,j]}+\cC_{\bt}^{[i,j]}
$$
of the particle--hole bubble propagator $\cC^{[i,j]}$
with
$$
\cC_{\tp}^{[i,j]}
=\sum_{i\le i_t\le j\atop i_b>j}C^{(i_t)}_v\otimes C^{(i_b)\,t}_v,
\qquad\cC_{\md}^{[i,j]}=
\sum_{i\le i_t\le j\atop i\le i_b\le j}C^{(i_t)}_v\otimes C^{(i_b)\,t}_v,
\qquad\cC_{\bt}^{[i,j]}=
\sum_{i_t> j\atop i\le i_b\le j}C^{(i_t)}_v\otimes C^{(i_b)\,t}_v
$$
and recall from (\eqnLADDelta) that
$$
\De=\set{\de\in\bbbn_0\times\bbbn_0^2}{\de_{0}\le r_0,\ \de_1+\de_2\le r_e}
$$
where $r_e+3$ is the degree of differentiability of the dispersion relation $e(\k)$ and $r_0$ is the number of $k_0$ derivatives that we wish to control.

\theorem{\STM\rebbound}{Let $1\le i\le j\le r$ and $\ell\ge 1$ and 
let $g$ and $h$ be sectorized,
 translation invariant functions on $\fY_{\ell,i}$ and $\fY_{i,r}$ 
respectively. Let $\ka_1,\ka_2\in\fK_r$. \hfill\break
i)
For any $\be\in\De$
$$\eqalign{
\sfrac{1}{M^{|\be|j}}\big\|g\bullet \rD_{1,3}^{\be}\cC^{[i,j]}_{\tp}
\bullet h\big\|_{\ka_1,\ka_2}
&\le\const \big|g\big|_{\ell,i}\big|h\big|_{i,r}\cr
\sfrac{1}{M^{|\be|j}}\big\|g\bullet \rD_{2,4}^{\be}\cC^{[i,j]}_{\bt}
\bullet h\big\|_{\ka_1,\ka_2}
&\le\const \big|g\big|_{\ell,i}\big|h\big|_{i,r}\cr
}$$
\noindent 
ii) 
$$\eqalign{
\big\|g\bullet\cC^{[i,j]}_\md\bullet h\big\|_{\ka_1,\ka_2}
&\le\const |j-i+1|\ 
\big|g\big|_{\ell,i}\big|h\big|_{i,r}\cr
}$$
and for any $\be\in\De$ with $|\be|\ge 1$ and
$(\mu,\mu')=(1,3),(2,4)$
$$\eqalign{
\sfrac{1}{M^{|\be|j}}\big\|g\bullet \rD_{\mu,\mu'}^{\be}\cC^{[i,j]}_{\md}
\bullet h\big\|_{\ka_1,\ka_2}
&\le\const \big|g\big|_{\ell,i}\big|h\big|_{i,r}\cr
}$$
The constant $\const$ depends on $e(\k)$, $M$ and $\De$, but not on
$i,\ell,j,r,g,h,\ka_1$ or $\ka_2$.
}
\noindent
The proof of Theorem \rebbound\ follows Lemma \:\lemXV.
\vskip\baselineskip
\noindent{\bf Proof  of Theorem \bbound b,c} (assuming Theorem \rebbound): \hfill\break
As pointed out above, we may assume without loss of generality that 
$\de_\li=\de_\ri=0$. Then parts (b) and (c) of Theorem \bbound\ follow directly from  Remark \remLADKnorm\ and parts (i) and (ii) of Theorem \rebbound, with 
$r=j$, respectively.
\endproof

\definition{\STM\defLADtdiff}{
 For any subset $d\subset \bbbm$, let $\cR(d)$ be the set of all
functions $R(t)$ that are identically one on $d$.

}
\lemma{\STM\lemXIV}{ 
Let $1\le i\le j$ and $\ell,r\ge 1$.
Let $\ka_1,\ka_2\in\fK_r$ and $g$ and $h$ be
sectorized, translation invariant functions on $\fY_{\ell,i}$ and $\fY_{i,r}$ 
respectively.  Let $W$ be a particle--hole bubble propagator whose total
Fourier transform is of the form
$$
\check W(p_1,k_1,p_2,k_2)=\sum_{m=i}^j\ \sum_{s_1,s_2\in\Si_m}
W^{(m)}_{s_1,s_2}(p_1,k_1,p_2,k_2)
\qquad\hbox{ if\ \ }p_2-k_2\in\ka_1-\ka_2
$$
with $W^{(m)}_{s_1,s_2}(p_1,k_1,p_2,k_2)$ vanishing unless $p_1,p_2\in
s_1$ and $k_1,k_2\in s_2$. Then 
$$\eqalign{
\|g\bullet W\bullet h\|_{\ka_1,\ka_2}
&\le 81|g|_{\ell,i}|h|_{i,r}\  
\sum_{m=i}^j\sum_{s_1,s_2\in\Si_m\atop(s_1-s_2)\cap(\ka_1-\ka_2)\ne\emptyset}
\inf_{R\in\cR(\ka_1-\ka_2)}\big\|W^{(m)}_{s_1,s_2,R}\big\|_{\rm bubble}\cr
&\le 81|g|_{\ell,i}|h|_{i,r}\ 
\sum_{m=i}^j\sum_{s_1,s_2\in\Si_m\atop(s_1-s_2)\cap(\ka_1-\ka_2)\ne\emptyset}
\big\|W^{(m)}_{s_1,s_2}\big\|_{\rm bubble}\cr
}$$
}
\prf Consider the case in which all of the external arguments of
$g\bullet W\bullet h$ are $({\rm position,sector})$'s.  Fix  (external) 
sectors  $\si'_1,\si'_2\in\Si_\ell$ and call $\si_1=\ka_1,\si_2=\ka_2\in\Si_r$ and $d=\si_1-\si_2$.
With the sector names
 
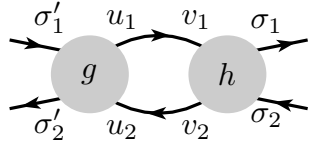
\centerline{\figplace{wwbubble3}{0 in}{0 in}}

\noindent we have
$$
g\bullet W\bullet h
=\sum_{m=i}^j\ 
\sum_{{u_1,v_1\in\Si_i\atop u_2,v_2\in\Si_i}\atop s_1,s_2\in \Si_m}
g({\sst(\,\cdot\,,\si'_1),(\,\cdot\,,\si'_2),(\,\cdot\,,u_1),(\,\cdot\,,u_2)})\fcirc
W^{(m)}_{s_1,s_2}\fcirc
 h({\sst(\,\cdot\,,v_1),(\,\cdot\,,v_2),(\,\cdot\,,\si_1),(\,\cdot\,,\si_2)})
$$
For each choice of $u_1,v_1, u_2,v_2, s_1,s_2$,
by conservation of momentum at the vertex $h$,
$$\eqalign{
&g({\sst(\,\cdot\,,\si'_1),(\,\cdot\,,\si'_2),(\,\cdot\,,u_1),(\,\cdot\,,u_2)})\fcirc
W^{(m)}_{s_1,s_2}\fcirc
 h({\sst(\,\cdot\,,v_1),(\,\cdot\,,v_2),(\,\cdot\,,\si_1),(\,\cdot\,,\si_2)})\cr
&=g({\sst(\,\cdot\,,\si'_1),(\,\cdot\,,\si'_2),(\,\cdot\,,u_1),(\,\cdot\,,u_2)})\fcirc
W^{(m)}_{s_1,s_2,R}\fcirc
 h({\sst(\,\cdot\,,v_1),(\,\cdot\,,v_2),(\,\cdot\,,\si_1),(\,\cdot\,,\si_2)})\cr
}$$
for all $R\in\cR(d)$ and the convolution vanishes
identically unless $(s_1-s_2)\cap d\ne\emptyset$.  
The convolution also vanishes identically unless 
$$\meqalign{
u_1\cap s_1&\ne\emptyset && s_1\cap v_1&\ne\emptyset \cr
u_2\cap s_2&\ne\emptyset && s_2\cap v_2&\ne\emptyset \cr
}$$
For each fixed $s_1,s_2$ there are only 81 quadruples $(u_1,u_2,v_1,v_2)$
satisfying these conditions. The same is true, by a similar argument, if one or more of the external arguments of $g\bullet W\bullet h$ are momenta.
Just replace, for example, $\si'_1$ by $\{k'\}$. Hence
$$
\|g\bullet W\bullet h\|_{\ka_1,\ka_2}
\le 81|g|_{\ell,i}|h|_{i,r}\  
\sum_{m=i}^j\sum_{s_1,s_2\in\Si_m\atop(s_1-s_2)\cap d\ne\emptyset}
\inf_{R\in\cR(d)}\big\|W^{(m)}_{s_1,s_2,R}\big\|_{\rm bubble}
$$
The second inequality follows by choosing an $R(t)$ that is identically
one on a large enough ball.
\endproof

\remark{\STM\remLADtdiff}{ 
Let $\ka_1,\ka_2\in\fK_r$.
\Item i) The set $\ka_1-\ka_2$ is contained in  a ball of radius $2\fl_r$.

\Item ii) Let $m\le r$. Then,
$$
\#\set{(s_1,s_2)\in\Si_m\times\Si_m}{(s_1-s_2)\cap (\ka_1-\ka_2)\ne\emptyset}
\le\sfrac{\const}{\fl_m}
$$
\Item iii) The set $\set{t_0\in\bbbr}{(t_0,\t)\in\ka_1-\ka_2\hbox{ for some }
\t\in\bbbr^2}$ is contained in an interval of length $\sfrac{4\sqrt{2M}}{M^r}$.
}
\prf Part (i) is an immediate consequence of the facts that $\ka_1$ and $\ka_2$
are each contained in a ball of radius $\fl_r$.
Given any fixed $s_1\in \Sigma_m$, 
$(s_1-s_2)\cap (\ka_1-\ka_2) \ne \emptyset$ only if 
$s_2$ intersects $s_1-\ka_1+\ka_2$. As $s_1-\ka_1+\ka_2$ is contained in a 
ball of radius at most $ 3 \fl_m$, there are at most eight sectors $s_2\in \Sigma_m$ that intersect it. This proves part (ii). Part (iii) follows from the 
fact that, for $\nu=1,2$, $\set{k_0\in\bbbr}{(k_0,\k)\in\ka_\nu\hbox{ for some }
\k\in\bbbr^2}$ is contained in an interval of length $\sfrac{2\sqrt{2M}}{M^r}$.
\endproof

\remark{\STM\remXIVector}{ 
Let $\bpi:k=(k_0,\k)\mapsto\k$ be the projection of $\bbbm=\bbbr\times\bbbr^2$
onto its second factor. If we retain all of the hypotheses of Lemma \lemXIV,
except that we only require  $\check W^{(m)}_{s_1,s_2}(p_1,k_1,p_2,k_2)$ 
to vanish unless $\bpi(p_1),\bpi(p_2)\in\bpi(s_1)$ 
and $\bpi(k_1),\bpi(k_2)\in\bpi(s_2)$, then we still have
$$\eqalign{
\|g\bullet W\bullet h\|_{\ka_1,\ka_2}
&\le 81|g|_{\ell,i}|h|_{i,r}\  
\sum_{m=i}^j\sum_{s_1,s_2\in\Si_m\atop\bpi(s_1-s_2)\cap
\bpi(\ka_1-\ka_2)\ne\emptyset}
\inf_{R\in\cR(\ka_1-\ka_2)}\big\|W^{(m)}_{s_1,s_2,R}\big\|_{\rm bubble}\cr
&\le 81|g|_{\ell,i}|h|_{i,r}\ 
\sum_{m=i}^j\sum_{s_1,s_2\in\Si_m\atop\bpi(s_1-s_2)\cap
\bpi(\ka_1-\ka_2)\ne\emptyset}
\big\|W^{(m)}_{s_1,s_2}\big\|_{\rm bubble}\cr
}$$

}

We are particularly interested in the particle--hole bubble propagator
$$
\cC^{[i,j]}(p,k)=\cC_\tp^{[i,j]}(p,k)+\cC^{[i,j]}_\md(p,k)+\cC_\bt^{[i,j]}(p,k)
$$
where
$$\eqalign{
\cC_\tp^{[i,j]}(p,k)
    &=\sum_{i\le m_1\le j\atop m_2>j}C^{(m_1)}_v(p)C^{(m_2)}_v(k)\cr
\cC_\md^{[i,j]}(p,k)
    &=\sum_{i\le m_1\le j\atop i\le m_2\le j}C^{(m_1)}_v(p)C^{(m_2)}_v(k)\cr
\cC_\bt^{[i,j]}(p,k)
    &=\sum_{m_1> j\atop i\le m_2\le j}C^{(m_1)}_v(p)C^{(m_2)}_v(k)\cr
}$$
We split $\cC_\tp^{[i,j]}$, $\cC_\md^{[i,j]}$ and $\cC_\bt^{[i,j]}$ into 
scales and we split each scale contribution
into pieces with additional sector restrictions on the momenta $p$ and
$k$ and the transfer momentum $p-k$. Recall, from just before Definition
\defLADresectoriz, that $\sum_{s\in\Si_m}\chi_s(k)$
is a partition of unity of the $m^{\rm th}$ neighbourhood
 subordinate to $\Si_m$. For any scale $i\le m\le j$ and 
sectors $s_1,s_2\in\Si_m$, set
$$\eqalign{
\cC^{(m)}_{\tp,j,s_1,s_2}(p,k)
     &=\sum_{m_2>j}C^{(m)}_v(p)\chi_{s_1}(p)C^{(m_2)}_v(k)\chi_{s_2}(k)\cr
\cC^{(m)}_{\md,j,s_1,s_2}(p,k)
&=\sum_{m_1,m_2\le j\atop \min(m_1,m_2)=m}
C^{(m_1)}_v(p)\chi_{s_1}(p)C^{(m_2)}_v(k)\chi_{s_2}(k)\cr
\cC^{(m)}_{\bt,j,s_1,s_2}(p,k)&=\sum_{m_1>j}
C^{(m_1)}_v(p)\chi_{s_1}(p)C^{(m)}_v(k)\chi_{s_2}(k)\cr
}$$
Then, for each of ${\rm loc}=\tp,\md,\bt$
$$
\cC^{[i,j]}_{{\rm loc}}
=\sum_{m=i}^j\ \sum_{s_1,s_2\in\Si_m}\cC^{(m)}_{{\rm loc},j,s_1,s_2}
$$

\lemma{\STM\lemXV}{ Let $1\le m\le j$ and $s_1,s_2\in\Si_m$. 
If $\be\in\De$ and $(\mu,\mu')\in\{(1,3),(2,4)\}$, then
$$\eqalign{ 
\|\rD_{1,3}^{\be}\cC^{(m)}_{\tp,j,s_1,s_2}\|_{\rm bubble}  
&\le\const\fl_m \sfrac{M^m}{M^j}M^{|\be|m}\cr 
\|\rD_{\mu,\mu'}^{\be}\cC^{(m)}_{\md,j,s_1,s_2}\|_{\rm bubble}
&\le\const\fl_m \cases{M^mM^{(|\be|-1)j}& if $|\be|\ge2$\cr
                         M^m( j-m+1)             & if $|\be|=1$\cr
                         1             & if $|\be|=0$\cr}\cr 
\|\rD_{2,4}^{\be}\cC^{(m)}_{\bt,j,s_1,s_2}\|_{\rm bubble}
&\le\const\fl_m \sfrac{M^m}{M^j}M^{|\be|m}\cr
}$$
}
\prf 
Set, for $s\in\Si_m$ and $n\ge1$,
$$
c^{(n)}_s(k)=C^{(n)}_v(k)\chi_s(k)
\EQN\eqnLADcsn$$
and denote by $c^{(n)}_s(x)$ its Fourier transform.
By Lemma \:\lemLADsimplepropbnd, for all $\be\in\De$,
$$\eqalignno{
\big\|x^\be c^{(m)}_s(x)\big\|_{L^1}&\le\const M^{(1+|\be|)m} 
&\EQNO\eqnLADlonepropbnd\cr
\big\|c^{(n)}_s(x)\big\|_{L^\infty}&\le \const\sfrac{\fl_m}{M^n}
&\EQNO\eqnLADlinftypropbnd\cr
\big\|x^\be c^{(n)}_s(x)\big\|_{L^\infty}&\le \const \fl_m M^{(|\be|-1)n}
\qquad\hbox{if $n\ge m$}
&\EQNO\eqnLADderivlinftypropbnd\cr
}$$
Recall that
$$
\rD_{1,3}^{\be}\cC^{(m)}_{\tp,j,s_1,s_2}(x_1,x_2,y_1,y_2) 
= \sum_{n> j}
(y_1-x_1)^{\be}c^{(m)}_{s_1}(y_1-x_1)\ c^{(n)}_{s_2}(x_2-y_2)
$$
Hence, by the triangle inequality and (\eqnBubbleTensor),
$$\eqalign{
\big\|\rD_{1,3}^{\be}\cC^{(m)}_{\tp,j,s_1,s_2}\big\|_{\rm bubble} 
&\le \sum_{n> j}
\big\|x^{\be}c^{(m)}_{s_1}\big\|_{L^1}\big\|c^{(n)}_{s_2}\big\|_{L^\infty}\cr
&\le \sum_{n> j}\const M^{(1+|\be|)m}\sfrac{\fl_m}{M^n}\cr
&\le\const \sfrac{M^m}{M^j}\fl_m M^{|\be|m}\cr
}$$
The bound on $\big\|\rD_{2,4}^{\be}\cC^{(m)}_{\bt,j,s_1,s_2}\big\|_{\rm bubble}$
 is proven similarly. As
$$\eqalign{
\rD_{1,3}^{\be}\cC^{(m)}_{\md,j,s_1,s_2}(x_1,x_2,y_1,y_2)
=& \sum_{m\le n\le j}
(y_1-x_1)^{\be}c^{(m)}_{s_1}(y_1-x_1)\ c^{(n)}_{s_2}(x_2-y_2)\cr
&+ \sum_{m<n\le j}
(y_1-x_1)^{\be}c^{(n)}_{s_1}(y_1-x_1)\ c^{(m)}_{s_2}(x_2-y_2)\cr
}$$
we have
$$\eqalign{
&\big\|\rD_{1,3}^{\be}\cC^{(m)}_{\md,j,s_1,s_2}\big\|_{\rm bubble} 
\le \sum_{m\le n\le j}
\big\|x^{\be}c^{(m)}_{s_1}\big\|_{L^1}\big\|c^{(n)}_{s_2}\big\|_{L^\infty}
+ \sum_{m< n\le j}
\big\|x^{\be} c^{(n)}_{s_1}\big\|_{L^\infty}\big\|c^{(m)}_{s_2}\big\|_{L^1}\cr
&\hskip.5in\le  \sum_{m\le n\le j}\const M^{(1+|\be|)m}\sfrac{\fl_m}{M^n}
+  \sum_{m< n\le j}\const \sfrac{\fl_m}{M^n}M^{|\be|n}M^{m}\cr
&\hskip.5in\le\const \fl_m \smsum_{m\le n\le j}\sfrac{M^m}{M^n}M^{|\be|n}\cr
}$$
To bound  $\big\|x^{\be} c^{(n)}_{s_1}\big\|_{L^\infty}$,
  we used (\eqnLADderivlinftypropbnd).
\endproof
\proof{ of Theorem \rebbound.i}
We prove the bound for $\cC^{[i,j]}_\tp$. The proof for 
$\cC^{[i,j]}_\bt$ is virtually identical. By Lemma \lemXIV, Remark \remLADtdiff.ii\ and Lemma \lemXV,
$$\eqalign{
\big\|g\bullet \rD^{\be}_{1,3}\cC^{[i,j]}_\tp\bullet h\big\|_{\ka_1,\ka_2}
&\le\const\big|g\big|_{\ell,i}\big|h\big|_{i,r}
\sum_{m=i}^j\sfrac{\const}{\fl_m}\max_{s_1,s_2\in \Si_m}
\big\|\rD^{\be}_{1,3}\cC^{(m)}_{\tp,j,s_1,s_2}\big\|_{\rm bubble}\cr
&\le\const\big|g\big|_{\ell,i}
\big|h\big|_{i,r}\sum_{m=i}^j\sfrac{\const}{\fl_m}
\fl_m\sfrac{M^m}{M^j} M^{|\be| m}\cr
&\le\const M^{|\be| j}
\big|g\big|_{\ell,i}\big|h\big|_{i,r}\cr
}$$
\endproof
\goodbreak
\proof{ of Theorem \rebbound.ii}
By Lemma \lemXIV, followed by Lemma \lemXV\ and Remark \remLADtdiff.ii, we have
$$\eqalign{
\big\|g\bullet\cC^{[i,j]}_\md\bullet h\big\|_{\ka_1,\ka_2}
&\le\const\big|g\big|_{\ell,i}
\big|h\big|_{i,r}\sum_{m=i}^j\sfrac{\const}{\fl_m}\max_{s_1,s_2\in \Si_m}
\big\|\cC^{(m)}_{\md,s_1,s_2}\big\|_{\rm bubble}\cr
&\le\const\big|g\big|_{\ell,i}
\big|h\big|_{i,r}\sum_{m=i}^j\sfrac{\const}{\fl_m}\fl_m\cr
&\le\const |j-i+1|\big|g\big|_{\ell,i}\big|h\big|_{i,r}\cr
}$$
For $|\be|\ge 1$ and $(\mu,\mu')=(1,3),(2,4)$, by Lemma \lemXV,
$$\eqalign{
\big\|g\bullet \rD^{\be}_{\mu,\mu'}\cC^{[i,j]}_\md\bullet h\big\|_{\ka_1,\ka_2}
&\le\const\big|g\big|_{\ell,i}
\big|h\big|_{i,r}
\sum_{m=i}^j\sfrac{\const}{\fl_m}\max_{s_1,s_2\in \Si_m}
\big\|\rD^{\be}_{\mu,\mu'}\cC^{(m)}_{\md,j,s_1,s_2}\big\|_{\rm bubble}\cr
&\le\const\big|g\big|_{\ell,i}
\big|h\big|_{i,r}\sum_{m=i}^j\sfrac{1}{\fl_m}
\fl_mM^m\!\cases{M^{(|\be|-1)j}& $|\be|\ge2$\cr
                          (j-m+1)             & $|\be|=1$\cr}\cr
&\le\const M^{|\be| j}
\big|g\big|_{\ell,i}\big|h\big|_{i,r}\cr
}$$
since, for $|\be|\ge 2$,
$$
\sum_{m=i}^jM^mM^{(|\be|-1)j}\le\const M^{|\be|j}
$$
and, for $|\be|=1$,
$$
\sum_{m=i}^jM^m(j-m+1)=M^j\sum_{m=i}^jM^{-(j-m)}(j-m+1)\le\const M^{j}
$$

\endproof

\vskip.25in
We now start the proof of part (a) of Theorem \bbound. We shall prove,
at the end of this subsection, the following bound on the small transfer momentum
contributions to $g\bullet\cC^{[i,j]}\bullet h$.
\theorem{\STM\rebboundII}{Let $1\le i\le j\le r$ and $\ell\ge 1$
and let $\ka_1,\ka_2\in\fK_r$.
Set $d=\ka_1-\ka_2$ and denote by $\bd$ the projection of $d$ onto
$\{0\}\times\bbbr^2$ identified with $\bbbr^2$.
By Remark \remLADtdiff, the set $\bd$ is contained in a disc of 
radius $2\fl_r$. Fix such a disk and denote by $\btau$ its centre.
Furthermore, set $\tau_0=\inf\set{|t_0|}{(t_0,\t)\in d\hbox{ for some }
\t\in\bbbr^2}$. Assume that
$$
\tau_0\le\sfrac{1}{M^{j-1}}\qquad
|\btau|\le\max\big\{\sfrac{1}{M^j},r^3\fl_r\big\}\qquad
M^i\le \fl_j M^j
$$
Also assume that $p^{(i)}$ vanishes for all $i>j+1$.
For any sectorized, translation invariant functions $g$ and $h$ 
on $\fY_{\ell,i}$ and $\fY_{i,r}$ respectively,
$$
\big\|g\bullet\cC^{[i,j]}\bullet h\big\|_{\ka_1,\ka_2}
\le\const \max_{\al_\ri,\al_\li\in\bbbn_0\times\bbbn_0^2
            \atop|\al_\ri|+|\al_\li|\le 3}
\big|g\big|_{\ell,i}^{[0,0,\al_\ri]}\big|h\big|_{i,r}^{[\al_\li,0,0]}
$$
The constant $\const$ depends on $e(\k)$, $M$ and $\De$, but not on
$i,\ell,j,r,g,h,\ka_1$ or $\ka_2$.
}
\noindent
Theorem \rebboundII\ is proven following Proposition \:\propXXXII.

\vskip\baselineskip
\noindent{\bf Proof of Theorem \bbound a} (assuming Theorem 
\rebboundII):\hfill\break
As pointed out above, we may assume without loss of generality that 
$\de_\li=\de_\ri=0$. Fix $0\le i,\ell\le j$ and sectorized, translation invariant functions $g$ and $h$ on $\fY_{\ell,i}$ and $\fY_{i,j}$
as in Theorem \bbound. By Remark \remLADKnorm, it suffices to prove that
$$
\big\|g\bullet\cC^{[i,j]}\bullet h\big\|_{\ka_1,\ka_2}
\le\const\,i \max_{\al_\ri,\al_\li\in\bbbn_0\times\bbbn_0^2
            \atop|\al_\ri|+|\al_\li|\le 3}
\big|g\big|_{\ell,i}^{[0,0,\al_\ri]}\big|h\big|_{i,j}^{[\al_\li,0,0]}
\EQN\eqnLADbboundZ$$
for all $\ka_1,\ka_2\in\fK_j$. Fix $\ka_1,\ka_2\in\fK_j$. 
Set $d=\ka_1-\ka_2$ and denote by $\bd$ the projection of $d$ onto
$\{0\}\times\bbbr^2$ identified with $\bbbr^2$.
By Remark \remLADtdiff, the set $\bd$ is contained in a disc of 
radius $2\fl_j$. We fix such a disk and denote by $\btau$ its centre.
Furthermore, we define $\tau_0=\inf\set{|t_0|}{(t_0,\t)\in d\hbox{ for some }
\t\in\bbbr^2}$. Define
$$\eqalign{
j_0&=\cases{\max\set{n\in\bbbn_0}{\tau_0\le\sfrac{1}{M^{n-1}}}& 
                if $0<\tau_0\le M$\cr
             0 & if $\tau_0\ge M$\cr
             \infty & if $\tau_0=0$\cr
      }\cr
j_1&=\cases{\max\set{n\in\bbbn_0}{|\btau|\le\sfrac{1}{M^n}}& 
                if $j^3\fl_j<|\btau|\le 1$\cr
             0 & if $|\btau|\ge 1$\cr
             \infty & if $|\btau|\le j^3\fl_j$\cr
      }\cr
\bar\jmath&=\max\Big\{i-1,\min\{j,j_0,j_1\}\Big\}
}$$
One of the tools that we use in the proof that 
Theorem \rebboundII\ implies Theorem \bbound.a is
\proposition{\STM\propLADlarget (Large Transfer Momentum)}{ 
$$
\big\|g\bullet\cC^{[\bar\jmath+1,j]}\bullet h\big\|_{\ka_1,\ka_2}\le\const|g|_{\ell,i}|h|_{i,j}
$$
}
\prf
If $\min\{j,j_0,j_1\}=j$, then $\bar\jmath=j$ and $\cC^{[\bar\jmath+1,j]}=0$ so
that there is nothing to prove.
So we may assume that $\min\{j_0,j_1\}<j$.
\Item{\it Case 1: $j_0\le j_1$.} In this case, 
$\big\|g\bullet\cC^{[\bar\jmath+1,j]}\bullet h\big\|_{\ka_1,\ka_2}=0$,
because $\cC^{[\bar\jmath+1,j]}(p,k)$ vanishes unless
$|p_0|,|k_0|\le\sfrac{\sqrt{2M}}{M^{\bar\jmath+1}}$ and hence unless
$|p_0-k_0|\le\sfrac{2\sqrt{2M}}{M^{\bar\jmath+1}}
< \sfrac{1}{M^{\bar\jmath}}<\tau_0$, while $|t_0|\ge\tau_0$ for all $t\in d$.
\Item{\it Case 2: $j_1< j_0$.} In this case $|\btau|> j^3\fl_j$. Let
$\de_F$ be the constant of Lemma \:\lemAI. By Lemma \:\lemAI.a, 
with $\ep=2\fl_j$ and $m\le j$,
$$
\#\set{(s_1,s_2)\in\Si_m\times\Si_m}{(s_1-s_2)\cap d\ne\emptyset}
\le\const\cases{
\sfrac{1}{\sqrt{\fl_m}} & if $|\btau|\ge\de_F$\cr
1+\sfrac{1}{|\btau|\fl_m}\big(\sfrac{1}{M^m}+\fl_j\big) & otherwise\cr
}$$
Hence, by Lemma \lemXIV\ and Lemma \lemXV,
$$\eqalign{
\big\|g\bullet\cC^{[\bar\jmath+1,j]}\bullet h\big\|_{\ka_1,\ka_2}
&\le \const|g|_{\ell,i}|h|_{i,j}\hskip-3pt
\sum_{m=\bar\jmath+1}^j\hskip-5pt \fl_m\ 
\#\set{(s_1,s_2)\in\Si_m\times\Si_m}{(s_1-s_2)\cap d\ne\emptyset}\cr
&\le \const|g|_{\ell,i}|h|_{i,j}
\cases{
\sum\limits_{m=\bar\jmath+1}^j\sqrt{\fl_m} & if $|\btau|\ge\de_F$\cr
1+\sfrac{1}{|\btau|}\sum\limits_{m=j_1+1}^j\big(\sfrac{1}{M^m}+\fl_j\big) & otherwise\cr
}\cr
&\le \const|g|_{\ell,i}|h|_{i,j}\cr
}$$
since, by the definition of $j_1$,
$$
\sfrac{1}{|\btau|}\sum_{m=j_1+1}^j\big(\sfrac{1}{M^m}+\fl_j\big)
\le\sfrac{1}{|\btau|}\big(\sfrac{1}{M^{j_1}}+j\fl_j\big)
\le\const
$$
\endproof

\noindent{\bf Continuation of the proof of Theorem \bbound a} (assuming 
Theorem \rebboundII):\ \hfill\break
When $M^i\ge \fl_{\bar\jmath}M^{\bar\jmath}=M^{(1-\aleph)\bar\jmath}$,
we have $|\bar\jmath-i+1|\le\const i$. In this case Theorem \rebbound,
with $r=j$ and $j=\bar\jmath$, gives
$$\eqalign{
\big\|g\bullet\cC^{[i,\bar\jmath]}\bullet h\big\|_{\ka_1,\ka_2}
&\le\const i\ 
\big|g\big|_{\ell,i}\big|h\big|_{i,j}\cr
}$$
This together with Proposition \propLADlarget\ yields (\eqnLADbboundZ).
Therefore, we may assume that 
$$
M^i\le \fl_{\bar\jmath}M^{\bar\jmath}
\EQN\eqnLADsmalli$$
Furthermore, if $\bar\jmath= i-1$, $\cC^{[i,\bar\jmath]}=0$ and there is
nothing more to prove. So we may also assume that $j_0,j_1\ge i$ and 
$\bar\jmath\le j,j_0,j_1$.

Set $v'=\sum_{i=2}^{\bar\jmath+1}p^{(i)}$. 
 Recall that
$
\cC^{[i,\bar\jmath]}=\cC_{\tp}^{[i,\bar\jmath]}+\cC_{\md}^{[i,\bar\jmath]}
+\cC_{\bt}^{[i,\bar\jmath]}
$
with
$$
\cC_{\tp}^{[i,\bar\jmath]}
=\sum_{i\le i_t\le \bar\jmath\atop i_b>\bar\jmath}C^{(i_t)}_v\otimes C^{(i_b)\,t}_v,
\qquad\cC_{\md}^{[i,\bar\jmath]}=
\sum_{i\le i_t\le \bar\jmath\atop i\le i_b\le \bar\jmath}C^{(i_t)}_v\otimes C^{(i_b)\,t}_v,
\qquad\cC_{\bt}^{[i,\bar\jmath]}=
\sum_{i_t> \bar\jmath\atop i\le i_b\le \bar\jmath}C^{(i_t)}_v\otimes C^{(i_b)\,t}_v
$$ 
and set 
$
\cC^{'[i,\bar\jmath]}=\cC_{\tp}^{'[i,\bar\jmath]}+\cC_{\md}^{'[i,\bar\jmath]}+\cC_{\bt}^{'[i,\bar\jmath]}
$
with
$$
\cC_{\tp}^{'[i,\bar\jmath]}
=\sum_{i\le i_t\le \bar\jmath\atop i_b>\bar\jmath}C^{(i_t)}_{v'}\otimes C^{(i_b)\,t}_{v'},
\qquad\cC_{\md}^{'[i,\bar\jmath]}=
\sum_{i\le i_t\le \bar\jmath\atop i\le i_b\le \bar\jmath}C^{(i_t)}_{v'}\otimes C^{(i_b)\,t}_{v'},
\qquad\cC_{\bt}^{'[i,\bar\jmath]}=
\sum_{i_t> \bar\jmath\atop i\le i_b\le \bar\jmath}C^{(i_t)}_{v'}\otimes C^{(i_b)\,t}_{v'}
$$ 
As $v-v'$ is supported on the $(\bar\jmath+2)^{\rm nd}$ extended neighbourhood, $\cC_{\md}^{[i,\bar\jmath]}=\cC_{\md}^{'[i,\bar\jmath]}$. Hence,
by Theorem \rebbound.i,
with $\be=0$, $r=j$ and $j=\bar\jmath$,
$$\eqalign{
\big\|g\bullet\big[\cC^{[i,\bar\jmath]}-\cC^{'[i,\bar\jmath]}\big]\bullet h\big\|_{\ka_1,\ka_2}
&\le\const \big|g\big|_{\ell,i}\big|h\big|_{i,j}\cr
}\EQN\eqnLADctocprime$$
By (\eqnLADsmalli) and
the Definitions of $\bar\jmath$ and $\cC^{'[i,\bar\jmath]}$, the 
hypotheses of Theorem \rebboundII,
with $r=j$ and $j=\bar\jmath$, apply to
$g\bullet\cC^{'[i,\bar\jmath]}\bullet h$. Hence 
$$
\big\|g\bullet\cC^{'[i,\bar\jmath]}\bullet h\big\|_{\ka_1,\ka_2}
\le\const \max_{\al_\ri,\al_\li\in\bbbn_0\times\bbbn_0^2
            \atop|\al_\ri|+|\al_\li|\le 3}
\big|g\big|_{\ell,i}^{[0,0,\al_\ri]}\big|h\big|_{i,j}^{[\al_\li,0,0]}
$$
This together with (\eqnLADctocprime) and 
 Proposition \propLADlarget\ yields (\eqnLADbboundZ). 
This completes the proof that Theorem \rebboundII\ implies Theorem \bbound.a.
\endproof

\vskip.25in
The rest of this subsection is devoted to the proof of Theorem \rebboundII.
So we fix $\ell\ge 1$, $1\le i\le j\le r$ and sectorized, translation 
invariant functions, $g$ and $h$, on $\fY_{\ell,i}$ and $\fY_{i,j}$ 
respectively. We also fix $\ka_1,\ka_2\in\fK_r$ and
assume that
$$
\tau_0\le\sfrac{1}{M^{j-1}}\qquad
|\btau|\le\max\big\{\sfrac{1}{M^j},r^3\fl_r\big\}\qquad
M^i\le\fl_j M^j
\EQN\eqnLADassume$$
and that $p^{(i)}$ vanishes for all $i>j+1$.

We shall not need to decompose $
\cC^{[i,j]}=\cC_\tp^{[i,j]}+\cC^{[i,j]}_\md+\cC_\bt^{[i,j]}
$
but we still split $\cC^{[i,j]}$ into scales and split each scale contribution
into pieces with additional sector restrictions.  For
any scale $i\le m\le j$ and sectors $s_1,s_2\in\Si_m$, set
$$
\cC^{(m)}_{s_1,s_2}(p,k)=\sum_{m_1,m_2\ge0\atop \min(m_1,m_2)=m}
c_{s_1}^{(m_1)}(p)\ c_{s_2}^{(m_2)}(k)
$$
where $c_{s}^{(n)}$ was defined in (\eqnLADcsn).
Then
$$
\cC^{[i,j]}=\sum_{m=i}^j\ \sum_{s_1,s_2\in\Si_m}\cC^{(m)}_{s_1,s_2}
$$
By Lemmas \lemXIII\ and \lemXV,
$$\eqalign{
\|\cC^{(m)}_{s_1,s_2,R}\|_{\rm bubble}&\le\const\fl_m\|\hat R(x)\|_{L^1}\cr
\|\cC^{(m)}_{s_1,s_2}\|_{\rm bubble}&\le\const\fl_m\cr
}\EQN\eqLemXV$$

\goodbreak
\titled{Reduction to the Model Bubble Propagator}

The above argument for large transfer momentum 
implicitly exploited the fact that the particle--hole bubble is H\"older
continuous in the transfer momentum $t$ when $t$ is nonzero. 
As was pointed out in the introduction, 
this is false for $t=0$. However, if one restricts to transfer momenta
with $t_0=0$ then, at least for the delta function interaction, $C^\infty$
dispersion relation and a model
propagator with suitable cutoff procedure, the particle--hole bubble is 
in fact $C^\infty$ for $\t$ near zero. This was seen in
Lemma \lemLADprimitivemanfred.

Lemma \lemLADprimitivemanfred\  applied to the particle--hole bubble with a delta function
interaction and choice of cutoff different from that used in this paper.
In the present situation, we have general interaction kernels $g$ and $h$
rather than delta functions and cutoffs that do not treat $k_0$ and $e(\k)$
independently. Furthermore, the time component $t_0$ of the transfer momentum
need not be zero. We now perform three reduction steps leading to a situation
similar to that of Lemma \lemLADprimitivemanfred.

\noindent{\it Step 1 (Decoupling of the $k_0$ integral.)}

Define the zero component localization operator
$$
\cZ(x_1,x_2,y_1,y_2)=\de(x_1-y_1)
\de(\x_2-\y_2)\de(y_{1,0}-y_{2,0})
\EQN\eqnVII$$
where $x_i=(x_{i,0},\x_i)$ and $y_i=(y_{i,0},\y_i)$.
The transpose of this operator has kernel
$$
\cZ^t(x_1,x_2,y_1,y_2)=\de(x_1-y_1)
\de(\x_2-\y_2)\de(x_{1,0}-x_{2,0})
$$
\remark{\STM\remXXII}{
If $W(x_1,x_2,y_1,y_2)$ is a particle--hole propagator
$$
\big(\cZ\fcirc W\fcirc\cZ^t\big)(x_1,x_2,y_1,y_2)
=W\big(x_1,(x_{1,0},\x_2),y_1,(y_{1,0},\y_2)\big)
$$
If $W(x_1,x_2,y_1,y_2)$ is associated to $W(p,k)$ as in (\eqnVI), then
$$
\big(\cZ\fcirc W\fcirc\cZ^t\big)(x_1,x_2,y_1,y_2)
=\int\sfrac{d^3t}{(2\pi)^3}\sfrac{d^2\k}{(2\pi)^3}\ 
e^{\imath \k\cdot(\x_1-\y_1+\y_2-\x_2)}e^{\imath <t,x_1-y_1>_-}
\int dk_0\  W(k+t,k)
$$
That is, $\big(\cZ\fcirc W\fcirc\cZ^t\big)$ is associated to
$\de(k_0)\int d\om\ W\big((\om,\0)+p,(\om,\k)\big)$. 
}
\lemma{\STM\lemLADzprops}{
Let $W$ be a particle--hole bubble propagator.
\Item i) 
Let $R(t)$ be any cutoff function for the transfer momentum. Then,
$$
\big(\cZ\fcirc W\fcirc\cZ^t\big)_R=\cZ\fcirc W_R\fcirc\cZ^t
$$ 
\Item ii) For any translation invariant kernels $G$ 
on $\fY^2\times\big(\bbbr\times\bbbr^2\big)^2$ and $H$ on
$\big(\bbbr\times\bbbr^2\big)^2\times\fY^2$,
$$
\tn G\fcirc\cZ\tn\le \tn G\tn\qquad{\rm and}\qquad
\tn \cZ^t\fcirc H\tn\le \tn H\tn
$$
\Item iii)
$$
\|\cZ\fcirc W\|_{\rm bubble}\le\|W\|_{\rm bubble}\qquad\hbox{and}\qquad
\|W\fcirc \cZ^t\|_{\rm bubble}\le\|W\|_{\rm bubble}
$$
\Item iv)
$$\eqalign{
\|\cZ\fcirc W\fcirc\cZ^t\|_{\rm bubble}
\le\min\Big\{
&\min_{n=1,2}\sup_{x_1,x_2}\int d\y_ndy_{1,0}\sup_{\y_{\bar n}}
\big|W\big(x_1,(x_{1,0},\x_2),y_1,(y_{1,0},\y_2)\big)\big|,\cr
&\hskip-3pt\min_{n=1,2}\sup_{y_1,y_2}\int d\x_ndx_{1,0}\sup_{\x_{\bar n}}
\big|W\big(x_1,(x_{1,0},\x_2),y_1,(y_{1,0},\y_2)\big)\big|
\Big\}
}$$
where $\bar n=2$ if $n=1$ and $\bar n=1$ if $n=2$.

}
\prf i)
This is obvious from Remark \remLADOR,
since $\cZ^t\fcirc O_R=O_R\fcirc \cZ^t$.
\Item ii) This is obvious since
$$\eqalign{
(G\fcirc\cZ)(\ \cdot\ ,\ \cdot\ ,x_3,x_4)
&=\de(x_{3,0}-x_{4,0})\int d\om \ 
G\big(\ \cdot\ ,\ \cdot\ ,x_3,(\om,\x_4)\big)\cr
(\cZ^t\fcirc H)(x_1,x_2,\ \cdot\ ,\ \cdot\ )
&=\de(x_{1,0}-x_{2,0})\int d\om \ 
H\big(x_1,(\om,\x_2),\ \cdot\ ,\ \cdot\ \big)\cr
}$$
\Item iii)
By part (ii), for any translation invariant $G,H$
$$
\tn G\fcirc\cZ\fcirc W\fcirc H\tn
\le \tn G\fcirc\cZ\tn\ \|W\|_{\rm bubble}\ \tn H\tn
\le\tn G\tn\ \|W\|_{\rm bubble}\ \tn H\tn
$$
and similarly for $\tn G\fcirc W\fcirc\cZ^t\fcirc H\tn$.
\Item iv)
The bounds with $n=1,\bar n=2$ are direct consequences of Remark \remXXII\
and Lemma \lemLADbubblenorm. We prove
$$
\|\cZ\fcirc W\fcirc\cZ^t\|_{\rm bubble}
\le
\sup_{x_1,x_2}\int d\y_2\,dy_{1,0}\,\sup_{\y_1}
\big|W\big(x_1,(x_{1,0},\x_2),y_1,(y_{1,0},\y_2)\big)\big|
$$
The remaining case is similar. Let $G(y_1,y_2,u_1,u_2)$ and $H(v_1,v_2,y_3,y_4)$
be translation invariant four--legged kernels obeying
$$
G=G\big|_{(i_1,i_2,1,1)}\qquad
H=H\big|_{(1,1,i_3,i_4)}
$$
for some ${i_1,i_2,i_3,i_4\in\{0,1\}}$.
 By (\eqnLADbubblenormA) and
(\eqnLADbubblenormB), with $P$ replaced by $\cZ\fcirc W\fcirc\cZ^t$,
$$\eqalign{
&\tn G\fcirc \cZ\fcirc W\fcirc\cZ^t\fcirc H\tn\cr
&\hskip.2in\le\tn G\tn
\sup_{y_\nu\in\bbbm\atop {\nu=3,4\atop{\rm\ with\ }i_\nu=0}}\hskip-4pt
\sup_{u_1,u_2}
\int\hskip-7pt \smprod_{\nu=3,4\atop{\rm\ with\ }i_\nu=1}\hskip-10pt dy_\nu\ 
dv_1 dv_2\ 
\big|W\big(u_1,(u_{1,0},{\bf u}_2),v_1,(v_{1,0},{\bf v}_2)\big)
H(v_1,v_2,y_3,y_4)\big|\cr
&\hskip.2in\le\tn G\tn 
\sup_{y_\nu\in\bbbm\atop {\nu=3,4\atop{\rm\ with\ }i_\nu=0}}\hskip-4pt
\sup_{u_1,u_2}
\int\! dv_{1,0}\, d{\bf v}_2 \Big\{\sup_{{\bf v}_1}
\big|W\big(u_1,(u_{1,0},{\bf u}_2),v_1,(v_{1,0},{\bf v}_2)\big)\big|\cr
&\hskip2.5in
\int\hskip-7pt \smprod_{\nu=3,4\atop{\rm\ with\ }i_\nu=1}\hskip-10pt dy_\nu\ 
 dv_{2,0}\,d{\bf v}_1
\big|H(v_1,v_2,y_3,y_4)\big|\Big\}\cr
&\hskip.25in\le\tn G\tn \Big[\sup_{u_1,u_2}\int\! dv_{1,0} d{\bf v}_2\ \sup_{{\bf v}_1}
\big|W\big(u_1,(u_{1,0},{\bf u}_2),v_1,(v_{1,0},{\bf v}_2)\big)\big|\Big]\cr
&\hskip2.4in
\sup_{y_\nu\in\bbbm\atop {\nu=3,4\atop{\rm\ with\ }i_\nu=0}}\hskip-6.5pt
\sup_{v_{1,0}, {\bf v}_2}
\int\hskip-10pt \smprod_{\nu=3,4\atop{\rm\ with\ }i_\nu=1}\hskip-10pt dy_\nu\ 
 dv_{2,0}\,d{\bf v}_1
\big|H(v_1,v_2,y_3,y_4)\big|\cr
}$$
By translation invariance
$$\eqalign{
&\sup_{v_{1,0}, {\bf v}_2}
\int\hskip-10pt \smprod_{\nu=3,4\atop{\rm\ with\ }i_\nu=1}\hskip-10pt dy_\nu\ 
 dv_{2,0}\,d{\bf v}_1
\big|H(v_1,v_2,y_3,y_4)\big|\cr
&\hskip1in=\sup_{v_1}\int 
\hskip-10pt \smprod_{\nu=3,4\atop{\rm\ with\ }i_\nu=1}\hskip-10pt dy_\nu\ 
dv_2\ 
\big|H\big((v_{1,0},{\bf v}_2),(v_{2,0},{\bf v}_1),y_3,y_4\big)\big|\cr
&\hskip1in=\sup_{v_1}\int 
\hskip-10pt \smprod_{\nu=3,4\atop{\rm\ with\ }i_\nu=1}\hskip-10pt dy_\nu\ dv_2\ 
\big|H\big(v_1,(v_{2,0},2{\bf v}_1-{\bf v}_2),y_3,y_4\big)\big|\cr
&\hskip1in=\sup_{v_1}\int 
\hskip-10pt \smprod_{\nu=3,4\atop{\rm\ with\ }i_\nu=1}\hskip-10pt dy_\nu\ dv_2\ 
\big|H(v_1,v_2,y_3,y_4)\big|\le\tn H\tn\cr
}$$

\endproof

\proposition{\STM\propLADzdiff}{
$$
\big\|g\bullet\big(\cC^{[i,j]}-\cZ\bullet\cC^{[i,j]}\bullet\cZ^t \big)
\bullet h\big\|_{\ka_1,\ka_2}
\le\const\max_{\al_\li,\al_\ri\in\De\atop  \al_\li+\al_\ri=(1,0,0)}
|g|_{\ell,i}^{[0,0,\al_\ri]}|h|_{i,r}^{[\al_\li,0,0]}
$$
}
\noindent
In preparation for the proof, which follows Lemma \:\lemXXV,
we define
$$\eqalign{
g_\ri({\sst(x_1,\si'_1),(x_2,\si'_2),(x_3,u_1),(x_4,u_2)})
&=(x_{4,0}-x_{3,0})
g({\sst(x_1,\si'_1),(x_2,\si'_2),(x_3,u_1),(x_4,u_2)})\cr
h_\li({\sst(x_1,v_1),(x_2,v_2),(x_3,\si_1),(x_4,\si_2)})
&=(x_{2,0}-x_{1,0})h({\sst(x_1,v_1),(x_2,v_2),(x_3,\si_1),(x_4,\si_2)})
}$$
For a particle--hole bubble propagator $W(x_1,x_2,y_1,y_2)$ set
$$\eqalign{
(D_\li W)(x_1,x_2,y_1,y_2)
&=\int_0^1 d\om\ \sfrac{\partial\ W}{\partial x_{2,0}}
\big(x_1,(\om x_{2,0}+(1-\om)x_{1,0},\x_2),y_1,y_2\big)\cr
(D_\ri W)(x_1,x_2,y_1,y_2)
&=\int_0^1 d\om\ \sfrac{\partial\ W}{\partial y_{2,0}}
\big(x_1,x_2,y_1,(\om y_{2,0}+(1-\om)y_{1,0},\y_2)\big)\cr
}$$
\lemma{\STM\lemXXIV}{
$$
g\bullet(W-\cZ W\cZ^t)\bullet h=g_\ri\bullet D_\li W\bullet h
+g\fcirc\cZ\bullet D_\ri W\bullet h_\li
$$
}
\prf By Remark \remXXII,
$$\eqalign{
&g\bullet(W-\cZ W\cZ^t)\bullet h
=g\bullet\big\{W(x_1,x_2,y_1,y_2)-
W\big(x_1,(x_{1,0},\x_2),y_1,(y_{1,0},\y_2)\big)\big\}\bullet h\cr
&\hskip.5in=g_\ri\bullet\big[\sfrac{1}{x_{2,0}-x_{1,0}}
\big\{W(x_1,x_2,y_1,y_2)-
W\big(x_1,(x_{1,0},\x_2),y_1,y_2\big)\big\}
\big]\bullet h\cr
&\hskip.5in\ \ +g\bullet\big[\big\{W\big(x_1,(x_{1,0},\x_2),y_1,y_2\big)-
W\big(x_1,(x_{1,0},\x_2),y_1,(y_{1,0},\y_2)\big)\big\}\sfrac{1}{y_{2,0}-y_{1,0}}
\big]\bullet h_\li\cr
&\hskip.5in=g_\ri\bullet D_\li W\bullet h+g\fcirc\cZ\bullet D_\ri W\bullet h_\li
}$$
by the Fundamental Theorem of Calculus.
\endproof
\lemma{\STM\lemXXV}{
Let $i\le m\le j$ and $s_1,s_2\in\Si_m$. Then
$$\eqalign{
\|D_\li\cC^{(m)}_{s_1,s_2}\|_{\rm bubble}
\le\const\sfrac{\fl_m}{M^m}\cr
\|D_\ri\cC^{(m)}_{s_1,s_2}\|_{\rm bubble}
\le\const\sfrac{\fl_m}{M^m}\cr
}$$
}
\prf
We treat $D_\li\cC^{(m)}_{s_1,s_2}$. The other case is similar. For
each fixed $0\le\om\le 1$
$$\eqalign{
& \Big(\sfrac{\partial\hfill}{\partial x_{2,0}}\cC^{(m)}_{s_1,s_2}\Big)
\big(x_1,(\om x_{2,0}+(1-\om)x_{1,0},\x_2),y_1,y_2\big)\cr
&\hskip.5in=\hskip-5pt\sum_{m_1,m_2\ge 0\atop \min\{m_1,m_2\}=m}
c^{(m_1)}_{s_1}({\sst y_1-x_1}) 
\Big(\sfrac{\partial\hfill}{\partial x_{2,0}}c^{(m_2)}_{s_2}\Big)
({\sst(\om x_{2,0}+(1-\om)x_{1,0}-y_{2,0},\x_2-\y_2)})\cr
}$$
We bound the bubble norm of each term separately. For $m_1\ge m_2=m$,
by Lemma \lemLADbubblenorm,
$$\eqalign{
&\Big\|c^{(m_1)}_{s_1}
\Big(\sfrac{\partial\hfill}{\partial x_{2,0}}c^{(m_2)}_{s_2}\Big)
\Big\|_{\rm bubble}\cr
&\hskip.5in\le \big\|c^{(m_1)}_{s_1}\big\|_{L^\infty}
\sup_{x_1,x_2}
\int dy_2 \ \Big| 
\Big(\sfrac{\partial\hfill}{\partial x_{2,0}}c^{(m_2)}_{s_2}\Big)
({\sst(\om x_{2,0}+(1-\om)x_{1,0}-y_{2,0},\x_2-\y_2)})\Big|\cr
&\hskip.5in= \big\|c^{(m_1)}_{s_1}\big\|_{L^\infty}
\Big\|\sfrac{\partial\hfill}{\partial x_{2,0}}c^{(m_2)}_{s_2}\Big\|_{L^1}\cr
&\hskip.5in\le 
\const\sfrac{\fl_m}{M^{m_1}}\sfrac{1}{M^{m_2}}M^{m_2}
\le\const\sfrac{\fl_m}{M^{m_1}}\cr
}$$
by parts (iii) and (iv) of Lemma \:\lemLADsimplepropbnd. For $m=m_1\le m_2$
$$\eqalign{
\Big\|c^{(m_1)}_{s_1}
\Big(\sfrac{\partial\hfill}{\partial x_{2,0}}c^{(m_2)}_{s_2}\Big)
\Big\|_{\rm bubble}
&\le \Big\|\sfrac{\partial\hfill}{\partial x_{2,0}}c^{(m_2)}_{s_2}\Big\|_{L^\infty}\big\|c^{(m_1)}_{s_1}\big\|_{L^1}
\le \const\sfrac{\fl_m}{M^{2m_2}}M^{m_1}\cr
}$$
Hence
$$
\|D_\li\cC^{(m)}_{s_1,s_2}\|_{\rm bubble}
\le\const\sum_{n=m}^\infty\Big[\sfrac{\fl_m}{M^{n}}
+\sfrac{\fl_m}{M^{2n}}M^{m}\Big]
\le\const \sfrac{\fl_m}{M^{m}}
$$
\endproof
\proof{of Proposition \propLADzdiff} By Lemma \lemXXIV\ followed by Remark
\remXIVector\footnote{$^{(1)}$}{The operators $D_\li,D_\ri$ can enlarge
supports in the $k_0$ direction. So we cannot apply Lemma \lemXIV\ directly.}
$$\eqalign{
&\big\|g\bullet\big(\cC^{[i,j]}-\cZ\bullet\cC^{[i,j]}
\bullet\cZ^t \big)\bullet h\big\|_{\ka_1,\ka_2}\cr
&\hskip.5in\le \big\|g_\ri\bullet D_\li\cC^{[i,j]}\bullet h\big\|_{\ka_1,\ka_2}
+\big\|g\fcirc\cZ\bullet D_\ri\cC^{[i,j]}\bullet 
                     h_\li\big\|_{\ka_1,\ka_2}\cr
&\hskip.5in\le\const|g_\ri|_{\ell,i}|h|_{i,r}
\sum_{m=i}^{j}\sum_{s_1,s_2\in\Si_m\atop\bpi(s_1-s_2)\cap \bpi(d)\ne\emptyset}
\big\|D_\li\cC^{(m)}_{s_1,s_2}\big\|_{\rm bubble}\cr
&\hskip.5in\ +\const|g\fcirc\cZ|_{\ell,i}|h_\li|_{i,r}
\sum_{m=i}^{j}\sum_{s_1,s_2\in\Si_m\atop\bpi(s_1-s_2)\cap \bpi(d)\ne\emptyset}
\big\|D_\ri\cC^{(m)}_{s_1,s_2}\big\|_{\rm bubble}\cr
}$$
By the spatial projection of Remark \remLADtdiff.ii,
$$
\#\set{(s_1,s_2)\in\Si_m\times\Si_m}{\bpi(s_1-s_2)\cap \bpi(d)\ne\emptyset}
\le\sfrac{\const}{\fl_m}
\EQN\eqnLADspatcnt$$
Using this, Lemma \lemXXV, Lemma \lemLADzprops.ii  and the definitions
of $g_\ri,\ h_\li$, we have
$$\eqalign{
&\big\|g\bullet\big(\cC^{[i,j]}-\cZ\bullet\cC^{[i,j]}
\bullet\cZ^t \big)\bullet h\big\|_{\ka_1,\ka_2}\cr
&\hskip.4in\le\const M^i|g|_{\ell,i}^{[0,0,(1,0,0)]}|h|_{i,r}
\sum_{m=i}^{j}\sfrac{1}{\fl_m}\sfrac{\fl_m}{M^m}
+\const|g|_{\ell,i}M^i|h|_{i,r}^{[(1,0,0),0,0]}
\sum_{m=i}^{j}\sfrac{1}{\fl_m}\sfrac{\fl_m}{M^m}\cr
}$$
\endproof

\noindent{\it Step 2 (Reduction to $t_0=0$.)}

For any particle--hole bubble propagator $W(x_1,x_2,y_1,y_2)$ set
$$
\widetilde W(x_1,x_2,y_1,y_2)
=\de(y_{1,0}-x_{1,0})\int dz_0\  W\big(x_1,(x_{1,0},\x_2),(z_0,\y_1),(z_0,\y_2)\big)
\EQN\eqnVIII$$
If $W(x_1,x_2,y_1,y_2)$ is associated to $W(p,k)$ as in (\eqnVI), then
$\widetilde W(x_1,x_2,y_1,y_2)$ is associated to
$$
\widetilde W(p,k)=\de(k_0)\int d\om\ W\big((\om,\p),(\om,\k)\big)
$$
By Remark \remXXII, $\widetilde W=\cZ\fcirc\widetilde W\fcirc\cZ^t$
for all particle--hole bubble propagators $W$.
\proposition{\STM\propXXVIII}{
$$
\big\|g\bullet\big(\cZ\bullet\cC^{[i,j]}\bullet\cZ^t
-\widetilde\cC^{[i,j]}\big)\bullet h\big\|_{\ka_1,\ka_2}
\le\const|g|_{\ell,i}|h|_{i,r}
$$
}
\prf
Choose a $C^\infty_0$ function $\phi(t_0)$ that takes values in $[0,1]$,
is supported in the interval  $|t_0|\le2\sfrac{M+4\sqrt{2M}}{M^{j}}$,
is identically one for $|t_0|\le\sfrac{M+4\sqrt{2M}}{M^{j}}$ and obeys
$
\big|\sfrac{d^n\hfill}{dt_0^n}\phi(t_0)\big|\le\const M^{{j}n}
$
for $n\le 2$. By Remark \remLADtdiff.iii,
$$
\set{|t_0|}{(t_0,\t)\in d\hbox{ for some }\t\in\bbbr^2}
\subset\Big[\tau_0,\tau_0+\sfrac{4\sqrt{2M}}{M^j}\Big]
$$
Hence, by (\eqnLADassume),  $\phi$ is in $\cR(d)$.
By Remark \remXIVector\ and (\eqnLADspatcnt),
$$\eqalign{
&\big\|g\bullet\big(\cZ\fcirc\cC^{[i,j]}\fcirc\cZ^t
-\widetilde\cC^{[i,j]}\big)\bullet h\big\|_{\ka_1,\ka_2}\cr
&\hskip1in\le\const|g|_{\ell,i}|h|_{i,r} 
\sum_{m=i}^{j}\sfrac{1}{\fl_m}\max_{s_1,s_2\in\Si_m}
\big\|\cZ\fcirc\cC^{(m)}_{s_1,s_2,\phi}\fcirc\cZ^t
-\big(\widetilde\cC^{(m)}_{s_1,s_2}\big)_\phi
\big\|_{\rm bubble}
}$$
Here, we used $(\cZ\fcirc W\fcirc\cZ^t)_R=\cZ\fcirc W_R\fcirc\cZ^t$, which
was proven in Lemma \lemLADzprops.i.
The proposition follows from the next Lemma.

\endproof
\lemma{\STM\lemXXIX}{
Let $m\le {j}$ and $s_1,s_2\in\Si_m$. Then
$$
\big\|\cZ\fcirc\cC^{(m)}_{s_1,s_2,\phi}\fcirc\cZ^t
-\big(\widetilde\cC^{(m)}_{s_1,s_2}\big)_\phi
\big\|_{\rm bubble}\le\const\fl_m\sfrac{M^m}{M^{j}}
$$
}
\prf
For any $m_1,m_2\ge 0$, set
$$\eqalign{
W^{(m_1,m_2)}
&=\cZ\fcirc\big(c^{(m_1)}_{s_1}\otimes c^{(m_2)\,t}_{s_2}\big) \fcirc\cZ^t
-\big(c^{(m_1)}_{s_1}\otimes c^{(m_2)\,t}_{s_2}\big)^{\widetilde{}}\cr
&=\cZ\fcirc\big(c^{(m_1)}_{s_1}\otimes c^{(m_2)\,t}_{s_2}\big) \fcirc\cZ^t
-\cZ\fcirc\big(c^{(m_1)}_{s_1}\otimes c^{(m_2)\,t}_{s_2}\big)^{\widetilde{}}
\fcirc\cZ^t\cr
}$$
Observe that
$$
\cZ\fcirc\cC^{(m)}_{s_1,s_2,\phi}\fcirc\cZ^t
-\big(\widetilde\cC^{(m)}_{s_1,s_2}\big)_\phi
=\sum_{m_1,m_2\in\bbbn_0\atop\min\{m_1,m_2\}=m }
W^{(m_1,m_2)}_{\phi}
\EQN\eqnLADwsum$$
We now fix any $m_1,m_2\ge 0$ with $\min\{m_1,m_2\}=m$ and bound
$\ 
\big\|W^{(m_1,m_2)}_{\phi}\big\|_{\rm bubble}
\ $.
By definition
$$\eqalign{
&W^{(m_1,m_2)}(x_1,x_2,y_1,y_2)\cr
&\hskip.4in=c^{(m_1)}_{s_1}(y_1-x_1)c^{(m_2)}_{s_2}\big((x_{1,0}-y_{1,0},\x_2-\y_2)\big)
\cr
&\hskip.8in-\de(y_{1,0}-x_{1,0})\int du\  c^{(m_1)}_{s_1}\big((u-x_{1,0},\y_1-\x_1)\big)
c^{(m_2)}_{s_2}\big((x_{1,0}-u,\x_2-\y_2)\big)\cr
}$$
and
$$\eqalign{
&W^{(m_1,m_2)}_{\phi}(x_1,x_2,y_1,y_2)\cr
&\hskip.3in=
\int dz_0\ \hat \phi(z_0)\Big[ 
c^{(m_1)}_{s_1}({\sst (y_{1,0}-x_{1,0}-z_0,\y_1-\x_1)})
c^{(m_2)}_{s_2}({\sst (x_{1,0}-y_{1,0}+z_0,\x_2-\y_2)})
\cr
&\hskip.6in-\de(y_{1,0}-x_{1,0}-z_0)\int du\  
c^{(m_1)}_{s_1}({\sst(u-x_{1,0},\y_1-\x_1)})
c^{(m_2)}_{s_2}({\sst(x_{1,0}-u,\x_2-\y_2)})\Big]\cr
&\hskip.3in=
\int dz_0\ 
c^{(m_1)}_{s_1}({\sst (y_{1,0}-x_{1,0}-z_0,\y_1-\x_1)})
c^{(m_2)}_{s_2}({\sst (x_{1,0}-y_{1,0}+z_0,\x_2-\y_2)})\hat \phi({\sst z_0})
\cr
&\hskip.6in-\int du \  
c^{(m_1)}_{s_1}({\sst (u-x_{1,0},\y_1-\x_1)})
c^{(m_2)}_{s_2}({\sst(x_{1,0}-u,\x_2-\y_2)})
\hat \phi({\sst y_{1,0}-x_{1,0}})\cr
&\hskip.3in=\int dz_0 \  
c^{(m_1)}_{s_1}({\sst (z_0-x_{1,0},\y_1-\x_1)})
c^{(m_2)}_{s_2}({\sst(x_{1,0}-z_0,\x_2-\y_2)})
\big[\hat \phi({\sst y_{1,0}-z_0})-\hat \phi({\sst y_{1,0}-x_{1,0}})\big]\cr
}$$
The last factor
$$
\hat\phi(y_{1,0}-z_0)-\hat\phi(y_{1,0}-x_{1,0})
=(x_{1,0}-z_0)\int_0^1\!dt\ 
\hat\phi'\big(y_{1,0}-x_{1,0}+t(x_{1,0}-z_0)\big)
$$
Observe that
$$
\int dy_{1,0}\int_0^1\!dt\ 
 \big|\hat\phi'\big(y_{1,0}-x_{1,0}+t(x_{1,0}-z_0)\big)\big|
=\int dy_{1,0}\ |\hat\phi'(y_{1,0})|\le\sfrac{\const}{M^{j}}
$$
since
$$
|\hat\phi'(y_{1,0})|
\le\const\sfrac{1/M^{2{j}}}{{[1+|y_{1,0}/M^{j}|]}^2}
$$
If $m_2\ge m_1=m$, we apply
Lemma \lemLADzprops.iv\footnote{$^{(2)}$}{Note that in the bound on 
the right hand side of Lemma \lemLADzprops.iv, $W$ only appears in the
form $W(x_1,(x_{1,0},\x_2),y_1,(y_{1,0},\y_2))
=(\cZ\circ W\circ\cZ^t)(x_1,x_2,y_1,y_2)$.},  (\eqnLADlinftypropbnd) and 
(\eqnLADlonepropbnd), giving
$$\eqalign{
&\big\|W^{(m_1,m_2)}_{\phi}\big\|_{\rm bubble}\cr
&\hskip.3in\le\const\sup_{x_1,x_2}\int dy_1\sup_{\y_2}\int dz_0\ 
|x_{1,0}-z_0|\cr
&\hskip2in\big|c^{(m_1)}_{s_1}({\sst (z_0-x_{1,0},\y_1-\x_1)})\big|
\,\big|c^{(m_2)}_{s_2}({\sst(x_{1,0}-z_0,\x_2-\y_2)})\big|\cr
&\hskip2in\int_0^1\!dt\ 
\big|\hat\phi'({\sst y_{1,0}-x_{1,0}+t(x_{1,0}-z_0)})\big|\cr
&\hskip.3in\le\const\big\|c^{(m_2)}_{s_2}\big\|_{L^\infty}\sup_{x_1}
\int d\y_1 dz_0\ |x_{1,0}-z_0|
\big|c^{(m_1)}_{s_1}({\sst (z_0-x_{1,0},\y_1-\x_1)})\big|\cr
&\hskip2in\int dy_{1,0}\int_0^1\!dt\ 
\big|\hat\phi'({\sst y_{1,0}-x_{1,0}+t(x_{1,0}-z_0)})\big|\cr
&\hskip.3in\le\const\sfrac{1}{M^{j}}\sfrac{\fl_m}{M^{m_2}}
\Big[\sup_{x_1}\int d\y_1 dz_0\ |x_{1,0}-z_0|
\big|c^{(m_1)}_{s_1}({\sst (z_0-x_{1,0},\y_1-\x_1)})\big|\Big]
\cr
&\hskip.3in=\const\sfrac{1}{M^{j}}\sfrac{\fl_m}{M^{m_2}}
\big\|x_0c^{(m_1)}_{s_1}(x)\big\|_{L^1}\,\cr
&\hskip.3in\le\const\sfrac{1}{M^{j}}\sfrac{\fl_m}{M^{m_2}}M^{2m}
\,
}$$
Similarly, if $m_1\ge m_2=m$,
$$\eqalign{
&\big\|W^{(m_1,m_2)}_{\phi}\big\|_{\rm bubble}\cr
&\hskip.3in\le\const\sup_{x_1,x_2}\int dy_{1,0}d\y_2\sup_{\y_1}\int dz_0\ 
|x_{1,0}-z_0|\cr
&\hskip2in\big|c^{(m_1)}_{s_1}({\sst (z_0-x_{1,0},\y_1-\x_1)})\big|
\,\big|c^{(m_2)}_{s_2}({\sst(x_{1,0}-z_0,\x_2-\y_2)})\big|\cr
&\hskip2in\int_0^1\!dt\ 
\big|\phi'({\sst y_{1,0}-x_{1,0}+t(x_{1,0}-z_0)})\big|\cr
&\hskip.3in\le\const\sfrac{1}{M^{j}}\big\|c^{(m_1)}_{s_1}\big\|_{L^\infty}
\sup_{x_{1,0},\x_2}
\int d\y_2 dz_0\ |x_{1,0}-z_0|
\big|c^{(m_2)}_{s_2}({\sst (x_{1,0}-z_0,\x_2-\y_2)})\big|\,
\cr
&\hskip.3in\le\const\sfrac{1}{M^{j}}\sfrac{\fl_m}{M^{m_1}}M^{2m}
}$$
Consequently, by (\eqnLADwsum),
$$\eqalign{
\big\|\cZ\fcirc\cC^{(m)}_{s_1,s_2,\phi}\fcirc\cZ^t
-\big(\widetilde\cC^{(m)}_{s_1,s_2}\big)_\phi\big\|_{\rm bubble}
&\le\const\hskip-15pt
\sum_{m_1,m_2\ge 0\atop\min\{m_1,m_2\}=m}\hskip-15pt
\sfrac{M^{2m}}{M^{j}}\fl_m\min\big\{\sfrac{1}{M^{m_1}},
\sfrac{1}{M^{m_2}}\big\}\cr
&\le\const 
\sfrac{M^{2m}}{M^{j}}\fl_m\sum_{m'\ge m}\sfrac{1}{M^{m'}}\cr
&\le\const \sfrac{M^m}{M^{j}}\fl_m\ 
}$$
\endproof

\noindent{\it Step 3 (Introduction of Factorized Cutoff.)}

Define
$$\eqalign{
\nu_0(\om)&=\sum_{m=i+1}^{{j}-1}\nu\big(M^{2m}\om^2\big)\cr
\nu_1(\p,\k)&=\Big[\sum_{m_1=i+1}^{\infty}\nu\big(M^{2m_1}e(\p)^2\big)\Big]
\Big[\sum_{m_2=i+1}^{\infty}\nu\big(M^{2m_2}e(\k)^2\big)\Big]\cr
}$$
where $\nu$ is the single scale cutoff introduced in Definition
\defLADscales. Recall that $\nu(x)$ is identically one on 
$\big[\sfrac{2}{M},M]$ and is supported on $\big[\sfrac{1}{M},2M]$. Define 
$$
e'(k)=e(\k)-v(k)
$$
and the model particle--hole bubble propagator
$$
\cM(p,k)=\de(k_0)\int d\om\ 
\frac{\nu_0(\om)\nu_1(\p,\k)}
{[i\om-e'(\om,\p)][i\om-e'(\om,\k)]}
\EQN\eqnIX$$
Observe that $\cZ\fcirc\cM\fcirc\cZ^t=\cM$.
\proposition{\STM\propXXX}{
$$
\big\|g\bullet\big(\widetilde\cC^{[i,{j}]}-\cM\big)
               \bullet h\big\|_{\ka_1,\ka_2}
\le\const|g|_{\ell,i}|h|_{i,r}
$$
}
\prf The cutoff $\nu_0(\om)\nu_1(\p,\k)$ is supported on
$\Big\{(\om,\p,\k)\,\Big|\,|\om|,|e(\p)|,|e(\k)|\le\sqrt{\sfrac{2}{M}}\sfrac{1}{M^i}
\Big\}$. Since $\nu^{(m)}(k)$ vanishes 
for all $|ik_0-e(\k)|\le\sfrac{1}{\sqrt{M}}\sfrac{1}{M^{i-1}}$ when 
$m\le i-1$ 
$$
\nu_0(\om)\nu_1(\p,\k)=\sum_{m_1,m_2\ge i}
\nu_0({\sst\om})\nu_1({\sst\p,\k})\nu^{(m_1)}({\sst (\om,\p)})
\nu^{(m_2)}({\sst (\om,\k)})
$$
if $M$ is large enough. Since every $k$ in the support of $\nu^{(m)}(k)$ 
obeys $|ik_0-e(\k)|\le\sqrt{2M}\sfrac{1}{M^{j+1}}$ for all 
$m\ge j+1$ and since $\nu_0(\om)$ is supported on 
$|\om|\ge \sqrt{M}\sfrac{1}{M^j}$ 
$$
\nu_0(\om)\nu_1(\p,\k)
=\sum_{i\le \min\{m_1,m_2\}\le j}
\nu_0({\sst\om})\nu_1({\sst\p,\k})\nu^{(m_1)}({\sst (\om,\p)})
\nu^{(m_2)}({\sst (\om,\k)})
$$
Recall that
$$\eqalign{
\widetilde \cC^{[i,{j}]}(p,k)
&=\de(k_0)\int d\om\ \cC^{[i,{j}]}\big((\om,\p),(\om,\k)\big)\cr
&=\de(k_0)\int d\om\ \sfrac{\nu^{(\ge i)}((\om,\p))\nu^{(\ge i)}((\om,\k))
-\nu^{(\ge j+1)}((\om,\p))\nu^{(\ge j+1)}((\om,\k))}
{[i\om-e'(\om,\p)][i\om-e'(\om,\k)]}\cr
}$$
The difference of cutoff functions
$$\eqalign{
&\nu^{(\ge i)}({\sst (\om,\p)})\nu^{(\ge i)}({\sst (\om,\k)})
-\nu^{(\ge j+1)}({\sst (\om,\p)})\nu^{(\ge j+1)}({\sst (\om,\k)})
-\nu_0(\om)\nu_1(\p,\k)\cr
&\hskip3in
=\sum_{m=i}^j\sum_{m_1,m_2\ge 1\atop\min\{m_1,m_2\}=m}\mu^{(m_1,m_2)}(\om,\k,\p)
}$$
where
$$
\mu^{(m_1,m_2)}(\om,\k,\p)=\Big[1-\nu_0({\sst\om})\nu_1({\sst\p,\k})\Big]
\nu^{(m_1)}({\sst (\om,\p)})\nu^{(m_2)}({\sst (\om,\k)})
$$
Define, for $i\le m\le j$, $m_1,m_2\ge m$ and $s_1,s_2\in \Si_m$
$$
\cD^{m_1,m_2}_{s_1,s_2}(p,k)=\de(k_0)\phi(p_0)\De^{m_1,m_2}_{s_1,s_2}(\p,\k)
$$ 
where
$$
\De^{m_1,m_2}_{s_1,s_2}(\p,\k)
=\int d\om\ \frac{\mu^{(m_1,m_2)}(\om,\k,\p)\chi_{s_1}(0,\p)\chi_{s_2}(0,\k)}
{[i\om-e'(\om,\p)][i\om-e'(\om,\k)]}
$$
and $\phi$ was defined at the beginning of the proof of Proposition \propXXVIII.
Define
$$
\cD^{(m)}_{s_1,s_2}(p,k)
=\sum_{m_1,m_2\ge 1\atop\min\{m_1,m_2\}=m}\cD^{m_1,m_2}_{s_1,s_2}(p,k)
\EQN\eqnX$$
As $\phi(p_0)=1$ for all $|p_0|\le\sfrac{M+4\sqrt{2M}}{M^{j}}$,
$$
\widetilde\cC^{[i,{j}]}-\cM=\sum_{m=i}^{j}\ \sum_{s_1,s_2\in\Si_m}
\cD^{(m)}_{s_1,s_2}
\qquad\hbox{if }|p_0|\le\sfrac{M+4\sqrt{2M}}{M^{j}}
$$
Observe that the kernels of $\widetilde\cC^{[i,{j}]}$, $\cM$ and 
$\cD^{(m)}_{s_1,s_2}(p,k)$ each contain a factor of $\de(k_0)$. Hence,
in the product $g\bullet\big(\widetilde\cC^{[i,{j}]}-\cM
-\sum_{m=i}^{j}\cD^{(m)}_{s_1,s_2}\big)\bullet h$, $\check h$ is 
restricted to $k_0=0$, so that $p_0=t_0$, where $t$ is the transfer momentum.
But $t\in d$, so that, by Remark \remLADtdiff\ and (\eqnLADassume),
$|t_0|\le\tau_0+\sfrac{4\sqrt{2M}}{M^r}\le \sfrac{M+4\sqrt{2M}}{M^{j}}$. 
Hence, by Remark \remXIVector\ and (\eqnLADspatcnt),
$$
\big\|g\bullet\big(\widetilde\cC^{[i,{j}]}-\cM\big)
            \bullet h\big\|_{\ka_1,\ka_2}
\le\const |g|_{\ell,i}|h|_{i,r}\sum_{m=i}^{j}\sfrac{1}{\fl_m}
\max_{s_1,s_2\in\Si_m}\big\|\cD^{(m)}_{s_1,s_2}\big\|_{\rm bubble}
$$
The Proposition follows from Lemma \:\lemXXXI\ below.
\endproof
\lemma{\STM\lemXXXI}{
Let $i\le m\le j$ and $s_1,s_2\in\Si_m$. Then
$$
\big\|\cD^{(m)}_{s_1,s_2}\big\|_{\rm bubble}
\le\const\cases{\fl_m& if $m=i,i+1$\cr
\noalign{\vskip.1in}
(j-m+1)\sfrac{M^m}{M^{j}}\fl_m& if $m\ge i+2$\cr}
$$
}
\prf
Fix any $m_1,m_2$ with $\min\{m_1,m_2\}=m$. If $\om$ is in the support
of $\mu^{(m_1,m_2)}(\om,\k,\p)$ for some $\k,\ \p$ then $|\om|\le\sfrac{\const}
{M^{\max\{m_1,m_2\}}}$. In the case when $m>i+1$, $|\om|$ is restricted
even farther. Then, in the support of $\mu^{(m_1,m_2)}(\om,\k,\p)$, both
$|i\om-e(\p)|\le\sfrac{\sqrt{2M}}{M^{m_1}}\le\sfrac{\sqrt{2M}}{M^{i+2}}$
and $|i\om-e(\k)|\le\sfrac{\sqrt{2M}}{M^{m_2}}\le\sfrac{\sqrt{2M}}{M^{i+2}}$
and hence $|\om|,\ |e(\p)|,\ |e(\k)|\le \sfrac{\sqrt{2M}}{M^{i+2}}$. 
But $\nu_1(\p,\k)=1$ whenever 
$|e(\p)|,\ |e(\k)|\le \sfrac{1}{M^{i+1/2}}$. Hence on the support 
of $\mu^{(m_1,m_2)}(\om,\k,\p)$, $|\om|\le \sfrac{\sqrt{2M}}{M^{i+2}}$
and $\nu_0(\om)\ne 1$. This forces $|\om|\le \sfrac{\sqrt{2}}{M^{{j}-1/2}}\le \sfrac{1}{M^{{j}-3/4}}$.
Set
$$
b(m_1,m_2)=\cases{\sfrac{\const}{M^{\max\{m_1,m_2\}}}& if $m=i, i+1$\cr
\noalign{\vskip.1in} \min\Big\{\sfrac{1}{M^{{j}-3/4}},\sfrac{\const}{M^{\max\{m_1,m_2\}}}\Big\}& if $m\ge i+2$\cr}
$$
Thus
$$
\De^{m_1,m_2}_{s_1,s_2}(\p,\k)
=\int_{-b(m_1,m_2)}^{b(m_1,m_2)}
 d\om\ \frac{\mu^{(m_1,m_2)}(\om,\k,\p)\chi_{s_1}(\p)\chi_{s_2}(\k)}
{[i\om-e'(\om,\p)][i\om-e'(\om,\k)]}
$$
By Lemma \:\lemLADdeltabnd, the Fourier transform of 
$\De^{m_1,m_2}_{s_1,s_2}(\p,\k)$ obeys
$$\eqalign{
\int d\z_1\sup_{\z_2}\big|\hat \De^{m_1,m_2}_{s_1,s_2}(\z_1,\z_2)\big|
&\le\const b(m_1,m_2)\ \fl^2_{m}\sfrac{M^{m_1}}{\fl_{m_1}}\cr
\int d\z_2\sup_{\z_1}\big|\hat \De^{m_1,m_2}_{s_1,s_2}(\z_1,\z_2)\big|
&\le\const b(m_1,m_2)\ \fl^2_{m}\sfrac{M^{m_2}}{\fl_{m_2}}\cr
}$$
As the particle--hole bubble propagator associated to 
$\cD^{(m)}_{s_1,s_2}(p,k)$ by (\eqnVI), namely
$$
\cD^{m_1,m_2}_{s_1,s_2}(x_1,x_2,y_1,y_2)=\hat\phi(y_{1,0}-x_{1,0})
\hat\De^{m_1,m_2}_{s_1,s_2}(\y_1-\x_1,\x_2-\y_2),
$$
is independent of $x_{2,0}$ and $y_{2,0}$, 
$\cZ\circ\cD^{m_1,m_2}_{s_1,s_2}\circ\cZ^t=\cD^{m_1,m_2}_{s_1,s_2}$ and
we have, by Lemma \lemLADzprops.iv,
$$\eqalign{
\|\cD^{m_1,m_2}_{s_1,s_2}\|_{\rm bubble}
&\le\const\min\Big\{
\sup_{x_1,x_2}\int dy_1\sup_{\y_2}
\big|\hat\phi(y_{1,0}-x_{1,0})
\hat\De^{m_1,m_2}_{s_1,s_2}(\y_1-\x_1,\x_2-\y_2)\big|,\cr
&\hskip.8in\sup_{x_1,x_2}\int dy_{1,0}d\y_2\sup_{\y_1}
\big|\hat\phi(y_{1,0}-x_{1,0})
\hat\De^{m_1,m_2}_{s_1,s_2}(\y_1-\x_1,\x_2-\y_2)\big|
\Big\}\cr
&\le\const\min\Big\{\int d\z_1\sup_{\z_2}
\big|\hat\De^{m_1,m_2}_{s_1,s_2}(\z_1,\z_2)\big|,
\int d\z_2\sup_{\z_1}
\big|\hat\De^{m_1,m_2}_{s_1,s_2}(\z_1,\z_2)\big|\Big\}\cr
&\le\const b(m_1,m_2)\fl_m^2\min\big\{\sfrac{M^{m_1}}{\fl_{m_1}},\sfrac{M^{m_2}}{\fl_{m_2}}\big\}\cr
&=\const b(m_1,m_2)\fl_m M^{m}\cr
}$$
If $m\in\{i,i+1\}$
$$
\big\|\cD^{(m)}_{s_1,s_2}\big\|_{\rm bubble}
\le\sum_{\min\{m_1,m_2\}=m}\|\cD^{m_1,m_2}_{s_1,s_2}\|_{\rm bubble}
\le\const\sum_{m'\ge m}\sfrac{1}{M^{m'}}\fl_m M^{m}\le\const\fl_m
$$
and if $m\ge i+2$
$$\eqalign{
\big\|\cD^{(m)}_{s_1,s_2}\big\|_{\rm bubble}
&\le\sum_{\min\{m_1,m_2\}=m}\|\cD^{m_1,m_2}_{s_1,s_2}\|_{\rm bubble}\cr
&\le\const\sum_{m'= m}^{j}\sfrac{1}{M^{{j}}}\fl_m M^{m}
+\const\sum_{m'> j}\sfrac{1}{M^{m'}}\fl_m M^{m}\cr
&\le\const(j-m+1)\sfrac{M^m}{M^j}\fl_m
}$$
\endproof
\titled{Model Propagator in Position Space}
\lemma{\STM\lemLADrectangle}{
 The set $\bd=\set{\p_1-\p_2}{\p_1\in\bka_1,\ \p_2\in\bka_2}$ is contained in a 
rectangle  with two sides of length 
$\const \fl_{j}$ and two sides of length $\const\sfrac{1}{M^{j}}$.

}
\prf For every sector $\si\in\Si_r$, let $\k_\si$ be the centre of 
$\si\cap F$, $\Ln_\si$ a unit tangent vector to $F$ at $\k_\si$ and 
$$
\bsi=\set{\k\in\bbbr^2}{(k_0,\k)\in\si\hbox{ for some }k_0\in\bbbr}
$$ 
Then $\bsi$ is contained in a rectangle $R_\si$ centered at $\k_\si$ with two sides parallel to $\Ln_\si$ of length $\const \fl_r$ and two 
sides perpendicular to $\Ln_\si$ of length $\const\sfrac{1}{M^r}$.
If at least one of $\ka_1$ and $\ka_2$ are in $\bbbm$, the claim follows
since $j\le r$.
So assume that $\ka_1,\ka_2\in\Si_r$. Then the distance between $\k_{\ka_1}$ and $\k_{\ka_2}$ is at most $|\btau|+5\fl_r$ and therefore the angle between
$\Ln_{\ka_1}$ and $\Ln_{\ka_2}$ is at most $\const\big(|\btau|+\fl_r\big)$.
Consequently $\bd$
is contained in a rectangle with two sides parallel to $\Ln_{\ka_1}$ of 
length $\const \fl_r$ and two sides perpendicular to $\Ln_{\ka_1}$ of 
length 
$$
\const\Big(\sfrac{1}{M^r}+\big(|\btau|+\fl_r\big)\fl_r\Big)
\le \const\big(\sfrac{1}{M^r}+|\btau|\fl_r\big)
$$
By (\eqnLADassume), $|\btau|\le\max\{\sfrac{1}{M^{j}},
r^3\fl_r\}$, so that $|\btau|\fl_r\le\const\sfrac{1}{M^{j}}$.
\endproof

Fix two mutually perpendicular unit vector $\Ln$ and $\Sh$ and a rectangle 
$R$, with two sides parallel to $\Ln$ of length $\const \fl_{j}$ 
and two sides parallel to $\Sh$ of length $\const\sfrac{1}{M^{j}}$,
such that $\bd\subset R$. By Lemma \lemLADrectangle, such a rectangle exists.
Let $\rho(\t)$ be identically one on $R$, be supported on a set of area twice
that of $R$ and obey 
$$
\Big|\big(\Sh\cdot\partial_{\t}\big)^{\al_1}
\big(\Ln\cdot\partial_{\t}\big)^{\al_2}\rho(\t)\Big|
\le\const M^{\al_1j}\sfrac{1}{\fl_{j}^{\al_2}}
$$ 
for all $\al_1,\al_2\le 2$. Define $\cM_\rho$ as in Definition \defLADtrasfermomcutoff. Then
$$
\big\|g\bullet\cM\bullet h\big\|_{\ka_1,\ka_2}
=\big\|g\bullet\cM_\rho\bullet h\big\|_{\ka_1,\ka_2}
\EQN\eqnLADmeqmrho$$
\proposition{\STM\propXXXII}{
$$
\big\|g\bullet\cM\bullet h\big\|_{\ka_1,\ka_2}
\le\const\max_{\al_\ri,\al_\li\in\bbbn_0\times\bbbn_0^2
            \atop|\al_\ri|+|\al_\li|\le 3}
\big|g\big|_{\ell,i}^{[0,0,\al_\ri]}\big|h\big|_{i,r}^{[\al_\li,0,0]}
$$

}
\prf
Write
$$
\cM_\rho(p,k)=\sum_{s_1,s_2\in\Si_i}\cM_{s_1,s_2}(p,k)
$$
where, for $s_1,s_2\in\Si_i$
$$
\cM_{s_1,s_2}(p,k)=\cM(p,k)\rho(\p-\k)\chi_{s_1}(\p)\chi_{s_2}(\k)
$$
By (\eqnLADspatcnt),
$$\eqalign{
\big\|g\bullet\cM_\rho\bullet h\big\|_{\ka_1,\ka_2}
&\le\sum_{s_1,s_2\in\Si_i\atop (s_1-s_2)\cap d\ne \emptyset}
\big\|g\bullet\cM_{s_1,s_2}\bullet h\big\|_{\ka_1,\ka_2}\cr
&\le \sfrac{\const}{\fl_i}\max_{s_1,s_2\in\Si_i}
\big\|g\bullet\cM_{s_1,s_2}\bullet h\big\|_{\ka_1,\ka_2}\cr
}\EQN\eqnLADmrhotoms$$
Define, for $v_1,v_2\in\Si_i$,
$$\eqalign{
h_{v_1,v_2}(x_1,x_2,y_3,y_4)
&=h\big((x_1,v_1),(x_2,v_2),y_3,y_4\big)\quad
\hbox{where\ }y_\nu=\cases{\ka_{\nu-2}& if $\ka_{\nu-2}\in\bbbm$\cr
                           (x_\nu,\ka_{\nu-2})& if $\ka_{\nu-2}\in\Si_r$\cr}\cr
}$$
Observe that $h_{v_1,v_2}$
is a function on $\big(\bbbr\times\bbbr^2\big)^{2+n_\ri}$
where $n_\ri=\#\set{\nu\in\{1,2\}}{\ka_\nu\in\Si_r}$.
Also define, for $u_3,u_4\in\Si_i$ and $\la_1,\la_2\in\bbbm\dunion\Si_\ell$,
$$\eqalign{
g_{\la_1,\la_2;u_3,u_4}(y_1,y_2,x_3,x_4)
&=g\big(y_1,y_2,(x_3,u_3),(x_4,u_4)\big)\quad
\hbox{where\ }y_\nu=\cases{\la_\nu& if $\la_\nu\in\bbbm$\cr
                           (x_\nu,\la_\nu)& if $\la_\nu\in\Si_r$\cr}\cr
}$$
Observe that $g_{\la_1,\la_2;u_3,u_4}$ is a function on $\big(\bbbr\times\bbbr^2\big)^{2+n_\li}$ 
where $n_\li=\#\set{\nu\in\{1,2\}}{\la_\nu\in\Si_\ell}$.
Then, for all $s_1,s_2\in\Si_i$,
$$
\big\|g\bullet\cM_{s_1,s_2}\bullet h\big\|_{\ka_1,\ka_2}
\le\sup_{\la_1,\la_2}\sum_{u_3,u_4\in\Si_i\atop v_1,v_2\in\Si_i}
\TN g_{\la_1,\la_2;u_3,u_4}\fcirc\cM_{s_1,s_2}\fcirc h_{v_1,v_2}\TN_{1,\infty}
\EQN\eqnLADmsup$$
By conservation of momentum, the convolution
$\ 
g_{\la_1,\la_2;u_3,u_4}\fcirc\cM_{s_1,s_2}\fcirc h_{v_1,v_2}
\ $
vanishes identically unless
$$\meqalign{
&u_3\cap s_1\ne\emptyset && s_1\cap v_1&\ne\emptyset \cr
&u_4\cap s_2\ne\emptyset && s_2\cap v_2&\ne\emptyset \cr
}$$
There only $3^4$ quadruples 
$(u_3,u_4,v_1,v_2)$ satisfying these conditions, so that, by (\eqnLADmeqmrho),
(\eqnLADmrhotoms) and (\eqnLADmsup),
$$
\big\|g\bullet \cM\bullet h\big\|_{\ka_1,\ka_2}
\le\const\sfrac{1}{\fl_i}\sup_{\la_1,\la_2}
\max_{s_1,s_2\in \Si_i\atop{u_3,u_4\in\Si_i\atop v_1,v_2\in\Si_i}}
\TN g_{\la_1,\la_2;u_3,u_4}\fcirc\cM_{s_1,s_2}\fcirc h_{v_1,v_2}\TN_{1,\infty}
\EQN\eqnLADmsupmax$$
Fix $\la_1,\la_2,s_1,s_2,u_3,u_4,v_1,v_2$ and denote $g'=g_{\la_1,\la_2;u_3,u_4}$ and $h'=h_{v_1,v_2}$. 
Write the convolution
$$\eqalign{
&g'\fcirc\cM_{s_1,s_2}\fcirc h'
=\int {\sst d^3z_1d^3z_2d^3y_1d^3y_2}
\sfrac{d^3p}{(2\pi)^3}\sfrac{d^3k}{(2\pi)^3}d\om
\ e^{\imath <p,z_1-y_1>_-}e^{\imath <k,y_2-z_2>_-}\ 
g'(\,\cdot\,,\,\cdot\,,z_1,z_2)\cr
&\hskip1.4in
\de(k_0)\frac{\nu_0(\om)\nu_1(\p,\k)\chi_{s_1}(\p)\chi_{s_2}(\k)}
{[i\om-e'(\om,\p)][i\om-e'(\om,\k)]}\rho(\p-\k)
h'(y_1,y_2,\,\cdot\,,\,\cdot\,)\cr
&\hskip.25in=\int {\sst d^3z_1d^3z_2d^2\y_1d^3y_2}
\sfrac{d^2\p}{(2\pi)^2}\sfrac{d^2\k}{(2\pi)^2}
\sfrac{d\om}{2\pi}
\ e^{\imath \p\cdot(\z_1-\y_1)}e^{\imath \k\cdot(\y_2-\z_2)}\ 
g'(\,\cdot\,,\,\cdot\,,z_1,z_2)\cr
&\hskip1.4in
\frac{\nu_0(\om)\nu_1(\p,\k)\chi_{s_1}(\p)\chi_{s_2}(\k)}
{[i\om-e'(\om,\p)][i\om-e'(\om,\k)]}\rho(\p-\k)
h'\big((z_{1,0},\y_1),y_2,\,\cdot\,,\,\cdot\,\big)\cr
&\hskip.25in=\int {\sst d^3z_1d^3z_2d^2\y_1d^3y_2}
\sfrac{d^2\t}{(2\pi)^2}\sfrac{d^2\k}{(2\pi)^2}
\sfrac{d\om}{2\pi}
\ e^{\imath \t\cdot(\z_1-\y_1)}e^{\imath \k\cdot(\z_1-\z_2+\y_2-\y_1)}\ 
g'(\,\cdot\,,\,\cdot\,,z_1,z_2)\cr
&\hskip1.4in
\frac{\nu_0(\om)\nu_1(\k+\t,\k)\chi_{s_1}(\k+\t)\chi_{s_2}(\k)}
{[i\om-e'(\om,\k+\t)][i\om-e'(\om,\k)]}\rho(\t)
h'\big((z_{1,0},\y_1),y_2,\,\cdot\,,\,\cdot\,\big)\cr
&\hskip.25in=\int {\sst d^3z_1d^3z_2d^2\y_1d^3y_2}
\ g'(\,\cdot\,,\,\cdot\,,z_1,z_2)\ \widehat B_{s_1,s_2}(\z_1,\z_2,\y_1,\y_2)
\ h'\big((z_{1,0},\y_1),y_2,\,\cdot\,,\,\cdot\,\big)\cr
}$$
where
$$
\widehat B_{s_1,s_2}(\z_1,\z_2,\y_1,\y_2)
=\int \sfrac{d^2\t}{(2\pi)^2}\ e^{\imath \t\cdot(\z_1-\y_1)}
B_{s_1,s_2}(\t,\z_1-\z_2+\y_2-\y_1)\,\rho(\t)
$$
with
$$
B_{s_1,s_2}(\t,\w)
=\int \frac{d^2\k}{(2\pi)^2}\frac{d\om}{2\pi}
\ e^{\imath \k\cdot\w}
\frac{\nu_0(\om)\nu_1(\k+\t,\k)\chi_{s_1}(\k+\t)\chi_{s_2}(\k)}
{[i\om-e'(\om,\k+\t)][i\om-e'(\om,\k)]}
$$
Recall that $M^i\le\fl_j M^j$ and that $p^{(i)}=0$ for $i>j+1$.
By Theorem \:\thmLADBposn, with 
$u(\k,\t)=e^{\imath \k\cdot\w}\nu_1(\k+\t,\k)\chi_{s_1}(\k+\t)\chi_{s_2}(\k)$, 
there is an $a>1$ such that
$$
\big|\widehat B_{s_1,s_2}(\z_1,\z_2,\y_1,\y_2)\big|
\le \const \sfrac{\fl_i\fl_{j}}{M^{j}}
\sfrac{1+|\z_2-\z_1+\y_1-\y_2|^3/M^{3i}}
{[1+(\Sh\cdot(\z_1-\y_1)/M^{j})^{3/2}]
       [1+|\fl_{j}\Ln\cdot(\z_1-\y_1)|^{a}]}
$$
Since
$\ 
\sup_{\z_1}\int d^2\y_1\ 
\sfrac{\fl_{j}}{M^{j}}
\sfrac{1}{[1+(\Sh\cdot(\z_1-\y_1)/M^{j})^{3/2}]
       [1+|\fl_{j}\Ln\cdot(\z_1-\y_1)|^{a}]}
\le\const
\ $,
we have
$$
\TN g'\fcirc\cM_{s_1,s_2}\fcirc h'\TN_{1,\infty}
\le  \const\fl_i\max_{\al_\ri,\al_\li\in\bbbn_0\times\bbbn_0^2
            \atop|\al_\ri|+|\al_\li|\le 3}
\big|g\big|_{\ell,i}^{[0,0,\al_\ri]}\big|h\big|_{i,r}^{[\al_\li,0,0]}
\EQN\eqnLADgMh$$
and the Proposition follows by (\eqnLADmsupmax).
\endproof

\goodbreak
\proof{ of Theorem \rebboundII}
By Propositions \propLADzdiff, \propXXVIII, \propXXX\ 
and \propXXXII
$$\eqalign{
&\big\|g\bullet\cC^{[i,j]}\bullet h\big\|_{\ka_1,\ka_2}\cr
&\hskip0.5in\le \big\|g\bullet\big(\cC^{[i,j]}-\cZ\bullet\cC^{[i,j]}
\bullet\cZ^t \big)\bullet h\big\|_{\ka_1,\ka_2}
+\big\|g\bullet\big(\cZ\bullet\cC^{[i,j]}\bullet\cZ^t
-\widetilde\cC^{[i,j]}\big)\bullet h\big\|_{\ka_1,\ka_2}\cr
&\hskip1in+\big\|g\bullet\big(\widetilde\cC^{[i,{j}]}-\cM\big)
               \bullet h\big\|_{\ka_1,\ka_2}+\big\|g\bullet\cM\bullet h\big\|_{\ka_1,\ka_2}\cr
&\hskip.5in\le\const
 \max_{\al_\ri,\al_\li\in\bbbn_0\times\bbbn_0^2
            \atop|\al_\ri|+|\al_\li|\le 3}
\big|g\big|_{\ell,i}^{[0,0,\al_\ri]}\big|h\big|_{i,r}^{[\al_\li,0,0]}
}$$
as desired.
\endproof
\vskip.25in\null
\titlec{The Infrared Limit -- Nonzero Transfer Momentum}\PG\pgLADIIIa

\lemma{\STM\lemBubTnonzero}{ 
Let $\ka_1,\ka_2\in\bbbm$ and $g$ and $h$ be
sectorized, translation invariant functions on $\fY_{\Si_j}^4$ and
$\fY_{\Si_j}^2\times\bbbm^2$, respectively.  
Then 
$$
\|g\bullet \cC^{(j)}\bullet h\|_{\ka_1,\ka_2}
\le\const|g|_{j,j}\,\|h\|_{\ka_1,\ka_2}\  
\Big\{\sqrt{\fl_j}+\min\big\{1,\sfrac{1}{|\ka_1-\ka_2|M^j} \big\}\Big\}
$$
Furthermore, for 
$({\rm loc},\mu,\mu')\in\big\{(\tp,1,3),(\bt,2,4),(\md,1,3),(\md,2,4)\big\}$
and $\be\in\De$,
$$
\|g\bullet \rD_{\mu,\mu'}^{\be}\cC^{[j,j]}_{{\rm loc}}\bullet h\|_{\ka_1,\ka_2}
\le\const|g|_{j,j}\,\|h\|_{\ka_1,\ka_2}\  M^{|\be|j}
\Big\{\sqrt{\fl_j}+\min\big\{1,\sfrac{1}{|\ka_1-\ka_2|M^j} \big\}\Big\}
$$
}
\prf 
Write $\cC^{(j)}=\sum\limits_{s_1,s_2\in\Si_j}
\big[\cC^{(j)}_{\tp,j,s_1,s_2}+\cC^{(j)}_{\md,j,s_1,s_2}
                                +\cC^{(j)}_{\bt,j,s_1,s_2}\big]$.
As in Lemma \lemXIV,
$$
\|g\bullet \rD_{\mu,\mu'}^{\be}\cC^{[j,j]}_{{\rm loc}}\bullet h\|_{\ka_1,\ka_2}
\le\const|g|_{j,j}\,\|h\|_{\ka_1,\ka_2}\  
\sum_{s_1,s_2\in\Si_j\atop\ka_1-\ka_2\in s_1-s_2}
\big\|\rD_{\mu,\mu'}^{\be}\cC^{(j)}_{{\rm loc},j,s_1,s_2}\big\|_{\rm bubble}
$$
By Lemma \lemXV, for 
$({\rm loc},\mu,\mu')\in\big\{(\tp,1,3),(\bt,2,4),(\md,1,3),(\md,2,4)\big\}$
and $\be\in\De$,
$$
\big\|\rD_{\mu,\mu'}^{\be}\cC^{(j)}_{{\rm loc},j,s_1,s_2}\big\|_{\rm bubble}
\le\const \fl_j M^{|\be| j}
$$
and by Lemma \lemAI,
$$
\#\set{(s_1,s_2)\in\Si_j}{\ka_1-\ka_2\in s_1-s_2}
\le \const\Big\{\sfrac{1}{\sqrt{\fl_j}}+\sfrac{1}{\fl_j}
\min\big\{1,\sfrac{1}{|\ka_1-\ka_2|M^j} \big\}\Big\}
$$
so that the desired bounds follow.
\endproof

Recall, from just before Lemma \lemLADlebupbnd, that
$$\eqalign{
&\fL^{(j)}_{\ell,\ii_1,\cdots,\ii_{\ell+1}}(q,q',t,\si_1,\cdots\si_4)\cr
&\hskip.1in=
\Big[\big(F^{(\ii_1)}+L^{(\ii_1)\,f} \big)  \bullet 
{\cal C}^{[\max \{\ii_1,\ii_2\},j]}\!\bullet\!
\big(F^{(\ii_2)}+L^{(\ii_2)\,f} \big) \!\bullet \cdots \cr
&\hskip.3in\cdots \bullet\! 
{\cal C}^{[\max \{\ii_\ell,\ii_{\ell+1}\},j]}\! \bullet\!
\big(F^{(\ii_{\ell+1})}+L^{(\ii_{\ell+1})\,f} \big)\Big]_{i_1,i_2,i_3,i_4=0}
({\sst(q+{t\over 2},\si_1),(q-{t\over 2},\si_2),
(q'+{t\over 2},\si_3),(q'-{t\over 2},\si_4)})\cr
}$$

\lemma{\STM\lemLADtnonzero}{
For $t\ne 0$, the limit
$$
\fL_{\ell,\ii_1,\cdots,\ii_{\ell+1}}(q,q',t,\si_1,\cdots\si_4)
=\lim_{j\rightarrow\infty}
\fL^{(j)}_{\ell,\ii_1,\cdots,\ii_{\ell+1}}(q,q',t,\si_1,\cdots\si_4)
$$
exists. The limit is continuous in $(q,q',t)$ for $t\ne 0$.

}
\prf
It suffices to consider separately the spin and charge parts,
in the sense of Lemma \lemLADchargespin,  of $\fL^{(j)}_{\ell,\ii_1,\cdots,\ii_{\ell+1}}(q,q',t,\si_1,\cdots\si_4)$.
We denote them $\fL^{(j)}_{X,\ell,\ii_1,\cdots,\ii_{\ell+1}}(q,q',t)$ 
with $X=S,C$. 
Define $L_{X,j_1,\cdots,j_\ell}(q,q',t)$ to be the $X$ part of
$$\eqalign{
&\Big[\big(F^{(\ii_1)}+L^{(\ii_1)\,f} \big)  \bullet 
{\cal C}^{(j_1)}\!\bullet\!
\big(F^{(\ii_2)}+L^{(\ii_2)\,f} \big) \!\bullet \cdots \cr
&\hskip0.7in\cdots \bullet\! 
{\cal C}^{(j_\ell)}\! \bullet\!
\big(F^{(\ii_{\ell+1})}+L^{(\ii_{\ell+1})\,f} \big)\Big]_{i_1,i_2,i_3,i_4=0}
({\sst(q+{t\over 2},\si_1),(q-{t\over 2},\si_2),
(q'+{t\over 2},\si_3),(q'-{t\over 2},\si_4)})\cr
}$$
and
$$
\De\fL^{(p)}_X(q,q',t)
            =\sum_{  {{{
                    \max \{\ii_1,\ii_2\}\le j_1\le j
                    \atop\vdots}
                    \atop \max \{\ii_\ell,\ii_{\ell+1}\}\le j_\ell\le j}
                    \atop \max\{j_1,\cdots,j_\ell\}=p}
                  }
                  L_{X,j_1,\cdots,j_\ell}(q,q',t)
$$
Then
$$
\fL^{(j)}_{X,\ell,\ii_1,\cdots,\ii_{\ell+1}}(q,q',t,\si_1,\cdots\si_4)
=\sum_{p=2}^j \De\fL^{(p)}_X(q,q',t,\si_1,\cdots\si_4)
$$
By repeated use of Lemma \lemBubTnonzero, (\eqnLADKbnd) and its $X=C$ analog,
$$
\| L_{X,j_1,\cdots,j_\ell} \|_{\ka_1,\ka_2}
\le\const\prod_{m=1}^\ell 
\Big\{\sqrt{\fl_{j_m}}+\min\big\{1,\sfrac{1}{|\ka_1-\ka_2|M^{j_m}} \big\}\Big\}
$$
with the constant depending on $\ell$ and $i_1,\cdots,i_{\ell+1}$.
Summing this bound over $j_1,\cdots,j_\ell$ with $\max\{j_1,\cdots,j_\ell\}=p$
$$
\|\De\fL^{(p)}_X \|_{\ka_1,\ka_2}
\le\const 
\Big\{\sqrt{\fl_p}+\min\big\{1,\sfrac{1}{|\ka_1-\ka_2|M^p} \big\}\Big\}
$$
The constant now depends on $|\ka_1-\ka_2|$ as well but remains finite
as long as $\ka_1\ne\ka_2$. The existence of the limit $j\rightarrow\infty$
is now a consequence of the Lebesgue dominated convergence theorem.
Similarly, if $\rD$ is any first order differential operator in $q,q'$
or $t$,
$$\eqalign{
\|\rD L_{X,j_1,\cdots,j_\ell} \|_{\ka_1,\ka_2}
&\le\const M^{\max\{j_1,\cdots,j_\ell\}}\prod_{m=1}^\ell 
\Big\{\sqrt{\fl_{j_m}}+\min\big\{1,\sfrac{1}{|\ka_1-\ka_2|M^{j_m}} \big\}\Big\}
\cr
\|\rD\De\fL^{(p)}_X \|_{\ka_1,\ka_2}
&\le\const M^{p}
\Big\{\sqrt{\fl_p}+\min\big\{1,\sfrac{1}{|\ka_1-\ka_2|M^p} \big\}\Big\}\cr
}$$
Continuity now follows from
$$
\sfrac{|g(x)-g(y)|}{|x-y|^\ep}
=|g(x)-g(y)|^{1-\ep}\  \big[\sfrac{|g(x)-g(y)|}{|x-y|}\big]^\ep
\le\big[2\sup|g(x)|\big]^{1-\ep}\ \big[\sup|\nabla g(x)|\big]^\ep
$$
and the observation that 
$$
\sum_{p=1}^\infty M^{\ep p}
\Big\{\sqrt{\fl_p}+\min\big\{1,\sfrac{1}{|\ka_1-\ka_2|M^p} \big\}\Big\}
<\infty\qquad\hbox{if $\ep<\sfrac{\aleph}{2}$ and $\ka_1\ne \ka_2$}
$$
\endproof

\titlec{The Infrared Limit -- Reduction to Factorized Cutoffs}\PG\pgLADIIIb

Recall from (\eqnMBPmab) that
$$\eqalign{
\cC^{[i,j]}(p,k)
&=\sfrac{\nu^{(\ge i)}(p)\nu^{(\ge i)}(k)
-\nu^{(\ge j+1)}(p)\nu^{(\ge j+1)}(k)}
{[ip_0-e'(p)][ik_0-e'(k)]}\cr
\cA_{i,j}(p,k)&=\sfrac{\nu^{(\ge i)}(p)\nu^{(\ge i)}(k)
[1-\nu_j(e(\p))\nu_j(e(\k))]}
{[ip_0-e'(p)][ik_0-e'(k)]}\cr
\cB_{i,j}(p,k)&=\sfrac{\nu^{(\ge i)}(p)\nu^{(\ge i)}(k)
[1-\nu_j(p_0)\nu_j(k_0)]}
{[ip_0-e'(p)][ik_0-e'(k)]}\cr
}$$
where $e'(k)=e(\k)-v(k)$
and 
$$\eqalign{
\nu_j(\om)&=\sum_{m=j}^\infty\nu\big(M^{2m}\om^2\big)\cr
}$$
with $\nu$ being the single scale cutoff introduced in Definition
\defLADscales.

\proposition{\STM\propMPsamelim}{ 
Let $1\le i\le j$ and $\ell,r\ge 1$.
Let $\ka_1,\ka_2\in\bbbm$ with  and $g$ and $h$ be
sectorized, translation invariant functions on $\fY_{\ell,i}$ and $\fY_{i,r}$ 
respectively.
If $|\ka_1-\ka_2|>0$, then
$$
\big\|g\bullet\big(\cC^{[i,j]}-\cA_{i,j}\big)
               \bullet h\big\|_{\ka_1,\ka_2},
\big\|g\bullet\big(\cC^{[i,j]}-\cB_{i,j}\big)
               \bullet h\big\|_{\ka_1,\ka_2}
\le\const|g|_{\ell,i}\|h\|_{\ka_1,\ka_2}\,j\,
\big\{\sqrt{\fl_j}+\sfrac{1}{|\ka_1-\ka_2|M^j\fl_j}\big\}
$$
}
\prf 
Let
$$
\cD(p,k)=
\sfrac{\nu^{(\ge j+1)}(p)\nu^{(\ge j+1)}(k)}
{[ip_0-e'(p)][ik_0-e'(k)]}
\hbox{ or }
\sfrac{\nu^{(\ge i)}(p)\nu^{(\ge i)}(k)\nu_j(e(\p))\nu_j(e(\k))}
{[ip_0-e'(p)][ik_0-e'(k)]}
\hbox{ or }
\sfrac{\nu^{(\ge i)}(p)\nu^{(\ge i)}(k)\nu_j(p_0)\nu_j(k_0)}
{[ip_0-e'(p)][ik_0-e'(k)]}
$$
It suffices to prove that
$$
\big\|g\bullet\cD\bullet h\big\|_{\ka_1,\ka_2}
\le\const|g|_{\ell,i}\|h\|_{\ka_1,\ka_2}\,j\,
\big\{\sqrt{\fl_j}+\sfrac{1}{|\ka_1-\ka_2|M^j\fl_j}\big\}
$$
Define, for $m\ge i$, $m_1,m_2\ge m$ and $s_1,s_2\in \Si_m$
$$
\cD^{m_1,m_2}_{s_1,s_2}(p,k)
=\sfrac{\nu^{(m_1)}(p)\nu^{(m_2)}(k)\chi_{s_1}(p)\chi_{s_2}(k)}
{[ip_0-e'(p)][ik_0-e'(k)]}
\cases{\nu^{(\ge j+1)}(p)\nu^{(\ge j+1)}(k)&or\cr
     \noalign{\vskip.05in}
        \nu_j(e(\p))\nu_j(e(\k))&or\cr
     \noalign{\vskip.05in}
        \nu_j(p_0)\nu_j(k_0)&\cr}
$$ 
and
$$
\cD^{(m)}_{s_1,s_2}(p,k)
=\sum_{m_1,m_2\ge 1\atop\min\{m_1,m_2\}=m}\cD^{m_1,m_2}_{s_1,s_2}(p,k)
$$
We have
$$
\cD=\sum_{m=i}^\infty\ \sum_{s_1,s_2\in\Si_m}
\cD^{(m)}_{s_1,s_2}
$$
Hence, as in Lemma \lemXIV,
$$
\big\|g\bullet\cD\bullet h\big\|_{\ka_1,\ka_2}
\le\const |g|_{\ell,i}\|h\|_{\ka_1,\ka_2}\sum_{m=i}^\infty
\sum_{s_1,s_2\in\Si_m\atop\ka_1-\ka_2\in s_1-s_2}
\max_{s_1,s_2\in\Si_m}\big\|\cD^{(m)}_{s_1,s_2}\big\|_{\rm bubble}
$$
To bound $\big\|\cD^{(m)}_{s_1,s_2}\big\|_{\rm bubble}$,
set, for $s\in\Si_m$ and $n\ge m$,
$$
c^{(n,j)}_s(k)
=\sfrac{\nu^{(n)}(k)\chi_{s}(k)}{ik_0-e'(k)}
\cases{\nu^{(\ge j+1)}(k)&or\cr
     \noalign{\vskip.05in}
        \nu_j(e(\k))&or\cr
     \noalign{\vskip.05in}
        \nu_j(k_0)&\cr}
$$
and denote by $c^{(n,j)}_s(x)$ its Fourier transform.
As in Lemma \:\lemLADsimplepropbnd,
$$\eqalignno{
\big\|c^{(m,j)}_s(x)\big\|_{L^1}
&\le\const M^m\sfrac{\fl_m}{\fl_{\max\{m,j\}}} \cr
\big\|c^{(n,j)}_s(x)\big\|_{L^\infty}
&\le \const\fl_m M^n\sfrac{1}{M^n}\sfrac{1}{M^{\max\{n,j\}}}
\le \const \sfrac{\fl_m}{M^{\max\{n,j\}}}
\cr
}$$
The factor $\sfrac{\fl_m}{M^{\max\{n,j\}}}$ in the first inequality
arises, when $j>m$, from splitting $s$ up into sectors of length $\fl_j$.
Hence, by the triangle inequality and (\eqnBubbleTensor),
$$\eqalign{
\big\|\cD^{(m)}_{s_1,s_2}\big\|_{\rm bubble} 
&\le \sum_{n\ge m}
\Big[
\big\|c^{(m,j)}_{s_1}\big\|_{L^1}\big\|c^{(n,j)}_{s_2}\big\|_{L^\infty}
+\big\|c^{(n,j)}_{s_1}\big\|_{L^\infty}\big\|c^{(m,j)}_{s_2}\big\|_{L^1}
\Big]\cr
&\le \sum_{n> m}\const M^m\sfrac{\fl_m}{\fl_{\max\{m,j\}}}
             \sfrac{\fl_m}{M^{\max\{n,j\}}}\cr
&\le \const \sfrac{M^m\fl^2_m}{\fl_{\max\{m,j\}}}
            \sum_{n> m} \sfrac{1}{M^{\max\{n,j\}}}\cr
&\le \const \sfrac{M^m\fl^2_m}{\fl_{\max\{m,j\}}}
            \sfrac{j}{M^{\max\{m,j\}}}\cr
}$$
By Lemma \lemAI,
$$
\#\set{(s_1,s_2)\in\Si_m}{\ka_1-\ka_2\in s_1-s_2}
\le \const\Big\{\sfrac{1}{\sqrt{\fl_m}}+\sfrac{1}{\fl_m}
\min\big\{1,\sfrac{1}{|\ka_1-\ka_2|M^m} \big\}\Big\}
$$
so that
$$\eqalign{
\|g\bullet \cD\bullet h\|_{\ka_1,\ka_2}
&\le\const |g|_{\ell,i}\|h\|_{\ka_1,\ka_2}\sum_{m=i}^\infty
\Big\{\sfrac{1}{\sqrt{\fl_m}}\!+\!\sfrac{1}{\fl_m}
\min\big\{1,\sfrac{1}{|\ka_1-\ka_2|M^m} \big\}\Big\}
\sfrac{M^m\fl^2_m}{\fl_{\max\{m,j\}}}
            \sfrac{j}{M^{\max\{m,j\}}}\cr
&\le\const |g|_{\ell,i}\|h\|_{\ka_1,\ka_2}\ \sum_{m=1}^j
\Big\{\sfrac{1}{\sqrt{\fl_m}}+\sfrac{1}{|\ka_1-\ka_2|M^m\fl_m}\Big\}
\sfrac{M^m\fl^2_m}{\fl_j}\sfrac{j}{M^j}\cr
&\hskip.5in +\const |g|_{\ell,i}\|h\|_{\ka_1,\ka_2}\ j\sum_{m=j}^\infty
\Big\{\sfrac{1}{\sqrt{\fl_m}}+\sfrac{1}{|\ka_1-\ka_2|M^m\fl_m}\Big\}\fl_m\cr
&\le\const |g|_{\ell,i}\|h\|_{\ka_1,\ka_2}\  j
\Big\{\sfrac{1}{M^j\fl_j}M^j\fl_j^{3/2}
+\sfrac{1}{M^j\fl_j}\sfrac{1}{|\ka_1-\ka_2|}
+\sqrt{\fl_j}+\sfrac{1}{|\ka_1-\ka_2|M^j}\Big\}\cr
&\le\const |g|_{\ell,i}\|h\|_{\ka_1,\ka_2}\  j
\Big\{\sqrt{\fl_j}+\sfrac{1}{|\ka_1-\ka_2|M^j\fl_j}\Big\}\cr
}$$
\endproof
\corollary{\STM\corMPsamelim}{ Let $\ell\ge 1$, $i_1,\cdots,i_{\ell+1}\ge 2$, 
$j\ge \max\{i_1,\cdots,i_{\ell+1}\}$ and $|\ka_1-\ka_2|>0$. Define $I_0=i_1$,
$I_{\ell+1}=i_{\ell+1}$ and, for $1\le m\le \ell$, $I_m=\max\{i_m,i_{m+1}\}$.
For each $1\le m\le \ell+1$, let $g_m$  be a
sectorized, translation invariant function on $\fY_{I_{m-1},I_m}$ with
$|g_m|_{I_{m-1},I_m}<\infty$
Then
$$
\big\|g_1\bullet\cC^{[I_1,j]}\bullet g_2\bullet\cdots\bullet\cC^{[I_\ell,j]}
\bullet g_{\ell+1}
-g_1\bullet\cA_{I_1,j}\bullet g_2\bullet\cdots\bullet\cA_{I_\ell,j}
\bullet g_{\ell+1}
\big\|_{\ka_1,\ka_2}
$$
and
$$
\big\|g_1\bullet\cC^{[I_1,j]}\bullet g_2\bullet\cdots\bullet\cC^{[I_\ell,j]}
\bullet g_{\ell+1}
-g_1\bullet\cB_{I_1,j}\bullet g_2\bullet\cdots\bullet\cB_{I_\ell,j}
\bullet g_{\ell+1}
\big\|_{\ka_1,\ka_2}
$$
both converge to zero as $j\rightarrow\infty$.
}
\prf Use Proposition \propMPsamelim\ and the bounds
$$
\big\|g\bullet\cA_{i,j}\bullet h\big\|_{\ka_1,\ka_2},
\big\|g\bullet\cB_{i,j}\bullet h\big\|_{\ka_1,\ka_2},
\big\|g\bullet\cC^{i,j}\bullet h\big\|_{\ka_1,\ka_2}
\le\const |j-i+1| \,\big|g\big|_{\ell,i} \big\|h\big\|_{\ka_1,\ka_2}
$$
which follow from Proposition \propMPsamelim\ and Theorem \rebbound, 
with $\be=0$, to prove that
$$\eqalign{
&\big\|g_1\!\bullet\!\cA_{I_1,j}\!\bullet\!g_2
\!\bullet\!\cdots\!\bullet\!\cA_{I_{m-1},j}
\bullet\! g_m\!\bullet\!
\big[\cC^{[I_m,j]}-\cA_{I_m,j}\big]\!\bullet\! g_{m+1}\!\bullet
\!\cC^{[I_{m+1},j]}\!\bullet\!\cdots\!\bullet\!\cC^{[I_\ell,j]}
\!\bullet\! g_{\ell+1}
\big\|_{\ka_1,\ka_2}\ ,\ \cr
&\big\|g_1\!\bullet\!\cB_{I_1,j}\!\bullet\!g_2
\!\bullet\!\cdots\!\bullet\!\cB_{I_{m-1},j}
\bullet\! g_m\!\bullet\!
\big[\cC^{[I_m,j]}-\cB_{I_m,j}\big]\!\bullet\! g_{m+1}\!\bullet
\!\cC^{[I_{m+1},j]}\!\bullet\!\cdots\!\bullet\!\cC^{[I_\ell,j]}
\!\bullet\! g_{\ell+1}
\big\|_{\ka_1,\ka_2}\cr
&\hskip.5in
\le\cst{\ell}{}|j|^\ell
\big\{\sqrt{\fl_j}+\sfrac{1}{|\ka_1-\ka_2|M^j\fl_j}\big\}
\smprod_{m=1}^{\ell+1}|g_m|_{I_{m-1},I_m}
}$$
\endproof

\vfill\eject

\chap{Double Bubbles}\PG\pgLADIV

\EDEF\CHdbubbles{\caproman\chapno}

In this section we prove the ``double bubble bound'', Theorem \dbbound.
The techniques we use are essentially those of \S\CHbubbles\ with one
additional wrinkle -- volume improvement due to overlapping loops.

To illustrate the effect of overlapping loops 
 we consider one double bubble, namely
$$
 D(q,q',t)\ = \figplace{dbubble6}{0 in}{-0.5in}
$$
with the kernels of all vertices being identically one in momentum space
and all lines having propagator $C^{(j)}$.
By the Feynman rules
$$
D(q,q',t)=\int  dk\,dp\ \ 
C^{(j)}(p+q)\,C^{(j)}(k+p)\,C^{(j)}(k+t)\,C^{(j)}(k)
$$
The naive power counting bound is, for each $q,q',t\in\bbbr\times \bbbr^2$,
$$\eqalign{
|D(q,q',t)| 
&\le \int  dk\,dp\ \ 
|C^{(j)}(p+q)|\,|C^{(j)}(k+p)|\,|C^{(j)}(k+t)|\,|C^{(j)}(k)| \cr
&\le \|C^{(j)}\|^2_\infty\int  dk\,dp\ \ |C^{(j)}(k)|\,|C^{(j)}(p+q)|\cr
& = \|C^{(j)}\|^2_\infty\,\|C^{(j)}\|_1^2 \cr
&\le  \const \cr
}$$
since, denoting the $j^{\rm th}$ shell by $S_j$ (see Definition \defLADscales),
$$\eqalign{
\|C^{(j)}\|_\infty & =\sup_{k\in S_j} \sfrac{1}{|\imath k_o-e(\k)|}
\le\sqrt{M} \,M^j \cr
\|C^{(j)}\|_1 & \le \|C^{(j)}\|_\infty \,\big( {\rm volume\ of\ } S_j \big)  
\le  \const M^j\,\sfrac{1}{M^{2j}} = \sfrac{\const}{M^j}
}$$
In the naive bound, we ignored the constraint that
$|e(\k+\p)|\le\sfrac{\sqrt{2M}}{M^{j}}$. Taking it into account, 
one has the better estimate\goodbreak
$$\eqalign{
|D(q,q',t)| 
&\le \int  dk\,dp\ \ 
|C^{(j)}(p+q)|\,|C^{(j)}(k+p)|\,|C^{(j)}(k+t)|\,|C^{(j)}(k)| \cr
&\hskip-.5in\le \const  M^{4j}
\int  dk\,dp\  
\nu^{(j)}(p+q)\ \nu^{(j)}(k+q)\ \nu^{(j)}(k) \cr
&\hskip-.5in\le \const M^{4j}\ 
{\rm vol} \Big\{(k,p)\in \big(\bbbr\times\bbbr^2\big)^2\, \Big|\,
  |\imath (p_0+q_0)-e(\p+\q)|\le \sfrac{\sqrt{2M}}{M^j}, \cr 
& \hskip 1.8cm |\imath (k_0+p_0)-e(\k+\p)|\le \sfrac{\sqrt{2M}}{M^j},\ 
 |\imath k_0-e(\k)|\le \sfrac{\sqrt{2M}}{M^j} \Big\} \cr
& \hskip-.5in\le  \const M^{4j}\sfrac{2M}{M^{2j}}\  
{\rm vol} \Big\{(\k,\p)\in \bbbr^2\times \bbbr^2\, \Big|\,
  |e(\k)|, |e(\k+\p)|, |e(\p+\q)|\le \sfrac{\sqrt{2M}}{M^j} \Big\} \cr 
& \hskip-.5in \le \const M^{2j}\int_{|e(\p+\q)|\le{\sqrt{2M}\over M^{j}}} d\p\ 
  \ {\rm vol} \Big\{ \k\in \bbbr^2\,\Big|\,|e(\k)|,|e(\k+\p)| \le \sfrac{\sqrt{2M}}{M^j} \Big\} 
}$$
There is $\veps>0$ such that for $\p$ outside a ball of radius 
$\sfrac{\const}{M^{j(1-\veps)}}$ around the origin
$$
{\rm vol} \Big(
\big\{ \k\in \bbbr^2\,\big|\,|e(\k)| \le \sfrac{\sqrt{2M}}{M^j} \big\}
\cap \big\{ \k\in \bbbr^2\,\big|\,|e(\k+\p)| \le \sfrac{\sqrt{2M}}{M^j} \big\} \Big)
\ \le\ \sfrac{const}{M^{(1+\veps)j}}
$$

\centerline{\figput{overlapg}}

\noindent
because, roughly speaking, 
$\big\{ \k\in \bbbr^2\,\big|\,|e(\k)| \le \sfrac{\sqrt{2M}}{M^j} \big\}$
and $\big\{ \k\in \bbbr^2\,\big|\,|e(\k+\p)| \le \sfrac{\sqrt{2M}}{M^j} \big\}$
cross at an angle of about $\const|\p|\ge\sfrac{\const}{M^{j(1-\veps)}}$.
Therefore
$$\eqalign{
\|D\|_\infty &\le \const M^{2j}
\big( \sfrac{1}{M^{2j(1-\veps)}}\,\sfrac{\sqrt{2M}}{M^j} + 
\sfrac{\sqrt{2M}}{M^j}\,\sfrac{1}{M^{(1+\veps)j}} \big) \cr
&\le \const \sfrac{1}{M^{\veps j}} 
}$$

This ``volume improvement'' is realized in terms of sector counting 
in Lemma \:\lemAI. Sector counting and simple propagator estimates
(Lemma \:\lemXV\ and Lemma \:\lemXVbis) are combined using Corollary \:\corLADgDgWh\
(an analog of Lemma \lemXIV) to prove Theorem \:\redbbound\ (which is 
essentially a reformulation of Theorem \dbbound\ parts b and c in terms of 
the $\big\|\ \cdot\ \big\|_{\ka_1,\ka_2}$ norm of Definition \defLADKspace)
and to treat the large transfer momentum part of the reformulation,
Theorem \:\redbboundII, of Theorem \dbbound a (Proposition \:\propLADdlarget).
Theorem \dbbound\ parts b and c are proven following Theorem \:\redbbound.
The treatment of the small transfer momentum part of Theorem \:\redbboundII\
closely parallels the corresponding argument in \S\CHbubbles.
Theorem \dbbound a is proven following Theorem \:\redbboundII.

We first prove a general bound on
$(g_1\bullet\cD\bullet g_2)^f\bullet W\bullet h$ similar to Lemma \lemXIV.
\lemma{\STM\lemXXXIV}{ 
Let $1\le \ell\le i\le j$ and $r\ge 1$.
Let $\ka_1,\ka_2\in\fK_r$ and $g_1$, $g_2$ and $h$ be
sectorized, translation invariant functions on $\fY_{\ell,\ell}$, 
$\fY_{\ell,\ell}$ and $\fY_{i,r}$ respectively.  Let $W$ be a particle--hole 
bubble propagator whose total Fourier transform is of the form
$$
\check W(p_1,k_1,p_2,k_2)=\sum_{m=i}^j\ \sum_{s_1,s_2\in\Si_m}
W^{(m)}_{s_1,s_2}(p_1,k_1,p_2,k_2)
\qquad\hbox{ if\ \ }p_2-k_2\in\ka_1-\ka_2
$$
with $W^{(m)}_{s_1,s_2}(p_1,k_1,p_2,k_2)$ vanishing unless $\bpi(p_1),
\bpi(p_2)\in \bpi(s_1)$ and $\bpi(k_1),\bpi(k_2)\in\bpi(s_2)$. 
Here $\bpi:k=(k_0,\k)\mapsto\k$ is the projection 
of $\bbbm=\bbbr\times\bbbr^2$ onto its second factor. Let $\cV$
be another particle--hole bubble propagator with
$$
\check\cV(p_1,k_1,p_2,k_2)=\sum_{u_1,u_2\in\Si_\ell}
\cV_{u_1,u_2}(p_1,k_1,p_2,k_2)
$$
and $\cV_{u_1,u_2}(p_1,k_1,p_2,k_2)=0$
unless $\bpi(p_1),\bpi(p_2)\in\bpi(u_1)$
and $\bpi(k_1),\bpi(k_2)\in\bpi(u_2)$. Then 
$$\eqalign{
\big\|(g_1\bullet\cV\bullet g_2)^f\bullet W\bullet h\big\|_{\ka_1,\ka_2}
&\le 3^8\big|g_1\big|_{\ell,\ell}\big|g_2\big|_{\ell,\ell}\big|h\big|_{i,r}
\max_{(u_1,u_2)\in \Si_\ell}\|\cV_{u_1,u_2}\|_{\rm bubble}\cr
&\hskip.4in\sup_{\ka'\in\fK_\ell}\ 
\sum_{m=i}^j\sum_{s_1,s_2\in\Si_m\atop\bpi(s_1-s_2)\cap\bpi(\ka_1-\ka_2)\ne\emptyset}
\hskip-9pt
\inf_{R\in\cR(\ka_1-\ka_2)}\big\|W^{(m)}_{s_1,s_2,R}\big\|_{\rm bubble}\cr
&\hskip.5in
 \#\set{(u_1,u_2)\in\Si_\ell\times\Si_\ell}
{\bpi(u_1-u_2)\cap\bpi(\ka'-s_1)\ne\emptyset}\cr
}$$
}
\prf
 Consider the case in which all of the external arguments of
$(g_1\bullet\cV\bullet g_2)^f\bullet W\bullet h$ are 
$({\rm position,sector})$'s.  Set $\si'_1=\ka'$ and fix  an external 
sector  $\si'_2\in\Si_\ell$. With the sector names

\centerline{\figplace{dbubble3}{0 in}{0 in}}

\noindent we have
$$\eqalign{
&(g_1\bullet\cV\bullet g_2)^f\bullet W\bullet h\cr
&\hskip.5in=
\sum_{u_1,u_2\in\Si_\ell\atop 
     {w_{1,1},w_{1,2}\in\Si_\ell\atop w_{2,1},w_{2,2}\in\Si_\ell}}
\sum_{m=i}^j
\sum_{{w_{1,3},w_{2,3}\in\Si_\ell\atop s_1,s_2\in\Si_m}\atop v_1,v_2\in \Si_i}\cr
&\hskip.9in\Big(
g_1({\sst(\,\cdot\,,\si'_1),(\,\cdot\,,w_{1,3}),(\,\cdot\,,w_{1,1}),
(\,\cdot\,,w_{1,2})})\fcirc\cV_{u_1,u_2}\fcirc 
g_2({\sst(\,\cdot\,,w_{2,1}),(\,\cdot\,,w_{2,2}),(\,\cdot\,,\si'_2),
(\,\cdot\,,w_{2,3})})\Big)^f\cr
&\hskip3in\fcirc W^{(m)}_{s_1,s_2}\fcirc
 h({\sst(\,\cdot\,,v_1),(\,\cdot\,,v_2),(\,\cdot\,,\ka_1),(\,\cdot\,,\ka_2)})
}$$
For each choice of sectors, by conservation of momentum at the vertex $h$,
we may replace the $W^{(m)}_{s_1,s_2}$ above by $W^{(m)}_{s_1,s_2,R}$ 
with any $R\in\cR(\ka_1-\ka_2)$. Furthermore the multiple convolution vanishes
unless
$$
\bpi(s_1-s_2)\cap\bpi(\ka_1-\ka_2)\ne\emptyset \qquad
\bpi(u_1-u_2)\cap\bpi(\si'_1-s_1)\ne\emptyset\qquad
\EQN\eqnXIa$$
and 
$$\meqalign{
\bpi(w_{1,1})\cap \bpi(u_1)&\ne\emptyset && 
              \bpi(w_{2,1})\cap \bpi(u_1)&\ne\emptyset && 
\bpi(w_{1,2})\cap \bpi(u_2)&\ne\emptyset && 
              \bpi(w_{2,2})\cap \bpi(u_2)&\ne\emptyset\cr
\bpi(w_{1,3})\cap \bpi(s_1)&\ne\emptyset && 
                \bpi(v_{1})\cap \bpi(s_1)&\ne\emptyset && 
\bpi(w_{2,3})\cap \bpi(s_2)&\ne\emptyset && 
                \bpi(v_{2})\cap \bpi(s_2)&\ne\emptyset\cr
}\EQN\eqnXIb$$
For each fixed $(u_1,u_2,s_1,s_2)$, at most $3^8$ 8--tuples 
$(w_{1,1},w_{2,1},w_{1,2},w_{2,2}, w_{1,3}, v_1, w_{2,3},v_2)$
can satisfy (\eqnXIb). Hence
$$\deqalign{
\big\|(g_1\!\bullet\!\cV\!\bullet\! g_2)^f\!\bullet\! W\!\bullet\! h\big\|_{\ka_1,\ka_2}
\le 3^8\big|h\big|_{i,r}
\max_{\si'_1\in\Si_\ell}\ 
\!\sum_{m=i}^j\!
&\sum_{s_1,s_2\in\Si_m\atop\bpi(s_1-s_2)\cap\bpi(\ka_1-\ka_2)\ne\emptyset}
\hskip-11pt
\inf_{R\in\cR(\ka_1-\ka_2)}\big\|W^{(m)}_{s_1,s_2,R}\big\|_{\rm bubble}\cr
&\sum_{u_1,u_2\in\Si_\ell\atop\bpi(u_1-u_2)\cap\bpi(\si'_1-s_1)\ne\emptyset}
\big|(g_1\!\bullet\!\cV_{u_1,u_2}\!\bullet\! g_2)^f\big|_{\ell,\ell}
}$$
By definition of the $\|\ \cdot\ \|_{\rm bubble}$ norm
$$
\big|(g_1\!\bullet\!\cV_{u_1,u_2}\!\bullet\! g_2)^f\big|_{\ell,\ell}
\le\big|g_1\big|_{\ell,\ell}\big|g_2\big|_{\ell,\ell}
\|\cV_{u_1,u_2}\|_{\rm bubble}
$$
Inserting this gives the Lemma. 
\endproof
Recall, for the statement of Theorem \dbbound, that, for $\nu\in\bbbn_0\times\bbbn_0^2$,
$$\eqalign{
\cD^{(\ell)}_{\nu,\upl}(x_1,x_2,x_3,x_4)
&=\sfrac{1}{M^{|\nu| \ell}}
\sum_{m=\ell}^\infty \rD^{\nu}_{1;3}C_v^{(\ell)}(x_1,x_3)C_v^{(m)}(x_4,x_2)\cr
\cD^{(\ell)}_{\nu,\dnl}(x_1,x_2,x_3,x_4)
&=\sfrac{1}{M^{|\nu| \ell}}
\sum_{m=\ell+1}^\infty C_v^{(m)}(x_1,x_3)\rD^{\nu}_{2;4}C_v^{(\ell)}(x_4,x_2)\cr
}\EQN\eqnLADdupdown$$
Express 
$\cD^{(\ell)}_{\nu,\upl}
=\sum_{u_1,u_2\in\Si_\ell}\cD^{(\ell)}_{\nu,\upl,u_1,u_2}$ with
$$
\cD^{(\ell)}_{\nu,\upl,u_1,u_2}(x_1,x_2,x_3,x_4) 
=\sfrac{1}{M^{|\nu| \ell}}
\sum_{m\ge\ell}(x_1-x_3)^{\nu}c^{(\ell)}_{u_1}(x_3-x_1)
c^{(m)}_{u_2}(x_2-x_4)
$$
where $c^{(n)}_u(x)$ was defined in (\eqnLADcsn). Do the same for 
$\cD^{(\ell)}_{\nu,\upl}$.

\lemma{\STM\lemXVbis}{ Let $\ell\ge1$, $u_1,u_2\in\Si_\ell$ 
and $\nu\in\De$. Then
$$\eqalign{ 
\|\cD^{(\ell)}_{\nu,\upl,u_1,u_2}\|_{\rm bubble}  
&\le\const\fl_\ell \cr 
\|\cD^{(\ell)}_{\nu,\dnl,u_1,u_2}\|_{\rm bubble}
&\le\const\fl_\ell \cr
}$$
}
\goodbreak
\prf  By the triangle inequality, Lemma \lemLADbubblenorm\ and 
Lemma \:\lemLADsimplepropbnd,
$$\eqalign{
\|\cD^{(\ell)}_{\nu,\upl,u_1,u_2}\|_{\rm bubble} 
&\le \sfrac{1}{M^{|\nu| \ell}}\sum_{m\ge\ell}
\|x^{\nu}c^{(\ell)}_{u_1}\|_{L^1}\|c^{(m)}_{u_2}\|_{L^\infty}
\le \sum_{m\ge\ell}\const \sfrac{1}{M^{|\nu| \ell}}
M^{(1+|\nu|)\ell}\sfrac{\fl_\ell}{M^m}\cr
&\le\const \fl_\ell \cr
}$$
The bound on $\|\cD^{(\ell)}_{\nu,\dnl,u_1,u_2}\|_{\rm bubble}$
 is proven similarly. 
\endproof
\corollary{\STM\corLADgDgWh}{
Let $1\le \ell\le i\le j$ and $r\ge 1$.
Let $\ka_1,\ka_2\in\fK_r$ and $g_1$, $g_2$ and $h$ be
sectorized, translation invariant functions on $\fY_{\ell,\ell}$, 
$\fY_{\ell,\ell}$ and $\fY_{i,r}$ respectively.  Let $W$ be a particle--hole 
bubble propagator of the form
$$
W(p_1,k_1,p_2,k_2)=\sum_{m=i}^j\ \sum_{s_1,s_2\in\Si_m}
W^{(m)}_{s_1,s_2}(p_1,k_1,p_2,k_2)
\qquad\hbox{ if\ \ }p_2-k_2\in\ka_1-\ka_2
$$
with $W^{(m)}_{s_1,s_2}(p_1,k_1,p_2,k_2)$ vanishing unless 
$\bpi(p_1),\bpi(p_2)\in\bpi(s_1)$ and $\bpi(k_1),\bpi(k_2)\in \bpi(s_2)$. 
Let $\cD$ be either $\cD^{(\ell)}_{\nu,\upl}$
or $\cD^{(\ell)}_{\nu,\dnl}$, with $\nu\in\De$. Then
$$\eqalign{
&\big\|(g_1\bullet\cD\bullet g_2)^f\bullet W\bullet h\big\|_{\ka_1,\ka_2}\cr
&\hskip.26in\le\const\big|g_1\big|_{\ell,\ell}\big|g_2\big|_{\ell,\ell}
              \big|h\big|_{i,r}\ \fl_\ell\ 
\sup_{\ka'\in\fK_\ell}\ 
\sum_{m=i}^j\sum_{s_1,s_2\in\Si_m\atop
                      \bpi(s_1-s_2)\cap\bpi(\ka_1-\ka_2)\ne\emptyset}
\hskip-13pt\inf_{R\in\cR(\ka_1-\ka_2)}
\big\|W^{(m)}_{s_1,s_2,R}\big\|_{\rm bubble} \cr
&\hskip2.2in\#\set{(u_1,u_2)\in\Si_\ell\times\Si_\ell}
{\bpi(u_1-u_2)\cap\bpi(\ka'-s_1)\ne\emptyset}\cr
&\hskip.26in\le\const\big|g_1\big|_{\ell,\ell}\big|g_2\big|_{\ell,\ell}
              \big|h\big|_{i,r}\ \fl_\ell\ 
\sum_{m=i}^j\sup_{s_1,s_2\in\Si_m}
\big\|W^{(m)}_{s_1,s_2}\big\|_{\rm bubble} \cr
&\hskip1.45in
\sup_{\ka'\in\fK_\ell}\ 
\#\bigg\{(u_1,u_2,s_1,s_2)\in\Si_\ell^2\times\Si_m^2\,\bigg|\,
{\bpi(u_1-u_2)\cap\bpi(\ka'-s_1)\ne\emptyset\atop\bpi(s_1-s_2)\cap\bpi(\ka_1-\ka_2)\ne\emptyset}
\bigg\}\cr
}$$
}
\prf The first inequality follows directly from Lemmas \lemXXXIV\ and 
\lemXVbis. The second inequality follows by choosing an $R$ which is one on a large ball.
\endproof
\theorem{\STM\redbbound}{Let $1\le \ell\le i\le j\le r$,
$\ka_1,\ka_2\in\fK_r$ and 
let $g_1,\ g_2$ and $h$ be sectorized, translation invariant functions on $\fY_{\ell,\ell}$,  $\fY_{\ell,\ell}$ and $\fY_{i,r}$  respectively.
Let $\nu\in\De$ and $\cD$ be either $\cD^{(\ell)}_{\nu,\upl}$ or $\cD^{(\ell)}_{\nu,\dnl}$. \hfill\break
i)
For any $\be\in\De$
$$\eqalign{
\sfrac{1}{M^{|\be|j}}\big\|(g_1\bullet \cD\bullet g_2)^f\bullet 
\rD_{1;3}^{\be}\cC^{[i,j]}_{\tp}\bullet h\big\|_{\ka_1,\ka_2}
&\le\const \sqrt{\fl_\ell}\,\big|g_1\big|_{\ell,\ell}\big|g_2\big|_{\ell,\ell}\big|h\big|_{i,r}\cr
\sfrac{1}{M^{|\be|j}}\big\|(g_1\bullet \cD\bullet g_2)^f\bullet 
\rD_{2;4}^{\be}\cC^{[i,j]}_{\bt}\bullet h\big\|_{\ka_1,\ka_2}
&\le\const \sqrt{\fl_\ell}\,\big|g_1\big|_{\ell,\ell}\big|g_2\big|_{\ell,\ell}\big|h\big|_{i,r}\cr
}$$
\noindent 
ii) 
$$\eqalign{
\big\|(g_1\bullet \cD\bullet g_2)^f\bullet\cC^{[i,j]}_\md\bullet h\big\|_{\ka_1,\ka_2}
&\le\const |j-i+1|\ \sqrt{\fl_\ell}\,\big|g_1\big|_{\ell,\ell}\big|g_2\big|_{\ell,\ell}\big|h\big|_{i,r}\cr
}$$
and for any $\be\in\De$ with $|\be|\ge 1$ and
$(\mu,\mu')=(1,3),(2,4)$
$$\eqalign{
\sfrac{1}{M^{|\be|j}}\big\|(g_1\bullet \cD\bullet g_2)^f\bullet \rD_{\mu;\mu'}^{\be}\cC^{[i,j]}_{\md}
\bullet h\big\|_{\ka_1,\ka_2}
&\le\const \sqrt{\fl_\ell}\,\big|g_1\big|_{\ell,\ell}\big|g_2\big|_{\ell,\ell}\big|h\big|_{i,r}\cr
}$$
}

\prf i)
We treat $\cC_\tp^{[i,j]}$. The proof for $\cC_\bt^{[i,j]}$ is similar.
By Corollary \corLADgDgWh, followed by Lemma \:\lemXV\ and Lemma \:\lemAIII
$$\eqalign{
&\sfrac{1}{M^{|\be|j}}\big\|(g_1\bullet\cD\bullet g_2)^f\bullet 
  \rD_{1;3}^{\be}\cC_\tp^{[i,j]}\bullet h\big\|_{\ka_1,\ka_2}\cr
&\hskip.4in\le\const\big|g_1\big|_{\ell,\ell}\big|g_2\big|_{\ell,\ell}
              \big|h\big|_{i,r}\ \fl_\ell\ 
\sum_{m=i}^j\sup_{s_1,s_2\in\Si_m}
\sfrac{1}{M^{|\be|j}}
\big\|\rD_{1;3}^{\be}\cC_{\tp,j,s_1,s_2}^{(m)}\big\|_{\rm bubble} \cr
&\hskip1.45in
\sup_{\ka'\in\fK_\ell}\ 
\#\bigg\{(u_1,u_2,s_1,s_2)\in\Si_\ell^2\times\Si_m^2\,\bigg|\,
{\bpi(u_1-u_2)\cap\bpi(\ka'-s_1)\ne\emptyset\atop
          \bpi(s_1-s_2)\cap\bpi(\ka_1-\ka_2)\ne\emptyset}
\bigg\}\cr
&\hskip.4in
\le\const\big|g_1\big|_{\ell,\ell}\big|g_2\big|_{\ell,\ell}
\big|h\big|_{i,r}\ \fl_\ell\ 
\sum_{m=i}^j\sfrac{1}{M^{|\be|j}}\fl_m \sfrac{M^m}{M^j}M^{|\be| m}
\sfrac{1}{\fl_m\sqrt{\fl_\ell}}
\cr
&\hskip.4in
\le\const\big|g_1\big|_{\ell,\ell}\big|g_2\big|_{\ell,\ell}
\big|h\big|_{i,r}\ \sqrt{\fl_\ell}
\cr
}$$
\Item ii)
By Corollary \corLADgDgWh, followed by Lemma \lemXV\ and Lemma \lemAIII
$$\eqalign{
&\sfrac{1}{M^{|\be|j}}\big\|(g_1\bullet\cD\bullet g_2)^f\bullet 
  \rD_{\mu;\mu'}^{\be}\cC_\md^{[i,j]}\bullet h\big\|_{\ka_1,\ka_2}\cr
&\hskip.4in\le\const\big|g_1\big|_{\ell,\ell}\big|g_2\big|_{\ell,\ell}
              \big|h\big|_{i,r}\ \fl_\ell\ 
\sum_{m=i}^j\sup_{s_1,s_2\in\Si_m}
\sfrac{1}{M^{|\be|j}}
\big\|\rD_{\mu;\mu'}^{\be}\cC_{\md,j,s_1,s_2}^{(m)}\big\|_{\rm bubble} \cr
&\hskip1.45in
\sup_{\ka'\in\fK_\ell}\ 
\#\bigg\{(u_1,u_2,s_1,s_2)\in\Si_\ell^2\times\Si_m^2\,\bigg|\,
{\bpi(u_1-u_2)\cap\bpi(\ka'-s_1)\ne\emptyset
           \atop\bpi(s_1-s_2)\cap\bpi(\ka_1-\ka_2)\ne\emptyset}
\bigg\}\cr
&\hskip.4in
\le\const\big|g_1\big|_{\ell,\ell}
\big|g_2\big|_{\ell,\ell}
\big|h\big|_{i,r}\ \fl_\ell\ 
\sum_{m=i}^j
\sfrac{1}{\fl_m\sqrt{\fl_\ell}}
\cases{
\fl_m& $\be= 0$\cr
\fl_m\sfrac{M^m}{M^j}(j-m+1)
&$|\be|=1$\cr
\fl_m\sfrac{M^m}{M^j}&$|\be|\ge2$\cr}
\cr
}$$
For $\be=0$,
$\ 
\sum_{m=i}^j\sfrac{1}{\fl_m}\fl_m=|j-i+1|
\ $
as desired. For $\be\ne 0$,
$$
\sum_{m=i}^j\sfrac{1}{\fl_m}\fl_m\sfrac{M^m}{M^j}(j-m+1)
=\sum_{m=i}^jM^{-(j-m)}(j-m+1)
\le\const
$$
again, as desired.
\endproof
\vskip\baselineskip
\noindent{\bf Proof  of Theorem \dbbound b,c}: 
Replacing $h$ by $\sfrac{1}{M^{j|\de_\ri|}}\rD^{\de_\ri}_{3,4}h$ 
reduces consideration to $\de_\ri=0$. Suppose that 
$\cD=\cD^{(\ell)}_{\nu,\upl}$.
Observe that, by Leibniz (Lemma \leibniz) 
$$\eqalign{
\rD_{1,2}^{\de_\li}(g_1\bullet\cD\bullet g_2)^f
&=\big(\rD_{1;3}^{\de_\li}g_1\bullet\cD^{(\ell)}_{\nu,\upl}
\bullet g_2\big)^f\cr
&=\sum_{\be_1,\be_2\be_3\in\bbbn_0^3\atop
\be_1+\be_2+\be_3=\de_\li}
{\tst{\de_\li\choose\be_1,\be_2,\be_3}}
\big(\rD_{1;3}^{\be_1}g_1\bullet\rD_{1;3}^{\be_2}\cD^{(\ell)}_{\nu,\upl}
\bullet \rD_{1;3}^{\be_3}g_2\big)^f\cr
&=\sum_{\be_1,\be_2\be_3\in\bbbn_0^3\atop
\be_1+\be_2+\be_3=\de_\li}
{\tst{\de_\li\choose\be_1,\be_2,\be_3}}M^{|\be_2\ell|}
\big(\rD_{1;3}^{\be_1}g_1\bullet\cD^{(\ell)}_{\nu+\be_2,\upl}
\bullet \rD_{1;3}^{\be_3}g_2\big)^f\cr
}$$
Replacing $\sfrac{1}{M^{|\be_1|\ell}}\rD_{1;3}^{\be_1}g_1$ by $g_1$, $\nu+\be_2$ by $\nu$
and $\sfrac{1}{M^{|\be_1|\ell}}\rD_{1;3}^{\be_3}g_2$ by $g_2$,
Theorem \redbbound, with $r=j$, gives bounds on
$$
\sfrac{1}{M^{|\be|j}}\big\|\sfrac{1}{M^{|\de_\li|\ell}}M^{|\be_2\ell|}
\big(\rD_{1;3}^{\be_1}g_1\bullet\cD^{(\ell)}_{\nu+\be_2,\upl}
\bullet \rD_{1;3}^{\be_3}g_2\big)^f\bullet \rD_{\mu;\mu'}^{\be}\cC^{[i,j]}_{\rm loc}
\bullet h\big\|_{\ka_1,\ka_2}
$$
for each of ${\rm loc}=\tp,\md,\bt$. Theorem \dbbound b,c now follows by
Remark \remLADKnorm.
\endproof

\vskip.25in
\theorem{\STM\redbboundII}{
Let $1\le \ell\le i\le j\le r$ and  $\ka_1,\ka_2\in\fK_r$. 
Set $d=\ka_1-\ka_2$ and let $\bd$, the projection of $d$ onto
$\{0\}\times\bbbr^2$ identified with $\bbbr^2$, be contained
in a disc of radius $2\fl_r$ and centre $\btau$.
Furthermore, set $\tau_0=\inf\set{|t_0|}{(t_0,\t)\in d\hbox{ for some }
\t\in\bbbr^2}$. Assume that
$$
\tau_0\le\sfrac{1}{M^{j-1}}\qquad
|\btau|\le\max\big\{\sfrac{1}{M^j},r^3\fl_r\big\}\qquad
M^i\le \fl_j M^j
$$
Also assume that $p^{(i)}$ vanishes for all $i>j+1$.
Let $\nu\in\bbbn_0\times\bbbn_0^2$, with $\nu+\al\in\De$ for all $|\al|\le 3$
and let $\cD$ be either $\cD^{(\ell)}_{\nu,\upl}$ or $\cD^{(\ell)}_{\nu,\dnl}$.
For any sectorized, translation invariant functions
$g_1,\ g_2$ and $h$ on $\fY_{\ell,\ell}$,  $\fY_{\ell,\ell}$ and 
$\fY_{i,r}$  respectively,
$$
\big\|(g_1\bullet \cD\bullet g_2)^f\bullet\cC^{[i,j]}\bullet h\big\|_{\ka_1,\ka_2}
\le\const \sqrt{\fl_\ell}\ \max_{\al_\upl,\al_\dnl,\al_\li\in\bbbn_0\times\bbbn_0^2
            \atop|\al_\upl|+|\al_\dnl|+|\al_\li|\le 3} \big|g_1\big|_{\ell}^{\[\al_\upl\]} \big|g_2\big|_{\ell}^{\[\al_\dnl\]}
\big|h\big|_{i,r}^{[\al_\li,0,0]}
$$
}
\noindent
Theorem \redbboundII\ is proven at the end of this section.

\vskip\baselineskip
\noindent{\bf Proof of Theorem \dbbound a} (assuming Theorem \redbboundII):\ \hfill\break
As in the proof  of Theorem \dbbound b,c, we may assume without 
loss of generality that 
$\de_\li=\de_\ri=0$. Fix $1\le \ell\le i\le j$, $\nu\in\bbbn_0\times\bbbn_0^2$,
$\cD$ and sectorized, translation invariant functions $g_1,\ g_2$ and $h$ on 
$\fY_{\ell,\ell}$,  $\fY_{\ell,\ell}$ and $\fY_{i,j}$ as in Theorem \dbbound. 
By Remark \remLADKnorm, it suffices to prove that
$$
\big\|(g_1\bullet\cD\bullet g_2)^f \bullet \cC^{[i,j]} 
                      \bullet h\big\|_{\ka_1,\ka_2}
\le\const\ i\ \sqrt{\fl_\ell}\max_{\al_\upl,\al_\dnl,\al_\li\in\bbbn_0\times\bbbn_0^2
            \atop|\al_\upl|+|\al_\dnl|+|\al_\li|\le 3} \big|g_1\big|_{\ell}^{\[\al_\upl\]} \big|g_2\big|_{\ell}^{\[\al_\dnl\]}
\big|h\big|_{i,j}^{[\al_\li,0,0]}
\EQN\eqnLADdbboundZ$$
for all $\ka_1,\ka_2\in\fK_j$. Fix $\ka_1,\ka_2\in\fK_j$. 
Set $d=\ka_1-\ka_2$ and denote by $\bd$ the projection of $d$ onto
$\{0\}\times\bbbr^2$ identified with $\bbbr^2$.
By Remark \remLADtdiff, the set $\bd$ is contained in a disc of 
radius $2\fl_j$. We fix such a disk and denote by $\btau$ its centre.
Furthermore, as in the proof of Theorem \bbound a, 
we define $\tau_0=\inf\set{|t_0|}{(t_0,\t)\in d\hbox{ for some }
\t\in\bbbr^2}$ and
$$\eqalign{
j_0&=\cases{\max\set{n\in\bbbn_0}{\tau_0\le\sfrac{1}{M^{n-1}}}& 
                if $0<\tau_0\le M$\cr
             0 & if $\tau_0\ge M$\cr
             \infty & if $\tau_0=0$\cr
      }\cr
j_1&=\cases{\max\set{n\in\bbbn_0}{|\btau|\le\sfrac{1}{M^n}}& 
                if $j^3\fl_j<|\btau|\le 1$\cr
             0 & if $|\btau|\ge 1$\cr
             \infty & if $|\btau|\le j^3\fl_j$\cr
      }\cr
\bar\jmath&=\max\Big\{i-1,\min\{j,j_0,j_1\}\Big\}
}$$
The analog of Proposition \propLADlarget\ in the current double bubble
setting is
\proposition{\STM\propLADdlarget (Large Transfer Momentum)}{ 
$$
\big\|(g_1\bullet\cD\bullet g_2)^f\bullet\cC^{[\bar\jmath+1,j]}\bullet h\big\|_{\ka_1,\ka_2}\le
\const \sqrt{\fl_\ell}\, \big|g_1\big|_{\ell,\ell}\big|g_2\big|_{\ell,\ell}\big|h\big|_{i,j}
$$
}
\prf
If $\min\{j,j_0,j_1\}=j$, then $\bar\jmath=j$ and $\cC^{[\bar\jmath+1,j]}=0$ so
that there is nothing to prove.
So we may assume that $\min\{j_0,j_1\}<j$.
\Item{\it Case 1: $j_0\le j_1$.} In this case, 
$\big\|(g_1\bullet\cD\bullet g_2)^f\bullet\cC^{[\bar\jmath+1,j]}\bullet h\big\|_{\ka_1,\ka_2}=0$,
because $\cC^{[\bar\jmath+1,j]}(p,k)$ vanishes unless
$|p_0|,|k_0|\le\sfrac{\sqrt{2M}}{M^{\bar\jmath+1}}$ and hence unless
$|p_0-k_0|\le\sfrac{2\sqrt{2M}}{M^{\bar\jmath+1}}
< \sfrac{1}{M^{\bar\jmath}}<\tau_0$, while $|t_0|\ge\tau_0$ for all $t\in d$.
\Item{\it Case 2: $j_1< j_0$.} In this case $|\btau|\ge j^3\fl_j$. By Corollary 
 \corLADgDgWh,  Lemma \lemXV\ and Lemma \:\lemAIV
$$\eqalign{
&\big\|(g_1\bullet\cD\bullet g_2)^f\bullet \cC^{[\bar\jmath+1,j]}\bullet h\big\|_{\ka_1,\ka_2}\cr
&\hskip.26in\le\const\big|g_1\big|_{\ell,\ell}\big|g_2\big|_{\ell,\ell}
              \big|h\big|_{i,j}\ \fl_\ell\ 
\sum_{m=\bar\jmath+1}^j\sup_{s_1,s_2\in\Si_m}
\big\|\cC^{(m)}_{s_1,s_2}\big\|_{\rm bubble} \cr
&\hskip1.45in
\sup_{\ka'\in\fK_\ell}\ 
\#\bigg\{(u_1,u_2,s_1,s_2)\in\Si_\ell^2\times\Si_m^2\,\bigg|\,
{\bpi(u_1-u_2)\cap\bpi(\ka'-s_1)\ne\emptyset
         \atop\bpi(s_1-s_2)\cap\bpi(\ka_1-\ka_2)\ne\emptyset}
\bigg\}\cr
&\hskip.26in\le\const\big|g_1\big|_{\ell,\ell}\big|g_2\big|_{\ell,\ell}
              \big|h\big|_{i,j}\ 
\sum_{m=\bar\jmath+1}^j\Big[
\min \big(\ell\fl_\ell, \sfrac{1+M^m\fl_j}{M^m |\btau|} \big) +
\sfrac{\sqrt \fl_\ell}{|\btau|}\big(\sfrac{1}{M^m}+\fl_j\big) + \sqrt{\fl_m} \Big]\cr
}$$
The last sum
$$
\sum_{m=\bar\jmath+1}^j\sqrt{\fl_m}
\le\sqrt{\fl_{\bar\jmath}}
\le\sqrt{\fl_{i-1}}
\le\const\sqrt{\fl_\ell}
$$
As, by the definition of $j_1$,
$$
\sfrac{1}{|\btau|}\sum_{m=j_1+1}^j\big(\sfrac{1}{M^m}+\fl_j\big)
\le\sfrac{1}{|\btau|}\big(\sfrac{1}{M^{j_1}}+j\fl_j\big)
\le\const
$$
the middle sum
$$
\sum_{m=\bar\jmath+1}^j
\sfrac{\sqrt \fl_\ell}{|\btau|}\big(\sfrac{1}{M^m}+\fl_j\big)
\le \const\sqrt{\fl_\ell}
$$
As $
\min \big(\ell\fl_\ell, \sfrac{1+M^m\fl_j}{M^m |\btau|} \big)
\le \min \big(\ell\fl_\ell, \sfrac{1}{M^m |\btau|} \big)
+\min \big(\ell\fl_\ell, \sfrac{\fl_j}{|\btau|} \big)$,
the first sum
$$\eqalign{
\sum_{m=\bar\jmath+1}^j
\min \big(\ell\fl_\ell, \sfrac{1+M^m\fl_j}{M^m |\btau|} \big)
&\le j\min \big(\ell\fl_\ell, \sfrac{\fl_j}{|\btau|} \big)
+\sum_{m=\bar\jmath+1}^j
\min \big(\ell\fl_\ell, \sfrac{1}{M^m |\btau|} \big)\cr
&\le j\min \big(\ell\fl_\ell, \sfrac{\fl_j}{|\btau|} \big)
+\sum_{m=\bar\jmath+1}^j
(\ell\fl_\ell)^{2/3}
\sfrac{1}{(M^m |\btau|)^{1/3}}\cr
&\le j\min \big(\ell\fl_\ell, \sfrac{\fl_j}{|\btau|} \big)
+(\ell\fl_\ell)^{2/3}
\sfrac{1}{(M^{\bar\jmath} |\btau|)^{1/3}}\cr
&\le j\min \big(\ell\fl_\ell, \sfrac{\fl_j}{|\btau|} \big)
+\const(\ell\fl_\ell)^{2/3}\cr
}$$
If $j\le\sfrac{1}{\fl_\ell^{1/3}}$, then 
$$
j\ell\fl_\ell\le \ell\fl_\ell^{2/3}\le\const\sqrt{\fl_\ell}
$$
while, if $j\ge\sfrac{1}{\fl_\ell^{1/3}}$, then 
$$
\sfrac{j\fl_j}{|\btau|}\le \sfrac{1}{j^2}\le\fl_\ell^{2/3}
$$
\endproof

\noindent{\bf Continuation of the proof of Theorem \dbbound a} (assuming Theorem \redbboundII):\ \hfill\break
When $M^i\ge \fl_{\bar\jmath}M^{\bar\jmath}=M^{(1-\aleph)\bar\jmath}$,
we have $|\bar\jmath-i+1|\le\const i$. In this case Theorem \redbbound,
with $r=j$ and $j=\bar\jmath$, gives
$$\eqalign{
\big\|(g_1\bullet \cD\bullet g_2)^f\bullet\cC^{[i,\bar\jmath]}\bullet h\big\|_{\ka_1,\ka_2}
&\le\const i\ \sqrt{\fl_\ell}\,\big|g_1\big|_{\ell,\ell}\big|g_2\big|_{\ell,\ell}\big|h\big|_{i,j}\cr
}$$
This together with Proposition \propLADdlarget\ yields (\eqnLADdbboundZ).
Therefore, we may assume that 
$$
M^i\le \fl_{\bar\jmath}M^{\bar\jmath}
\EQN\eqnLADdsmalli$$
Furthermore, if $\bar\jmath= i-1$, $\cC^{[i,\bar\jmath]}=0$ and there is
nothing more to prove. So we may also assume that $j_0,j_1\ge i$ and 
$\bar\jmath\le j,j_0,j_1$.

Set $v'=\sum_{i=2}^{\bar\jmath+1}p^{(i)}$ and
$
\cC^{'[i,\bar\jmath]}=\cC_{\tp}^{'[i,\bar\jmath]}+\cC_{\md}^{'[i,\bar\jmath]}+\cC_{\bt}^{'[i,\bar\jmath]}
$
with
$$
\cC_{\tp}^{'[i,\bar\jmath]}
=\sum_{i\le i_t\le \bar\jmath\atop i_b>\bar\jmath}C^{(i_t)}_{v'}\otimes C^{(i_b)\,t}_{v'},
\qquad\cC_{\md}^{'[i,\bar\jmath]}=
\sum_{i\le i_t\le \bar\jmath\atop i\le i_b\le \bar\jmath}C^{(i_t)}_{v'}\otimes C^{(i_b)\,t}_{v'},
\qquad\cC_{\bt}^{'[i,\bar\jmath]}=
\sum_{i_t> \bar\jmath\atop i\le i_b\le \bar\jmath}C^{(i_t)}_{v'}\otimes C^{(i_b)\,t}_{v'}
$$ 
as in \S III. Again, $v-v'$ is supported on the $(\bar\jmath+2)^{\rm nd}$ 
extended neighbourhood
and $\cC_{\md}^{[i,\bar\jmath]}=\cC_{\md}^{'[i,\bar\jmath]}$. 
Hence, by Theorem \redbbound.i,
with $\be=0$, $r=j$ and $j=\bar\jmath$,
$$\eqalign{
\big\|(g_1\bullet \cD\bullet g_2)^f\bullet\big[\cC^{[i,\bar\jmath]}-\cC^{'[i,\bar\jmath]}\big]\bullet h\big\|_{\ka_1,\ka_2}
&\le\const \sqrt{\fl_\ell}\,\big|g_1\big|_{\ell,\ell}\big|g_2\big|_{\ell,\ell}\big|h\big|_{i,j}\cr
}\EQN\eqnLADdctocprime$$
By (\eqnLADdsmalli) and
the Definitions of $\bar\jmath$ and $\cC^{'[i,\bar\jmath]}$, the 
hypotheses of Theorem \redbboundII,
with $r=j$ and $j=\bar\jmath$, apply to
$(g_1\bullet \cD\bullet g_2)^f\bullet\cC^{'[i,\bar\jmath]}\bullet h$. Hence 
$$
\big\|(g_1\bullet \cD\bullet g_2)^f\bullet\cC^{'[i,\bar\jmath]}\bullet h\big\|_{\ka_1,\ka_2}
\le\const \sqrt{\fl_\ell}\ \max_{\al_\upl,\al_\dnl,\al_\li\in\bbbn_0\times\bbbn_0^2
            \atop|\al_\upl|+|\al_\dnl|+|\al_\li|\le 3} \big|g_1\big|_{\ell}^{\[\al_\upl\]} \big|g_2\big|_{\ell}^{\[\al_\dnl\]}
\big|h\big|_{i,j}^{[\al_\li,0,0]}
$$
This together with (\eqnLADdctocprime) and 
 Proposition \propLADdlarget\ yields (\eqnLADdbboundZ). 
This completes the proof that Theorem \redbboundII\ implies Theorem \dbbound.a.
\endproof

\vskip.25in
The rest of this section is devoted to the proof of Theorem \redbboundII.
So we fix $\nu\in\De$, 
$1\le \ell\le i\le j\le r$,  $\cD=\cD^{(\ell)}_{\nu,\upl}$ or $\cD^{(\ell)}_{\nu,\dnl}$
 and sectorized, translation invariant functions, $g_1,\ g_2$ and $h$, on $\fY_{\ell,\ell}$,  $\fY_{\ell,\ell}$ 
and $\fY_{i,r}$ respectively. We also fix $\ka_1,\ka_2\in\fK_r$ and
assume that
$$
\tau_0\le\sfrac{1}{M^{j-1}}\qquad
|\btau|\le\max\big\{\sfrac{1}{M^j},r^3\fl_r\big\}\qquad
M^i\le\fl_j M^j
\EQN\eqnLADdassume$$
and that $p^{(i)}$ vanishes for all $i>j+1$. As in \S III, we 
reduce the particle--hole bubble propagator $\cC^{[i,j]}$
to the model bubble propagator of (\eqnIX). This is done in the following
two lemmata.

\lemma{\STM\lemXXXVI}{ Let $\cZ$ be the operator defined in (\eqnVII). Then
$$
\big\|(g_1\bullet\cD\bullet g_2)^f\bullet (\cZ\bullet\cC^{[i,j]}\bullet\cZ^t-\cM)\bullet h\big\|_{\ka_1,\ka_2}
\le\const\sqrt{\fl_\ell}\,\big|g_1\big|_{\ell,\ell}
\big|g_2\big|_{\ell,\ell}\big|h\big|_{i,r}
$$
}
\prf
Expand
$$
\cZ\bullet\cC^{[i,j]}\bullet\cZ^t-\cM=
(\widetilde\cC^{[i,j]}-\cM)+(\cZ\bullet\cC^{[i,j]}\bullet\cZ^t
-\widetilde\cC^{[i,j]})
$$
where $\widetilde\cC^{[i,j]}$ was defined in (\eqnVIII). Then,
for $|p_0|\le\sfrac{M+4\sqrt{2M}}{M^j}$,
$$
(\widetilde\cC^{[i,j]}-\cM)+(\cZ\bullet\cC^{[i,j]}\bullet\cZ^t
-\widetilde\cC^{[i,j]})
=\sum_{m=i}^j\sum_{s_1,s_2\in\Si_m}\hskip-4pt\cD^{(m)}_{s_1,s_2}
+\sum_{m=i}^j\sum_{s_1,s_2\in\Si_m}\hskip-4pt
(\cZ\bullet\cC^{(m)}_{s_1,s_2}\bullet\cZ^t
-\widetilde\cC^{(m)}_{s_1,s_2})
$$
where $\cD^{(m)}_{s_1,s_2}$  was defined in (\eqnX). So, by 
Corollary \corLADgDgWh,
$$\eqalign{
&\big\|(g_1\bullet\cD\bullet g_2)^f\bullet \big(
\widetilde\cC^{[i,j]}-\cM+\cZ\bullet\cC^{[i,j]}\bullet\cZ^t
-\widetilde\cC^{[i,j]}\big)
\bullet h\big\|_{\ka_1,\ka_2}\cr
&\hskip.2in\le\const\big|g_1\big|_{\ell,\ell}\big|g_2\big|_{\ell,\ell}\big|h\big|_{i,r}\ \fl_\ell\ 
\sum_{m=i}^j
\cr&\hskip.35in
\ \sup_{\ka'\in \fK_\ell}\#\set{(s_1,s_2,u_1,u_2)}
{\bpi(s_1-s_2)\cap\bpi(\ka_1-\ka_2)\ne\emptyset,
\ \bpi(u_1-u_2)\cap\bpi(\ka'-s_1)\ne\emptyset}
\cr&\hskip.35in
\max_{s_1,s_2\in\Si_m}\Big[
\big\|\cD^{(m)}_{s_1,s_2}\big\|_{\rm bubble}
+\big\|\cZ\bullet\cC^{(m)}_{s_1,s_2,\phi}\bullet\cZ^t
-\big(\widetilde\cC^{(m)}_{s_1,s_2}\big)_\phi\big\|_{\rm bubble}\Big]
}$$
where $\phi$ was defined at the beginning of the proof of Proposition \propXXVIII.
Then by Lemma \:\lemAIII, Lemma \lemXXXI\ and Lemma \lemXXIX
$$\eqalign{
&\big\|(g_1\bullet\cD\bullet g_2)^f\bullet \big(
\widetilde\cC^{[i,j]}-\cM+\cZ\bullet\cC^{[i,j]}\bullet\cZ^t
-\widetilde\cC^{[i,j]}\big)
\bullet h\big\|_{\ka_1,\ka_2}\cr
&\le\const\big|g_1\big|_{\ell,\ell}\big|g_2\big|_{\ell,\ell}\big|h\big|_{i,r}\ \fl_\ell
\Big[\sum_{m=i}^{i+1}\!\sfrac{1}{\fl_m\sqrt{\fl_\ell}}\fl_m
+\sum_{m=i+2}^{j}\!\sfrac{j-m+1}{\fl_m\sqrt{\fl_\ell}}\sfrac{M^m}{M^j}\fl_m
+\sum_{m=i}^j\sfrac{1}{\fl_m\sqrt{\fl_\ell}}\fl_m\sfrac{M^m}{M^j}\Big]\cr
&\le\const\sqrt{\fl_\ell}\ \big|g_1\big|_{\ell,\ell}\big|g_2\big|_{\ell,\ell}\big|h\big|_{i,r}\cr
}$$
\endproof
\lemma{\STM\lemXXXVII}{ 
$$
\big\|(g_1\bullet\cD\bullet g_2)^f\bullet 
(\cC^{[i,j]}-\cZ\bullet\cC^{[i,j]}\bullet\cZ^t)\bullet h\big\|_{\ka_1,\ka_2}
\le\const\sqrt{\fl_\ell} \max_{|\al_\upl|+|\al_\dnl|\le1}
\big|g_1\big|_{\ell}^{\[\al_\upl\]}
\big|g_2\big|_{\ell}^{\[\al_\dnl\]}
\big|h\big|_{i,r}^{[1,0,0]}
$$
}
\prf
By Lemma \lemXXIV
$$\eqalign{
&(g_1\bullet\cD\bullet g_2)^f\bullet
(\cC^{[i,j]}-\cZ\bullet\cC^{[i,j]}\bullet\cZ^t)\bullet h\cr
&\hskip1in=(g_1\bullet\cD\bullet g_2)^f_\ri\bullet D_\li\cC^{[i,j]}\bullet h
+(g_1\bullet\cD\bullet g_2)^f\bullet\cZ\bullet D_\ri\cC^{[i,j]}\bullet h_\li\cr
}$$
Both terms are bounded as in the previous lemma, using Lemma \:\lemAIII\ to 
bound the number of allowed 4--tuples $(s_1,s_2,u_1,u_2)$
by $\sfrac{1}{\fl_m\sqrt{\fl_\ell}}$. 
Lemma \lemXXV, (and, for the second term, Lemma \lemLADzprops.iii)
are used to bound 
$\|D_\li\cC^{(m)}_{s_1,s_2}\|_{\rm bubble}\le\const\sfrac{\fl_m}{M^m}$
and $\|\cZ\bullet D_\ri\cC^{(m)}_{s_1,s_2}\|_{\rm bubble}
\le\|D_\ri\cC^{(m)}_{s_1,s_2}\|_{\rm bubble}\le\const\sfrac{\fl_m}{M^m}$.
The right derivative in $(g_1\bullet\cD\bullet g_2)^f_\ri$ acts as a central
derivative on $g_1\bullet\cD\bullet g_2$ and may be written, using Leibniz's
rule, as a sum of three terms with the first containing a central derivative
acting on $g_1$, the second a central derivative acting on $g_2$ and the
third having one component of $\cD$'s index $\nu$ increased by
one. Lemma \lemXVbis\ is used to bound the bubble norms of the sectorized 
contributions to $\cD$. All together,
$$\eqalign{
&\big\|(g_1\bullet\cD\bullet g_2)^f\bullet
\big(\cC^{[i,j]}-\cZ\bullet\cC^{[i,j]}\bullet\cZ^t \big)\bullet h\big\|_{\ka_1,\ka_2}\cr
&\hskip.1in\le 
\big\|(g_1\bullet\cD\bullet g_2)^f_\ri\bullet D_\li\cC^{[i,j]}\bullet h\big\|_{\ka_1,\ka_2}
+\big\|(g_1\bullet\cD\bullet g_2)^f\bullet
\cZ\bullet D_\ri\cC^{[i,j]}\bullet h_\li\|_{\ka_1,\ka_2}\cr
&\hskip.1in\le\const\Big[
M^\ell\big|g_1\big|_{\ell}^{\[(1,0,0)\]}\ \fl_\ell\ \big|g_2\big|_{\ell,\ell}
+\big|g_1\big|_{\ell,\ell}\ M^\ell\fl_\ell\ \big|g_2\big|_{\ell,\ell}
+\big|g_1\big|_{\ell,\ell}\ \fl_\ell\ M^\ell\big|g_2\big|_{\ell}^{\[(1,0,0)\]}\Big]\cr
&\hskip2in
\big|h\big|_{i,r}
\sum_{m=i}^j\sfrac{1}{\fl_m\sqrt{\fl_\ell}}\sfrac{\fl_m}{M^m}\cr
&\hskip.3in\ +\const\big|g_1\big|_{\ell,\ell}\ \fl_\ell\ \big|g_2\big|_{\ell,\ell}|h_\li|_{i,r}
\sum_{m=i}^j\sfrac{1}{\fl_m\sqrt{\fl_\ell}}\sfrac{\fl_m}{M^m}\cr
&\hskip.1in\le\const\sqrt{\fl_\ell}\max_{|\al_\upl|+|\al_\dnl|\le1}
\big|g_1\big|_{\ell}^{\[\al_\upl\]}
\big|g_2\big|_{\ell}^{\[\al_\dnl\]}
\big|h\big|_{i,r}^{[1,0,0]}\sum_{m=i}^j\sfrac{M^i}{M^m}\cr
&\hskip.1in\le\const\sqrt{\fl_\ell}\max_{|\al_\upl|+|\al_\dnl|\le1}
\big|g_1\big|_{\ell}^{\[\al_\upl\]}
\big|g_2\big|_{\ell}^{\[\al_\dnl\]}
\big|h\big|_{i,r}^{[1,0,0]}\cr
}$$
\endproof

\proposition{\STM\propXXXVIII}{
$$
\big\|(g_1\bullet\cD\bullet g_2)^f\bullet\cM\bullet h\big\|_{\ka_1,\ka_2}
\le\const\sqrt{\fl_\ell}\   
\max_{\al_\upl,\al_\dnl,\al_\li\in\bbbn_0\times\bbbn_0^2
            \atop|\al_\upl|+|\al_\dnl|+|\al_\li|\le 3} \big|g_1\big|_{\ell}^{\[\al_\upl\]} \big|g_2\big|_{\ell}^{\[\al_\dnl\]}
\big|h\big|_{i,r}^{[\al_\li,0,0]}
$$
}
\prf
Write
$$
\cM(p,k)=\sum_{s_1,s_2\in\Si_i}\cM_{s_1,s_2}(p,k)
$$
where
$$
\cM_{s_1,s_2}(p,k)=\cM(p,k)\rho(\p-\k)\chi_{s_1}(\p)\chi_{s_2}(\k)
$$
and $\rho$ was defined just before (\eqnLADmeqmrho).
Then
$$\eqalign{
&(g_1\bullet\cD\bullet g_2)^f\bullet \cM\bullet h\cr
&\hskip.5in=
\sum_{u_1,u_2\in\Si_\ell\atop 
     {w_{1,1},w_{1,2}\in\Si_\ell\atop w_{2,1},w_{2,2}\in\Si_\ell}}
\sum_{{w_{1,3},w_{2,3}\in\Si_\ell\atop s_1,s_2\in\Si_i}\atop v_1,v_2\in \Si_i}\cr
&\hskip.9in\Big(
g_1({\sst y_1,(\,\cdot\,,w_{1,3}),(\,\cdot\,,w_{1,1}),
(\,\cdot\,,w_{1,2})})
\fcirc\cD_{u_1,u_2}\fcirc 
g_2({\sst(\,\cdot\,,w_{2,1}),(\,\cdot\,,w_{2,2}),y_2,
(\,\cdot\,,w_{2,3})})
\Big)^f\cr
&\hskip2in\fcirc \cM_{s_1,s_2}\fcirc
 h({\sst(\,\cdot\,,v_1),(\,\cdot\,,v_2),y_3,y_4})
}$$
where, for $\nu=1,2$,
$$
y_\nu=\cases{\ka'_{\nu}& if $\ka'_{\nu}\in\bbbm$\cr
            (x_\nu,\ka'_{\nu})& if $\ka'_{\nu}\in\Si_\ell$\cr}
\qquad
y_{\nu+2}=\cases{\ka_{\nu}& if $\ka_{\nu}\in\bbbm$\cr
                (x_{\nu+2},\ka_{\nu})& if $\ka_{\nu}\in\Si_r$\cr}
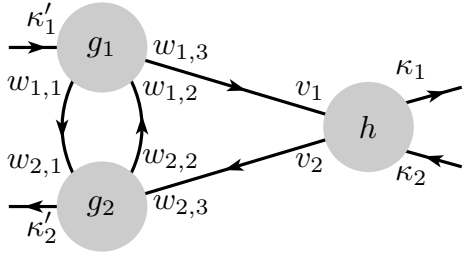
$$
\centerline{\figplace{dbubble5}{0 in}{0 in}}

\noindent 
The multiple convolution vanishes unless
$$
\bpi(s_1-s_2)\cap\bpi(\ka_1-\ka_2)\ne\emptyset \qquad
\bpi(u_1-u_2)\cap\bpi(\ka'_1-s_1)\ne\emptyset\qquad
$$
and 
$$\meqalign{
w_{1,1}\cap u_1&\ne\emptyset && 
                  w_{2,1}\cap u_1&\ne\emptyset && 
w_{1,2}\cap u_2&\ne\emptyset && 
                   w_{2,2}\cap u_2&\ne\emptyset\cr
\bpi(w_{1,3})\cap \bpi(s_1)&\ne\emptyset && 
                   \bpi(v_{1})\cap \bpi(s_1)&\ne\emptyset && 
\bpi(w_{2,3})\cap \bpi(s_2)&\ne\emptyset && 
                    \bpi(v_{2})\cap \bpi(s_2)&\ne\emptyset\cr
}\EQN\eqnXII$$
Fix the external sectors/momenta $(\ka_1,\ka_2,\ka'_1,\ka'_2)$. Then, for each fixed $(s_1,s_2,u_1,u_2)$, 
there are at most $3^8$ 8--tuples 
$(w_{1,1},w_{2,1},w_{1,2},w_{2,2}, w_{1,3}, v_1, w_{2,3},v_2)$
satisfying (\eqnXII). 
By Lemma \lemAIII, the number of allowed 4--tuples $(s_1,s_2,u_1,u_2)$
is bounded by $\sfrac{1}{\fl_i\sqrt{\fl_\ell}}$. Set, for each 
$\varsigma=(\ka_1',\ka_2',w_{1,1}, w_{1,2}, w_{1,3}, w_{2,1}, w_{2,2}, w_{2,3},u_1,u_2)\in\fK_\ell^2\times\Si_\ell^{8}$
$$
g'_\varsigma
=g_1({\sst y_1,(\,\cdot\,,w_{1,3}),(\,\cdot\,,w_{1,1}),
(\,\cdot\,,w_{1,2})})
\fcirc\cD_{u_1,u_2}\fcirc 
g_2({\sst(\,\cdot\,,w_{2,1}),(\,\cdot\,,w_{2,2}),y_2,
(\,\cdot\,,w_{2,3})})
$$
for each 
$\tau=(w_{1,1}, w_{1,2}, w_{2,1}, w_{2,2},,u_1,u_2)\in\Si_\ell^{6}$,
$$
g_\tau
=g_1({\sst \,\cdot\,,\,\cdot\,,(\,\cdot\,,w_{1,1}),
(\,\cdot\,,w_{1,2})})
\fcirc\cD_{u_1,u_2}\fcirc 
g_2({\sst(\,\cdot\,,w_{2,1}),(\,\cdot\,,w_{2,2}),\,\cdot\,,\,\cdot\,})
$$
and, for each $v_1,v_2\in\Si_i$,
$$
h_{v_1,v_2}(x_1,x_2,y_3,y_4)
=h\big((x_1,v_1),(x_2,v_2),y_3,y_4\big)
$$
Then
$$
\big\|(g_1\bullet\cD\bullet g_2)^f\bullet\cM\bullet h\big\|_{\ka_1,\ka_2}
\le\const\sfrac{1}{\fl_i\sqrt{\fl_\ell}}
\sup_{\varsigma\in\fK_\ell^2\times\Si_\ell^{8}}
\max_{s_1,s_2\in\Si_i\atop v_1,v_2\in\Si_i}\!
\TN {g'_\varsigma}^f\fcirc\cM_{s_1,s_2}\fcirc h_{v_1,v_2}\TN_{1,\infty}
$$
By (\eqnLADgMh)
$$
\TN {g'_\varsigma}^f\fcirc\cM_{s_1,s_2}\fcirc h_{v_1,v_2}\TN_{1,\infty}
\le  \const\fl_i\max_{\tau\in\Si_\ell^6}
\max_{\al_\ri,\al_\li\in\bbbn_0\times\bbbn_0^2
            \atop|\al_\ri|+|\al_\li|\le 3}
\big|g_\tau^f\big|_{\ell,i}^{[0,0,\al_\ri]}\big|h\big|_{i,r}^{[\al_\li,0,0]}
$$
Bounding
$$\eqalign{
\big|g_\tau^f\big|_{\ell,i}^{[0,0,\al_\ri]}
&\le |g_\tau|_{\ell,\ell}^{[0,\al_\ri,0]}\cr
&\le\const\fl_\ell
\max_{\al_\upl,\al_\dnl\in\bbbn_0\times\bbbn_0^2
            \atop \al_\upl+\al_\dnl\le \al_\ri} |g_1|_{\ell}^{\[\al_\upl\]}|g_2|_{\ell}^{\[\al_\dnl\]}\cr
}$$
by Leibniz and Lemma \lemXVbis, yields
$$\eqalign{
\big\|(g_1\bullet\cD\bullet g_2)^f\bullet\cM\bullet h\big\|_{\ka_1,\ka_2}
&\le\const\sfrac{1}{\fl_i\sqrt{\fl_\ell}}\ \fl_i\fl_\ell\ 
\max_{\al_\upl,\al_\dnl,\al_\li\in\bbbn_0\times\bbbn_0^2
            \atop|\al_\upl|+|\al_\dnl|+|\al_\li|\le 3} \big|g_1\big|_{\ell}^{\[\al_\upl\]} \big|g_2\big|_{\ell}^{\[\al_\dnl\]}
\big|h\big|_{i,r}^{[\al_\li,0,0]}\cr
&\le\const \sqrt{\fl_\ell}\ 
\max_{\al_\upl,\al_\dnl,\al_\li\in\bbbn_0\times\bbbn_0^2
            \atop|\al_\upl|+|\al_\dnl|+|\al_\li|\le 3} \big|g_1\big|_{\ell}^{\[\al_\upl\]} \big|g_2\big|_{\ell}^{\[\al_\dnl\]}
\big|h\big|_{i,r}^{[\al_\li,0,0]}\cr
}$$
\endproof
\goodbreak
\proof{ of Theorem \redbboundII}
By Lemmas \lemXXXVI,\ \lemXXXVII\ 
and Proposition \propXXXVIII,
$$\eqalign{
\big\|(g_1\bullet \cD\bullet g_2)^f\bullet\cC^{[i,j]}\bullet h\big\|_{\ka_1,\ka_2}
&\le \big\|(g_1\bullet\cD\bullet g_2)^f\bullet (\cZ\bullet\cC^{[i,j]}\bullet\cZ^t-\cM)\bullet h\big\|_{\ka_1,\ka_2}\cr
&\hskip.5in+\big\|(g_1\bullet\cD\bullet g_2)^f\bullet 
(\cC^{[i,j]}-\cZ\bullet\cC^{[i,j]}\bullet\cZ^t)\bullet h\big\|_{\ka_1,\ka_2}
\cr
&\hskip.5in+\big\|(g_1\bullet\cD\bullet g_2)^f\bullet\cM\bullet h\big\|_{\ka_1,\ka_2}
\cr
&\le\const \sqrt{\fl_\ell}\ 
\max_{\al_\upl,\al_\dnl,\al_\li\in\bbbn_0\times\bbbn_0^2
            \atop|\al_\upl|+|\al_\dnl|+|\al_\li|\le 3} \big|g_1\big|_{\ell}^{\[\al_\upl\]} \big|g_2\big|_{\ell}^{\[\al_\dnl\]}
\big|h\big|_{i,r}^{[\al_\li,0,0]}
}$$
as desired.
\endproof

\vfill\eject

\appendix{\APpropbnd}{Bounds on Propagators }\PG\pgLADA
Fix, as in Theorem \thmLADmodcompLadder, a sequence, $p^{(2)},p^{(3)},\cdots$, 
of sectorized, translation invariant functions $p^{(i)}$ on 
$\Big(\big(\bbbr\times\bbbr^2\big)\times\Si_i\Big)^2$ obeying
$$
\v p^{(i)}\v_{1,\Sigma_i} \le \sfrac{\rho\,\fl_i}{M^i}\cb_i\qquad
\check p^{(i)}(0,\k)=0
$$
and set, for all $j\ge 1$,
$$\meqalign{
v(k) &=\smsum\limits_{i=2}^\infty \check p^{(i)}(k) \qquad&&
e'(k)&=e(\k)-v(k)  \cr
v_j(k) &=\smsum\limits_{i=2}^j \check p^{(i)}(k) &&
e'_j(k)&=e(\k)-v_j(k)  \cr
\cr
}$$
\lemma{\STM\lemLADeprime}{
There is a $\tilde\rho_0>0$ such that for all $\rho\le\tilde\rho_0$ 
and $(k_0,\k)$ in the first neighbourhood
{\parindent=.25in
\item{ i)}
$$\meqalign{
& \nabla_{\k} e'(k_0,\k)\ne 0 &&
& \nabla_{\k} e'_j(k_0,\k)\ne 0 
\cr
&\big|\partial_{k_0} e'(k_0,\k)\big|\le\rho<\half &&
&\big|\partial_{k_0} e'_j(k_0,\k)\big|\le\rho<\half 
\cr
&\big|\imath k_0-e'(k)\big|\ge\half\big|\imath k_0-e(\k)\big|\qquad&&
&\big|\imath k_0-e'_j(k)\big|\ge\half\big|\imath k_0-e(\k)\big|
\cr
}$$
\medskip
\item{ii)}\hskip.75in
$
\big|\partial_k^\al e'_j(k)\big|\le
\abcst\cases{1&if $|\al|\le 1$\cr
       \noalign{\vskip.05in}
             \fl_jM^{(|\al|-1)j}& if $\al\in\De, |\al|\ge 2$}
$
\medskip
\item{iii)} If $|\ga|=1$, then
$$\eqalign{
\big|\partial_{k}^\ga e'_j(p_0,\k)-\partial_{k}^\ga e'_j(k_0,\k)\big|,
\big|\partial_{k}^\ga e'(p_0,\k)-\partial_{k}^\ga e'(k_0,\k)\big|
&\le\abcst\,\rho\,|p_0-k_0|^\aleph\cr
\big|\partial_{k_0} e'_j(k)-\partial_{k_0} e'_j(k')\big|,
\big|\partial_{k_0} e'(k)-\partial_{k_0} e'(k')\big|
&\le\abcst\,\rho\,|k-k'|^\aleph\cr
\big|\partial_{k}^\ga e'_j(k)-\partial_{k}^\ga e'_j(k')\big|,
\big|\partial_{k}^\ga e'(k)-\partial_{k}^\ga e'(k')\big|
&\le\abcst\,|k-k'|^\aleph\cr
}$$
}}
\prf i)
By setting $p^{(i)}=0$ for all $i>j$, it suffices to prove the statements 
regarding $e'$. All statements follow from
$$
\sup_k\big|\partial_k v(k)\big|
\le \sum_{i=2}^\infty\sup_k\big|\partial_k \check p^{(i)}(k)\big|
\le \sum_{i=2}^\infty2\rho\fl_i\le \rho
$$
For the second inequality, we used Lemma \lemOSNormMom\ of [FKTo3].
\Item ii) Again by Lemma \lemOSNormMom\ of [FKTo3],
$$\eqalign{
\big|\partial_k^\al e'_j(k)\big|
&\le \big|\partial_k^\al e(\k)\big|+
\sum_{i=2}^j\big|\partial_k^\al \check p^{(i)}(k)\big|
\le \big|\partial_k^\al e(\k)\big|+
\sum_{i=2}^j2\al!\rho\fl_i M^{(|\al|-1)i}\cr
&\le
\abcst\cases{1&if $|\al|\le 1$\cr
       \noalign{\vskip.05in}
             \fl_jM^{(|\al|-1)j}& if $\al\in\De, |\al|\ge 2$}\cr
}$$

\Item iii) 
Apply Lemma \lemNPhoelder\ of [FKTf3] with $C_0=C_1=\abcst\,\rho$, $\al=\aleph$ 
and $\be=1-\aleph$. 
When dealing with $e'$, use
$f_i(t)=\partial_{k}^\ga \check p_i(t,\k)$ for the first bound,
$f_i(t)=\partial_{k_0} \check p_i\big(k+t\sfrac{k'-k}{|k'-k|}\big)$ for the second bound and
$f_i(t)=\partial_{k}^\ga \check p_i\big(k+t\sfrac{k'-k}{|k'-k|}\big)$ for the third bound.
When dealing with $e'_j$ use the above $f_i$'s for $2\le i\le j$ and zero otherwise. 
The contribution from $e(\k)$ vanishes in the first two bounds and 
is bounded by $\abcst\,|k-k'|$ in the third.
\endproof

Recall that, for $j\ge 1$, $C^{(j)}_v(k)=\sfrac{\nu^{(j)}(k)}{ik_0-e'(k)}$.
Set, for $m\ge 1$ and $s\in\Si_m$,
$$
c^{(j)}_s(k)=C^{(j)}_v(k)\chi_s(k)
$$
and denote by $c^{(j)}_s(x)$ its Fourier transform.
\lemma{\STM\lemLADsimplepropbnd}{ There are $M$--dependent constants 
$\const\!$ and $\tilde \rho_0$ such that the following holds for all 
$\rho\le\tilde\rho_0$, $\be\in\De$ and $j\ge 1$.
\Item i) For $s\in\Si_j$,
$$
\big\|x^\be c^{(j)}_s(x)\big\|_{L^1}\le\const M^{(1+|\be|)j} 
$$
\Item ii) For $1\le m\le j$ and $s\in\Si_m$,
$$
\big\|x^\be c^{(j)}_s(x)\big\|_{L^\infty}\le \const \fl_m M^{(|\be|-1)j}
$$
\Item iii) For $m\ge 1$ and $s\in\Si_m$, 
$$
\big\|c^{(j)}_s(x)\big\|_{L^\infty}\le \const\sfrac{\fl_m}{M^j}\qquad
\Big\|\sfrac{\partial\hfill}{\partial x_{0}}c^{(j)}_{s}(x)\Big\|_{L^\infty}
\le \const\sfrac{\fl_m}{M^{2j}}
$$
\Item iv)  For $s\in\Si_j$,
$$
\Big\|\sfrac{\partial\hfill}{\partial x_{0}}c^{(j)}_{s}(x)\Big\|_{L^1}
\le\const
$$

}
\prf For any sector $s'$ of any scale,
 $c^{(j)}_{s'}(k)$ is supported on the $j^{\rm th}$ shell and
$\check p^{(i)}(k)$   is supported on the $i^{\rm th}$ extended neighbourhood.
If $i>j+1$, the $j^{\rm}$ shell and $i^{\rm th}$ extended
neighbourhood do not intersect,
so we may assume that $p^{(i)}=0$ for all $i> j+1$. Therefore,
by Corollary \corOSresectorvanishkzero\ and Proposition \propOSresectorI.iii of [FKTo4],
$$
\v v\v_{1,\Si_j}\le \sum_{i=2}^{j+1}\v p^{(i)}\v_{1,\Si_j}
\le \const\sum_{i=2}^{j+1}\sfrac{\rho\,\fl_i}{M^j}\cb_j
\le\const\sfrac{\rho}{M^j}\cb_j
$$
Hence the hypotheses of Proposition \propOSrealpropbound\ of [FKTo3]
are fulfilled. To apply this Proposition, we let, for $s',s''\in\Si_j$,
$\tilde c\big((\xi,s'),(\xi',s''))$ be the Fourier transform, as in Definition 
\defOSftcov\ of [FKTo2], of $\chi_{s'}(k)\,C^{(j)}_v(k)\,\chi_{s''}(k)$.
Comparing this Fourier transform with that specified before Definition
\defLADresectoriz, we see that
$$
c^{(j)}_{s'}(x)=\sum_{s''\in\Si_j}
\tilde c\big((0,\uparrow,0,s'),(x,\uparrow,a,s''))
$$
By conservation of momentum, only three $s''$'s give nonzero contributions
to the right hand side for each $s'$.
Hence, by parts (ii) and (iii) of Proposition \propOSrealpropbound\ of [FKTo3] 
and Corollary \corOSappMonoidIV.i of [FKTo1],
$$\eqalignno{
\sum_{\de\in\bbbn_0\times\bbbn_0^2}\sfrac{1}{\de!}
\big\|x^\de c^{(j)}_{s'}(x)\big\|_{L^1}t^\de
&\le\abcst\,\V \tilde c\V_{1,\Si_j}
\le\const \sfrac{M^j\cb_j}{1-\const\rho\cb_j}\le\const M^j\cb_j
&\EQNO\eqnLADcspI\cr
\sum_{\de\in\bbbn_0\times\bbbn_0^2}\sfrac{1}{\de!}
\big\|x^\de c^{(j)}_{s'}(x)\big\|_{L^\infty}t^\de
&\le\const \sfrac{\fl_j}{M^j}\sfrac{\cb_j}{1-\const\rho\cb_j}
\le\const \sfrac{\fl_j}{M^j}\cb_j
&\EQNO\eqnLADcspII\cr
}$$
\Item i) follows from (\eqnLADcspI) by choosing $s'=s$.

\Item ii) By (\eqnLADcspII),
$$
\big\|x^\de c^{(j)}_{s'}(x)\big\|_{L^\infty}
\le \const \sfrac{\fl_j}{M^j}M^{|\de|j}
$$
for all $s'\in\Si_j$ and $\de\in\De$. Write $c^{(j)}_s(k)
=\sum_{s'\in\Si_j\atop s\cap s'\ne\emptyset}c^{(j)}_{s'}(k)\chi_s(k)$.
By Lemma \lemOSsectpartunit.iii of [FKTo3]
$$\eqalign{
\big\|x^\be c^{(j)}_{s}(x)\big\|_{L^\infty}
&\le \abcst\sum_{s'\in\Si_j\atop s\cap s'\ne\emptyset}
\sum_{\de,\de'\in\bbbn_0\times\bbbn_0^2\atop\de+\de'=\be}
\big\|x^{\de} c^{(j)}_{s'}(x)\big\|_{L^\infty}
\big\|x^{\de'} \hat\chi_{s}(x)\big\|_{L^1}\cr
&\le \const
\sum_{s'\in\Si_j\atop s\cap s'\ne\emptyset} 
\sum_{\de,\de'\in\bbbn_0\times\bbbn_0^2\atop\de+\de'=\be}
\sfrac{\fl_j}{M^j}M^{|\de|j}M^{|\de'|j}
\le \const
\sfrac{\fl_m}{\fl_j} \sfrac{\fl_j}{M^j}M^{|\be|j}
}$$
\Item iii) follows from the observations that
$$
\sup_k\big|c^{(j)}_s(k)|\le\const M^j\qquad
\sup_k\big|k_0c^{(j)}_s(k)|\le\const
$$
 and $c^{(j)}_s(k)$ is supported
in a region of volume $\const\sfrac{\fl_m}{M^{2j}}$.
\Item iv) follows from part (iv) of Proposition \propOSrealpropbound\ 
of [FKTo3].

\endproof

Recall from Lemma \lemXXXI\ that
$$
\De^{m_1,m_2}_{s_1,s_2}(\p,\k)
=\int_{-b(m_1,m_2)}^{b(m_1,m_2)}
 d\om\ \frac{\mu^{(m_1,m_2)}(\om,\k,\p)\chi_{s_1}(\p)\chi_{s_2}(\k)}
{[i\om-e'(\om,\p)][i\om-e'(\om,\k)]}
$$
where
$$
\mu^{(m_1,m_2)}(\om,\k,\p)=\Big[1-\nu_0({\sst\om})\nu_1({\sst\p,\k})\Big]
\nu^{(m_1)}({\sst (\om,\p)})\nu^{(m_2)}({\sst (\om,\k)})
$$
and
$$
b(m_1,m_2)=\cases{\sfrac{\const}{M^{\max\{m_1,m_2\}}}& if $m=i, i+1$\cr
\noalign{\vskip.1in} \min\Big\{\sfrac{1}{M^{{j}-3/4}},\sfrac{\const}{M^{\max\{m_1,m_2\}}}\Big\}& if $m\ge i+2$\cr}
$$
The functions $\nu_0$ and $\nu_1$ were defined just before (\eqnIX).
\lemma{\STM\lemLADdeltabnd}{
Let $i\le m\le j$, $\min\{m_1,m_2\}=m$ and $s_1,s_2\in\Si_m$. Then
$$\eqalign{
\int d\z_1\sup_{\z_2}\big|\hat\De^{m_1,m_2}_{s_1,s_2}(\z_1,\z_2)\big|
&\le\const b(m_1,m_2)\ \fl^2_{m}\sfrac{M^{m_1}}{\fl_{m_1}}\cr
\int d\z_2\sup_{\z_1}\big|\hat\De^{m_1,m_2}_{s_1,s_2}(\z_1,\z_2)\big|
&\le\const b(m_1,m_2)\ \fl^2_{m}\sfrac{M^{m_2}}{\fl_{m_2}}\cr
}$$

}
\prf
We may write $\De^{m_1,m_2}_{s_1,s_2}(\p,\k)$ as a sum of two pieces,
each of the form
$$
\sum_{u_1,v_1\in\Si_{m_1}\atop
         {u_1\cap s_1\ne\emptyset\atop u_1\cap v_1\ne\emptyset}}
\sum_{u_2,v_2\in\Si_{m_2}\atop
         {u_2\cap s_2\ne\emptyset\atop u_2\cap v_2\ne\emptyset}}
\De_{u_1,v_1,u_2,v_2}(\p,\k)
$$
with
$$
\De_{u_1,v_1,u_2,v_2}(\p,\k)
=\pm\int d\om\ \ze_{m_1,m_2}(\om) c^{(m_1)}_{u_1}(\om,\p)  c^{(m_2)}_{u_2}(\om,\k)
\chi_{1,v_1}(\om,\p)\chi_{2,v_2}(\om,\k)
$$
where 
$$\eqalign{
\chi_{1,v_1}(\om,\p)
&=\left.\cases{1& or\cr
\noalign{\vskip.05in}
\sum\limits_{\ell=i+1}^{\infty}\nu\big(M^{2\ell}e(\p)^2\big)&\cr}
\right\}
\chi_{v_1}(\om,\p)\chi_{s_1}(0,\p)
\cr
\noalign{\vskip.1in}
\chi_{2,v_2}(\om,\k)
&=\left.\cases{1& or\cr
\noalign{\vskip.05in}
\sum\limits_{\ell=i+1}^{\infty}\nu\big(M^{2\ell}e(\k)^2\big)&\cr}\right\}
\chi_{v_2}(\om,\k)\chi_{s_2}(0,\k)
\cr
\noalign{\vskip.1in}
\ze_{m_1,m_2}(\om)
&=\left.\cases{1& or\cr
\noalign{\vskip.05in}
\nu_0(\om)&\cr}\right\}
\chi_{m_1,m_2}(\om)
\cr
}$$
and
$\chi_{m_1,m_2}(\om)$ is the characteristic function of the interval
$\big[-b(m_1,m_2),b(m_1,m_2)\big]$.
The Fourier transform of $\De_{u_1,v_1,u_2,v_2}(\p,\k)$ is then
$$\eqalign{
\hat\De_{u_1,v_1,u_2,v_2}(\z_1,\z_2)
=\int d\x_1\,d\x_2\,\smprod_{\ell=1}^4 dt_\ell\ 
&c^{(m_1)}_{u_1}(t_1,\x_1) 
\hat \chi_{1,v_1}(t_2-t_1,\z_1-\x_1) \cr
&c^{(m_2)}_{u_2}(t_3-t_2,\x_2)
\hat \chi_{2,v_2}(t_4-t_3,\z_2-\x_2)
\hat \ze_{m_1,m_2}(-t_4) 
}$$
This is bounded using
$$\eqalign{
\sup_{t_4}\big|\hat \ze_{m_1,m_2}(-t_4)\big|&\le 2b(m_1,m_2)\cr
\sup_{t_3,\x_2,\z_2}\int dt_4\ \big|\hat \chi_{2,v_2}(t_4-t_3,\z_2-\x_2)\big|
&\le\const\sfrac{\fl_{m_2}}{M^{m_2}} \cr
\sup_{t_2}\int dt_3\,d\x_2\ \big|c^{(m_2)}_{u_2}(t_3-t_2,\x_2)\big|
&\le\const M^{m_2} \cr
\sup_{t_1,\x_1}\int dt_2\,d\z_1\ \big|\hat \chi_{1,v_1}(t_2-t_1,\z_1-\x_1)\big|
&\le\const \cr
\int dt_1\,d\x_1\ \big|c^{(m_1)}_{u_1}(t_1,\x_1)\big|
&\le\const M^{m_1} \cr
}\EQN\eqnLADdepartbnds$$
The supremum of $|\hat \ze_{m_1,m_2}(-t_4)|$ was bounded by the
$L^1$ norm of $\ze_{m_1,m_2}(\om)$. The bounds on $c^{(m_\ell)}_{u_\ell}$, 
$\ell=1,2$ are immediate consequences of Lemma \lemLADsimplepropbnd.i with
$\be=0$. The bounds on $\hat \chi_{\ell,v_\ell}$, $\ell=1,2$ are proven
much as Lemma \lemOSmorepartunity\ of [FKTo3]. Indeed (\eqnOSsecpropboundII)
of [FKTo3] applies with $j=m_\ell$ and $\fl=\fl_{m_\ell}$. Denoting
by $\x_{\perp}$ and $\x_{\|}$ the components of $\x$ perpendicular and
parallel, respectively, to the Fermi curve at the centre of $v_\ell$,
Proposition \propOSGenDecay.i gives
$$
\big|\hat \chi_{\ell,v_\ell}(t,\x)\big|
\le \const\sfrac{\fl_{m_\ell}}{M^{2m_\ell}}
\sfrac{1}{\big[1+M^{-m_\ell}|t|]^2\,
[1+M^{-m_\ell}|\x_{\perp} | +\fl_{m_\ell} |\x_{\|}|]^3}
$$
from which the desired bounds follow.
Applying, in order, the bounds of (\eqnLADdepartbnds)  yields
$$\eqalign{
\int d\z_1\sup_{\z_2}\big|\hat\De_{u_1,v_1,u_2,v_2}(\z_1,\z_2)\big|
&\le\const b(m_1,m_2)\ \fl_{m_2}M^{m_1}\cr
}$$
Similarly,
$$\eqalign{
\int d\z_2\sup_{\z_1}\big|\hat\De_{u_1,v_1,u_2,v_2}(\z_1,\z_2)\big|
&\le\const b(m_1,m_2)\ \fl_{m_1} M^{m_2}\cr
}$$
For each fixed $s_1\in\Si_m$, there are at most 
$\const\sfrac{\fl_m}{\fl_{m_1}}$ pairs $(u_1,v_1)\in\Si_{m_1}^2$
obeying $u_1\cap s_1\ne\emptyset$, $u_1\cap v_1\ne\emptyset$ and
for each fixed $s_2\in\Si_m$, there are at most 
$\const\sfrac{\fl_m}{\fl_{m_2}}$ pairs $(u_2,v_2)\in\Si_{m_2}^2$
obeying $u_2\cap s_2\ne\emptyset$, $u_2\cap v_2\ne\emptyset$.
Hence
$$\deqalign{
\int d\z_1\sup_{\z_2}\big|\hat\De^{m_1,m_2}_{s_1,s_2}(\z_1,\z_2)\big|
&\le \const b(m_1,m_2)\ \sfrac{\fl_m}{\fl_{m_1}}\sfrac{\fl_m}{\fl_{m_2}}
\ \fl_{m_2}M^{m_1}
&\le\const b(m_1,m_2)\ \fl^2_{m}\sfrac{M^{m_1}}{\fl_{m_1}}\cr
\int d\z_2\sup_{\z_1}\big|\hat\De^{m_1,m_2}_{s_1,s_2}(\z_1,\z_2)\big|
&\le \const b(m_1,m_2)\ \sfrac{\fl_m}{\fl_{m_1}}\sfrac{\fl_m}{\fl_{m_2}}
\ \fl_{m_1}M^{m_2}
&\le\const b(m_1,m_2)\ \fl^2_{m}\sfrac{M^{m_2}}{\fl_{m_2}}\cr
}$$

\endproof

\vfill\eject

\appendix{\APmodelbnd}{Bound on the Generalized Model Bubble}\PG\pgLADB
Fix, as in Theorem \thmLADmodcompLadder, a sequence, $p^{(2)},p^{(3)},\cdots$, 
of sectorized, translation invariant functions $p^{(i)}$ on 
$\Big(\big(\bbbr\times\bbbr^2\big)\times\Si_i\Big)^2$ obeying
$$
\v p^{(i)}\v_{1,\Sigma_i} \le \sfrac{\rho\,\fl_i}{M^i}\cb_i\qquad
\check p^{(i)}(0,\k)=0
$$
Let $I$ be an interval of length $\fl$ on the Fermi
surface $F$ and $u(\k,\t)$ a function that vanishes unless $\pi_F(\k)\in I$,
where $\pi_F$ is projection on the Fermi curve $F$.
Set, for  $1\le i\le j$,
$$
B_{i,j}(\t)=
\int dk\ \frac{\nu^{(i,j)}_0(k_0)u(\k,\t)}
{[\imath k_0-e'(k_0,\k)][\imath k_0-e'(k_0,\k+\t)]}
$$
where
$$
\nu^{(i,j)}_0(k_0)=\sum_{\ell=i+1}^{j-1}\nu(M^{2\ell}k_0^2)\qquad
e'(k)=e(\k)-\smsum_{\ell=2}^{j+1}\check p^{(\ell)}(k)
$$

\lemma{\STM\lemLadBmom}{  Let $1\le i\le j$ obey $M^i\le \fl_jM^j$.
Then
$$
\big|\partial_\t^\al B_{i,j}(\t)\big|\le \const\fl\max\{1,\,j\,\fl_jM^{|\al|j}\}
\max_{|\be+\ga|\le|\al|}\sup_{\k,\t} \sfrac{1}{M^{i|\be+\ga|}}
\big|\partial^{\be}_\t\partial^{\ga}_\k u(\k,\t)\big|
$$
for all $|\al|\le 4$ and all $\t$ in a neighbourhood of the origin.
}
\prf Let
$$\eqalign{
E'(k_0,\k,\t,s)&=s e'(k_0,\k)+(1-s)e'(k_0,\k+\t)\cr
E(\k,\t,s)&=E'(0,\k,\t,s)=s e(\k)+(1-s)e(\k+\t)\cr
w(\k,\t,s)&=1-\sfrac{1}{i}\sfrac{\partial E'}{\partial k_0}(0,\k,\t,s)\cr
\tilde E(k_0,\k,\t,s)&=E'(k_0,\k,\t,s)-E'(0,\k,\t,s)
-k_0\sfrac{\partial E'}{\partial k_0}(0,\k,\t,s)\cr
}$$
Then
$$\eqalignno{
B_{i,j}(\t)
&=\int\! dk\ \frac{\nu^{(i,j)}_0(k_0)u(\k,\t)}
{[\imath k_0-e'(k_0,\k)][\imath k_0-e'(k_0,\k+\t)]}\cr
&=\int\! dk\
\frac{\nu^{(i,j)}_0(k_0)u(\k,\t)}{e'(k_0,\k)-e'(k_0,\k+\t)}
\Big[\frac{1}{\imath k_0-e'(k_0,\k)}-\frac{1}{\imath k_0-e'(k_0,\k+\t)}\Big]\cr
&=\int dk\int_0^1ds\ 
\frac{\nu^{(i,j)}_0(k_0)u(\k,\t)}{[\imath k_0-E'(k_0,\k,\t,s)]^2}\cr
&=\int dk\int_0^1ds\ 
\frac{\nu^{(i,j)}_0(k_0)u(\k,\t)}{[\imath w(\k,\t,s) k_0- E(\k,\t,s)
-\tilde E(k_0,\k,\t,s)]^2}&\EQNO\eqnMana\cr
}$$

\goodbreak\Item {\it Case i: $|\al|\ge 1$}
Make, for each fixed $s$, the change of variables from $\k$ to $E$ 
and  an ``angular'' variable $\th$. Denote by $J(E,\t,\th,s)$ the Jacobian 
of this change of variables. Then
$$
B_{i,j}(\t)
=\int_0^1ds\int dk_0 \int d\th dE\ \frac{\nu^{(i,j)}_0(k_0)u\big(\k(E,\t,\th,s),\t\big)J(E,\t,\th,s)}
{[\imath w(\k(E,\t,\th,s),\t,s)k_0-E-\tilde E(k_0,\k(E,\t,\th,s),\t,s)]^2}
\EQN\eqnManb
$$
Since $E(\k,\t,s)=s e(\k)+(1-s)e(\k+\t)$ and $\t$ is restricted to a small
neighbourhood of the origin,
$$
\big|\nabla_\k E(\k,\t,s)\big|\ge \abcst >0\qquad
\big|\partial_k^\al\partial_\t^\be E(\k,\t,s)\big|\le \abcst'
\EQN\eqnMancm$$
for all $\al+\be$ having spatial component at most $r$.
Using
$$
\partial_{\t_i}\k_\ell(E,\t,\th,s)=-\sfrac
{\partial_{\k_2}\th(\k)\de_{\ell,1}
-\partial_{\k_1}\th(\k)\de_{\ell,2}}
{\partial_{\k_1}E(\k,\t,s)\,\partial_{\k_2}\th(\k)
-\partial_{\k_2}E(\k,\t,s)\,\partial_{\k_1}\th(\k)}
\partial_{\t_i}E(\k,\t,s)\Big|_{\k=\k(E,\t,\th,s)}
$$
one proves, by induction on $|\be|$, that, for $|\be|\le r$,
$$
\big|\partial_\t^\be\k(E,\t,\th,s)\big|
\le\abcst''
\EQN\eqnMand$$
Using this bound and
$$
J(E,\t,\th,s)=\sfrac{1}
{|\partial_{\k_1}E(\k,\t,s)\,\partial_{\k_2}\th(\k)
-\partial_{\k_2}E(\k,\t,s)\,\partial_{\k_1}\th(\k)|}
\Big|_{\k=\k(E,\t,\th,s)}
$$
one proves that, for $|\be|\le r-1$,
$$
\big|\partial_\t^\be J(E,\t,\th,s)\big|
\le\abcst''
\EQN\eqnMane$$
By Lemma \lemLADeprime.ii, for $\al+\be+(1,0,0)\in\De$,
$$\eqalign{
\big|\partial_k^\al\partial_\t^\be w(\k,\t,s)\big|
&\le\abcst\cases{1&if $|\al|+ |\be|= 0$\cr
             \noalign{\vskip.02in}
             \fl_jM^{(|\al|+|\be|)j\phantom{-1}}& if $|\al|+ |\be|\ge 1$}\cr
}\EQN\eqnManc$$
Since
$$\eqalign{
\tilde E(k_0,\k,\t,s)
&=E'(k_0,\k,\t,s)-E'(0,\k,\t,s)
-k_0\sfrac{\partial E'}{\partial k_0}(0,\k,\t,s)\cr
&=\int_0^{k_0}\!\!\!d\ka\ \big[
\sfrac{\partial E'}{\partial k_0}(\ka,\k,\t,s)
-\sfrac{\partial E'}{\partial k_0}(0,\k,\t,s)\big]
}$$
parts (ii) and (iii) of Lemma \lemLADeprime\ imply that, again
for $\al+\be+(1,0,0)\in\De$,
$$\eqalign{
\big|\partial_k^\al\partial_\t^\be \tilde E(k_0,\k,\t,s)\big|
&\le\abcst\,\rho\cases{|k_0|^{1+\aleph}&if $|\al|+ |\be|= 0$\cr
             \noalign{\vskip.02in}
             |k_0|^\aleph+|k_0|\fl_jM^{j}& if $|\al|+ |\be|= 1$\cr
             \noalign{\vskip.02in}
             \fl_jM^{(|\al|+|\be|-1)j}+|k_0|\fl_jM^{(|\al|+|\be|)j}& if $|\al|+ |\be|> 1$\cr}\cr
}$$
For $k_0$ in the support of $\nu^{(i,j)}_0(k_0)$, $|k_0|\ge\const \sfrac{1}{M^j}$
and 
$$\eqalign{
\big|\partial_k^\al\partial_\t^\be \tilde E(k_0,\k,\t,s)\big|
&\le\const\,\rho\cases{|k_0|^{1+\aleph}&if $|\al|+ |\be|= 0$\cr
             \noalign{\vskip.02in}
             |k_0|\fl_jM^{(|\al|+|\be|)j}& if $|\al|+ |\be|\ge 1$\cr}\cr
}\EQN\eqnMante$$

 Applying $\partial_\t^\al$ to
(\eqnManb) yields an integral whose integrand is a sum of terms (whose
number is bounded by a universal constant) of the form
a combinatorial factor (which is bounded by a universal constant)
times $\nu^{(i,j)}_0(k_0)$ times
$$\eqalign{
\sfrac{1}{[\imath wk_0-E-\tilde E]^{m+2}}
\smprod_{p=1}^m\Big[
-\partial^{\ga^{(p)}}_\k\partial^{\be^{(p)}}_\t (\imath k_0 w-\tilde E)
\smprod_{\ell=1}^{|\ga^{(p)}|}\partial^{\al^{(p,\ell)}}_\t \k_{i_\ell}
\Big]\!\quad
\Big[\partial^{\ga'}_\k\partial^{\be'}_\t u
\smprod_{\ell'=1}^{|\ga'|}\partial^{{\al'}^{(\ell')}}_\t \k_{i_{\ell'}}\big]\quad
\partial^{\be''}_\t J
\cr
}$$
with the various degrees obeying
$$\deqalign{
&\be''+\be'+\smsum_{\ell'=1}^{|\ga'|}{\al'}^{(\ell')}
+\sum_{p=1}^m\Big[\be^{(p)}+\smsum_{\ell=1}^{|\ga^{(p)}|}\al^{(p,\ell)}\Big]
=\al
\hidewidth\cr
&|\ga^{(p)}|+|\be^{(p)}|\ge 1\qquad&\hbox{for all }1\le p\le m\cr
&|\al^{(p,\ell)}|\ge 1\qquad&\hbox{for all }1\le p\le m,\ 1\le \ell\le
|\ga^{(p)}|\cr
&|{\al'}^{(\ell')}|\ge 1\qquad&\hbox{for all }1\le \ell'\le |\ga'|\cr
}$$
Using the fact, from Lemma \lemLADeprime.i, 
that $|w(\k,\t,s)-1|\le\rho\le\half$
and the bounds on the derivatives of $\k$, $J$, $w$ and $\tilde E$
of (\eqnMand--\eqnMante), we may bound this term by
$$\eqalign{
&\abcst'\ \sfrac{1}{|\imath k_0-E|^{m+2}}
\smprod_{p=1}^m\bigg[|k_0|\fl_jM^{(|\ga^{(p)}|+|\be^{(p)}|)j}
\bigg]
\sup_{\k,\t}\big|\partial^{\ga'}_\k\partial^{\be'}_\t u\big|\cr
&\le \abcst'\, \sfrac{|k_0|^m}{|\imath k_0-E|^{m+2}}
\smprod_{p=1}^m\Big[\fl_jM^{(|\ga^{(p)}|+|\be^{(p)}|)j}\Big]\,
M^{i(|\be'|+|\ga'|)}\max_{|\be+\ga|\le|\al|}\sup_{\k,\t} \sfrac{1}{M^{i|\be+\ga|}}
\big|\partial^{\be}_\t\partial^{\ga}_\k u(\k,\t)\big|
\cr
&\le \abcst'\, \sfrac{|k_0|^m}{|\imath k_0-E|^{m+2}}M^{|\al-\be''|j}
M^{|\be'+\ga'|(i-j)}\,\fl_j^m
\max_{|\be+\ga|\le|\al|}\sup_{\k,\t} 
\sfrac{1}{M^{i|\be+\ga|}}
\big|\partial^{\be}_\t\partial^{\ga}_\k u(\k,\t)\big|
\cr
&\le \abcst'\, \sfrac{1}{|\imath k_0-E|^2}M^{|\al|j}\fl_j
\max_{|\be+\ga|\le|\al|}\sup_{\k,\t} 
\sfrac{1}{M^{i|\be+\ga|}}
\big|\partial^{\be}_\t\partial^{\ga}_\k u(\k,\t)\big|
\cr
}$$
For the second inequality, we used that 
$|\be'|+|\ga'|+\smsum_{p=1}^m[|\ga^{(p)}|+|\be^{(p)}|]\le |\al-\be''|$.
For the final inequality we used that one of $m$, $|\ga'|$, $|\be'|$, $|\be''|$ 
must be nonzero for $|\al|$ to be nonzero and we also used the hypothesis that
$M^{i-j}\le\fl_j$. The bound is completed by applying
$$
\int_{\const\over M^j}^\const dk_0 \int_{I}d\th
\int_{-\const}^\const dE \sfrac{1} {|\imath k_0-E|^2}
\le \abcst'\,\fl \int_{\const\over M^j}^\const dR\ \sfrac{1}{R}\le 
\const\,j\fl
$$
\goodbreak
\Item {\it Case ii: $\al=0$.}
Recall from (\eqnMana) that
$$
B_{i,j}(\t)
=\int dk\int_0^1ds\ 
\frac{\nu^{(i,j)}_0(k_0)u(\k,\t)}{[\imath w(\k,\t,s) k_0- E(\k,\t,s)
- \tilde E(k_0,\k,\t,s)]^2}
$$
and set
$$
B'_{i,j}(\t)
=\int dk\int_0^1ds\ 
\frac{\nu^{(i,j)}_0(k_0)u(\k,\t)}{[\imath w(\k,\t,s) k_0- E(\k,\t,s)]^2}
$$
By Lemma \lemLADeprime.i, (\eqnMante) and the reality of $k_0$ and $E(\k,\t,s)$,
$$
\sfrac{1}{4}\big|\imath k_0- E(\k,\t,s)\big|
\le \big|\imath w(\k,\t,s) k_0- E(\k,\t,s)- \tilde E(k_0,\k,\t,s)\big|
\le 2\big|\imath k_0- E(\k,\t,s)\big|
$$
Hence, by (\eqnMante),
$$\eqalign{
&\big|B_{i,j}(\t)-B'_{i,j}(\t)\big|\cr
&\hskip.25in\le\int dk\int_0^1ds\ |\nu^{(i,j)}_0(k_0)u(\k,\t)|\Big|
\sfrac{
[\imath w(\k,\t,s) k_0- E(\k,\t,s)]^2
       -[\imath w(\k,\t,s) k_0- E(\k,\t,s)- \tilde E(k_0,\k,\t,s)]^2}
{[\imath w(\k,\t,s) k_0- E(\k,\t,s)]^2
           [\imath w(\k,\t,s) k_0- E(\k,\t,s)- \tilde E(k_0,\k,\t,s)]^2}
\Big|\cr
&\hskip.25in=\int dk\int_0^1ds\ |\nu^{(i,j)}_0(k_0)u(\k,\t)|\Big|
\sfrac{
\tilde E(k_0,\k,\t,s)
       [2\imath w(\k,\t,s) k_0- 2E(\k,\t,s)- \tilde E(k_0,\k,\t,s)]}
{[\imath w(\k,\t,s) k_0- E(\k,\t,s)]^2
           [\imath w(\k,\t,s) k_0- E(\k,\t,s)- \tilde E(k_0,\k,\t,s)]^2}
\Big|\cr
&\hskip.25in\le\int dk\int_0^1ds\ \const |k_0|^{\aleph}
|\nu^{(i,j)}_0(k_0)u(\k,\t)|
\sfrac{|\tilde E(k_0,\k,\t,s)|}{|\imath  k_0- E(\k,\t,s)|^3}
\cr
&\hskip.25in\le\int dk\int_0^1ds\ \const |k_0|^{\aleph}
|\nu^{(i,j)}_0(k_0)u(\k,\t)|
\sfrac{1}{|\imath  k_0- E(\k,\t,s)|^2}
\cr
&\hskip.25in\le\const\sup_{\k,\t}|u(\k,\t)|\int_I d\th
\int_{-const}^{\const} dk_0\int_{-\const}^{\const} dE\ 
\sfrac{|k_0|^\aleph}{|\imath  k_0- E|^2}
\cr
&\hskip.25in\le\const\sup_{\k,\t}|u(\k,\t)|\int_I d\th
\int_{-const}^{\const} dk_0\ |k_0|^{\aleph-1}\int dE'\ 
\sfrac{1}{|\imath- E'|^2}\cr
&\hskip.25in\le\const\fl\,\sup_{\k,\t}|u(\k,\t)|
}$$
and it suffices to consider $B'_{i,j}(\t)$.

Make, for each fixed $s$, the change of variables from $\k$ to $E=E(\k,\t,s)$ 
and  an ``angular'' variable $\th$. Then
$$\eqalign{
B'_{i,j}(\t)
&=\int_0^1ds\int d\th
\int dk_0 dE\ \nu^{(i,j)}_0(k_0)
\frac{u\big(\k(E,\t,\th,s),\t\big)J(E,\t,\th,s)}
{[\imath w(\k(E,\t,\th,s),\t,s)k_0-E]^2}\cr
}$$
Integrating by parts with respect to $k_0$,
$$\eqalignno{
B'_{i,j}(\t)
&=\int_0^1ds\int d\th dE
\int_{-\infty}^\infty dk_0 \ \nu^{(i,j)}_0(k_0)\,\sfrac{u\,J}{w}\,
\imath\sfrac{d\hfill}{dk_0}\sfrac{1}{\imath wk_0-E}\cr
&=\int_0^1ds\int d\th dE
\int_{-\infty}^\infty dk_0 \ \frac{{u\,J\over w}
\,(-\imath)\sfrac{d\hfill}{dk_0}\nu^{(i,j)}_0(k_0)}{\imath wk_0-E}\cr
}$$
Since $\sfrac{d\hfill}{dk_0}\nu^{(i,j)}_0(k_0)$ is odd under $k_0\rightarrow-k_0$, 
$$\eqalign{
 B'_{i,j}(\t)&=\int_0^1ds\int d\th dE
\int_0^\infty dk_0\ \sfrac{u\,J}{w}\,(-\imath)\Big[\sfrac{d\hfill}{dk_0}\nu^{(i,j)}_0(k_0)\Big]
\Big[\sfrac{1}{\imath wk_0-E}-\sfrac{1}{-\imath wk_0-E}\Big]\cr
&=\int_0^1ds\int d\th dE
\int_0^\infty dk_0
\sfrac{u\,J}{w}\,(-\imath)\Big[\sfrac{d\hfill}{dk_0}\nu^{(i,j)}_0(k_0)\Big]
\sfrac{-2\imath wk_0}{w^2k_0^2+E^2}\cr
&=-2\int_0^1ds\int d\th \int_{-\infty}^\infty\! dE
\int_0^\infty dk_0 \ 
\Big[\sfrac{d\hfill}{dk_0}\nu^{(i,j)}_0(k_0)\Big]
\sfrac{u(\k(E,\t,\th,s),\t)J(E,\t,\th,s)k_0}
{w(\k(E,\t,\th,s),\t,s)^2k_0^2+E^2}\cr
}$$
Hence, since $\big|w(\k(E,\t,\th,s),\t,s)-1\big|\le\rho\le\sfrac{1}{5}$,
$\big|w(\k(E,\t,\th,s),\t,s)^2-1\big|\le\sfrac{1}{2}$ and
$$\eqalign{
\big|B'_{i,j}(\t)\big|
&\le4\sup|uJ|\int_0^1ds\int_{I'} d\th \int_0^\infty dk_0\int_{-\infty}^\infty\! dE 
\ \Big|\sfrac{d\hfill}{dk_0}\nu^{(i,j)}_0(k_0)\Big|
\sfrac{k_0}{k_0^2+E^2}\cr
}$$
where $I'$ is some interval of length $\const\fl$. Finally,
$$\eqalign{
\big|B'_{i,j}(\t)\big|
&\le\const\fl\,\sup|u|
\int_0^\infty dk_0 \ \Big|\sfrac{d\hfill}{dk_0}\nu^{(i,j)}_0(k_0)\Big|\cr
&\le\const\fl\,\sup_{\k,\t}|u(\k,\t)|
}$$
since 
$\int_0^\infty dk_0 \ \big|\sfrac{d\hfill}{dk_0}\nu^{(i,j)}_0(k_0)\big|=2$.
\endproof

\theorem{\STM\thmLADBposn}{
Let $1\le i\le j$ obey $M^i\le \fl_jM^j$.
Let $\Ln$ and $\Sh$ be mutually perpendicular  unit vectors in $\bbbr^2$
and $\rho(\k)$ be a function that is supported in  a rectangle
in $\k$ having one side of length $\sfrac{\const}{M^j}$ parallel to $\Sh$ and  
one side of length $\const\fl_j$ parallel to $\Ln$. Furthermore
assume that, for all $\al_1,\al_2\le 2$,
$$
\Big|\big(\Sh\cdot\partial_{\k}\big)^{\al_1}
\big(\Ln\cdot\partial_{\k}\big)^{\al_2}\rho(\k)\Big|
\le\const M^{\al_1j}\sfrac{1}{\fl_j^{\al_2}}
$$
Let $a=\sfrac{1/2}{1-\aleph}$ and
$\hat B_{i,j}(\x)$ be the
Fourier transform of $\rho(\k) B_{i,j}(\k)$. Then 
$$
\big|\hat B_{i,j}(\x)\big|\le
\const \sfrac{\fl\,\fl_{j}}{M^{j}}
\sfrac{1}
{(1+|\Sh\cdot\x/M^{j}|^{3/2})
       (1+|\fl_{j}\Ln\cdot\x|^{(1+a)/2})}
\max_{|\be+\ga|\le3}\sup_{\k,\t} \sfrac{1}{M^{i|\be+\ga|}}
\big|\partial^{\be}_\t\partial^{\ga}_\k u(\k,\t)\big|
$$
}
\prf 
Denote
$$
U=\max_{|\be+\ga|\le3}\sup_{\k,\t} \sfrac{1}{M^{i|\be+\ga|}}
\big|\partial^{\be}_\t\partial^{\ga}_\k u(\k,\t)\big|
$$
Note that $1<a<\sfrac{3}{2}$.
The first step is to prove that for all $\al_1\in\{0,1\}$ and $0\le\al_2\le a$
$$
\Big|\big(\Sh\cdot\partial_{\k}\big)^{\al_1}
\big(\Ln\cdot\partial_{\k}\big)^{[\al_2]}\big[B_{i,j}(\k+q\Ln)-B_{i,j}(\k)\big]\Big|
\le C\,\fl U\, M^{\al_1j}\sfrac{1}{\fl_j^{\al_2}}|q|^{\al_2-[\al_2]}
\EQN\eqnManaa$$
where $[\al_2]$ is the integer part of $\al_2$. Here $C$ is a constant that
is independent of $i$, $j$, $\k$ and $q$. To prove (\eqnManaa) when 
$[\al_2]=0$, apply Lemma \lemLadBmom\ twice to obtain
$$\eqalign{
\Big|\big(\Sh\cdot\partial_{\k}\big)^{\al_1}\big[B_{i,j}(\k+q\Ln)
      \!-\!B_{i,j}(\k)\big]\Big|
&\le 
\Big|\big(\Sh\cdot\partial_{\k}\big)^{\al_1}B_{i,j}(\k+q\Ln)\Big|
\!+\!\Big|\big(\Sh\cdot\partial_{\k}\big)^{\al_1}B_{i,j}(\k)\Big|\cr
&\le 2\,\const\fl U\, M^{\al_1j}\cr
\Big|\big(\Sh\cdot\partial_{\k}\big)^{\al_1}
               \big[B_{i,j}(\k+q\Ln)\!-\!B_{i,j}(\k)\big]\Big|
&\le |q|\sup_{\p}
\Big|\big(\Sh\cdot\partial_{\p}\big)^{\al_1}\big(\Ln\cdot\partial_{\p}\big)
    B_{i,j}(\p)\Big|\cr
&\le \const\,\fl U\, |q|\,j\fl_j\,M^{(\al_1+1)j}\cr
}$$
Multiplying the $(1-\al_2)^{\rm th}$ power of the first bound by the
$\al_2^{\rm th}$ power of the second gives
$$\eqalign{
\Big|\big(\Sh\cdot\partial_{\k}\big)^{\al_1}
               \big[B_{i,j}(\k+q\Ln)\!-\!B_{i,j}(\k)\big]\Big|
&\le 2^{1-\al_2}\const\,\fl U\, |q|^{\al_2}M^{\al_1j}\,\big(j\fl_j\,M^{j}\big)^{\al_2}\cr
&= 2^{1-\al_2}\const\,\fl U\, M^{\al_1j}\sfrac{1}{\fl_j^{\al_2}}|q|^{\al_2}
\,\big(jM^{(1-2\aleph)j}\big)^{\al_2}\cr
&\le C\,\fl U\, M^{\al_1j}\sfrac{1}{\fl_j^{\al_2}}|q|^{\al_2}\cr
}$$
To prove (\eqnManaa) when 
$[\al_2]=1$, again apply Lemma \lemLadBmom\ twice to obtain
$$\eqalign{
\Big|\big(\Sh\cdot\partial_{\k}\big)^{\al_1}\big(\Ln\cdot\partial_{\k}\big)
    \big[B_{i,j}(\k+q\Ln)\!-\!B_{i,j}(\k)\big]\Big|
&\le 2\,\const\,\fl U\, j\fl_j\,M^{(\al_1+1)j}\cr
\Big|\big(\Sh\cdot\partial_{\k}\big)^{\al_1}\big(\Ln\cdot\partial_{\k}\big)
    \big[B_{i,j}(\k+q\Ln)\!-\!B_{i,j}(\k)\big]\Big|
&\le |q|\sup_{\p}
\Big|\big(\Sh\cdot\partial_{\p}\big)^{\al_1}\big(\Ln\cdot\partial_{\p}\big)^2
    B_{i,j}(\p)\Big|\cr
&\le \const\,\fl U\, |q|\,j\fl_j\,M^{(\al_1+2)j}\cr
}$$
Multiplying the $(2-\al_2)^{\rm th}$ power of the first bound by the
$(\al_2-1)^{\rm th}$ power of the second gives
$$\eqalign{
\Big|\big(\Sh\cdot\partial_{\k}\big)^{\al_1}\big(\Ln\cdot\partial_{\k}\big)
               \big[B_{i,j}(\k+q\Ln)\!-\!B_{i,j}(\k)\big]\Big|
&\le 2^{2-\al_2}\const\,\fl U\, |q|^{\al_2-1}M^{(\al_1+1)j}\,j\fl_j\,M^{(\al_2-1) j}\cr
&\hskip-1in= 2^{2-\al_2}\const\,\fl U\, M^{\al_1j}\sfrac{1}{\fl_j^{\al_2}}|q|^{\al_2-1}
\,\big(jM^{-(\aleph+\al_2\aleph-\al_2) j}\big)\cr
&\hskip-1in\le C\,\fl U\, M^{\al_1j}\sfrac{1}{\fl_j^{\al_2}}|q|^{\al_2-1}\cr
}$$
since $\aleph+\al_2\aleph-\al_2=\aleph-(1-\aleph)\al_2\ge \aleph-(1-\aleph)a=\aleph-\half>0$.
We also have, for all $\al_1\in\{0,1,2\}$ and $\al_2\in\{0,1\}$,
$$
\Big|\big(\Sh\cdot\partial_{\k}\big)^{\al_1}
\big(\Ln\cdot\partial_{\k}\big)^{\al_2}B_{i,j}(\k)\Big|
\le\const\fl U M^{\al_1j}\sfrac{1}{\fl_j^{\al_2}}
\EQN\eqnManaabis$$
since $j\fl_j\, M^{\al_2j}
=\sfrac{1}{\fl_j^{\al_2}}jM^{-(\aleph+\al_2\aleph-\al_2) j}
\le \const\sfrac{1}{\fl_j^{\al_2}}$ for $\al_2=0,1$.

The next step is to prove that for all $\al_1\in\{0,1\}$ 
and $0\le\al_2\le a$
$$
\Big|\big(\Sh\cdot\partial_{\k}\big)^{\al_1}
\big(\Ln\cdot\partial_{\k}\big)^{[\al_2]}
\big[\rho(\k+q\Ln)B_{i,j}(\k+q\Ln)-\rho(\k)B_{i,j}(\k)\big]\Big|
\le C'\,\fl U\, M^{\al_1j}\sfrac{1}{\fl_j^{\al_2}}|q|^{\al_2-[\al_2]}
\EQN\eqnManbb$$
Applying the hypothesis on $\rho$ twice,
$$\eqalign{
\Big|\big(\Sh\cdot\partial_{\k}\big)^{\al_1}
\big(\Ln\cdot\partial_{\k}\big)^{[\al_2]}
\big[\rho(\k+q\Ln)-\rho(\k)\big]\Big|
&\le 2\,\const M^{\al_1j}\sfrac{1}{\fl_j^{[\al_2]}}\cr
\Big|\big(\Sh\cdot\partial_{\k}\big)^{\al_1}
\big(\Ln\cdot\partial_{\k}\big)^{[\al_2]}
\big[\rho(\k+q\Ln)-\rho(\k)\big]\Big|
&\le \const M^{\al_1j}\sfrac{1}{\fl_j^{[\al_2]+1}}|q|\cr
}$$
Multiplying the $(1+[\al_2]-\al_2)^{\rm th}$ power of the first bound by the
$(\al_2-[\al_2])^{\rm th}$ power of the second gives
$$
\Big|\big(\Sh\cdot\partial_{\k}\big)^{\al_1}
\big(\Ln\cdot\partial_{\k}\big)^{[\al_2]}
\big[\rho(\k+q\Ln)-\rho(\k)\big]\Big|
\le 2^{1+[\al_2]-\al_2}
\const M^{\al_1j}\sfrac{1}{\fl_j^{\al_2}}|q|^{\al_2-[\al_2]}
\EQN\eqnManrho$$
The bound (\eqnManbb) follows from the product rule and
$$\eqalign{
&\Big|\big(\Sh\cdot\partial_{\k}\big)^{\al_1}
\big(\Ln\cdot\partial_{\k}\big)^{[\al_2]}
\big[\rho(\k+q\Ln)B_{i,j}(\k+q\Ln)-\rho(\k)B_{i,j}(\k)\big]\Big|\cr
&\hskip.5in\le\Big|\big(\Sh\cdot\partial_{\k}\big)^{\al_1}
\big(\Ln\cdot\partial_{\k}\big)^{[\al_2]}
\big[\rho(\k+q\Ln)B_{i,j}(\k+q\Ln)-\rho(\k)B_{i,j}(\k+q\Ln)\big]\Big|\cr
&\hskip1in+\Big|\big(\Sh\cdot\partial_{\k}\big)^{\al_1}
\big(\Ln\cdot\partial_{\k}\big)^{[\al_2]}
\big[\rho(\k)B_{i,j}(\k+q\Ln)-\rho(\k)B_{i,j}(\k)\big]\Big|\cr
}$$
by using (\eqnManrho) and (\eqnManaabis) to bound the first line and   
(\eqnManaa) and the hypothesis on the derivatives of $\rho$, 
to bound the second.

The Lemma will follow from
$$
\sup_\x\ \big|\Sh\cdot\x/M^j\big|^{\al_1}
|\fl_j\Ln\cdot\x|^{\al_2}\big|\hat B_{i,j}(\x)\big|\le \const\,\fl U\,\sfrac{\fl_j}{M^{j}}
$$
for all $\al_1\in\{0,\sfrac{3}{2}\}$ and $\al_2\in\{0,\sfrac{1+a}{2}\}$.
This in turn will follow from
$$
\sup_\x\ \big|\Sh\cdot\x/M^j\big|^{\al_1}
|\fl_j\Ln\cdot\x|^{\al_2}\big|\hat B_{i,j}(\x)\big|\le \const\,\fl U\,\sfrac{\fl_j}{M^{j}}
\EQN\eqnMancc$$
for all $\al_1\in\{0,1,2\},\ \al_2\in\{0,1,a\}$, $(\al_1,\al_2)\ne (2,a)$,
by taking various geometric means. In particular, to handle the case 
$(\al_1,\al_2)=\big(\sfrac{3}{2},\sfrac{1+a}{2}\big)$,
take the geometric mean of the bounds with $(\al_1,\al_2)=(1,a)$
and $(\al_1,\al_2)=(2,1)$.
For $\al_2=0,1$, (\eqnMancc) follows, by integration by parts, from
$$
\Big|\big(\Sh\cdot\partial_{\k}\big)^{\al_1}
\big(\Ln\cdot\partial_{\k}\big)^{\al_2}
\big[\rho(\k)B_{i,j}(\k)\big]\Big|
\le \const\,\fl U\, M^{\al_1j}\sfrac{1}{\fl_j^{\al_2}}
$$
and the fact that $\rho(\k)B_{i,j}(\k)$ is supported in a region
of volume $\const \sfrac{\fl_j}{M^{j}}$. Furthermore, if $|\fl_j\Ln\cdot\x|\le
1$,
$$
\big|\Sh\cdot\x/M^j\big|^{\al_1}
|\fl_j\Ln\cdot\x|^{a}\big|\hat B_{i,j}(\x)\big|
\le \big|\Sh\cdot\x/M^j\big|^{\al_1}\big|\hat B_{i,j}(\x)\big|
\le \const\,\fl U\, \sfrac{\fl_j}{M^{j}}
$$
so it suffices to consider $\al_1=0,1$, $\al_2=a$ and $|\Ln\cdot\x|\ge\sfrac{1}{\fl_j}$.
Let $\tilde D_j(\x)$ denote the Fourier transform of 
$$
D_{i,j}(\k)=\sfrac{1}{M^{\al_1j}}\big(\Sh\cdot\partial_{\k}\big)^{\al_1}
\big(\fl_j\Ln\cdot\partial_{\k}\big)\big[\rho(\k)B_{i,j}(\k)\big]
$$
Then
$$\eqalign{
\big|e^{-\imath q\Ln\cdot\x}-1\big|\big|\Sh\cdot\x/M^j\big|^{\al_1}
|\fl_j\Ln\cdot\x|\big|\hat B_{i,j}(\x)\big|
&=\big|e^{-\imath q\Ln\cdot\x}-1\big|\ 
\Big|\int\sfrac{d^2\k}{(2\pi)^2}e^{\imath \k\cdot \x}D_{i,j}(\k)\Big|\cr
&=\Big|\int\sfrac{d^2\k}{(2\pi)^2}\big[e^{\imath (\k-q\Ln)\cdot \x}
        -e^{\imath \k\cdot \x}\big]D_{i,j}(\k)\Big|\cr
&=\Big|\int\sfrac{d^2\k}{(2\pi)^2}
e^{\imath \k\cdot \x}\big[D_{i,j}(\k+q\Ln)-D_{i,j}(\k)\big]\Big|\cr
}$$
By (\eqnManbb)
$$
\big|D_{i,j}(\k+q\Ln)-D_{i,j}(\k)\big|\le C'\,\fl U\,\sfrac{|q|^{a-1}}{\fl_j^{a-1}}
$$
Furthermore, if $|q|<\fl_j$, $D_{i,j}(\k+q\Ln)-D_{i,j}(\k)$ is supported in a region
of volume $\const\sfrac{\fl_j}{M^{j}}$, so that
$$
\big|e^{-\imath q\Ln\cdot\x}-1\big|
\big|\Sh\cdot\x/M^j\big|^{\al_1}
|\fl_j\Ln\cdot\x|\big|\hat B_{i,j}(\x)\big|
\le \const\,\fl U\,\sfrac{\fl_j}{M^{j}}\sfrac{|q|^{a-1}}{\fl_j^{a-1}}
$$
To finish the proof of (\eqnMancc), and the Lemma, it now suffices to choose
$q=\sfrac{1}{10\,\Ln\cdot \x}$ and observe that then 
$\big|e^{-\imath q\Ln\cdot\x}-1\big|=\big|e^{-\imath/10}-1\big|>0$.
\endproof

\lemma{\STM\lemLadABdiff}{  
Let $I$ be an interval of length $\fl$ on the Fermi surface $F$,
$K$ be a compact subset of $\bbbr^2$, 
and $u_j(k,t_0,z)=u_j\big((k_0,\k),t_0,z\big)$ and 
$v_j(k,\t,z)=v_j\big((k_0,\k),\t,z\big)$,
$j>1$, be functions that vanish unless $\k\in K$ and
$\pi_F(\k)\in I$, where $\pi_F$ is 
projection on the Fermi surface. The variable $z$ runs over $\bbbr^n$ for
some $n\in\bbbn$. Let $n_j(\om)$, $j>1$, be functions that
take values in $[0,1]$, vanish in a ($j$--dependent) neighbourhood of $\om=0$,
are supported in a compact set that is independent of $j$ and converge
pointwise as $j\rightarrow\infty$.
Set, for  $j>1$,
$$\eqalign{
A_j(t_0,z)&=
\int dk\ \frac{n_j(e(\k))u_j(k,t_0,z)}
{[\imath k_0-e'(k_0,\k)][\imath (k_0+t_0)-e'(k_0+t_0,\k)]}\cr
B_j(\t,z)&=
\int dk\ \frac{n_j(k_0)v_j(k,\t,z)}
{[\imath k_0-e'(k_0,\k)][\imath k_0-e'(k_0,\k+\t)]}\cr
}$$
and let
$$\eqalign{
U_{\tilde\aleph} &= \sup_{j,k,t_0,z}|u_j(k,t_0,z)|
+\sup_{j,k,t_0,\ka,z}
       \sfrac{|u_j(k-(\ka,\0),t_0,z)-u_j(k,t_0,z)|}{|\ka|^{\tilde\aleph}}\cr
&\hskip1in
+\sup_{j,k,t_0,z}\sfrac{|u_j(k,t_0,z)-u_j(k,0,z)|}{|t_0|^{\tilde\aleph}}
+\sup_{j,k,t_0,z,z'}\sfrac{|u_j(k,t_0,z)-u_j(k,t_0,z')|}{|z-z'|^{\tilde\aleph}}
\cr
V_{\tilde\aleph} &= \sup_{j,k,\t,z}|v_j(k,\t,z)|
+\sup_{j,k_0,\k,\k',\t,z}
   \sfrac{|v_j((k_0,\k'),\t,z)-v_j((k_0,\k),\t,z)|}{|\k-\k'|^{\tilde\aleph}}\cr
&\hskip1in
+\sup_{j,k,\t,z}\sfrac{|v_j(k,\t,z)-v_j(k,\0,z)|}{|\t|^{\tilde\aleph}}
+\sup_{j,k,\t,z,z'}\sfrac{|v_j(k,\t,z)-v_j(k,\t,z')|}{|z-z'|^{\tilde\aleph}}
\cr
}$$
be finite.

\Item a) Let  $0\le\aleph'<\tilde\aleph<\aleph$. Then
$$\eqalign{
\big| A_j(t_0,z)-A_j(0,z)\big|\le 
&\const\fl\,U_{\tilde\aleph}\,|t_0|^{\aleph'}\cr
\big| A_j(t_0,z)-A_j(t_0,z')\big|\le 
&\const\fl\,U_{\tilde\aleph}\,|z-z'|^{\aleph'}\cr
}$$
for all $t_0$ in a ($j$--independent) neighbourhood of the origin.

\Item b) If, in addition, $u_j(k,t_0,z)$ converges pointwise as 
$j\rightarrow\infty$, the limit 
$A(t_0,z)=\lim\limits_{j\rightarrow\infty}A_j(t_0,z)$ 
exists for $t_0$ in a neighbourhood of the origin and obeys
$$\eqalign{
\big| A(t_0,z)-A(0,z)\big|\le 
&\const\fl\,U_{\tilde\aleph}\,|t_0|^{\aleph'}\cr
\big| A(t_0,z)-A(t_0,z')\big|\le 
&\const\fl\,U_{\tilde\aleph}\,|z-z'|^{\aleph'}\cr
}$$
for all $t_0$ in a neighbourhood of the origin.

\Item c) Let $0\le\aleph'<\tilde\aleph<\aleph$. Then
$$\eqalign{
\big| B_j(\t,z)-B_j(\0,z)\big|\le 
&\const\fl\,V_{\tilde\aleph}\,|\t|^{\aleph'}\cr
\big| B_j(\t,z)-B_j(\t,z')\big|\le 
&\const\fl\,V_{\tilde\aleph}\,|z-z'|^{\aleph'}\cr
}$$
for all $\t$ in a ($j$--independent) neighbourhood of the origin.

\Item d) If, in addition, $v_j(k,\t,z)$ converges pointwise as 
$j\rightarrow\infty$, the limit 
$B(\t,z)=\lim\limits_{j\rightarrow\infty}B_j(\t,z)$ 
exists for $\t$ in a neighbourhood of the origin and obeys
$$\eqalign{
\big| B(\t,z)-B(\0,z)\big|\le 
&\const\fl\,V_{\tilde\aleph}\,|\t|^{\aleph'}\cr
\big| B(\t,z)-B(\t,z')\big|\le 
&\const\fl\,V_{\tilde\aleph}\,|z-z'|^{\aleph'}\cr
}$$

}
\prf The proofs of parts (a) and (b) are very similar to the proofs of
parts (c) and (d) respectively. So we only give the latter.

\Item c) 
As in Lemma \lemLadBmom,
$$\eqalign{
B_j(\t,z)
&=\int dk\int_0^1ds\ 
\frac{n_j(k_0)v_j(k,\t,z)}{[\imath w(\k,\t,s) k_0- E(\k,\t,s)
- \tilde E(k_0,\k,\t,s)]^2}\cr
}$$
where
$$\eqalign{
E'(k_0,\k,\t,s)&=s e'(k_0,\k)+(1-s)e'(k_0,\k+\t)\cr
E(\k,\t,s)&=E'(0,\k,\t,s)=s e(\k)+(1-s)e(\k+\t)\cr
w(\k,\t,s)&=1-\sfrac{1}{i}\sfrac{\partial E'}{\partial k_0}(0,\k,\t,s)\cr
\tilde E(k_0,\k,\t,s)&=E'(k_0,\k,\t,s)-E'(0,\k,\t,s)
-k_0\sfrac{\partial E'}{\partial k_0}(0,\k,\t,s)\cr
}$$
Set
$$
B'_j(\t,z)
=\int dk\int_0^1ds\ 
\frac{n_j(k_0)v_j(k,\t,z)}{[\imath w(\k,\t,s) k_0- E(\k,\t,s)]^2}
$$
and
$$\eqalign{
I(k_0,\k,\t,s)&=\sfrac{1}{[\imath w(\k,\t,s) k_0- E(\k,\t,s)
- \tilde E(k_0,\k,\t,s)]^2}
-\sfrac{1}{[\imath w(\k,\t,s) k_0- E(\k,\t,s)]^2}\cr
&=\sfrac{
[\imath w(\k,\t,s) k_0- E(\k,\t,s)]^2
       -[\imath w(\k,\t,s) k_0- E(\k,\t,s)- \tilde E(k_0,\k,\t,s)]^2}
{[\imath w(\k,\t,s) k_0- E(\k,\t,s)]^2
           [\imath w(\k,\t,s) k_0- E(\k,\t,s)- \tilde E(k_0,\k,\t,s)]^2}\cr
&=\sfrac{ \tilde E(k_0,\k,\t,s)[2\imath w(\k,\t,s) k_0- 2E(\k,\t,s)- \tilde E(k_0,\k,\t,s)]}
{[\imath w(\k,\t,s) k_0- E(\k,\t,s)]^2
           [\imath w(\k,\t,s) k_0- E(\k,\t,s)- \tilde E(k_0,\k,\t,s)]^2}\cr
}$$
so that
$$\eqalign{
B_j(\t,z)-B'_j(\t,z)
=\int dk\int_0^1ds\ n_j(k_0)v_j(k,\t,z)I(k_0,\k,\t,s)
}\EQN\eqnLADBminusBp$$
By (\eqnMante) and Lemma \lemLADeprime, using
$\tilde E(k_0,\k,\t,s)
=\int_0^{k_0}\!d\ka\, \big[
\sfrac{\partial E'}{\partial k_0}(\ka,\k,\t,s)
-\sfrac{\partial E'}{\partial k_0}(0,\k,\t,s)\big]$,
$$\meqalign{
|\tilde E(k_0,\k,\t,s)|&\le\const\rho\, |k_0|^{1+\aleph} &&
|\tilde E(k_0,\k,\t,s)-\tilde E(k_0,\k,\0,s)|
&\le\const |k_0|\,|\t|^\aleph\cr
|w(\k,\t,s)-1|&\le\rho  &&
|w(\k,\t,s)-w(\k,\0,s)|
&\le\const\rho\,|\t|^\aleph\cr
& &&
|E(\k,\t,s)-E(\k,\0,s)|
&\le\const|\t|\cr
}$$
Consequently
$$
|I(k_0,\k,\t,s)|\le \const\sfrac{|k_0|^{1+\aleph}|\imath k_0-E(\k,\t,s)|}
{|\imath k_0-E(\k,\t,s)|^4}
\le \const\sfrac{|k_0|^{\aleph}}{|\imath k_0-E(\k,\t,s)|^2}
\EQN\eqnLADIbnd$$
and, we claim, for $0\le\aleph'\le\aleph$,
$$\eqalign{
|I(k_0,\k,\t,s)-I(k_0,\k,\0,s)|
&\le \const|k_0|^{\aleph-\aleph'}|\t|^{\aleph'}
\Big[\sfrac{1}{|\imath k_0-E(\k,\0,s)|^2}+\sfrac{1}{|\imath k_0-E(\k,\t,s)|^2}\Big]\cr
}$$
We prove the last bound in the case $|E(\k,\t,s)|\le |E(\k,\0,s)|$. The 
other case is similar. Set
$$\meqalign{
c&=\imath w(\k,\t,s) k_0- E(\k,\t,s) &&
C&=\imath w(\k,\0,s) k_0- E(\k,\0,s) \cr
d&=\imath w(\k,\t,s) k_0- E(\k,\t,s)- \tilde E(k_0,\k,\t,s) &&
D&=\imath w(\k,\0,s) k_0- E(\k,\0,s) - \tilde E(k_0,\k,\0,s)\cr
a&=\tilde E(k_0,\k,\t,s) &&
A&=\tilde E(k_0,\k,\0,s) \cr
b&=c+d &&
B&=C+D\cr
}$$
Then, for $|k_0|,|\t|\le\const$,
$$\eqalign{
|a|,|A|&\le\const |k_0|^{1+\aleph}\cr
|b|,|c|,|d|&\le \const |\imath k_0-E(\k,\t,s)|\le \const |\imath k_0-E(\k,\0,s)|\cr
|c|,|d|&\ge \const |\imath k_0-E(\k,\t,s)|\cr
|B|,|C|,|D|&\le \const |\imath k_0-E(\k,\0,s)|\cr
|C|,|D|&\ge \const |\imath k_0-E(\k,\0,s)|\cr
|a-A|&\le \const\min\big\{|k_0|^{1+\aleph},|k_0||\t|^{\aleph}\big\}
\le \const|k_0|^{1+\aleph-\aleph'}|\t|^{\aleph'}\cr
|c-C|, |d-D|, |b-B|&\le \const|k_0||\t|^{\aleph}+\const\min\{|E(\k,\0,s)|+|E(\k,\t,s)|,|\t|\}\cr
       &\le\const |\t|^{\aleph'}|\imath k_0-E(\k,\0,s)|^{1-\aleph'}\cr
}$$
Applying these to
$$\eqalign{
\sfrac{AB}{C^2D^2}-\sfrac{ab}{c^2d^2}
&=\sfrac{(A-a)B}{C^2D^2}+\sfrac{a(B-b)}{C^2D^2}
+\sfrac{ab}{c^2d^2}\sfrac{c^2d^2-C^2D^2}{C^2D^2}\cr
&=\sfrac{(A-a)B}{C^2D^2}+\sfrac{a(B-b)}{C^2D^2}
+\sfrac{ab}{c^2d^2}\sfrac{c^2(d-D)(d+D)}{C^2D^2}
+\sfrac{ab}{c^2d^2}\sfrac{(c-C)(c+C)}{C^2}\cr
}$$
gives
$$\eqalign{
|I(k_0,\k,\t,s)-I(k_0,\k,\0,s)|
&\le \const\sfrac{|k_0|^{\aleph-\aleph'}|\t|^{\aleph'}}{|\imath k_0-E(\k,\0,s)|^2}
+\const\sfrac{|k_0|^\aleph}{|\imath k_0-E(\k,\t,s)|^2}
\sfrac{|\t|^{\aleph'}}{|\imath k_0-E(\k,\0,s)|^{\aleph'}}\cr
&\le \const|k_0|^{\aleph-\aleph'}|\t|^{\aleph'}
\Big[\sfrac{1}{|\imath k_0-E(\k,\0,s)|^2}+\sfrac{1}{|\imath k_0-E(\k,\t,s)|^2}\Big]\cr
}$$
which is the desired bound.

Using these bounds gives, for $0\le\aleph'<\tilde \aleph<\aleph$,
$$\eqalign{
&\big|B_j(\t,z)-B'_j(\t,z)-B_j(\0,z)+B'_j(\0,z)\big|\cr
&\le\int\! dk\int_0^1\!\!ds\  n_j(k_0)
        \Big|v_j(k,\t,z)I(k_0,\k,\t,s)-v_j(k,\0,z)I(k_0,\k,\0,s)|\Big|\cr
&\le\const\!\!\!\int\! dk\!\int_0^1\!\!\!ds\,
n_j(k_0)\Big[
\sfrac{|v_j(k,\t,z)-v_j(k,\0,z)||k_0|^\aleph}{|\imath  k_0- E(\k,\t,s)|^2}
\!+\!\sfrac{|v_j(k,\0,z)||k_0|^{\aleph-\aleph'}|\t|^{\aleph'}}{|\imath k_0-E(\k,\0,s)|^2}
\!+\!\sfrac{|v_j(k,\0,z)||k_0|^{\aleph-\aleph'}|\t|^{\aleph'}}{|\imath k_0-E(\k,\t,s)|^2}
\Big]
\cr
&\le\const\int_I d\th
\int_{-\const}^{\const}\!\!\! dk_0\int_{-\const}^{\const} dE\ 
\Big[\sfrac{|k_0|^\aleph\sup_{k}|v_j(k,\t,z)-v_j(k,\0,z)|}{|\imath  k_0- E|^2}
+\sfrac{|k_0|^{\aleph-\aleph'}|\t|^{\aleph'}\sup_{k}|v_j(k,\0,z)|}
{|\imath  k_0- E|^2}
\Big]\cr
&\le\const\fl\,\Big[\sup_{k}|v_j(k,\t,z)-v_j(k,\0,z)|
+|\t|^{\aleph'}\sup_{k}|v_j(k,\0,z)|\Big]\cr
&\le\const\fl\,V_{\tilde\aleph}\,|\t|^{\aleph'}\cr
}$$
and
$$\eqalign{
&\big|B_j(\t,z)-B'_j(\t,z)-B_j(\t,z')+B'_j(\t,z')\big|\cr
&\hskip.5in\le\int\! dk\int_0^1\!\!ds\  n_j(k_0)
        \Big|v_j(k,\t,z)I(k_0,\k,\t,s)-v_j(k,\t,z')I(k_0,\k,\t,s)|\Big|\cr
&\hskip.5in\le\const\int dk\int_0^1ds\ n_j(k_0)
\sfrac{|v_j(k,\t,z)-v_j(k,\t,z')||k_0|^\aleph}{|\imath  k_0- E(\k,\t,s)|^2}
\cr
&\hskip.5in\le\const\int_I d\th
\int_{-\const}^{\const} dk_0\int_{-\const}^{\const} dE\ 
\sfrac{|k_0|^\aleph\sup_{k}|v_j(k,\t,z)-v_j(k,\t,z')|}{|\imath  k_0- E|^2}\cr
&\hskip.5in\le\const\fl\,\sup_{k}|v_j(k,\t,z)-v_j(k,\t,z')|\cr
&\hskip.5in\le\const\fl\,V_{\tilde\aleph}\,|z-z'|^{\aleph'}\cr
}$$
Hence it suffices to consider $B'_j(\t,z)$.

Making, as in Lemma \lemLadBmom, for each fixed $s$, the change of variables 
from $\k$ to $E=E(\k,\t,s)$ and  an ``angular'' variable $\th$
$$\eqalign{
B'_j(\t,z)
&=\int_0^1ds\int d\th
\int dk_0 dE\ n_j(k_0)
\frac{v_j\big((k_0,\k(E,\t,\th,s)),\t,z\big)J(E,\t,\th,s)}
{[\imath w(\k(E,\t,\th,s),\t,s)k_0-E]^2}\cr
}$$
Set
$$\eqalign{
B''_j(\t,z)
&=\int_0^1ds\int d\th
\int dk_0 dE\ n_j(k_0)
\frac{v_j\big((k_0,\k(0,\t,\th,s)),\t,z\big)J(0,\t,\th,s)}
{[\imath w(\k(0,\t,\th,s),\t,s)k_0-E]^2}\cr
}$$

Using
$$
\partial_E\k_\ell(E,\t,\th,s)=\sfrac
{\partial_{\k_2}\th(\k)\de_{\ell,1}
-\partial_{\k_1}\th(\k)\de_{\ell,2}}
{\partial_{\k_1}E(\k,\t,s)\,\partial_{\k_2}\th(\k)
-\partial_{\k_2}E(\k,\t,s)\,\partial_{\k_1}\th(\k)}
\Big|_{\k=\k(E,\t,\th,s)}
$$
one proves, by induction on $n$, that, for $n\le r$,
$$
\big|\partial_E^n\k(E,\t,\th,s)\big|
\le\abcst''
\EQN\eqnMEa$$
Using this bound and the observation that the Jacobian
$$
J(E,\t,\th,s)=\sfrac{1}
{|\partial_{\k_1}E(\k,\t,s)\,\partial_{\k_2}\th(\k)
-\partial_{\k_2}E(\k,\t,s)\,\partial_{\k_1}\th(\k)|}
\Big|_{\k=\k(E,\t,\th,s)}
$$
one proves that, for $n\le r-1$,
$$
\big|\partial_E^n J(E,\t,\th,s)\big|
\le\abcst''
\EQN\eqnMEb$$
Since 
$$\eqalign{
|w(\k,\t,s)-w(\k',\t,s)|
&\le s\big|\sfrac{\partial e'}{\partial k_0}(0,\k)
-\sfrac{\partial e'}{\partial k_0}(0,\k')\big|
+(1-s)\big|\sfrac{\partial e'}{\partial k_0}(0,\k+\t)
-\sfrac{\partial e'}{\partial k_0}(0,\k'+\t)\big|\cr
&\le\const|\k-\k'|^\aleph
}\EQN\eqnMEc$$ 
(\eqnMEa), (\eqnMEb) and the fact, from Lemma \lemLADeprime.i, 
that $|w(\k,\t,s)-1|\le\rho\le\half$ imply that
$$\eqalign{
&\Big|\sfrac{v_j((k_0,\k(E,\t,\th,s)),\t,z)J(E,\t,\th,s)}
{[\imath w(\k(E,\t,\th,s),\t,s)k_0-E]^2}
-\sfrac{v_j((k_0,\k(0,\t,\th,s)),\t,z)J(0,\t,\th,s)}
{[\imath w(\k(0,\t,\th,s),\t,s)k_0-E]^2}\Big|\cr
&\hskip0.05in\le \const\big|v_j\big((k_0,\k(E,\t,\th,s)),\t,z\big)
            \!-\!v_j\big((k_0,\k(0,\t,\th,s)),\t,z\big)\big|\sfrac{1}{k_0^2+E^2}\cr
&\hskip1in+\const\sup_{k,\t}\big|v_j(k,\t,z)\big|\sfrac{|E|}{k_0^2+E^2}
 + \const\sup_{k,\t}\big|v_j(k,\t,z)\big|
\sfrac{[|k_0|+|E|]|k_0||E|^\aleph}{[k_0^2+E^2]^2}\cr
&\hskip0.05in\le \const
\Big[\sup_{\k,\k'}\sfrac{\big|v_j((k_0,\k'),\t,z)-v_j((k_0,\k),\t,z)\big|}
        {|\k-\k'|^{\tilde\aleph}}
+\sup_{k,\t}\big|v_j(k,\t,z)\big|\Big]\sfrac{|E|^{\tilde\aleph}}{k_0^2+E^2}
\le\const V_{\tilde\aleph} \sfrac{|E|^{\tilde\aleph}}{k_0^2+E^2}\cr
}$$
for all small $E$. For $E$ bounded away from zero
$$\eqalign{
\Big|\sfrac{v_j((k_0,\k(E,\t,\th,s)),\t,z)J(E,\t,\th,s)}
{[\imath w(\k(E,\t,\th,s),\t,s)k_0-E]^2}
-\sfrac{v_j((k_0,\k(0,\t,\th,s)),\t,z)J(0,\t,\th,s)}
{[\imath w(\k(0,\t,\th,s),\t,s)k_0-E]^2}\Big|
\le \const
\sup_{k,\t}\big|v_j(k,\t,z)\big|\sfrac{1}{k_0^2+E^2}
}$$
so 
$$\eqalign{
\Big|\sfrac{v_j((k_0,\k(E,\t,\th,s)),\t,z)J(E,\t,\th,s)}
{[\imath w(\k(E,\t,\th,s),\t,s)k_0-E]^2}
-\sfrac{v_j((k_0,\k(0,\t,\th,s)),\t,z)J(0,\t,\th,s)}
{[\imath w(\k(0,\t,\th,s),\t,s)k_0-E]^2}\Big|
\le \const V_{\tilde\aleph}\sfrac{\min\{|E|^{\tilde\aleph},1\}}{k_0^2+E^2}
}\EQN\eqnMEd$$
Since 
$$
|w(\k,\t,s)-w(\k,\0,s)|=|1-s|\big|\sfrac{\partial e'}{\partial k_0}(0,\k+\t)
-\sfrac{\partial e'}{\partial k_0}(0,\k)\big|\le\const|\t|^\aleph
$$ 
(\eqnMEc), (\eqnMand) and (\eqnMane) imply that
$$\eqalign{
&\Big|\sfrac{v_j((k_0,\k(E,\t,\th,s)),\t,z)J(E,\t,\th,s)}
{[\imath w(\k(E,\t,\th,s),\t,s)k_0-E]^2}
-\sfrac{v_j((k_0,\k(E,\0,\th,s)),\0,z)J(E,\0,\th,s)}
{[\imath w(\k(E,\0,\th,s),\0,s)k_0-E]^2}\Big|\cr
&\hskip0.5in\le \const\big|v_j((k_0,\k(E,\t,\th,s)),\t,z)-v_j((k_0,\k(E,\0,\th,s)),\0,z)\big|
                \sfrac{1}{k_0^2+E^2}\cr
&\hskip1in +\const|\t|\sup_k\big|v_j(k,\0,z)\big|\sfrac{1}{k_0^2+E^2}
          + \const|\t|^\aleph
\sup_k\big|v_j(k,\0,z)\big|\sfrac{|k_0|[|k_0|+|E|]}{[k_0^2+E^2]^2}\cr
&\hskip0.5in\le \const
\Big[\hskip-3pt\sup_{\k,\k'\atop|\k-\k'|\le\const|\t|}\hskip-12pt
          \big|v_j\big((k_0,\k'),\t,z\big)-v_j\big((k_0,\k),\0,z\big)\big|
+|\t|^\aleph\sup_\k\big|v_j(k,\0,z)\big|\Big]\sfrac{1}{k_0^2+E^2}\cr
&\hskip0.5in\le \const V_{\tilde\aleph} |\t|^{\tilde\aleph}\sfrac{1}{k_0^2+E^2}\cr
}\EQN\eqnMEe$$
and that
$$\eqalign{
&\Big|\sfrac{v_j((k_0,\k(0,\t,\th,s)),\t,z)J(0,\t,\th,s)}
{[\imath w(\k(0,\t,\th,s),\t,s)k_0-E]^2}
-\sfrac{v_j((k_0,\k(0,\0,\th,s)),\0,z)J(0,\0,\th,s)}
{[\imath w(\k(0,\0,\th,s),\0,s)k_0-E]^2}\Big|\cr
&\hskip0.5in\le \const\big|v_j((k_0,\k(0,\t,\th,s)),\t,z)-v_j((k_0,\k(0,\0,\th,s)),\0,z)\big|
                \sfrac{1}{k_0^2+E^2}\cr
&\hskip1in +\const|\t|\sup_k\big|v_j(k,\0,z)\big|\sfrac{1}{k_0^2+E^2}
          + \const|\t|^\aleph
\sup_k\big|v_j(k,\0,z)\big|\sfrac{|k_0|[|k_0|+|E|]}{[k_0^2+E^2]^2}\cr
&\hskip0.5in\le \const V_{\tilde\aleph} |\t|^{\tilde\aleph}\sfrac{1}{k_0^2+E^2}\cr
}\EQN\eqnMEf$$
Combining, (\eqnMEd), a second copy of (\eqnMEd) with $\t=\0$,
(\eqnMEe) and (\eqnMEf), gives, for all $0\le\aleph'<\tilde\aleph<\aleph$,
$$\eqalign{
&\Big|\sfrac{v_j((k_0,\k(E,\t',\th,s)),\t',z)J(E,\t',\th,s)}
{[\imath w(\k(E,\t',\th,s),\t',s)k_0-E]^2}\big|_{\t'=\0}^{\t'=\t}
-\sfrac{v_j((k_0,\k(0,\t',\th,s)),\t',z)J(0,\t',\th,s)}
{[\imath w(\k(0,\t',\th,s),\t',s)k_0-E]^2}\big|_{\t'=\0}^{\t'=\t}
\Big|\cr
&\hskip0.5in\le \const V_{\tilde\aleph} |\t|^{\aleph'}
           \sfrac{\min\{|E|^{\tilde\aleph-\aleph'},1\}}{k_0^2+E^2}\cr
}$$
and 
$$\eqalign{
&\big|B'_j(\t,z)-B''_j(\t,z)-B'_j(\0,z)+B''_j(\0,z)\big|\cr
&\hskip.5in\le\const V_{\tilde\aleph} |\t|^{\aleph'}\int_I d\th
\int_{-\const}^{\const} dk_0\int dE\ 
\sfrac{\min\{|E|^{\tilde\aleph-\aleph'},1\}}{k_0^2+E^2}\cr
&\hskip.5in\le\const\fl\,V_{\tilde\aleph}\,|\t|^{\aleph'}\cr
}$$
Similarly, combining, (\eqnMEd), a second copy of (\eqnMEd) with 
$z\rightarrow z'$ and 
$$\eqalign{
&\Big|\sfrac{v_j((k_0,\k(E',\t,\th,s)),\t,z)J(E',\t,\th,s)}
{[\imath w(\k(E',\t,\th,s),\t,s)k_0-E]^2}
-\sfrac{v_j((k_0,\k(E',\t,\th,s)),\t,z')J(E',\t,\th,s)}
{[\imath w(\k(E',\t,\th,s),\t,s)k_0-E]^2}\Big|
\le \const V_{\tilde\aleph} |z-z'|^{\tilde\aleph}\sfrac{1}{k_0^2+E^2}\cr
}$$
gives, for all $0\le\aleph'<\tilde\aleph<\aleph$,
$$\eqalign{
&\Big|\sfrac{v_j((k_0,\k(E,\t,\th,s)),\t,\tilde z)J(E,\t,\th,s)}
{[\imath w(\k(E,\t,\th,s),\t,s)k_0-E]^2}\big|_{\tilde z=z'}^{\tilde z=z}
-\sfrac{v_j((k_0,\k(0,\t,\th,s)),\t,\tilde z)J(0,\t,\th,s)}
{[\imath w(\k(0,\t,\th,s),\t,s)k_0-E]^2}\big|_{\tilde z=z'}^{\tilde z=z}
\Big|\cr
&\hskip0.5in\le \const V_{\tilde\aleph} |z-z'|^{\aleph'}
           \sfrac{\min\{|E|^{\tilde\aleph-\aleph'},1\}}{k_0^2+E^2}\cr
}$$
and 
$$\eqalign{
&\big|B'_j(\t,z)-B''_j(\t,z)-B'_j(\t,z')+B''_j(\t,z')\big|\cr
&\hskip.5in\le\const V_{\tilde\aleph} |z-z'|^{\aleph'}\int_I d\th
\int_{-\const}^{\const} dk_0\int dE\ 
\sfrac{\min\{|E|^{\tilde\aleph-\aleph'},1\}}{k_0^2+E^2}\cr
&\hskip.5in\le\const\fl\,V_{\tilde\aleph}\,|z-z'|^{\aleph'}\cr
}$$
Fix $s$, $\t$ and $\th$ and write $w$ for $w(\k(0,\t,\th,s),\t,s)$. Then,
for all $k_0\ne 0$,
$$
\int_{-\infty}^\infty dE\ \sfrac{1}{[\imath w k_0-E]^2}
=\int_{-\infty}^\infty dE\ \sfrac{d\hfill}{dE}
\sfrac{1}{\imath w k_0-E}=0
$$
Thus $B''_j(\t,z)\equiv 0$.

\Item d)
By part (a), it suffices to prove that $B_j(\t,z)$ converges 
as $j\rightarrow\infty$. By (\eqnLADIbnd), $\sup_j\big|v_j(k,\t,z)I(k_0,\k,\t,s)\big|$
is locally $L^1$ in $k$ and $s$. Hence, by the Lebesgue dominated convergence
theorem applied to the integral in (\eqnLADBminusBp), 
$B_j(\t,z)-B'_j(\t,z)$ converges as $j\rightarrow\infty$.
So it suffices to prove that $B'_j(\t,z)$ converges.
By (\eqnMEd), the Lebesgue dominated convergence theorem also implies that
$B'_j(\t,z)-B''_j(\t,z)$ converges as $j\rightarrow\infty$.
We have already observed that  $B''_j(\t,z)\equiv 0$.

\endproof

The function $A(t_0,z)$ of Lemma \lemLadABdiff.b was constructed in such
a way that the cutoff in the $k_0$ direction was removed first (it does not even
appear in the definition of $A_j(t_0,z)$) and the cutoff in the $e(\k)$
direction was removed second (in the limit $j\rightarrow\infty$).
On the other hand, for the function $B(\t,z)$ of Lemma \lemLadABdiff.d,
the cutoff in the $e(\k)$ direction was removed before the cutoff in the $k_0$
direction. The following Lemma illustrates, in a simplified setting,
that the order of removal of the two cutoffs matters when $t=0$.
\lemma{\STM\lemLadBmodbub}{
Let, for $0<a,b<1$ and $w>0$,
$$
B^{(1)}_{a,b}
=\int_{a<|k_0|\le 1}dk_0\int_{b<|E|\le 1}dE\ \sfrac{1}{[\imath wk_0-E]^2}
$$
Then $B^{(1)}_{a,b}$ is bounded uniformly on $w\ge \ep>0$, $0<a,b<1$ and
$$
\lim_{a\rightarrow 0}\lim_{b\rightarrow 0}B^{(1)}_{a,b}
=-\sfrac{4}{w}\tan^{-1}w\qquad\qquad
\lim_{b\rightarrow 0}\lim_{a\rightarrow 0}B^{(1)}_{a,b}
=-\sfrac{4}{w}\big[\tan^{-1}w-\sfrac{\pi}{2}\big]
$$
}
\prf
$$\meqalign{
B^{(1)}_{a,b}
&=\int_{a<|k_0|\le 1}dk_0\ 
\Big[\sfrac{1}{\imath wk_0-E}\big|_b^1
     +\sfrac{1}{\imath wk_0-E}\big|_{-1}^{-b}\Big]
&=-2\int_{a<|k_0|\le 1} dk_0\ \Big[\sfrac{1}{1+w^2k_0^2}-
\sfrac{b}{b^2+w^2k_0^2}\Big]\cr
&=-4\int_a^1 dk_0\ \Big[\sfrac{1}{1+w^2k_0^2}-
\sfrac{b}{b^2+w^2k_0^2}\Big]
&=-\sfrac{4}{w}\int_{wa}^w dx\ \sfrac{1}{1+x^2}
+\sfrac{4}{w}\int_{wa}^w dx\ \sfrac{b}{b^2+x^2}\cr
&=-\sfrac{4}{w}\int_{wa}^w dx\ \sfrac{1}{1+x^2}
+\sfrac{4}{w}\int_{wa/b}^{w/b} dy\ \sfrac{1}{1+y^2}\cr
&=-\sfrac{4}{w}\Big[\tan^{-1}w-\tan^{-1}wa-\tan^{-1}\sfrac{w}{b}
+\tan^{-1}\sfrac{wa}{b}\Big]\hidewidth\cr
}$$

\endproof

\vfill\eject

\appendix{\APbubsec}{Sector  Counting with Specified}\PG\pgLADC
\null\vskip-.6in
\centerline{\tafontt Transfer Momentum}
\vskip 1cm
As pointed out in the introduction to \S\CHbubbles, we are interested in 
translates $F+\t=\set{\k+\t}{\k\in F}$ of the Fermi surface $F$, and in 
particular in the distances from points of $F+\t$ to $F$.

\lemma{\STM\lemAO}{ There are constants $\de,\abcst>0$ that depend only 
on the Fermi curve $F$ such that the following holds:
Let $\p\in F$,  $|\t| \le \de$ such that $\p-\t\in F$. Denote by $U$ the
disc of radius $\de$ around $\p$. Then for any  $\k \in F\cap U$. 

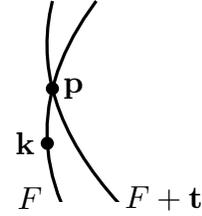
$$
|\k-\p| \le \sfrac{\abcst}{|\t|} {\rm dist}\big( \k, F+\t\big)\qquad\qquad
\smash{\figplace{distF6}{0 in}{-0.35 in}}
$$
\vskip.05in
}
\prf
If ${\rm dist}\big( \k, F+\t\big) \ge \half |\t|$ or if $\k=\p$ 
there is nothing to prove. So assume that ${\rm dist}\big( \k, F+\t\big) \le \half |\t|$. 
If $\de$ was chosen small enough, the angle between the chords
$-\t=(\p-\t)-\p$ and $\k-\p$ of $F$ is sufficiently small. In particular, $(\k-\p)\cdot \t \ne 0$.

\noindent
{\it Case 1:} $(\k-\p)\cdot \t >0$

\noindent We may assume without loss of generality that $\p=(0,0)$, 
that $\t=(0,t_2)$ with $t_2>0$ and that the tangent direction of $F$ 
at $\p$ is $(\al,1)$ with some $\al >0$.
As $F$ is strictly convex, and both $\p=(0,0)$ and $\t=(0,t_2)$ are on $F$, 
$\,F'=\set{\k'\in F\cap U}{(\k'-\p)\cdot \t \ge 0}\,$ 
is contained in the first quadrant, if $\de$ was chosen small enough. 
By the implicit function theorem, $F'$ can be parametrized in the form 
$\,F'=\set{\big(x,\,y(x)\big)}{0\le x<\abcst}\,$ 
with an $r_e+3$ times differentiable function satisfying $y'(0)=\sfrac{1}{\al}$,
$y''<0$.

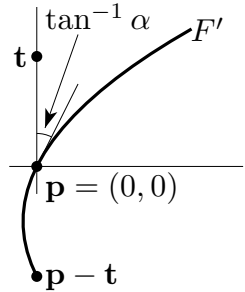
\centerline{\figput{distF1}}

\noindent
Since the curvature of $F$ is bounded above and below, there are constants
$\abcst_1, \abcst_2>0$ such that
$$
\abcst_1 |\t| \le \al \le \abcst_2 |\t| 
$$
If $\de$ was chosen small enough, $y'>1$.
Let $c_1$ resp. $c_2$ be the maximal resp. minimal curvature of $F$, and let 
$C_1$ resp. $C_2$ be the circles of curvature $c_1$ resp. $c_2$ that are tangent to $F$ at $\p$ and curved in the same direction  as $F$ at $\p$.
Then $F'$ lies between $C_1$ and $C_2$, and the slope of $F'$ at a point
$(x,\,y(x))$ lies between the slopes of $C_1$ resp. $C_2$ at the points with the same $x$--coordinate in the first quadrant. 

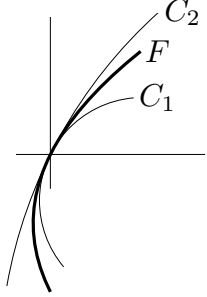
\centerline{\figput{distF2}}

\noindent
If $C$ is a circle of a radius $r>0$ that is tangent to $F$ at $\p$ and curved in the same direction as $F$ at $\p$ then the slope of $C$ at any of its points $(x,y)$ in the first quadrant is equal to 
$$
\frac{ \sfrac{r}{ \sqrt{1+\al^2} }-x }{ y+\sfrac{\al r}{ \sqrt{1+\al^2} } }
$$
Therefore, for any point $\big(x,\,y(x)\big)$ of $F'$
$$
y'(x)
 \le \frac{\sfrac{1}{ c_2 \sqrt{1+\al^2} }-x }
{ y(x)+\sfrac{\al }{ c_2 \sqrt{1+\al^2} } }
\le  \frac{ \abcst_3 }{ y(x)+\abcst_4|\t|}
$$

Let $F''$ be the union of $\t+ F'$ and the segment joining $\p=(0,0)$ to $\t$.

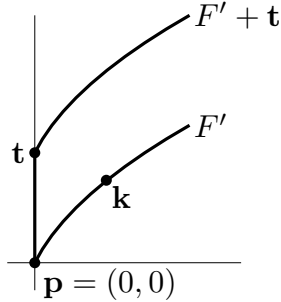
\centerline{\figput{distF3}}

\noindent
Then
$$
{\rm dist}(\k,F+\t) \ge {\rm dist}(\k,F'') 
  \ge \min\big\{ k_1, {\rm dist}(\k,F'+\t) \big\}
$$
As $k_1 \ge \al k_2 \ge \abcst |\t|\,|\k|$, we get that
$$
\sfrac{\abcst}{|\t|}{\rm dist}(\k,F+\t) 
  \ge \min\big\{ |\k-\p|, \sfrac{1}{|\t|}{\rm dist}(\k,F'+\t) \big\}
\EQN\eqnAOI$$

If $ k_2 \le  |\t|$ then the distance from $\k$ to $F'+\t$ is larger
than the distance from $\k$ to the ray through $\t$ in the direction $(1,1)$,
since $y'>1$. 

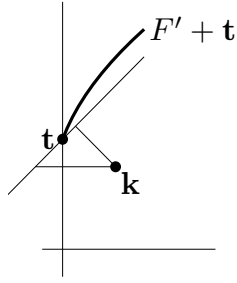
\centerline{\figput{distF4}}

\noindent
Consequently
$$
{\rm dist}(\k,F'+\t) \ge\sfrac{1}{\sqrt{2}}k_1 \ge  \abcst  |\t|\,|\k-\p|
$$
Together with (\eqnAOI) this gives the claim of the Lemma in the situation that 
$ k_2 \le  |\t|$ .
 
Now assume that $ k_2 \ge  |\t|$. Let $\k'=(k_1',k_2-|\t|)$ be the point of $F'$
with $y$--coordinate $k'_2=k_2-|\t|$ that lies to the left of $\k$, i.e. $k_1'<k_1$.

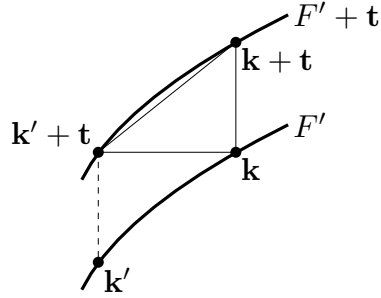
\centerline{\figput{distF5}}

\noindent
By the convexity of $F$, the distance of $\k$ to $\t+F$ is bounded below by the distance of $\k$ to the line segment joining $\k'+\t$ and $\k+\t$. Thus
$$
{\rm dist}(\k,F'+\t) \ge \sfrac{1}{\sqrt{2}}  \min\{ |\t|, k_1-k_1' \}
$$ 
Since $F'$ is strictly convex
$$\eqalign{
k_1-k_1' \ge \sfrac{|\t|}{y'(k_1')} 
         \ge \abcst_5 |\t| (y(k_1') + \abcst_4 |\t|)
          =  \abcst_5 |\t| (k_2' + \abcst_4 |\t|) 
}$$
If $k_2 \le 2|\t|$ then $|\t|\ge \abcst\,|\k|=\abcst\,|\k-\p|$ and 
$$
k_1-k_1' \ge \abcst |\t|^2\ge \abcst  |\t|\,|\k-\p|
$$
and if $k_2 \ge 2|\t|$
$$
k_1-k_1' \ge \abcst\,|\t|\,k'_2 \ge \abcst\,|\t|\,k_2  
\ge \abcst\,|\t|\,|\k|=\abcst\,|\t|\,|\k-\p|
$$
Therefore
$$
\sfrac{1}{|\t|} {\rm dist}(\k,F'+\t) \ge \abcst \min\{1, |\k-\p|\}
\ge \abcst  |\k-\p| 
$$
Again, (\eqnAOI) implies the claim of the Lemma in the situation that 
$ k_2 \ge  |\t|$ .

\vskip .2cm
\noindent
{\it Case 2:} $(\k-\p)\cdot \t <0$

\noindent Let $\k'\in F+\t$ such that ${\rm dist}(\k,F+\t) =|\k-\k'|$. Then
$$
{\rm dist}\big(\k',(F+\t)-\t\big) \le |\k'-\k|= {\rm dist}(\k,F+\t)
$$
and
$$
|\k-\p| \le |\k-\k'| +|\k'-\p| 
        \le {\rm dist}(\k,F+\t) +|\k'-\p| 
\EQN\eqnAOII$$
Observe that the point $\p$ of $F+\t$ has the property that $\p-(-\t)$ lies also in $F+\t$. Also, if $\de$ was chosen small enough, the
angle between the tangents to $F$ at $\p$ and to $F+\t$ at $\p$ (which
is parallel to the tangent to $F$ at $\p-\t$) is very small and
$(\k'-\p)\cdot(-\t)>0$.

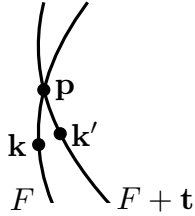
\centerline{\figput{distF7}}

\noindent
Thus we can apply the results of Case 1, with $F$ replaced by $F+\t$,
$\t$ replaced by $-\t$ and $\k$ replaced by $\k'$ and get
$$
|\k'-\p| \le \sfrac{\abcst}{|\t|}{\rm dist}\big(\k',(F+\t)-\t\big)
   \le  \sfrac{\abcst}{|\t|}{\rm dist}(\k,F+\t)
$$
This, together with (\eqnAOII), proves the Lemma in Case 2.
\endproof

\lemma{\STM\lemAI}{ 
There are constants $\de_F,\const$ that depend only on $F$ and $M$ such that the following holds: 

\noindent
Let $\btau \in \bbbr^2$,  $\epsilon>0$ and $D$ the disc centered at 
$\btau$ with radius $\epsilon$.
Let $m\ge 1$ be a scale with $\fl_m \ge \half\ep$. Define
$$
N= \# \set{(s_1,s_2) \in \Sigma_m\times \Sigma_m}
{(s_1- s_2) \cap D \ne \emptyset}
$$
where $D\subset \bbbr^2$ is viewed as $\set{(0,\t)}{\t\in D}\subset \bbbr\times\bbbr^2$.

\Item{a)}{ If $\ |\btau| \ge \de_F$, then 
$\ N\le  \sfrac{\const}{ \sqrt{ \fl_m}}\ $.   }

\Item{b)}{ If
$\ |\btau| \le \de_F$, then 
$$
N \le  \sfrac{ \const }{\fl_m |\btau| }\Big(\sfrac{1}{M^m}+\ep\Big) +\const 
$$
}
}

\prf
We first observe that, given any fixed $s_1\in \Sigma_m$, 
$\ (s_1-s_2)\cap D \ne \emptyset\ $ only if 
$\ s_2\cap (s_1-D) \ne \emptyset\ $. As $s_1-D$ is contained in a ball of radius at most $ 3 \fl_m$, there are at most five sectors $s_2\in \Sigma_m$ that intersect it. Hence
$$\eqalign{
N&\le \const\# \set{s_1\in \Sigma_m}
{\exists s_2\in \Sigma_m \ {\rm such\ that\ }
s_2\cap (s_1-D) \ne \emptyset} \cr
&\le \const\# \set{s_1\in \Sigma_m}
{\exists \k \in s_1\cap F \ {\rm such\ that\ }{\rm dist}\big( \k-\btau,F \big)
\le \cst{}{1} \big(\sfrac{1}{M^m}+\ep\big) }
}$$
Define
$$
I = \set{\k\in F}{ {\rm dist}(\k-\btau,F) \le 
\cst{}{1}\big(\sfrac{1}{M^m}+\ep\big)}
$$
Then
$$
N \le \const\# \set{s\in \Sigma_m}{s\cap I \ne \emptyset} 
\EQN\eqnAI$$
Clearly $I\subset I'$, where
$$
I'=\set{\k\in F}{ {\rm dist}(\k,F+\btau) \le \const\,\fl_m}
$$
and hence
$$
N\le \const\# \set{s\in \Sigma_m}{s\cap I' \ne \emptyset} \le \sfrac{\const}{\fl_m}\,{\rm length}(I') + \const
\EQN\eqnAIp$$
We choose $\de_F$ to be smaller than the constant $\de$ of Lemma \lemAO.

\Item{a)} If  $|\btau| \ge \de_F$, we use (\eqnAIp)
and that
$$ 
{\rm length} \big( I' \big)
\le \const \,\sqrt{\fl_m}
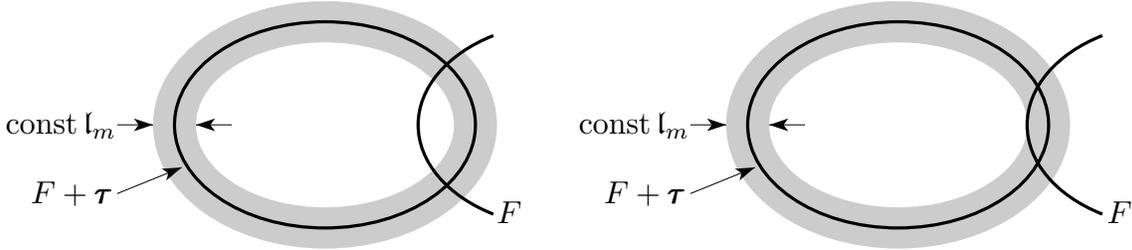
$$
\centerline{\figplace{FmR2}{0 in}{0 in}}

\Item{b)} Assume that $|\btau| \le \de_F$. 
Since $F$ is strictly convex, $F\cap(F+\btau)$ consists of two points, say $\p_1,\p_2$. Let $U_1$ and $U_2$ be the discs of radius $\de_F$ around $\p_1$ and $\p_2$, respectively.
If $\de_F$ is small enough, $U_1$ and $U_2$ are disjoint.
 By Lemma \lemAO, for $i=1,2$, $U_i\cap I$ is contained in an interval of length $\sfrac{\abcst}{|\btau|} \big(\sfrac{1}{M^m}+\ep\big)$. 
Also by Lemma \lemAO, for $i=1,2$, the distance between $F+\btau$ and any endpoint of $F\cap U_i$ is bigger than $\abcst\,\de_F|\btau|$.
If $\sfrac{1}{|\btau|} \big(\sfrac{1}{M^m}+\ep\big)\ge 
\sfrac{\abcst}{2\cst{}{1}}\de_F$, 
the desired bound follows immediately from (\eqnAI) and the fact that 
$\#\Si_m\le\sfrac{\abcst}{\fl_m}$.
So we may assume that,
for $i=1,2$ the distance between $F+\btau$ and any endpoint of $F\cap U_i$ is bigger than $2\cst{}{1} \big(\sfrac{1}{M^m}+\ep\big)$.

\noindent
We now show that $I\subset U_1\cup U_2$. For this purpose, let $\k\in I$. Then there is ${\bf v}\in\bbbr^2$ with $|{\bf v}| \le \cst{}{1} \big(\sfrac{1}{M^m}+\ep\big)$ such that 
$\k\in F+\btau+{\bf v}$. Now for $i=1,2$ there is point $\k_i \in F\cap U_i$ such that $\k_i-{\bf v} \in F+\btau$. (At the two endpoints $\k'$ of $F\cap U_i$, the points $\k'-{\bf v}$ lie on opposite sides of $ F+\btau$.) Since $F\cap (F+\btau +{\bf v})$ consists of only two points,
$\k=\k_1$ or $\k=\k_2$; in particular $\k\in U_1\cup U_2$.

Therefore $I$ is contained in two intervals of length 
$\sfrac{\abcst}{|\btau|} \big(\sfrac{1}{M^m}+\ep\big)$
and, by (\eqnAI),
$$
N\le \sfrac{\const}{\fl_m}\,\sfrac{\abcst}{|\btau|} \big(\sfrac{1}{M^m}+\ep\big) + \const
$$
\endproof

\noindent
Recall that $\bpi:k=(k_0,\k)\mapsto\k$ is the projection of 
$\bbbm=\bbbr\times\bbbr^2$ onto its second factor.

\lemma{\STM\lemAIII}{ Let $1\le \ell\le m \le r$ and 
$\,\ka'\in \fK_\ell, \ \ka_1,\ka_2 \in \fK_r\,$.
Then the number of $4$--tuples
$\,(u_1,u_2,s_1,s_2) \in \Si_\ell \times \Si_\ell \times 
\Sigma_m \times \Sigma_m\,$ fulfilling
$$\eqalign{
\bpi(s_1-s_2) \cap \bpi(\ka_1 -\ka_2) &\ne \emptyset \cr
\bpi(u_1-u_2) \cap \bpi(\ka'-s_1) &\ne \emptyset \cr
}\EQN\eqnAII$$
is bounded by $\,\sfrac{\const}{\fl_m \sqrt{\fl_\ell}}\,$ 
with the constant $\const$ independent of 
$\,\ka_1,\ka_2,\ka',\, \ell,m\,$ and $r$.
}

\prf
Observe that for each fixed $s_1\in \Sigma_m\,$ there are at most $\const$ sectors $s_2\in \Sigma_m\,$ fulfilling 
$\,\bpi(s_1-s_2) \cap \bpi(\ka_1 -\ka_2) \ne \emptyset\,$. Recall that for 
any sector $s$, $\k_s$ denotes the center of $F\cap s$.
When $\ka'\in\bbbm$, set $\k_{\ka'}=\bka'$.
We bound each of the three terms in
$$\eqalign{
&\# \{ (u_1,u_2,s_1,s_2)\big| {\rm (\eqnAII)}\ {\rm holds} \}\cr
&\hskip1in \le \# \{(u_1,u_2,s_1,s_2) \big|
     {\rm (\eqnAII)}\ {\rm holds},\ |\k_{\ka'}-\k_{s_1}| \le \const\fl_\ell \} \cr
&\hskip1.1in +   \# \{(u_1,u_2,s_1,s_2) \big|
     {\rm (\eqnAII)}\ {\rm holds},\ \const\fl_\ell \le |\k_{\ka'}-\k_{s_1}| 
         \le \de_F\} \cr
&\hskip1.1in +   \# \{(u_1,u_2,s_1,s_2) \big|
     {\rm (\eqnAII)}\ {\rm holds},\ \de_F \le |\k_{\ka'}-\k_{s_1}| 
        \} \cr
}$$
separately. For the first term observe that there are at most 
$\,\const \,\big[ \sfrac{\fl_\ell}{ \fl_m} +1 \big]\, $ sectors $s_1$ with
$\,|\k_{s_1}-\k_{\ka'}| \le \const\fl_\ell\,$, and that for any given $s_1$, there are at most $\sfrac{\const}{\fl_\ell}$ pairs $(u_1,u_2)$ such that 
$\,\bpi(u_1-u_2) \cap \bpi(\ka'-s_1) \ne \emptyset \,$. Hence
$$
\# \{(u_1,u_2,s_1,s_2) \big|
     {\rm (\eqnAII)}\ {\rm holds},\ |\k_{\ka'}-\k_{s_1}| \le \const\fl_\ell \}
\,\le\, \const \,\big[ \sfrac{\fl_\ell}{\fl_m} +1 \big]\, \sfrac{1}{\fl_\ell}
\,\le\, \sfrac{\const}{\fl_m}
$$

We next bound the third term. There are at most $\sfrac{\const}{\fl_m}$ pairs 
$(s_1,s_2)$ obeying  $\,\bpi(s_1-s_2) \cap \bpi(\ka_1 -\ka_2) \ne \emptyset\,$ 
and $\,|\k_{\ka'}-\k_{s_1}| \ge \de_F \,$. For each such a pair $(s_1,s_2)$,
$\bpi(\ka'-s_1)$ is contained in a disc of radius $\,2\fl_\ell\,$, centered
a distance at least $\de_F$ from the origin, so, by Lemma \lemAI a,
with $m$ replaced by $\ell$, there are at most
$\,\sfrac{\const}{\sqrt{\fl_\ell}}\,$ pairs $(u_1,u_2)$ such that 
(\eqnAII) holds.  Hence
$$
\# \{(u_1,u_2,s_1,s_2) \big|
     {\rm (\eqnAII)}\ {\rm holds},\ \de_F \le |\k_{\ka'}-\k_{s_1}|\}
\le \sfrac{\const}{\fl_m \sqrt{\fl_\ell}} 
$$

Finally, for the second term, we observe that, for each fixed $(s_1,s_2)$ 
satisfying
$\,\const\fl_\ell \le |\k_{\ka'}-\k_{s_1}|\le \de_F\,$
there are, by Lemma \lemAI b with $\ep=2\fl_\ell$ and $m$ replaced by 
$\ell$, at most 
$\,\sfrac{\const}{\fl_\ell |\k_{\ka'}-\k_{s_1}| }
         \big[\sfrac{1}{M^\ell}+\fl_\ell\big]+\const
\le \sfrac{\const}{|\k_{\ka'}-\k_{s_1}|}\,$ pairs $(u_1,u_2)$ 
such that (\eqnAII) holds. Furthermore, we may order the allowed $s_1$'s so that
the $\mu^{\rm th}$ obeys $|\k_{s_1}-\k_{\ka'}|\ge\const(\fl_\ell+\mu\fl_m)$.
Hence
$$\eqalign{
&\# \{(u_1,u_2,s_1,s_2)\ \big|\  
     {\rm (\eqnAII)}\ {\rm holds},\ \const\fl_\ell \le |\k_{\ka'}-\k_{s_1}| 
         \le \de_F\}  
\le \sum_{\mu=1}^{\const/\fl_m} 
   \sfrac{\const}{\fl_\ell +\mu \fl_m}\cr
&\hskip1in\le \sfrac{\const} {\fl_m}
    \smsum_{\mu=1}^{\const/\fl_m}  \sfrac{1}{ \sfrac{\fl_\ell}{\fl_m} +\mu }
\le \sfrac{\const} {\fl_m} 
    \ln \sfrac{ \sfrac{\fl_\ell}{\fl_m} + \sfrac{\const}{\fl_m} }
              { \sfrac{\fl_\ell}{\fl_m} } 
\le \sfrac{\const} {\fl_m}\ln\big(1+\sfrac{\const}{\fl_\ell}\big) \cr
&\hskip1in\le \sfrac{\const} {\fl_m}\ell
\le \sfrac{\const}{\fl_m \sqrt{\fl_\ell}} 
}$$
\endproof

\lemma{\STM\lemAIV}{ 
Let $1\le \ell\le m $ and $\,\ka'\in \fK_\ell\,$. 
Let $D$ be the disc of radius $\epsilon$ centered at $\btau$,
with $\btau \in \bbbr^2$ and $0\le\epsilon\le2\fl_m$. Let
$N$ be the number of $4$--tuples
$\,(u_1,u_2,s_1,s_2) \in \Si_\ell\times\Si_\ell \times\Sigma_m\times\Si_m\,$ fulfilling
$$\eqalign{
\bpi(s_1-s_2) \cap D &\ne \emptyset \cr
\bpi(u_1-u_2) \cap \bpi(\ka'-s_1) &\ne \emptyset \cr
}\EQN\eqnAIII$$
\Item a) If $\ |\btau|\ge\de_F$, 
then $N\le\sfrac{\const}{\fl_\ell\sqrt{\fl_m}}$.
\Item b) If $\ |\btau|\le\de_F$, then
$$
N\le \sfrac{\const}{\fl_m \fl_\ell} \Big[
\min \big(\ell\fl_\ell, \sfrac{1+M^m\ep}{M^m |\btau|} \big) +
\sfrac{\sqrt \fl_\ell}{|\btau|}\big(\sfrac{1}{M^m}+\ep\big) + \fl_m \Big]
$$
}

\prf a) 
By Lemma \lemAI.a,
$
\# \{ (s_1,s_2) \in \Sigma_m^2 \big| \bpi(s_1-s_2) \cap D \ne \emptyset \}
\le \sfrac{\const}{\sqrt{\fl_m}}
$.
For each fixed $(s_1,s_2)$ there are at most $\sfrac{\const}{\fl_\ell}$ pairs
$(u_1,u_2)$ such that ${\rm (\eqnAIII)}$ holds. The desired bound follows.
\Item b)
As in the proof of Lemma \lemAIII,
we bound each of the three terms in
$$\eqalign{
&\# \{ (u_1,u_2,s_1,s_2)\big| {\rm (\eqnAIII)}\ {\rm holds} \}\cr
&\hskip 1in \le \# \{(u_1,u_2,s_1,s_2) \big|
     {\rm (\eqnAIII)}\ {\rm holds},\ |\k_{\ka'}-\k_{s_1}| \le \const\fl_\ell \} \cr
&\hskip1.1in +   \# \{(u_1,u_2,s_1,s_2) \big|
     {\rm (\eqnAIII)}\ {\rm holds},\ \const\fl_\ell \le |\k_{\ka'}-\k_{s_1}| 
         \le \de_F\} \cr
&\hskip1.1in +   \# \{(u_1,u_2,s_1,s_2) \big|
     {\rm (\eqnAIII)}\ {\rm holds},\ \de_F \le |\k_{\ka'}-\k_{s_1}| 
        \} \cr
}$$
separately.

By Lemma \lemAI.b
$$
\# \{ (s_1,s_2) \in \Sigma_m^2 \big| \bpi(s_1-s_2) \cap D \ne \emptyset \}
\le \const \big[ 1+\sfrac{1+M^m\ep}{M^m \fl_m |\btau|} \big]
\EQN\eqnLADappCn$$
As well, for each fixed $s_1$ there are at most $\const$ $s_2\in \Sigma_m$ such 
that $\,\bpi(s_1-s_2) \cap D \ne \emptyset\,$. Hence
$$
\# \{ (s_1,s_2) \in \Sigma_m^2 \big| \bpi(s_1-s_2) \cap D \ne \emptyset,\
|\k_{s_1} -\k_{\ka'}| \le \const\fl_\ell \}
\le \const \,\min \big\{ 1+\sfrac{1+M^m\ep}{M^m \fl_m |\btau|},\ 
\sfrac{\fl_\ell}{\fl_m } \big\}
$$
Also, for each fixed $(s_1,s_2)$ there are at most $\sfrac{\const}{\fl_\ell}$ pairs
$(u_1,u_2)$ such that ${\rm (\eqnAIII)}$ holds. Hence
$$\eqalign{
\# \{(u_1,u_2,s_1,s_2) \big|
     {\rm (\eqnAIII)}\ {\rm holds},\ |\k_{\ka'}-\k_{s_1}| \le \const\fl_\ell \}
&\le\sfrac{\const}{\fl_m \fl_\ell}\,
  \min\big\{ \fl_m+ \sfrac{1+M^m\ep}{M^m |\btau|},\ \fl_\ell \big\}\cr
&\le\sfrac{\const}{\fl_m \fl_\ell}\,\big[
  \min\big\{\sfrac{1+M^m\ep}{M^m |\btau|},\ \fl_\ell \big\}+\fl_m\big]\cr
}$$
This gives the desired bound for the first term.

We next bound the third term. For each fixed $(s_1,s_2)$ with
$\,|\k_{\ka'}-\k_{s_1}| \ge \de_F \,$, there are, by Lemma \lemAI a, at most
$\,\sfrac{\const}{\sqrt{\fl_\ell}}\,$ pairs $(u_1,u_2)$ such that (\eqnAIII)
holds. Hence
$$\eqalign{
\# \{(u_1,u_2,s_1,s_2) \big|
     {\rm (\eqnAIII)}\ {\rm holds},\ \de_F \le |\k_{\ka'}-\k_{s_1}| \}
& \le \sfrac{\const}{\sqrt{\fl_\ell}} \big[1+ \sfrac{1+M^m\ep}{M^m \fl_m |\btau|} \big] \cr
& \le\sfrac{\const}{\fl_m \fl_\ell}\,
\big[ \fl_m \sqrt{\fl_\ell} + \sfrac{\sqrt{\fl_\ell}}{|\btau|}
\big(\sfrac{1}{M^m}+\ep\big) \big] \cr
}$$  
which is smaller than the desired bound.

Finally, for the second term, we observe that, for each fixed $(s_1,s_2)$ 
satisfying
$\,\const\fl_\ell \le |\k_{\ka'}-\k_{s_1}|\le \de_F\,$
there are, by Lemma \lemAI b, with $\ep=2\fl_\ell$ and $m$ replaced by $\ell$, 
at most $\,\sfrac{\const}{\fl_\ell |\k_{\ka'}-\k_{s_1}| }
         \big[\sfrac{1}{M^\ell}+\fl_\ell\big]
\le  \sfrac{\const}{ |\k_{\ka'}-\k_{s_1}| }
\le\sfrac{\const}{\fl_\ell}\,$ pairs $(u_1,u_2)$ such that ${\rm (\eqnAIII)}$ 
holds. Hence, by (\eqnLADappCn) and the last argument of Lemma \lemAIII,
$$\eqalign{
\# \{(u_1,u_2,s_1,s_2)&\big| 
     {\rm (\eqnAIII)}\ {\rm holds},\ \const\fl_\ell \le |\k_{\ka'}-\k_{s_1}| 
         \le \de_F\} \cr
&\le \min \Big\{ \sum_{\mu=1}^{\const/\fl_m} 
   \sfrac{\const}{\fl_\ell +\mu \fl_m },\ 
    \sfrac{\const}{\fl_\ell} \big[ 1+ \sfrac{1+M^m\ep}{M^m \fl_m |\btau|} \big]
      \Big\} \cr 
&\le \sfrac{\const} {\fl_m\fl_\ell} \min \Big\{ \fl_\ell
    \smsum_{\mu=1}^{\const/\fl_m}  \sfrac{1}{ \sfrac{\fl_\ell}{\fl_m} +\mu },
\ \fl_m+\sfrac{1+M^m\ep}{M^m |\btau|}  \Big\} \cr
&\le \sfrac{\const} {\fl_m\fl_\ell} \min \Big\{ \fl_\ell\ell,
\ \fl_m+\sfrac{1+M^m\ep}{M^m |\btau|}  \Big\} \cr
&\le \sfrac{\const} {\fl_m\fl_\ell} \big[ \fl_m + \min \{
\ell\fl_\ell ,\ \sfrac{1+M^m\ep}{M^m |\btau|} \} \big] \cr
}$$
\endproof

\vfill\eject

\titlea{ References}\PG\pgLADref
\vskip-.1in
\item{[FHN]}
H. Fukuyama, Y. Hasegawa and O. Narikiyo, {\bf Two--Dimensional Hubbard Model at Low Electron Density}, Journal of the Physical Society of Japan, {\bf 60} (1991),
2013--2030.
\smallskip%
\item{[FKLT1]} J. Feldman,  H. Kn\"orrer, D. Lehmann and E. Trubowitz,
{\bf Fermi Liquids in Two Space Dimensions}, in {\it ``Constructive Physics''},
 edited by V. Rivasseau, Springer LNP {\bf 446}, 267--299 (1995).
\smallskip%
\item{[FKLT2]} J. Feldman,  H. Kn\"orrer, D. Lehmann and E. Trubowitz,
{\bf A Class of Fermi Liquids}, in {\it ``Particles and Fields '94''},
 edited by G. Semenoff and L. Vinet, Springer 35-62 (1999).
\smallskip%
\item{[FMST]} J.Feldman, H. Kn\"orrer, R. Sinclair and E. Trubowitz, 
{\bf Superconductivity in a Repulsive Model}, Helvetica Physica Acta, {\bf 70} (1997)
154--191.
\smallskip%
\item{[FKTf1]} J. Feldman, H. Kn\"orrer, E. Trubowitz, 
{\bf A Two Dimensional Fermi Liquid, Part 1: Overview}, preprint.
\smallskip%
\item{[FKTf2]} J. Feldman, H. Kn\"orrer, E. Trubowitz, 
{\bf A Two Dimensional Fermi Liquid, Part 2: Convergence}, preprint.
\smallskip%
\item{[FKTf3]} J. Feldman, H. Kn\"orrer, E. Trubowitz, 
{\bf A Two Dimensional Fermi Liquid, Part 3: The Fermi Surface}, preprint.
\smallskip%
\item{[FKTo1]} J. Feldman, H. Kn\"orrer, E. Trubowitz, 
{\bf Single Scale Analysis of Many Fermion Systems, Part 1: Insulators}, preprint.
\smallskip%
\item{[FKTo2]} J. Feldman, H. Kn\"orrer, E. Trubowitz, 
{\bf Single Scale Analysis of Many Fermion Systems, Part 2: The First Scale}, preprint.
\smallskip%
\item{[FKTo3]} J. Feldman, H. Kn\"orrer, E. Trubowitz, 
{\bf Single Scale Analysis of Many Fermion Systems, Part 3: Sectorized Norms}, preprint.
\smallskip%
\item{[FKTo4]} J. Feldman, H. Kn\"orrer, E. Trubowitz, 
{\bf Single Scale Analysis of Many Fermion Systems, Part 4: Sector Counting}, preprint.
\smallskip%
\item{[FKTr1]} J. Feldman, H. Kn\"orrer, E. Trubowitz, 
{\bf Convergence of Perturbation Expansions in Fermionic Models, Part 1: Nonperturbative Bounds}, preprint.
\smallskip%
\item{[FKTr2]} J. Feldman, H. Kn\"orrer, E. Trubowitz, 
{\bf Convergence of Perturbation Expansions in Fermionic Models, Part 2: Overlapping Loops}, preprint.
\smallskip%
\item{[FMRT]} J.Feldman, J. Magnen, V. Rivasseau, E. Trubowitz, 
{\bf Two Dimensional Many Fermion Systems as Vector Models},
Europhysics Letters, {\bf 24} (1993) 521-526.
\smallskip%
\item{[N]} P. Nozi\`eres, {\it Theory of Interacting Fermi Systems}, Benjamin
(1964).
\smallskip%
\item{[S]}  M. Salmhofer, private communication.

\vfill\eject

\titlea{Notation}\PG\pgLADnot
\null\vskip-.7in
\titlec{Configuration Spaces}
\centerline{
\vbox{\offinterlineskip
\hrule
\halign{\vrule#&
         \strut\hskip0.05in\hfil#\hfil&
         \hskip0.05in\vrule#\hskip0.05in&
          #\hfil\hfil&
         \hskip0.05in\vrule#\hskip0.05in&
          #\hfil\hfil&
           \hskip0.05in\vrule#\cr
height2pt&\omit&&\omit&&\omit&\cr
&Symbol&&Interpretation&&Reference&\cr
height2pt&\omit&&\omit&&\omit&\cr
\noalign{\hrule}
height2pt&\omit&&\omit&&\omit&\cr
&$\bbbm$&&momentum&&after Definition \defLADsectors&\cr
height2pt&\omit&&\omit&&\omit&\cr
&$\fY$&&momentum or position&&before Definition \defLADbubblenorm&\cr
height2pt&\omit&&\omit&&\omit&\cr
&$\fY_\Si$&&momentum or (position, sector)&&after Definition \defLADsectors&\cr
height2pt&\omit&&\omit&&\omit&\cr
&$\fY_{0,\Si}$&&momentum&&(\eqnLADfourdunion)&\cr
height2pt&\omit&&\omit&&\omit&\cr
&$\fY_{1,\Si}$&&(position, sector)&&(\eqnLADfourdunion)&\cr
height2pt&\omit&&\omit&&\omit&\cr
&$\fY_{2,\Si}$&&(momentum, sector)&&Definition \defLADfourtrans&\cr
height2pt&\omit&&\omit&&\omit&\cr
&$\fY^\updownarrow_\Si$&&(momentum, spin) or (position, spin, sector)&&after Definition \defLADsectors&\cr
height2pt&\omit&&\omit&&\omit&\cr
&$\fX_\Si$&&(momentum, spin, creation/annihilation index)&&after Definition \defLADsectors&\cr
height2pt&\omit&&\omit&&\omit&\cr
& &&or (position, spin, creation/annihilation index, sector)&&after Definition \defLADsectors&\cr
height2pt&\omit&&\omit&&\omit&\cr
&$\cB^\updownarrow$&&(position, spin)&&after Definition \defLADsectors&\cr
height2pt&\omit&&\omit&&\omit&\cr
&$\check\cB^\updownarrow$&&(momentum, spin)&&after Definition \defLADsectors&\cr
height2pt&\omit&&\omit&&\omit&\cr
&$\cB$&&(position, spin, creation/annihilation index)&&after Definition \defLADsectors&\cr
height2pt&\omit&&\omit&&\omit&\cr
&$\check\cB$&&(momentum, spin, creation/annihilation index)&&after Definition \defLADsectors&\cr
height2pt&\omit&&\omit&&\omit&\cr
&$\fY^{(4)}_{\Si,\Si'}$&&$\fY^2_\Si\times\fY^2_{\Si'}$&&(\eqnLADfourdunion)&\cr
height2pt&\omit&&\omit&&\omit&\cr
&$\fY_{\ell,r}$&&$\fY^{(4)}_{\Si_\ell,\Si_r}$&&Convention \convLADlr&\cr
height2pt&\omit&&\omit&&\omit&\cr
height2pt&\omit&&\omit&&\omit&\cr
&$\fK_r$&&momentum or sector&&Definition \defLADKspace&\cr
}\hrule}}

\vfill
\titlec{Norms}
\centerline{
\vbox{\offinterlineskip
\hrule
\halign{\vrule#&
         \strut\hskip0.05in\hfil#\hfil&
         \hskip0.05in\vrule#\hskip0.05in&
          #\hfil\hfil&
         \hskip0.05in\vrule#\hskip0.05in&
          #\hfil\hfil&
           \hskip0.05in\vrule#\cr
height2pt&\omit&&\omit&&\omit&\cr
&Norm&&Characteristics&&Reference&\cr
height2pt&\omit&&\omit&&\omit&\cr
\noalign{\hrule}
height2pt&\omit&&\omit&&\omit&\cr
&$\tn\ \cdot\ \tn_{1,\infty}$&&no derivatives, external positions only&&Definition \defLADlonelinfty&\cr
height3pt&\omit&&\omit&&\omit&\cr
&$\tn\ \cdot\ \tn$&&no derivatives, external positions and momenta&&Definition  \defLADbubblenorm&\cr
height3pt&\omit&&\omit&&\omit&\cr
&$\|\ \cdot\ \|_{\rm bubble}$&&operator norm for bubble propagators&&Definition  \defLADbubblenorm&\cr
height3pt&\omit&&\omit&&\omit&\cr
&$\v\ \cdot\ \v^\de_{1,\Si}$&&two--legged kernel, $\de$ derivatives, sectors
&&Definition \defLADlonelinftyII&\cr
height3pt&\omit&&\omit&&\omit&\cr
&$\v\ \cdot\ \v^{(\de_\li,\de_\ci,\de_\ri)}_{\Si,\Si'} $&&four--legged kernel,
$(\de_\li,\de_\ci,\de_\ri)$ derivatives, sectors&&Definition \defLADlonelinftyIV&\cr
height3pt&\omit&&\omit&&\omit&\cr
&$\v\ \cdot\ \v_{1,\Si}$&&two--legged kernel, all derivatives, sectors&&Definition \defLADnormdomainnorm&\cr
height3pt&\omit&&\omit&&\omit&\cr
&$\v\ \cdot\ \v_{\Si}$&&four--legged kernel,
all derivatives, sectors&&Definition \defLADnormdomainnorm&\cr
height3pt&\omit&&\omit&&\omit&\cr
&$\|\ \cdot\ \|^{(\de_\li,\de_\ci,\de_\ri)}_{\ell,r}$&&
$(\de_\li,\de_\ci,\de_\ri)$ scaled derivatives, sectors $\Si_\ell,\Si_r$&&
Definition \secrepnorm&\cr
height3pt&\omit&&\omit&&\omit&\cr
&$|\ \cdot\ |^{[\de_\li,\de_\ci,\de_\ri]}_{\ell,r}$&&$\le(\de_\li,\de_\ci,\de_\ri)$ scaled derivatives, sectors $\Si_\ell,\Si_r$&&Def'ns \secrepnorm,\resecrepnorm&\cr
height3pt&\omit&&\omit&&\omit&\cr
&$|\ \cdot\ |^{\[\de\]}_j$&&$\de_\li+\de_\ci+\de_\ri\le\de$ scaled derivatives, sectors $\Si_j$&&Def'ns \secrepnorm,\resecrepnorm&\cr
height3pt&\omit&&\omit&&\omit&\cr
&$|\ \cdot\ |_{\ell,r}$&&no derivatives, sectors $\Si_\ell,\Si_r$&&Definition \defLADshortnorm&\cr
height3pt&\omit&&\omit&&\omit&\cr
&$\|\ \cdot\ \|_{\ka_1,\ka_2}$&&no derivatives, specified right hand momenta/sectors&&Definition \defLADKspace&\cr
height2pt&\omit&&\omit&&\omit&\cr
}\hrule}}

\vfill\eject
\titlec{Propagators and Ladders}
\centerline{
\vbox{\offinterlineskip
\hrule
\halign{\vrule#&
         \strut\hskip0.05in\hfil#\hfil&
         \hskip0.05in\vrule#\hskip0.05in&
          #\hfil\hfil&
         \hskip0.05in\vrule#\hskip0.05in&
          #\hfil\hfil&
           \hskip0.05in\vrule#\cr
height2pt&\omit&&\omit&&\omit&\cr
&Symbol&&Interpretation&&Reference&\cr
height2pt&\omit&&\omit&&\omit&\cr
\noalign{\hrule}
height2pt&\omit&&\omit&&\omit&\cr
&$C_v^{(j)}$&&$C_v^{(j)}(k)=\sfrac{\nu^{(j)}(k)}{ik_0-e(\k)-v(k)}$,
single scale propagator&&before Definition \defLADresectoriz&\cr
height2pt&\omit&&\omit&&\omit&\cr
&$C_v^{(\ge j)}$&&$C_v^{(\ge j)}(k)=\sfrac{\nu^{(\ge j)}(k)}{ik_0-e(\k)-v(k)}$,
multi scale propagator&&before Definition \defLADresectoriz&\cr
height2pt&\omit&&\omit&&\omit&\cr
&$\cC(A,B)$&&$A\otimes A^t +A\otimes B^t+ B\otimes A^t$, bubble propagator&&Definition \defLADphladder&\cr
height2pt&\omit&&\omit&&\omit&\cr
&$\cC^{(j)}$&&$\hskip-14pt\sum\limits_{i_1,i_2\ge 1 \atop {\hskip14pt \min(i_1,i_2)=j}}
\hskip-22pt
C_v^{(i_1)}\otimes C_v^{(i_2)\,t}$, single scale bubble propagator
&&before Convention \convLADresectorbullet&\cr
height2pt&\omit&&\omit&&\omit&\cr
&$\cC^{[i,j]}$&&$\sum_{i\le\ell\le j} {\cal C}^{(\ell)}$, 
multi scale bubble propagator&&before Convention \convLADresectorbullet&\cr
height2pt&\omit&&\omit&&\omit&\cr
&$\cC_{\tp}^{[i,j]}$
&&$\sum_{i\le i_t\le j\atop i_b>j}C^{(i_t)}_v\otimes C^{(i_b)\,t}_v$
&&Definition \topbottom&\cr
height2pt&\omit&&\omit&&\omit&\cr
&$\cC_{\md}^{[i,j]}$&&$\sum_{i\le i_t\le j\atop i\le i_b\le j}C^{(i_t)}_v$   &&Definition \topbottom&\cr
height2pt&\omit&&\omit&&\omit&\cr
&$\cC_{\bt}^{[i,j]}$
&&$\sum_{i_t> j\atop i\le i_b\le j}C^{(i_t)}_v
\otimes C^{(i_b)\,t}_v$
&&Definition \topbottom&\cr
height2pt&\omit&&\omit&&\omit&\cr
&$\cD^{(\ell)}_{\nu,\upl}$&&$\sfrac{1}{M^{|\nu| \ell}}
\sum_{m=\ell}^\infty \rD^{\nu}_{1;3}C_v^{(\ell)}\otimes C_v^{(m)\,t}$
&&Theorem \dbbound&\cr
height2pt&\omit&&\omit&&\omit&\cr
&$\cD^{(\ell)}_{\nu,\dnl}$&&$\sfrac{1}{M^{|\nu| \ell}}
\sum_{m=\ell+1}^\infty C_v^{(m)}\otimes\rD^{\nu}_{2;4}C_v^{(\ell)\,t}$
&&Theorem \dbbound&\cr
height2pt&\omit&&\omit&&\omit&\cr
&$\cM$&&model particle--hole bubble propagator&&(\eqnIX)&\cr
height2pt&\omit&&\omit&&\omit&\cr
&$\cL^{(j)}_v(\vec F)$&&compound particle--hole ladder
&&Definition \defLADcompoundphladder&\cr
height2pt&\omit&&\omit&&\omit&\cr
&$L^{(j)}$&&single scale compound particle--hole ladder
&& Definition \defLADcompoundphladderscalej&\cr
height2pt&\omit&&\omit&&\omit&\cr
}\hrule}}

\vfill
\titlec{Scales and Sectors}
\centerline{
\vbox{\offinterlineskip
\hrule
\halign{\vrule#&
         \strut\hskip0.05in\hfil#\hfil&
         \hskip0.05in\vrule#\hskip0.05in&
          #\hfil\hfil&
         \hskip0.05in\vrule#\hskip0.05in&
          #\hfil\hfil&
           \hskip0.05in\vrule#\cr
height2pt&\omit&&\omit&&\omit&\cr
&Symbol&&Interpretation&&Reference&\cr
height2pt&\omit&&\omit&&\omit&\cr
\noalign{\hrule}
height2pt&\omit&&\omit&&\omit&\cr
&$M$&&scale parameter, $M>1$, large enough&&Lemma \lemLADprimitivemanfred&\cr
height2pt&\omit&&\omit&&\omit&\cr
&$\nu,\varphi$&&used in constructing scale functions&&Definition \defLADscales &\cr
height2pt&\omit&&\omit&&\omit&\cr
&$\nu^{(j)},\ j\ge 1$&&partition of unity that implements scales&&Definition \defLADscales &\cr
height2pt&\omit&&\omit&&\omit&\cr
&$\nu^{(\ge j)}$&&basically $\sum_{i\ge j}\nu^{(i)}$
&&Definition \defLADscales &\cr
height2pt&\omit&&\omit&&\omit&\cr
&$\nu_0(\om)\nu_1(\p,\k)$&&factorized cutoff for model bubble propagator
&&before (\eqnIX)&\cr
height2pt&\omit&&\omit&&\omit&\cr
&$\aleph$&&$\half<\aleph<\sfrac{2}{3}$, parameter controlling sector length&& 
before Definition \defLADresectoriz&\cr
height2pt&\omit&&\omit&&\omit&\cr
&$\fl_j$&&$\fl_j=\sfrac{1}{M^{\aleph j}}$, sector length for scale $j$&& 
before Definition \defLADresectoriz&\cr
height2pt&\omit&&\omit&&\omit&\cr
&$\chi_s,\ s\in\Si$&&partition of unity that implements sectorization&& 
before Definition \defLADresectoriz&\cr
height2pt&\omit&&\omit&&\omit&\cr
&$\Si_j$&&set of sectors of scale $j$&&before Definition \defLADresectoriz&\cr
height2pt&\omit&&\omit&&\omit&\cr
&$p_{\Si},f_{\Si,\Si'}$&&resectorization&&Definition \defLADresectoriz&\cr
height2pt&\omit&&\omit&&\omit&\cr
}\hrule}}

\vfill\eject
\titlec{Miscellaneous}
\centerline{
\vbox{\offinterlineskip
\hrule
\halign{\vrule#&
         \strut\hskip0.05in\hfil#\hfil&
         \hskip0.05in\vrule#\hskip0.05in&
          #\hfil\hfil&
         \hskip0.05in\vrule#\hskip0.05in&
          #\hfil\hfil&
           \hskip0.05in\vrule#\cr
height2pt&\omit&&\omit&&\omit&\cr
&Symbol&&Interpretation&&Reference&\cr
height2pt&\omit&&\omit&&\omit&\cr
\noalign{\hrule}
height2pt&\omit&&\omit&&\omit&\cr
&$\const$&&generic constant, independent of scale&& &\cr
height2pt&\omit&&\omit&&\omit&\cr
&$\abcst$&&generic constant, independent of scale and $M$&& &\cr
height2pt&\omit&&\omit&&\omit&\cr
&$F$&&Fermi curve $=\set{\k\in\bbbr^2}{e(\k)=0}$
&&before Definition \defLADscales&\cr
height2pt&\omit&&\omit&&\omit&\cr
&$r_0,r_e$&&$r_0,r_e\ge 6$, number of derivatives controlled&& before Definition \defLADscales &\cr
height2pt&\omit&&\omit&&\omit&\cr
&$\pi_F$&&projection on the Fermi surface&&Definition \defLADsectors&\cr
height2pt&\omit&&\omit&&\omit&\cr
&$\bpi$&&$\bpi(k_0,\k)=\k$&&Remark \remXIVector&\cr
height2pt&\omit&&\omit&&\omit&\cr
&$<\!k,x\!>_-$&& $-k_0x_0+\k_1\x_1+\k_2\x_2$&&Definition \defLADtransinv&\cr
height2pt&\omit&&\omit&&\omit&\cr
&$b_\nu$&&$b_1=b_4=0$, $b_1=b_2=1$&&Definition \defLADtransinv &\cr
height2pt&\omit&&\omit&&\omit&\cr
&$K^f$&&flipped vertex&&(\eqnLADflipped)&\cr
height2pt&\omit&&\omit&&\omit&\cr
&$\cb_j$&&$\sum_{\de\in\De}M^{j|\de|}t^\de+\sum_{\de\notin\De}\infty t^\de$
&&(\eqnLADcj)&\cr
height2pt&\omit&&\omit&&\omit&\cr
&$\De$&&$\set{\de\in\bbbn_0\times\bbbn_0^2}{\de_0\le r_0,\ \de_1+\de_2\le r_e}$
&&(\eqnLADDelta)&\cr
height2pt&\omit&&\omit&&\omit&\cr
&$\vec\De$&&$\set{\vec\de=(\de_\li,\de_\ci,\de_\ri)\in
\big(\bbbn_0\times\bbbn_0^2\big)^3}{\de_\li+\de_\ci+\de_\ri\in\De}$
&&(\eqnLADDelta)&\cr
height2pt&\omit&&\omit&&\omit&\cr
&$\bullet$&&convolution with sector sums
&&Definition \defLADphladder&\cr
height2pt&\omit&&\omit&&\omit&\cr
&$\circ$&&convolution without sector sums
&&before (\eqnBubIa)&\cr
height2pt&\omit&&\omit&&\omit&\cr
&$f_C,\ f_S$&&charge and spin components&&Lemma \lemLADchargespin&\cr
height2pt&\omit&&\omit&&\omit&\cr
&$W_R$&&$W_R(p,k)=W(p,k)R(p-k)$, transfer momentum cutoff
&&Definition \defLADtrasfermomcutoff&\cr
height2pt&\omit&&\omit&&\omit&\cr
&$\cR(d)$&&set of functions $R(t)$ that are identically one on $d$
&&Definition \defLADtdiff&\cr
height2pt&\omit&&\omit&&\omit&\cr
&$\cZ,\ \cZ^t$&&zero component localization operator and transpose&&(\eqnVII)&\cr
height2pt&\omit&&\omit&&\omit&\cr
& &&$\big(\cZ\fcirc W\fcirc\cZ^t\big)(p,k)=\de(k_0)\int d\om\ W\big((\om,\0)+p,(\om,\k)\big)$&& &\cr
height2pt&\omit&&\omit&&\omit&\cr
&$\widetilde W$&&$\widetilde W(p,k)=\de(k_0)\int d\om\ W\big((\om,\p),(\om,\k)\big)$
&&(\eqnVIII)&\cr
height2pt&\omit&&\omit&&\omit&\cr
&$\bar\jmath$&&boundary between large and small transfer momentum
&&before Prop'n \propLADlarget &\cr
height2pt&\omit&&\omit&&\omit&\cr
}\hrule}}

\end